%% file: main.tex
\documentclass[%
 reprint,  
superscriptaddress,
 amsmath,amssymb,
 aps,
doi=true,
rmp,
floatfix,
raggedbottom,
]{revtex4-1}
\usepackage[colorlinks=true,linkcolor=blue,urlcolor=blue,citecolor=blue]{hyperref}
\usepackage{amsthm,amssymb,amsmath,epsfig,graphics,color,verbatim,booktabs,multirow,rotating}
\AtBeginDocument{
\heavyrulewidth=.08em
\lightrulewidth=.05em
\cmidrulewidth=.03em
\belowrulesep=.65ex
\belowbottomsep=0pt
\aboverulesep=.4ex
\abovetopsep=0pt
\cmidrulesep=\doublerulesep
\cmidrulekern=.5em
\defaultaddspace=.5em
}
\renewcommand{\arraystretch}{1.2}

\usepackage{sidecap}
\usepackage[normalem]{ulem} 
\usepackage{url} 
\usepackage{amsmath}
\usepackage{amsfonts}
\usepackage{dsfont}
\usepackage{relsize}
\usepackage{wrapfig}
\usepackage{subfigure}
\usepackage[scr]{rsfso}
\usepackage{upgreek}
\usepackage{xfrac}
\usepackage[usenames,dvipsnames]{xcolor}
\usepackage[margin=.58in]{geometry}
\usepackage[percent]{overpic}
\usepackage{chngcntr}
\counterwithin{figure}{section}
\counterwithin{table}{section}

\usepackage{lipsum} 
\usepackage{enumitem} 

\usepackage[]{nicefrac}

\usepackage[export]{adjustbox}

\usepackage{array}
\newcolumntype{L}[1]{>{\raggedright\let\newline\\\arraybackslash\hspace{0pt}}m{#1}}
\newcolumntype{C}[1]{>{\centering\let\newline\\\arraybackslash\hspace{0pt}}m{#1}}
\newcolumntype{R}[1]{>{\raggedleft\let\newline\\\arraybackslash\hspace{0pt}}m{#1}}

\usepackage{upgreek}

\newcommand{\appropto}{\mathrel{\vcenter{
  \offinterlineskip\halign{\hfil$##$\cr
    \propto\cr\noalign{\kern2pt}\sim\cr\noalign{\kern-2pt}}}}}
    
\usepackage[T1]{fontenc}

\usepackage{chngcntr}
\counterwithout{figure}{section}
\counterwithout{table}{section}

\usepackage[outline]{contour}

\definecolor{dandelion}{rgb}{0.94, 0.88, 0.19}
\definecolor{thepurple}{rgb}{0.65, 0.24, 0.59}
\definecolor{theorange}{rgb}{0.95, 0.40, 0.13}
\definecolor{thegreen}{rgb}{0.05, 0.50, 0.25}
\definecolor{mathematicagreen}{rgb}{0.0, 0.80, 0.0}
\definecolor{igoraquamarine}{rgb}{0.0, 0.6, 0.6}
\definecolor{igorgreen}{rgb}{0.0, 0.6, 0.0}
\definecolor{springgreen}{rgb}{0.0, 0.8, 0.0}
\definecolor{skyblue}{rgb}{0.0, 0.6, 1.0}
\definecolor{black}{rgb}{0.0 0.0 0.0}

\definecolor{inkscapeblue}{rgb}{0.0 0.0 1.0}
\definecolor{inkscapepurple}{rgb}{0.5 0.0 0.5}
\definecolor{inkscapegreen}{rgb}{0.0 0.5 0.0}
\definecolor{inkscapered}{rgb}{1 0.0 0.0}
\definecolor{black}{rgb}{0.0 0.0 0.0}

\begin{document}


\title{Sensitivity Optimization for NV-Diamond Magnetometry}

\author{John F. Barry}
\thanks{These authors contributed equally}
\affiliation{Lincoln Laboratory, Massachusetts Institute of Technology, Lexington, Massachusetts 02421, USA}
\affiliation{Harvard-Smithsonian Center for Astrophysics, Cambridge, Massachusetts 02138, USA}
\affiliation{Department of Physics, Harvard University, Cambridge, Massachusetts 02138, USA}
\affiliation{Center for Brain Science, Harvard University, Cambridge, Massachusetts 02138, USA}
\author{}
\email{john.barry@ll.mit.edu}
\noaffiliation
\author{Jennifer M. Schloss}
\thanks{These authors contributed equally}
\affiliation{Lincoln Laboratory, Massachusetts Institute of Technology, Lexington, Massachusetts 02421, USA}
\affiliation{Harvard-Smithsonian Center for Astrophysics, Cambridge, Massachusetts 02138, USA}
\affiliation{Center for Brain Science, Harvard University, Cambridge, Massachusetts 02138, USA}
\affiliation{Department of Physics, Massachusetts Institute of Technology, Cambridge, Massachusetts 02139, USA}
\author{}
\email{jennifer.schloss@ll.mit.edu}
\noaffiliation
\author{Erik Bauch}
\thanks{These authors contributed equally}
\affiliation{Department of Physics, Harvard University, Cambridge, Massachusetts 02138, USA}
\author{Matthew J. Turner}
\affiliation{Department of Physics, Harvard University, Cambridge, Massachusetts 02138, USA}
\affiliation{Center for Brain Science, Harvard University, Cambridge, Massachusetts 02138, USA}
\author{Connor A. Hart}
\affiliation{Department of Physics, Harvard University, Cambridge, Massachusetts 02138, USA}
\author{Linh M. Pham}
\affiliation{Lincoln Laboratory, Massachusetts Institute of Technology, Lexington, Massachusetts 02421, USA}
\affiliation{Harvard-Smithsonian Center for Astrophysics, Cambridge, Massachusetts 02138, USA}
\author{Ronald L. Walsworth}
\email{walsworth@umd.edu}
\affiliation{Harvard-Smithsonian Center for Astrophysics, Cambridge, Massachusetts 02138, USA}
\affiliation{Department of Physics, Harvard University, Cambridge, Massachusetts 02138, USA}
\affiliation{Center for Brain Science, Harvard University, Cambridge, Massachusetts 02138, USA}
\affiliation{Quantum Technology Center, University of Maryland, College Park, Maryland 20742, USA}
\affiliation{Department of Electrical and Computer Engineering, University of Maryland, College Park, Maryland 20742, USA}
\affiliation{Department of Physics, University of Maryland, College Park, Maryland 20742, USA}

\date{\today}

\begin{abstract}
    Solid-state spin systems including nitrogen-vacancy (NV) centers in diamond constitute an increasingly favored quantum sensing platform. However, present NV ensemble devices exhibit sensitivities orders of magnitude away from theoretical limits. The sensitivity shortfall both handicaps existing implementations and curtails the envisioned application space. This review analyzes present and proposed approaches to enhance the sensitivity of broadband ensemble-NV-diamond magnetometers. Improvements to the spin dephasing time, the readout fidelity, and the host diamond material properties are identified as the most promising avenues and are investigated extensively. Our analysis of sensitivity optimization establishes a foundation to stimulate development of new techniques for enhancing solid-state sensor performance.
\end{abstract}

\maketitle

\tableofcontents







\renewcommand{\thetable}{\arabic{table}}   
\renewcommand{\thefigure}{\arabic{figure}}

   
\input{sec01} 
\input{sec02} 
\input{sec03} 
\input{sec04} 
\input{sec05} 

\input{sec06} 
\input{sec07} 
\input{sec08} 


\clearpage\newpage
\bibliography{thebib2016.bib}

\end{document}

%% file: sec01.tex
\section{Introduction and background}\label{introintro}
\subsection{NV-diamond magnetometry overview}\label{intro}
Quantum sensors encompass a diverse class of devices that exploit quantum coherence to detect weak or nanoscale signals. As their behavior is tied to physical constants, quantum devices can achieve accuracy, repeatability, and precision approaching fundamental limits~\cite{Budker2007}. As a result, these sensors have shown utility in a wide range of applications spanning both pure and applied science~\cite{Degen2017}. A rapidly emerging quantum sensing platform employs atomic-scale defects in crystals. In particular, magnetometry using nitrogen vacancy (NV) color centers in diamond has garnered increasing interest.

The use of NV centers as magnetic field sensors was first proposed~\cite{Taylor2008,Degen2008} and demonstrated with single NVs~\cite{Maze2008,Balasubramanian2008} and NV ensembles~\cite{Acosta2009} circa 2008. In the decade following, both single- and ensemble-NV-diamond magnetometers~\cite{Doherty2013,Rondin2014} have found use for applications in condensed matter physics~\cite{Casola2018}, neuroscience and living systems biology~\cite{Schirhagl2014,Wu2016}, nuclear magnetic resonance (NMR)~\cite{Wu2016}, Earth and planetary science~\cite{Glenn2017}, and industrial vector magnetometry~\cite{Grosz2017}. 

\begin{figure}[ht]
\begin{minipage}[t]{4.4cm}
\begin{overpic}[height=1.65 in]{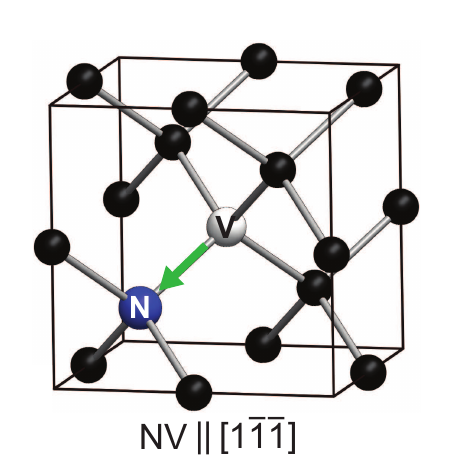}
\put (2.5, 94) {\large a}
\end{overpic}
\end{minipage}
\begin{minipage}[t]{4.4cm}
\begin{overpic}[height=1.75 in]{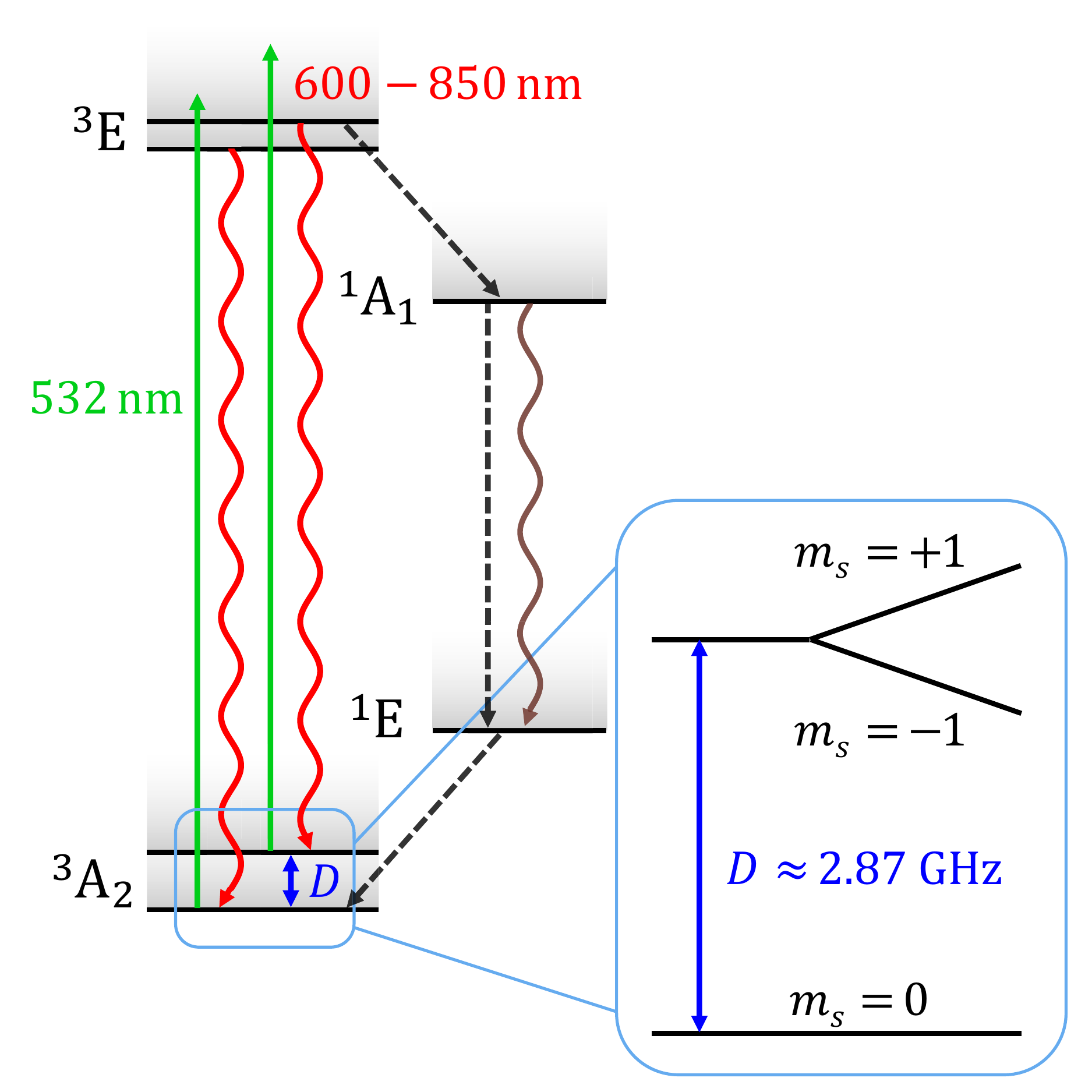}
\put (-10, 90) {\large b}
\end{overpic}
\end{minipage}
\caption[NV Center Overview]{Overview of the \textcolor{black}{nitrogen-vacancy (NV)} center quantum system. a) Diagram of diamond lattice containing \textcolor{black}{an NV} center, which consists of a substitutional nitrogen adjacent to a lattice vacancy. The green arrow marks the NV symmetry axis, oriented along the $[1\bar{1}\bar{1}]$ diamond crystallographic axis for the particular NV center shown here. From Ref.~\cite{Pham2013thesis}. b) Energy level diagram for the negatively charged NV$^\text{-}$ center in diamond, with zero-field splitting $D$ between the ground-state electronic spin levels $m_s \!=\!0$ and $m_s \!=\!\pm 1$. The $m_s \!=\!\pm 1$ energy levels experience a Zeeman shift in the presence of a magnetic field $\vec{B}$, \textcolor{black}{which forms the basis for NV$^\text{-}$ magnetometry}. Adapted from Ref.~\cite{Schloss2018}.}  \label{fig:nvlatticeandenergylevels}
\end{figure}

Solid-state defects such as NV centers exhibit quantum properties similar to traditional atomic systems yet confer technical and logistical advantages for sensing applications. NVs are point defects composed of a substitutional nitrogen fixed adjacent to a vacancy within the rigid carbon lattice (see Fig.~\ref{fig:nvlatticeandenergylevels}a). Each NV center's symmetry axis is constrained to lie along one of the four [111] crystallographic directions. While NVs are observed to exist in three charge states (NV$^\text{-}$, NV$^0$ and NV$^\text{+}$), the negatively charged NV$^\text{-}$ center is favored for quantum sensing and quantum information applications~\cite{Doherty2013}. The NV$^\text{-}$ defect exhibits a spin-1 triplet electronic ground state with long spin lifetimes at room temperature; longitudinal relaxation times $T_1 \approx 6$~ms~\cite{Jarmola2012,rosskopf2014investigation} are typical, and coherence times $T_2$ up to a few ms are achievable~\cite{Balasubramanian2009}. 
The defect's spin energy levels are sensitive to magnetic fields, electric fields, strain, and temperature variations~\cite{Doherty2013}, allowing NV$^\text{-}$ to operate as a multi-modal sensor. Coherent spin control is achieved by application of resonant microwaves (MWs) near 2.87~GHz. 
Upon optical excitation, \textcolor{black}{nonradiative decay through} a spin-state-dependent intersystem crossing~\cite{Goldman2015,Goldman2015b} \textcolor{black}{produces} both spin-state-dependent fluorescence contrast and optical spin initialization into the NV$^\text{-}$ center's $m_s=0$ ground state (see Fig.~\ref{fig:nvlatticeandenergylevels}b). 

Relative to alternative technologies~\cite{Grosz2017}, sensors employing NV$^\text{-}$ centers excel in technical simplicity and spatial resolution~\cite{Grinolds2014,Arai2015,jaskula2017superresolution}. Such devices may operate as broadband sensors, with bandwidths up to $\sim 100$~kHz~\cite{Acosta2010b,Barry2016,Schloss2018}, or as \textcolor{black}{high frequency} detectors for signals up to \textcolor{black}{$\sim$ GHz~\cite{Shin2012,Loretz2013,Boss2016,Wood2016,Cai2013, Steinert2013,Lovchinsky2016, Pham2016,Shao2016,Boss2017, Schmitt2017,Aslam2017,Casola2018, horsley2018microwave, Tetienne2013,pelliccione2014two,hall2016detection}
.} Importantly, effective optical initialization and readout of NV$^\text{-}$ spins does not require narrow-linewidth lasers; rather, a single free-running $532$~nm solid-state laser is sufficient. NV-diamond sensors operate at ambient temperatures, pressures, and magnetic fields, and thus require no cryogenics, vacuum systems, or \textcolor{black}{tesla-scale} applied bias fields. Furthermore, diamond is chemically inert, making NV$^\text{-}$ devices biocompatible. These properties allow sensors to be placed within $\sim 1$~nm of field sources~\cite{Pham2016}, which enables magnetic field imaging with nanometer-scale spatial resolution~\cite{Grinolds2014, Arai2015,jaskula2017superresolution}. NV-diamond sensors are also operationally robust and may function at pressures up to 60~GPa~\cite{Ivady2014,Doherty2014,Hsieh2018imaging} and temperatures from cryogenic to 600~K~\cite{Toyli2012,Toyli2013,Plakhotnik2014}. 

Although single NV$^\text{-}$ centers find numerous applications in ultra-high-resolution sensing due to their angstrom-scale size~\cite{Balasubramanian2008,Maze2008,Casola2018},  sensors employing ensembles of NV$^\text{-}$ centers provide improved signal-to-noise ratio (SNR) at the cost of spatial resolution by virtue of statistical averaging over multiple spins~\cite{Taylor2008,Acosta2009}. Diamonds may be engineered to contain concentrations of NV$^\text{-}$ centers as high as $10^{19}$~cm$^{\text{-}3}$~\cite{choi2017depolarization}, which facilitates high-sensitivity measurements from single-channel bulk detectors as well as wide-field parallel magnetic imaging~\cite{Taylor2008,Steinert2010,Pham2011,Steinert2013,LeSage2013,Glenn2015,Davis2018,Fescenko2018}. These engineered diamonds typically contain NV$^\text{-}$ centers with symmetry axes distributed along all four crystallographic orientations, each \textcolor{black}{primarily sensitive to the magnetic field projection along its axis}; thus, ensemble-NV$^\text{-}$ devices provide full vector magnetic field sensing without heading errors or dead zones~\cite{Maertz2010,Steinert2010,Pham2011,LeSage2013,Schloss2018}. NV$^\text{-}$ centers have also been employed for high-sensitivity imaging of temperature~\cite{Kucsko2013}, strain, and electric fields~\cite{Dolde2011,Barson2017}. Recent examples of ensemble-NV$^\text{-}$ sensing applications include magnetic detection of single-neuron action potentials~\cite{Barry2016}; magnetic imaging of living cells~\cite{LeSage2013,Steinert2013}, malarial hemozoin~\cite{Fescenko2018}, and biological tissue with subcellular resolution~\cite{Davis2018}; nanoscale thermometry~\cite{Kucsko2013,Neumann2013}; single protein detection~\cite{Shi2015single,Lovchinsky2016}; nanoscale and micron-scale NMR~\cite{Staudacher2013nuclear,Glenn2018,Devience2015,Kehayias2017,Loretz2014,bucher2018hyperpolarization,Rugar2015,sushkov2014alloptical}; and studies of meteorite composition~\cite{Fu2014} and paleomagnetism~\cite{Glenn2017, Farchi2017}.


Despite demonstrated utility in a number of applications, the present performance of ensemble-NV$^\text{-}$ sensors remains far from theoretical limits. Even the \textcolor{black}{most sensitive} ensemble-based devices demonstrated to date exhibit readout fidelities \textcolor{black}{$\mathcal{F}$ $\sim 0.01$}, \textcolor{black}{limiting sensitivity to at best} $\sim 100\times$ worse than the spin projection limit. Additionally, reported dephasing times $T_2^*$ in NV-rich diamonds remain $100$ to $1000\times$ shorter than the \textcolor{black}{theoretical maximum of $2 T_1$~\cite{Jarmola2012,Bauch2018,Bauch2019}.} As a result, whereas present state-of-the-art ensemble-NV$^\text{-}$ magnetometers  exhibit $\text{pT/}\sqrt{\text{Hz}}$-level sensitivities, competing technologies such as superconducting quantum interference devices (SQUIDs) and spin-exchange relaxation-free (SERF) magnetometers exhibit sensitivities at the $\text{fT/}\sqrt{\text{Hz}}$-level and below~\cite{kitching2018chipscale}. This $\sim 1000\times$ sensitivity discrepancy corresponds to a $\sim 10^6\times$ increase in required averaging time, which precludes many envisioned applications.  \textcolor{black}{In particular}, the sensing times required to detect weak static signals with an NV-diamond sensor may be unacceptably long; e.g., biological systems may have only \textcolor{black}{a short period of} viability. In addition, many applications, such as spontaneous event detection and time-resolved sensing of dynamic processes~\cite{Marblestone2013,Shao2016}, are incompatible with signal averaging. Realizing NV-diamond magnetometers with improved sensitivity could enable a new class of scientific and industrial applications poorly matched to bulkier SQUID and vapor-cell technologies. Examples include noninvasive, real-time magnetic imaging of neuronal circuit dynamics~\cite{Barry2016}, high throughput nanoscale and micron-scale NMR spectroscopy~\cite{Smits2019two,Glenn2018,bucher2018hyperpolarization}, nuclear quadrupole resonance (NQR)~\cite{Lovchinsky2017magnetic}, human magnetoencephalography~\cite{Hamalainen1993}, subcellular magnetic resonance imaging (MRI) of dynamic processes\textcolor{black}{~\cite{Davis2018}}, precision metrology, \textcolor{black}{tests of fundamental physics~\cite{rajendran2017}, and simulation of exotic particles~\cite{kirschner2018proposal}.}



This review accordingly focuses on understanding present sensitivity limitations for ensemble-NV$^\text{-}$ magnetometers to guide future research efforts. We survey and analyze methods for optimizing magnetic field sensitivity, which we divide into three broad categories: (i) improving spin dephasing and coherence times; (ii) improving readout fidelity; and (iii) improving quality and consistency of host diamond material properties. Given the square-root improvement of sensitivity with number of interrogated spins, we primarily concentrate on ensemble-based devices with $\gtrsim 10^{4}$ NV$^\text{-}$ centers~\cite{Acosta2009,LeSage2012,Wolf2015,Barry2016,Clevenson2015}. However, we also examine single-NV$^\text{-}$ magnetometry techniques in order to determine their applicability to ensembles. Moreover, while this work primarily treats broadband, time-domain magnetometry from DC up to $\sim 100$~kHz, narrowband AC sensing techniques are also analyzed when considered relevant to future DC and broadband magnetometry advances. \textcolor{black}{Alternative phase-insensitive AC magnetometry techniques, such as $T_1$ relaxometry~\cite{hall2016detection,Shao2016,Casola2018, Ariyaratne2018, pelliccione2014two, Tetienne2013,vanderSar2015nanometre,Romach2015}, are not discussed}.

This document is organized as follows: \textcolor{black}{the remainder of Sec.~\ref{introintro} provides introductory material on NV$^\text{-}$ magnetometry, with Sec.~\ref{magintro} introducing magnetic field sensing, Sec.~\ref{nvhamiltonian} presenting the NV$^\text{-}$ spin Hamiltonian and its magnetic-field-dependent transitions, Sec.~\ref{ramseyintro} describing quantum measurements using the NV$^\text{-}$ spin, Sec.~\ref{t2*t2intro} outlining how spin dephasing and decoherence limit magnetometry, and Sec.~\ref{ACDC} summarizing differences between DC and AC sensing approaches while focusing subsequent discussion on DC sensing. Section~\ref{sensitivityconsiderations} concentrates on magnetic field sensitivity}, with Sec.~\ref{magneticfieldsensitivity} introducing the mathematical formalism governing sensitivity of \textcolor{black}{Ramsey-based} ensemble-NV$^\text{-}$ magnetometers, Sec.~\ref{cwpulsed} reviewing \textcolor{black}{common alternatives to Ramsey protocols for DC magnetometry}, and Sec.~\ref{physicallimits} overviewing key parameters that determine magnetic field sensitivity. Section~\ref{T2starlimits} examines the NV$^\text{-}$ spin ensemble dephasing time, $T_2^*$, and coherence time, $T_2$. In particular, Sec.~\ref{T2*improvement} motivates efforts to extend $T_2^*$, Sec.~\ref{T2*ensemble} highlights relevant definitional differences of $T_2^*$ for ensembles and single spins, Sec.~\ref{T2*params} characterizes various mechanisms contributing to NV$^\text{-}$ ensemble $T_2^*$, and Secs.~\ref{nitrogenlimitT2*}-\ref{NVNVlimit} investigate limits to $T_2^*$ and $T_2$ from dipolar interactions with specific paramagnetic species within the diamond. Section~\ref{T2*extensionmethods} analyzes methods to extend the NV$^\text{-}$ ensemble dephasing and coherence times using DC and radiofrequency (RF) magnetic fields. Section~\ref{fidelityimprovementmethods} analyzes a variety of techniques demonstrated to improve the NV$^\text{-}$ ensemble readout fidelity. Section~\ref{sampleengineering} reviews progress in engineering diamond samples for high-sensitivity magnetometry, primarily focusing on increasing the NV$^\text{-}$ concentration while maintaining long $T_2^*$ times and good readout fidelity. Section~\ref{miscellaneoustechniques} analyzes several additional NV-diamond magnetometry techniques not covered in previous sections. Section~\ref{conclusion} provides concluding remarks and an outlook on areas where further study is needed. 
We note that this document aims to comprehensively cover relevant results reported through mid-2017 and provides limited coverage of results published thereafter.

\subsection{Magnetometry introduction}\label{magintro}
Magnetometry is the measurement of a magnetic field's magnitude, direction, or projection onto a particular axis. A simple magnetically-sensitive device is a compass needle, which aligns along \textcolor{black}{the planar projection of the ambient} magnetic field. \textcolor{black}{Regardless of sophistication}, all magnetometers exhibit one or more parameters dependent upon the external magnetic field.
For example, the voltage induced across a pickup coil varies with applied AC magnetic field, as does the resistance of a giant magnetoresistance sensor. In atomic systems such as gaseous alkali atoms, the Zeeman interaction causes the electronic-ground-state energy levels to shift with magnetic field. Certain color centers including NV$^\text{-}$ in diamond also exhibit magnetically sensitive energy levels. 
For both NV$^\text{-}$ centers and gaseous alkali atoms, magnetometry reduces to measuring transition frequencies between energy levels \textcolor{black}{that display a difference in response to magnetic fields}. Various approaches allow direct determination of a transition frequency; for example, frequency-tunable electromagnetic radiation may be applied to the system, and the transition frequency localized from recorded absorption, dispersion, or fluorescence features. Transition frequencies may also be measured via interferometric techniques, which record a transition-frequency-dependent phase~\cite{rabi1937space,ramsey1950molecular}.


\subsection{The NV$^\text{-}$ ground state spin}\label{nvhamiltonian}
The NV$^\text{-}$ center's electronic ground state Hamiltonian can be expressed as
\begin{equation}\label{eqn:fullham}
    \mathscr{H} = \mathscr{H}_{0}+\mathscr{H}_{\text{nuclear}}+\mathscr{H}_{\text{elec}|\text{str}},
    \end{equation}
where $\mathscr{H}_0$ encompasses the NV$^\text{-}$ electron spin interaction with external magnetic field $\vec{B}$ and zero-field-splitting parameter $D\approx 2.87$ GHz, which results from an electronic spin-spin interaction within the NV$^\text{-}$; $\mathscr{H}_{\text{nuclear}}$ characterizes interactions arising from the nitrogen's nuclear spin; and $\mathscr{H}_{\text{elec}|\text{str}}$ describes the electron spin interaction with electric fields and crystal strain. 
Defining $z$ to be along the NV$^\text{-}$ internuclear axis, $\mathscr{H}_0$ may be expressed as
\begin{equation}\label{eqn:h0}
\mathscr{H}_{0}/h =  D S_z^2  +  \frac{g_e\mu_B}{h}\left(\vec{B}\cdot \vec{S}\right),
\end{equation}
where $g_e\approx 2.003$ is the NV electronic g-factor, $\mu_B$ is the Bohr magneton, $h$ is Planck's constant, and $\vec{S}=(S_x,S_y,S_z)$ is the dimensionless electronic spin-1 operator. $\mathscr{H}_{0}$ is the simplest Hamiltonian sufficient to model basic NV$^\text{-}$ spin behavior in the presence of a magnetic field. 

The NV$^\text{-}$ center's nitrogen nuclear spin ($I = 1$ for $^{14}$N and $I = \nicefrac{1}{2}$ for $^{15}$N) creates additional coupling terms characterized by \begin{align}\label{eqn:nuclearham}
    \begin{split}
    \mathscr{H}_{\text{nuclear}}/h &= A_\parallel S_z I_z + A_\perp\!\left(S_x I_x + S_y I_y\right) \\ 
    &+ P\left(I_z^2 - I(I+1)/3\right) \\
    &-\frac{g_I \mu_N}{h}\left(\vec{B}\cdot \vec{I}\right),
    \end{split}
\end{align}
where $A_\parallel$ and $A_\perp$ are (respectively) the axial and transverse magnetic hyperfine coupling coefficients, $P$ is the nuclear electric quadrupole parameter, $g_I$ is the nuclear g-factor for the relevant nitrogen isotope, $\mu_N$ is the nuclear magneton,  and $\vec{I}=(I_x,I_y,I_z)$ is the dimensionless nuclear spin operator. Experimental values of $A_\parallel$, $A_\perp$, and $P$ are reported in Table~\ref{tab:hamconstants}. Note that the term proportional to $P$ vanishes for $I = \nicefrac{1}{2}$ in $^{15}$NV$^\text{-}$, as no quadrupolar moment exists for spins $I<1$. 

\begin{figure}[htbp]
    \centering
    \begin{overpic}[width = .6\columnwidth]{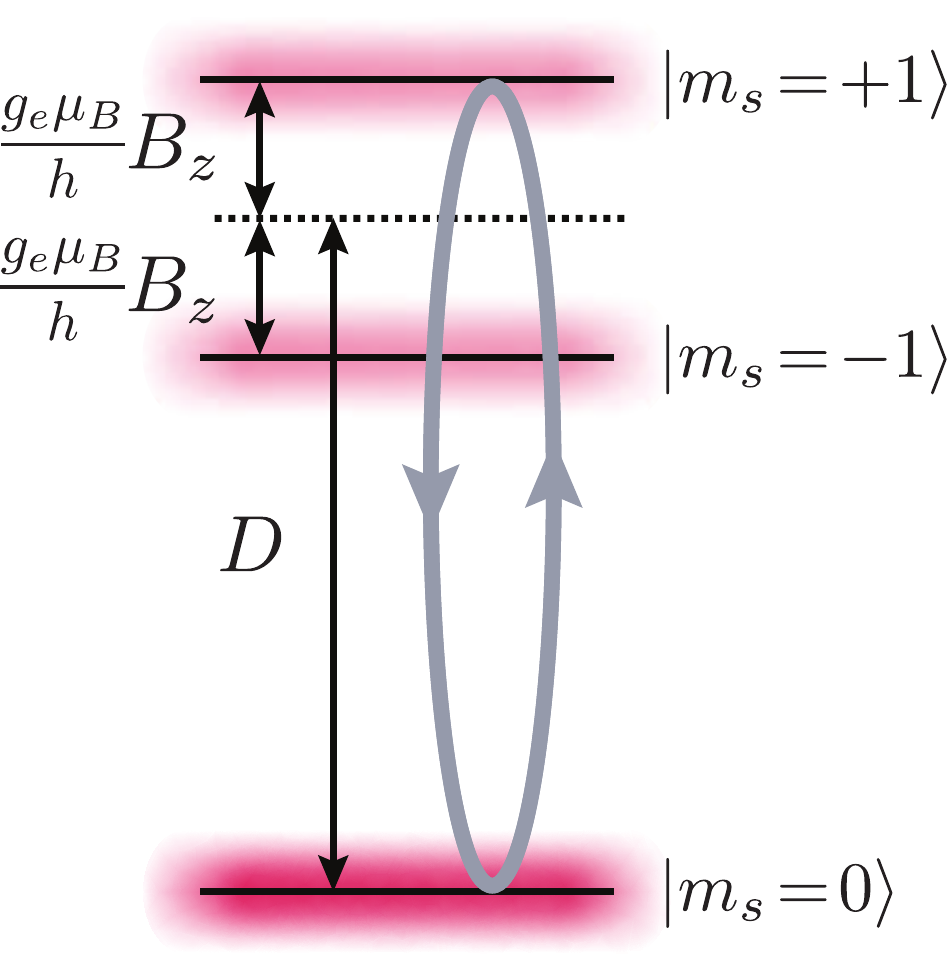}
    \end{overpic}
    \caption[]{\textcolor{black}{Energy level diagram for the NV$^\text{-}$ ground state spin in the presence of an axial magnetic field $B_z$ and ignoring nuclear spin, as described by Eqn.~\ref{eqn:spin1ham}. Population in the $|m_s\!=\!0\rangle$ state results in higher fluorescence under optical illumination than population in the $|m_s\!=\!\pm1\rangle$ states. In this diagram, resonant MWs (gray oval) address the $|m_s\!=\!0\rangle\rightarrow|m_s\!=\!+1\rangle$ transition. Eqn.~\ref{eqn:spinhalfham} describes the pseudo-spin-\nicefrac{1}{2} subspace occupied by these two levels.}}
    \label{fig:simpleleveldiagram}
\end{figure}

\begin{figure*}[htbp]
    \centering
    \begin{overpic}[width = \textwidth]{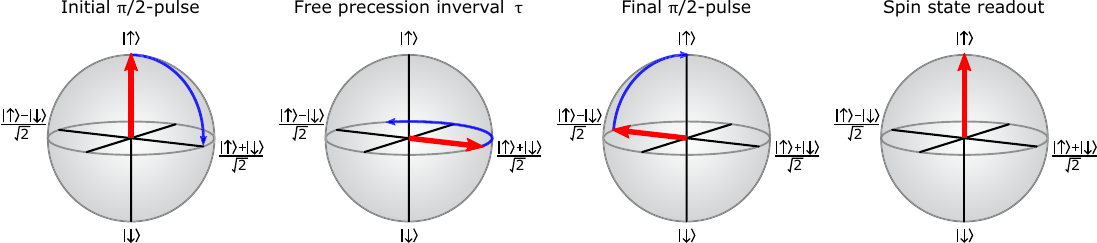}
    \end{overpic}
    \caption[Bloch sphere depiction of Ramsey sequence]{\textcolor{black}{Bloch sphere depiction of Ramsey sequence. After initialization to the spin state $|\!\!\uparrow\rangle$, a sinusoidally-varying magnetic field rotates the state vector by $\pi/2$, thus preparing a superposition of $|\!\!\uparrow \rangle$ and $|\!\!\downarrow \rangle$ spin states. Next, the Bloch vector undergoes free precession for duration $\tau$, accumulating a phase $\phi$ proportional to the static magnetic field being sensed. After time $\tau$, a second $\pi/2$-pulse maps the accumulated phase onto a population difference between the $|\!\!\uparrow \rangle$ and $|\!\!\downarrow \rangle$ states. Here, a $\phi = \pi$ phase accumulation is shown, which maps back to the state $|\!\!\uparrow \rangle$. Finally, a projective spin state measurement detects the population difference, allowing determination of the static magnetic field sensed by the spin.}}
    \label{fig:bloch}
\end{figure*}

The NV$^\text{-}$ electron spin also interacts with electric fields $\vec{E}$ and crystal stress (with associated strain)~\cite{kehayias2019diamond}. In terms of the axial dipole moment $d_\parallel$, transverse dipole moments $d_\perp$ and $d_\perp'$, and spin-strain coupling parameters \{$\mathscr{M}_z$, $\mathscr{M}_x$, $\mathscr{M}_y$, $\mathscr{N}_x$, $\mathscr{N}_y$\}, the interaction is presently best approximated by~\cite{VanOort1990, Doherty2012, udvarhelyi2018spinstrain,barfuss2018spinstress}
\begin{align}\label{eqn:elecstrham}
    \begin{split}
     \mathscr{H}_{\text{elec}|\text{str}}/h &= \left(d_\parallel E_z+\mathscr{M}_z\right)S_z^2 \\ &+ \left(d_\perp E_x +\mathscr{M}_x\right)\left(S_y^2-S_x^2\right)\\
     &+\left(d_\perp E_y + \mathscr{M}_y\right)\left(S_x S_y+S_y S_x\right) \\ &+ \left(d_\perp' E_x + \mathscr{N}_x\right)\left(S_x S_z+S_z S_x\right)\\
     &+\left(d_\perp' E_y + \mathscr{N}_y\right)\left(S_y S_z+S_z S_y\right).   
    \end{split}
\end{align}
Experimental values of $d_\perp$ and $d_\parallel$ are given in Table~\ref{tab:hamconstants}. In magnetometry measurements, the terms proportional to $(d_\perp' E_i$ +$\mathscr{N}_i)$ for $i = x,y$ are typically ignored, as \textcolor{black}{they are off-diagonal in the $S_z$ basis, and the energy level shifts they produce are thus suppressed by $D$~\cite{kehayias2019diamond}}. Furthermore, many magnetometry implementations operate with an applied bias field $\vec{B_0}$ satisfying $d_\perp E_i$ +$\mathscr{M}_i  \ll \frac{g_e \mu_B}{h} B_0 \ll D$ for $i = x,y$ in order to operate in the linear Zeeman regime, where the energy levels are maximally sensitive to magnetic field changes (see Appendix~\ref{StarkZeeman}). In the linear Zeeman regime, the terms in $\mathscr{H}_{\text{elec}|\text{str}}$ proportional to $(d_\perp E_i$ +$\mathscr{M}_i)$ can also be ignored. The sole remaining term in $\mathscr{H}_{\text{elec}|\text{str}}$ acts on the NV$^\text{-}$ spin in the same way as the temperature-dependent $D$ and is often combined into the parameter $D$ for a given NV$^\text{-}$ orientation~\cite{Glenn2017}. Except for extreme cases such as sensing in highly strained diamonds or in the presence of large electric fields, the values of all the electric field and strain parameters in $\mathscr{H}_{\text{elec}|\text{str}}$ are $\sim 1 ~\text{MHz}$ or lower. Consequently, for most magnetic sensing applications, $\mathscr{H}_{0}$ can be taken as the Hamiltonian describing the NV$^\text{-}$ ground state spin for each of the hyperfine states. 


In the presence of a magnetic field \textcolor{black}{oriented along the NV internuclear axis} $\vec{B} = (0, 0, B_z)$, $\mathscr{H}_{0}$ is given in matrix form by

\begin{equation}\label{eqn:spin1ham}
\mathscr{H}_{0}^{(z)}/h = \left(
\begin{array}{ccc}
 D+\frac{g_e \mu_B}{h}B_z & 0 & 0 \\
 0 & 0 & 0 \\
 0 & 0 & D-\frac{g_e \mu_B}{h}B_z \\
\end{array}
\right).
\end{equation}
with eigenstates \mbox{$|m_s\!=\!0\rangle$}, \mbox{$|m_s\!=\!-1\rangle$}, and \mbox{$|m_s\!=\!+1\rangle$} and magnetic-field-dependent transition frequencies 
\begin{equation}
\nu_\pm = D\pm\frac{g_e \mu_B}{h}B_z, 
\end{equation}
which are depicted in Fig.~\ref{fig:simpleleveldiagram}. For the general case of a magnetic field $\vec{B}$ with both axial and transverse components $B_z$ and $B_\perp$, the transition frequencies are given to third order in $\left(\frac{g_e \mu_B}{h}\frac{B}{D}\right)$ by
\small
\begin{align}
\begin{split}
\nu_\pm &= D\left[1 \pm \left(\!\frac{g_e \mu_B}{h}\frac{B}{D}\!\right)\cos{\theta_B} + \frac{3}{2}\left(\!\frac{g_e \mu_B}{h}\frac{B}{D}\!\right)^2\sin^2{\theta_B} \right. \\
  &\left.\;\;\;\;\pm \!\left(\!\frac{g_e \mu_B}{h}\frac{B}{D}\!\right)^3 \!\left(\frac{1}{8}\sin^3\!{\theta_B}\tan{\theta_B}-\frac{1}{2}\sin^2\!{\theta_B} \cos{\theta_B} \right)\!\right],
\end{split}
\end{align}
\normalsize
where $\tan{\theta_B} = B_\perp/B_z$.

Magnetic sensing experiments utilizing NV$^\text{-}$ centers often interrogate one of these two transitions, allowing the unaddressed state to be neglected. 
For example, choosing the $|0\rangle$ and $|\!+\!1\rangle$ states and subtracting a common energy offset allows $\mathscr{H}_{0}^{(z)}$ from Eqn.~\ref{eqn:spin1ham} to be reduced to the spin-\nicefrac{1}{2} Hamiltonian $H$ given by 
\begin{equation}\label{eqn:spinhalfham}
H/h = \left(
\begin{array}{cc}
 \frac{D}{2}+\frac{1}{2}\frac{g_e \mu_B}{h} B_z &  0 \\
 0 & -\frac{D}{2}-\frac{1}{2}\frac{g_e \mu_B}{h} B_z \\
\end{array}
\right).
\end{equation}
This simplification is appropriate when off-resonant excitation of the $|m_s\!=\!-1\rangle$ state can be ignored and operations on the spin system are short compared to $T_1$. From this simple picture, the full machinery typically employed for two-level systems can be leveraged.

\subsection{Spin-based measurements on NV$^\text{-}$}\label{ramseyintro}


We now outline Norman Ramsey's \textit{Method of separated oscillatory fields} when adapted for magnetic field measurement using one or more NV$^\text{-}$ centers in a two-level subspace, e.g., \{$|0\rangle$,$|\!+\!1\rangle$\}. 
After initialization of the spin state to $|0\rangle$, a periodically varying magnetic field $B_1(t)$, with polarization in the $x$-$y$ plane and frequency $\nu_+$ resonant with the $|0\rangle \leftrightarrow|\!+\!1\rangle$ transition, causes spin population to oscillate between the $|0\rangle $ and $|\!+\!1\rangle$ states at angular frequency $\Omega_R \propto B_1$, called the Rabi frequency.  The resonant field $B_1(t)$ is applied for a particular finite duration $\pi/(2\Omega_R)$ \textcolor{black}{known as a $\pi/2$-pulse}, which transforms the initial state $|0\rangle$ into an equal superposition of $|0\rangle$ and $|\!+\!1\rangle$. This state is then left to precess unperturbed for duration $\tau$, during which a magnetic-field-dependent phase $\phi$ accumulates between the two states. Next, a second $\pi/2$-pulse is applied, mapping the phase $\phi$ onto a population difference between $|0\rangle$ and $|\!+\!1\rangle$. 
Figure~\ref{fig:bloch} provides a Bloch sphere depiction of the Ramsey sequence, where the states $|0\rangle$ and $|\!+\!1\rangle$ are denoted by $|\!\!\uparrow\rangle$ and $|\!\!\downarrow\rangle$ respectively.  See Appendix~\ref{ramsey} for a full \textcolor{black}{mathematical description} of a Ramsey magnetic field measurement. 

The subsequent spin readout process is fundamentally limited by quantum mechanical uncertainty. 
If a measurement of the final state's spin projection $S_z$ is performed in the \{$|0\rangle$, $|\!+\!1\rangle$\} basis, 
only two measurement outcomes are possible: 0 and 1. The loss of information associated with this projective measurement is commonly referred to as spin projection noise~\cite{Itano1993}. 
A projection-noise-limited sensor is characterized by a spin readout fidelity $\mathcal{F} = 1$. Other considerations, such as the photon shot noise, may lead to reductions in the fidelity $\mathcal{F}$, which degrade magnetic field sensitivity. The sensitivities of $S = \nicefrac{1}{2}$ magnetometers at the spin-projection and shot-noise limits are discussed in Sec.~\ref{magneticfieldsensitivity} and treated in detail in Appendices~\ref{spinproj} and \ref{shot}.

\subsection{Spin dephasing and decoherence}\label{t2*t2intro}
Pulsed magnetometry measurements benefit from long sensing intervals $\tau$, as the accumulated magnetic-field-dependent phase $\phi$ typically increases with $\tau$. For example, in a Ramsey measurement, $\phi = \gamma_e B \tau$ with $\gamma_e = g_e \mu_B/\hbar$; maximal sensitivity of the observable $\phi$ to changes in $B$ is therefore achieved when $\frac{d\phi}{dB}= \gamma_e \tau$ is maximized. At the same time, contrast degrades with increasing $\tau$ due to \textcolor{black}{dephasing, decoherence, and spin-lattice interactions, with associated respective relaxation times $T_2^*$, $T_2$, and $T_1$}. The optimal interrogation time must therefore balance these two competing concerns. 

\begin{figure}[htbp]
    \centering
    \begin{overpic}[width = .99\columnwidth]{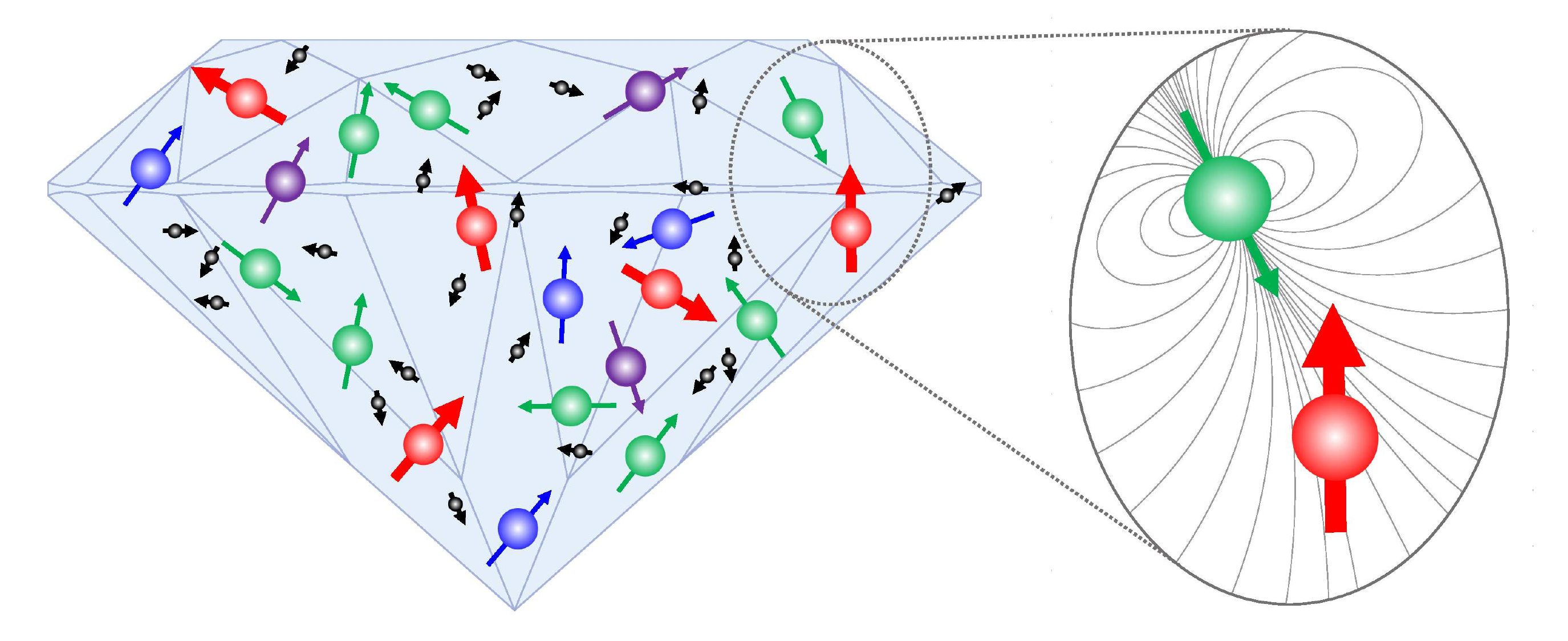}
    \end{overpic}
    \caption[]{\textcolor{black}{Diamond containing spin impurities. NV$^\text{-}$ centers [thick red arrows (\textcolor{red}{\contour{red}{$\rightarrow$}})] experience magnetic fields caused by other spin defects in the diamond, including substitutional nitrogen [thin green arrows (\textcolor{igorgreen}{$\rightarrow$})], $^{13}$C nuclei [small black arrows (\tiny\textcolor{black}{$\rightarrow$}\normalsize)], and other paramagnetic impurities [blue (\textcolor{blue}{$\rightarrow$}) and purple (\textcolor{violet}{$\rightarrow$}) arrows]. The inhomogeneous and time-varying dipolar magnetic fields generated by these spins dephase and decohere the NV$^\text{-}$ spin ensemble.}}
    \label{fig:diamondspinsdipole}
\end{figure}
\begin{figure*}[htbp]
\begin{overpic}[width = .9\textwidth]{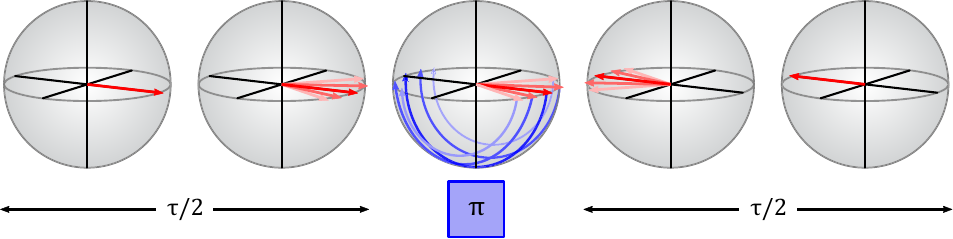}
\end{overpic}
\caption[Recovery of spin phase coherence following central $\pi$-pulse]{\textcolor{black}{Recovery of spin phase coherence with central $\pi$-pulse. Bloch sphere depiction of spin dephasing due to static field inhomogeneities (characterized by $T_2^*$) followed by application of a $\pi$ pulse at time $\tau/2$ and then spin rephasing at time $\tau$. The $\pi$-pulse cancels $T_2^*$ dephasing as well as sensitivity to static signal fields.}}  \label{fig:blochecho}
\end{figure*}

The parameter $T_2^*$ characterizes dephasing associated with static or slowly varying inhomogeneities in a spin system, e.g., dipolar fields from other spin impurities in the diamond, as depicted in Fig.~\ref{fig:diamondspinsdipole}. $T_2^*$ is the characteristic $1/e$ time of a free induction decay (FID) measurement, wherein a series of Ramsey sequences are performed with varying free precession interval $\tau$, and an exponential envelope decay is observed (see Fig.~\ref{fig:cwvsramsey}a). Inhomogeneous fields limit $T_2^*$ by causing spins within an ensemble to undergo Larmor precession at different rates. As depicted in the second Bloch sphere in Fig.~\ref{fig:blochecho}, the spins dephase from one another after free precession intervals $\tau \sim T_2^*$. 

Dephasing from fields that are static over the measurement duration can be reversed by application of a $\pi$-pulse halfway through the free precession interval. In this protocol~\cite{Hahn1950}, the $\pi$-pulse alters the direction of spin precession, such that the phase accumulated due to static fields during the second half of the sequence cancels the phase from the first half. Thus, spins in nonuniform fields rephase, producing a recovered signal termed a "spin echo" (Fig.~\ref{fig:blochecho}). The decay of this echo signal, due to fields that fluctuate over the course of the measurement sequence, is characterized by the coherence time $T_2$, also called the transverse or spin-spin relaxation time. In NV$^\text{-}$ ensemble systems, $T_2$ can exceed $T_2^*$ by orders of magnitude~\cite{deLange2010, Bauch2018, Bauch2019}. As the $T_2$-limited spin echo sequence is intrinsically insensitive to DC magnetic fields, it is frequently employed for detecting AC signals. Meanwhile, the $T_2^*$-limited Ramsey sequence is commonly employed for DC sensing experiments. 
 

\color{black}
\subsection{DC and AC sensing}\label{ACDC}

Quantum sensing approaches may be divided into two broad categories based on the spectral characteristics of the fields to be detected, summarized in Table~\ref{tab:ACDC}. In particular, \textcolor{black}{DC sensing protocols are sensitive to static, slowly-varying, or broadband near-DC signals, whereas AC sensing protocols typically detect narrowband, time-varying signals at frequencies up to $\sim 10$~MHz~\cite{Shin2012,Loretz2013,Boss2016,Wood2016,Cai2013, Steinert2013, Pham2016,Shao2016,Boss2017, Schmitt2017}, although AC sensing experiments of $\sim 100$~MHz signals have also been demonstrated for niche applications~\cite{Aslam2017}. Both DC and AC sensors employing NV$^\text{-}$ ensembles exhibit sensitivities limited, in part, by the relevant NV$^\text{-}$ spin relaxation times.} DC sensitivity is limited by the ensemble's inhomogeneous dephasing time $T_2^*$, \textcolor{black}{which is of order $1~\upmu\text{s}$ in most present implementations}. AC sensitivity is limited by the coherence time $T_2$, which\textcolor{black}{, as mentioned above,} is typically one to two orders of magnitude longer than $T_2^*$~\cite{deLange2010,Bauch2019}\textcolor{black}{, and which can be extended through use of dynamical decoupling protocols to approach the longitudinal spin relaxation time $T_1$ (see Sec.~\ref{DD}). Additionally, alternative forms of $T_1$-limited AC sensing such as $T_1$ relaxometry allow phase-insensitive detection of signals at frequencies in the $\sim$~GHz regime~\cite{Tetienne2013,hall2016detection,Casola2018,Shao2016,pelliccione2014two}. In general,} the enhanced field sensitivities afforded by longer AC sensor coherence times coincide with reduced sensing bandwidth as well as insensitivity to static fields, restricting the application space of \textcolor{black}{of sensors employing these techniques} (see Table~\ref{tab:ACDC}). This review concentrates primarily on DC sensing protocols with particular focus on sensors designed to detect broadband time-varying magnetic fields from DC to $\sim 100$~kHz.

\begin{table*}[ht]
\centering 
\begin{tabular}{l p{7.25cm} p{7.25cm}}
\toprule
 & \parbox[t]{7.25cm}{Broadband DC sensing} & \parbox[t]{7.25cm}{AC sensing} \\ [0.5ex] 
\midrule
Common techniques & Ramsey (Sec.~\ref{magneticfieldsensitivity}), CW-ODMR (Sec.~\ref{cwodmr}), pulsed ODMR (Sec.~\ref{pulsedodmr}) & Hahn echo, dynamical decoupling (Sec.~\ref{DD}) \\[0.8ex]
\textcolor{black}{Relevant relaxation}  & Inhomogeneous spin dephasing ($T_2^*$) & Homogeneous spin decoherence ($T_2$) and longitudinal relaxation ($T_1$)\\[0.8ex]
\parbox[t]{3cm}{\raggedright Frequency/bandwidth} & 0 to $\sim\!100$~kHz (pulsed), 0 to $\sim\!10$~kHz (CW) & Center frequency: $\sim\!1~\text{kHz}$ to $\sim\!10$~MHz; bandwidth: $\lesssim 100$~kHz \\[0.8ex]
\parbox[t]{3cm}{\raggedright Example magnetic sensing applications}  & Biocurrent detection, magnetic particle tracking, magnetic imaging of rocks and meteorites, imaging of magnetic nanoparticles in biological systems, magnetic imaging of electrical current flow in materials, magnetic anomaly detection, navigation & \textcolor{black}{Single biomolecule and protein detection, nanoscale nuclear magnetic resonance, nanoscale electron spin resonance, magnetic resonant phenomena in materials, noise spectroscopy} \\
\bottomrule 
\end{tabular}
\caption{\textcolor{black}{Operational regimes} and selected applications of \textcolor{black}{broadband DC and} AC sensing protocols employing NV$^\text{-}$ ensembles in diamond. \textcolor{black}{$T_1$ relaxometry methods are not considered.}}
\label{tab:ACDC} 
\end{table*}

\textcolor{black}{\section{Measurement sensitivity considerations}\label{sensitivityconsiderations}}

\subsection{Magnetic field sensitivity}\label{magneticfieldsensitivity}

\textcolor{black}{The spin-projection-limited sensitivity of an ensemble magnetometer consisting of $N$ non-interacting spins is approximately given by~\cite{Budker2007,Taylor2008}
\begin{equation}\label{eqn:spinprojlim}
\eta_\text{sp}^\text{ensemble} \approx \frac{\hbar}{\Delta m_s g_e \mu_B} \frac{1}{\sqrt{N \tau}},
\end{equation}
where $g_e \approx 2.003$ is the NV$^\text{-}$ center's electronic g-factor~\cite{Doherty2013}, $\mu_B$ is the Bohr magneton, $\hbar$ is the reduced Planck constant,  $\tau$ is the free precession (i.e., interrogation) time per measurement, and $\Delta m_s$ is the difference in spin quantum number between the two interferometry states (e.g., $\Delta m_s\!=\! 1$ for a spin $S\!=\!\nicefrac{1}{2}$ system, and $\Delta m_s\!=\!2$ for a $S\!=\!1$ system employing $m_s\!=\!+1$ and $m_s\!=\!-1$ states).} 
\textcolor{black}{Certain pulsed magnetometry schemes such as Ramsey-based protocols can allow sensitivities approaching the spin-projection limit, in part by ensuring that the spin state readout does not interfere with the magnetic field interrogation~\cite{ramsey1950molecular}.}
However, even when employing Ramsey protocols, NV$^\text{-}$ ensemble magnetometers suffer from at least three major experimental non-idealities, which deteriorate the achievable magnetic field sensitivity. 

First, for NV$^\text{-}$ ensemble magnetometers, the spin-state initialization time $t_I$ and readout time $t_R$ may be significant compared to the interrogation time $\tau$. By decreasing the fraction of time devoted to spin precession, the finite values of $t_I$ and $t_R$ deteriorate the sensitivity by the factor
\begin{equation}\label{eqn:duty}
\sqrt{\frac{t_I + \tau + t_R}{\tau}}.
\end{equation}



\begin{figure}[ht]
\centering
\begin{overpic}[width = 0.8\columnwidth]{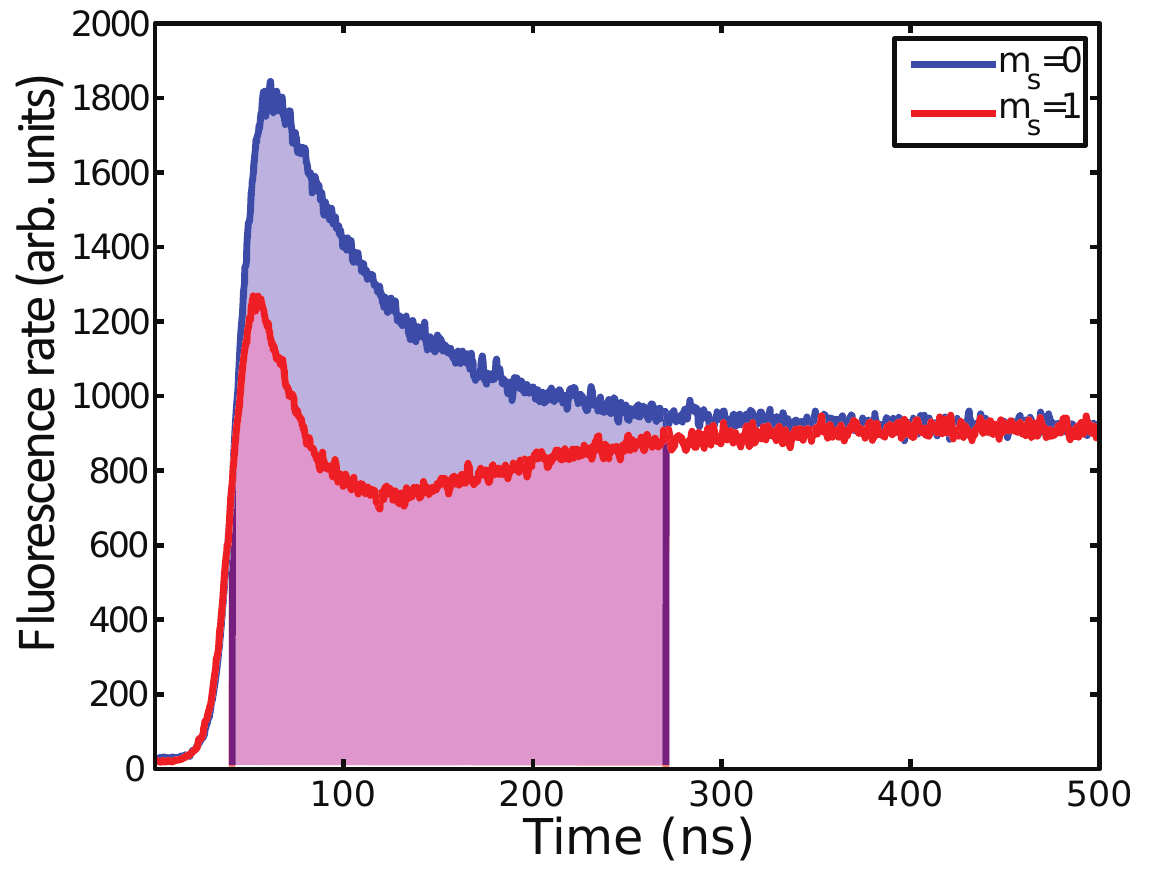}
\end{overpic}
\caption[Spin-dependent fluorescence]{Fluorescence of the NV$^\text{-}$ spin states. NV$^\text{-}$ centers prepared in the $m_s=0$ state emit photons at a higher rate than centers prepared in the $m_s=\pm 1$ states. This spin-dependent fluorescence forms the basis of conventional NV$^\text{-}$ readout. Data courtesy of Brendan Shields.}  \label{fig:spindependentfluorescencecontrast}
\end{figure}

Second, the conventional NV$^\text{-}$ optical readout technique~\cite{Doherty2013}, \textcolor{black}{which detects the spin-state-dependent fluorescence (also commonly referred to as photoluminescence or PL) in the 600\,-\,850~nm band, does not allow single-shot determination of the NV$^\text{-}$ spin state to the spin projection limit (i.e., the standard quantum limit)}\textcolor{black}{~\cite{Itano1993}}. An NV$^\text{-}$ center in the electronically excited spin-triplet state will decay either directly to the spin-triplet ground state or indirectly though a cascade of spin-singlet states~\cite{Rogers2008} via an inter-system crossing~\cite{Goldman2015,Goldman2015b,Thiering2018}. Conventional NV$^\text{-}$ optical readout exploits the $m_s\!=\!\pm 1$ states' higher likelihood to enter the singlet-state cascade more often than the $m_s\!=\!0$ state (see Table~\ref{tab:branchingratios}). An NV$^\text{-}$ center that enters the singlet state cascade does not fluoresce in the 600\,-\,850~nm band, whereas an NV$^\text{-}$ center decaying directly to the spin-triplet ground state can continue cycling between the ground and excited triplet states, producing fluorescence in the 600\,-\,850~nm band. The $m_s\! =\! \pm 1$ \textcolor{black}{states therefore} 
produce on average less PL in the 600\,-\,850~nm band, as shown in Fig.~\ref{fig:spindependentfluorescencecontrast}.
Unfortunately the $\sim 140-200$~ns~\cite{Gupta2016,Robledo2011,Acosta2010b} spin-singlet cascade lifetime and limited differences in $m_s = \pm 1$ and $m_s = 0$ decay behavior allows for only probabilistic determination of the NV$^\text{-}$ initial spin state. Following Ref.~\cite{Shields2015}, we quantify the added noise from imperfect readout with the parameter $\sigma_R \geq 1$, such that $\sigma_R=1$ corresponds to readout at the spin projection limit. This parameter is the inverse of the measurement fidelity: $\mathcal{F} \equiv 1/\sigma_R$. For imperfect readout schemes, the value of $\sigma_R$ can be calculated as~\cite{Taylor2008,Shields2015}
\begin{eqnarray}\label{eqn:readnoise}
\sigma_R &=& \sqrt{1+\frac{2(a+b)}{(a-b)^2}}\\
&=&\sqrt{1+\frac{1}{C^2 n_\text{avg}}},\label{eqn:readnoise2}
\end{eqnarray}
where $a$ and $b$ respectively denote the average numbers of photons detected from the $m_s\!=\!0 $ and $ m_s\!=\!\pm 1$ states of a single NV$^\text{-}$ center during a single readout. In Eqn.~\ref{eqn:readnoise2} we identify $C = \frac{a-b}{a+b}$ as the measurement contrast (i.e., the interference fringe visibility)  and $n_\text{avg} = \frac{a+b}{2}$ as the average number of photons collected per NV$^\text{-}$ center per measurement. Although sub-optimal initialization and readout times $t_I$ and $t_R$ can degrade the value of $C$, it is henceforth assumed that $t_I$ and $t_R$ are \textcolor{black}{chosen optimally}.

Third, the sensitivity $\eta$ is degraded for increased values of $\tau$ due to spin dephasing during precession. For Ramsey-type pulsed magnetometry (i.e., with no spin echo), the dephasing occurs with characteristic time $T_2^*$ so that $\eta$ is additionally deteriorated by the factor
\begin{equation}\label{eqn:dephasing}
\frac{1}{e^{-(\tau/T_2^*)^p}},
\end{equation}
where the value of the stretched exponential parameter $p$ depends on the origin of the dephasing (see Appendix~\ref{FIDexponent}). NV$^\text{-}$ spin resonance lineshapes with exactly Lorentzian profiles correspond to dephasing with $p = 1$, and spin resonance lineshapes with Gaussian profiles correspond to $p=2$  (see Appendix~\ref{linewidth}).

\textcolor{black}{Combining Eqns.~\ref{eqn:spinprojlim}, \ref{eqn:duty}, \ref{eqn:readnoise}, and \ref{eqn:dephasing} gives the sensitivity for a Ramsey-type NV$^\text{-}$ broadband ensemble magnetometer~\cite{Popa2004} as
\small
\begin{align}\label{eqn:ramseyshotexact}
\eta_\text{Ramsey} ^\text{ensemble} &\approx \\ \nonumber & \!\!\!\!\!\!\!\!\!\!\!\!\!\!\!\!\!\! \underbrace{\frac{\hbar}{\Delta m_s g_e \mu_B} \frac{1}{\sqrt{N\tau}}}_{\text{Spin projection limit}} \;   \underbrace{\frac{1}{e^{-(\tau/T_2^*)^p}}}_{\text{Spin dephasing}} \; \underbrace{\sqrt{1\!+\!\frac{1}{C^2 n_\text{avg}}}}_{\text{Readout}} \; \underbrace{\sqrt{\frac{t_I\! +\! \tau\! +\! t_R}{\tau}}}_{\text{Overhead time}}
\end{align}
\normalsize
where $N$ is the number of NV$^\text{-}$ centers in the ensemble \textcolor{black}{and $\Delta m_s\!=\!1$ for the effective $S\!=\!\nicefrac{1}{2}$ subspace employed for NV$^\text{-}$ magnetometry using the $m_s\!=\! 0$ and $m_s\!=\!\pm1$ basis}. However, in the limit of measurement contrast $C\ll1$ and when the number of photons collected per NV$^\text{-}$ center per optical readout is much less than one, the readout fidelity is limited by photon shot noise and can be approximated using $1/\mathcal{F} \!=\!\sigma_R \!\approx\! \frac{1}{C \sqrt{n_\text{avg}}}$.  Defining $\mathscr{N} \!=\! N n_\text{avg}$ to be the average number of photons detected per measurement from the ensemble of $N$ NV$^\text{-}$ centers yields the following shot-noise-limited sensitivity equation for a Ramsey scheme~\cite{Pham2013thesis}:}
\begin{equation}\label{eqn:ramseyshot}
\eta^\text{ensemble,shot}_\text{Ramsey} \approx \frac{\hbar}{\Delta m_sg_e \mu_B} \frac{1}{C e^{-(\tau/T_2^*)^p}\sqrt{\mathscr{N}}} \frac{\sqrt{t_I + \tau + t_R}}{\tau}.
\end{equation}
Hereafter, we assume broadening mechanisms produce Lorentzian lineshapes, so that $p=1$. For negligible $t_I$ and $t_R$, the optimal measurement time is $\tau =  T_2^*/2$, whereas for $t_I + t_R \gg T_2^*$, the optimal $\tau$ approaches $T_2^*$ (see Appendix~\ref{optimalprecession}). Equation~\ref{eqn:ramseyshot} illustrates the benefits attained by increasing the dephasing time $T_2^*$, the measurement contrast $C$, the number of NV$^\text{-}$ spin sensors $N$, and the \textcolor{black}{average number of photons detected per NV$^\text{-}$ per measurement  $n_\text{avg}$}. Table~\ref{tab:parameters} lists values of $\sigma_R$ and $\mathscr{N}$  achieved using conventional optical readout in pulsed and CW magnetometry measurements, with both single NV$^\text{-}$ centers and ensembles. At present, conventional optical readout is insufficient to reach the spin projection limit for both single- and ensemble-NV$^\text{-}$ sensors. Appendix~\ref{derivations} derives the sensitivity for a Ramsey-type magnetometer in both the spin projection and shot noise limits. 

In addition to Ramsey-type methods, other protocols allow \textcolor{black}{measurement of DC magnetic fields}.  These alternative methods, including continuous-wave and pulsed optically detected magnetic resonance, offer reduced sensitivity compared to Ramsey-type sequences (for a fixed number of NV$^\text{-}$ centers addressed), \textcolor{black}{as discussed in the following sections}. 

\begin{figure}[ht]
\centering
\begin{overpic}[height=3.75 in]{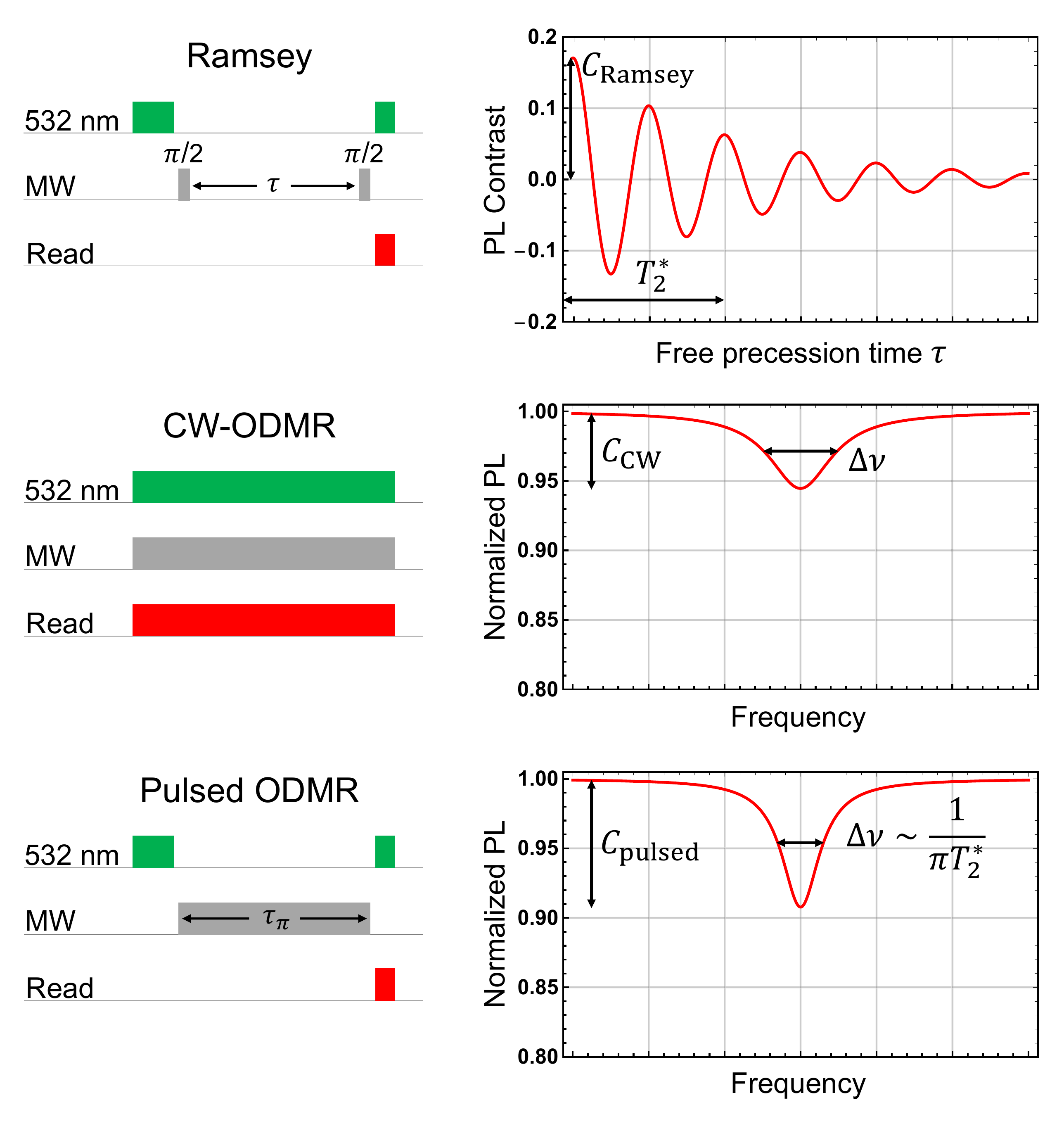}
\put (0, 97) {\large a}
\put (41.25, 97) {\large b}
\put (0, 64.5) {\large c}
\put (41.25, 64.5) {\large d}
\put (0, 32) {\large e}
\put (41.25, 32) {\large f}
\end{overpic}
\caption[CWODMRPulsedODMRRamseyFigure]{Overview of Ramsey, CW-ODMR, and pulsed ODMR magnetometry protocols. a) Schematic of Ramsey magnetometry protocol. b) Representation of free induction decay associated with a Ramsey protocol versus free precession time $\tau$. Fringes exhibit contrast $C_\text{Ramsey}$ and decay exponentially with dephasing time $T_2^*$. c) Schematic of CW-ODMR sensing protocol. d) Representation of CW-ODMR spectrum with contrast $C_\text{CW}$ and linewidth $\Delta \nu$. e) Schematic of pulsed ODMR sensing protocol with MW $\pi$-pulse time $\tau_\pi \sim T_2^*$. f) Representation of pulsed ODMR spectrum with contrast $C_\text{pulsed}$ and linewidth $\Delta \nu \sim 1/(\pi T_2^*)$.}  \label{fig:cwvsramsey}
\end{figure}
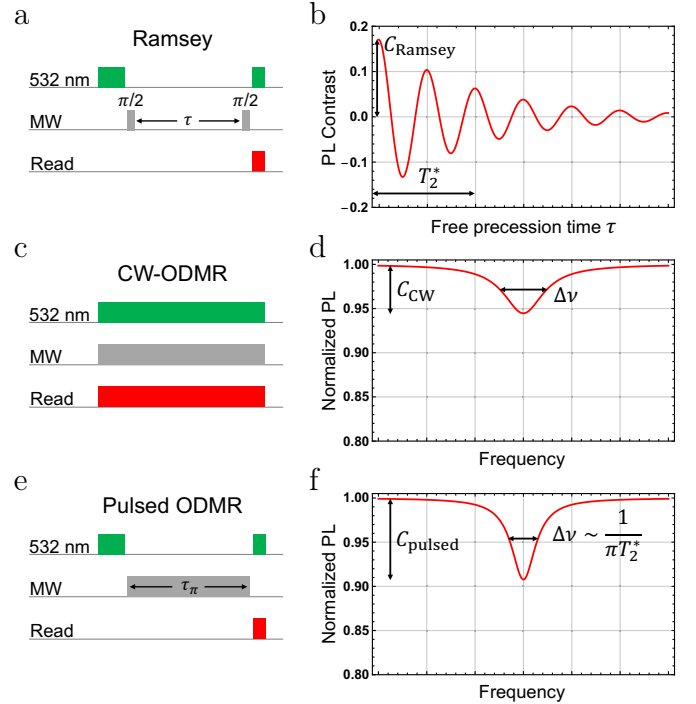




\begin{table*}[ht]
\centering 
\begin{tabular}{c c c c c c} 
\toprule
Reference & Readout method & Single NV$^\text{-}$/ensemble & $\sigma_R$ & $\mathscr{N}$ [counts/measurement] \\ [0.5ex] 

\midrule
\cite{Shields2015} & conventional & single & 10.6 & $9.45 \times 10^5 $~cps~$\times$~$t_R$  \\ 
\cite{Shields2015} & spin-to-charge conversion & single & 2.76 & - \\ 
\cite{Lovchinsky2016} & conventional & single & 35 & $\sim\!10^5  $~cps~$\times$~$t_R$ \\
\cite{Lovchinsky2016} & ancilla-assisted & single & 5 & - \\
\cite{Fang2013}  & conventional & single & 80 & 0.01  \\
\cite{Hopper2016}  & conventional & single & 48 & 0.04 \\
\cite{Hopper2016}  & spin-to-charge conversion & single & 3 & - \\
\cite{jaskula2017improved}  & conventional & single & 54 & 0.022 \\
\cite{jaskula2017improved}  & spin-to-charge conversion & single & 5 & -\\
\cite{Neumann2010} & ancilla-assisted & single & 1.1 & -\\
\cite{LeSage2012}  & conventional & ensemble & $67$ & $2\!\times\! 10^8$ \\
\cite{Wolf2015}  & conventional & ensemble & $\sim\!1000$ & $10^{12}$ \\
\cite{Chatzidrosos2017} & NIR absorption$^\dagger$ & ensemble & 65 & - \\
\cite{Barry2016} & conventional$^\dagger$ & ensemble & $\sim 5000$ & - \\ 
\cite{Schloss2018} & conventional$^\dagger$ & ensemble & $\sim 5000$ & - \\ 
\bottomrule 
\end{tabular}
\caption[Pulsed readout parameters from the literature]{Example literature values for readout schemes employing conventional optical readout or alternative techniques. \textcolor{black}{The parameter $\sigma_R$ characterizes the \textcolor{black}{factor above the spin projection limit} and $\mathscr{N}$ is the average number of photons collected per measurement.} Conventional NV$^\text{-}$ readout is unable to reach the spin projection limit ($\sigma_R=1$), whereas alternative schemes can allow readout to approach this limit. The best demonstrated pulsed readout methods with ensembles are presently $\sim 100\times$ away from the spin projection limit. The symbol $^\dagger$ denotes non-pulsed schemes for comparison, and dashed lines (-) indicate values not reported (or not applicable to non-pulsed schemes).}
\label{tab:parameters} 
\end{table*}

\subsection{Alternatives to Ramsey magnetometry}\label{cwpulsed}
\subsubsection{CW-ODMR}\label{cwodmr}

Continuous-wave optically detected magnetic resonance (CW-ODMR) is a simple, widely employed magnetometry method~\cite{Fuchs2008,Acosta2009,Dreau2011,Schoenfeld2011,Tetienne2012,Barry2016,Schloss2018} wherein the MW driving and the optical polarization and readout occur simultaneously (see Fig.~\ref{fig:cwvsramsey}c). Laser excitation continuously polarizes NV$^\text{-}$ centers into the \textcolor{black}{more} fluorescent $m_s = 0$ ground state, while MWs tuned near resonance with one of the $m_s = 0 \leftrightarrow m_s = \pm 1$ transitions drive NV$^\text{-}$ population into the less fluorescent $m_s = \pm 1$ state (reducing the emitted light). A change in the local magnetic field shifts the resonance feature with respect to the MW drive frequency, causing a change in the detected fluorescence, as illustrated in Fig.~\ref{fig:cwvsramsey}d.

\textcolor{black}{In the simplest CW-ODMR implementation, the MW frequency is swept across the entire NV$^\text{-}$ resonance spectrum, allowing all resonance line centers to be determined. Alternatively, the MW frequency may be tuned to a specific resonance feature's maximal slope, so that incremental changes in magnetic field result in maximal changes in PL. The sensitivity of this latter approach can be further improved by modulating the MW frequency to combat noise or by exciting multiple hyperfine transitions simultaneously to improve contrast~\cite{Barry2016,Schloss2018,elella2017optimised}.}

\textcolor{black}{CW-ODMR does not require pulsed optical excitation, MW phase control, fast photodetectors, multichannel timing generators, or switches; the technique is therefore technically easier to implement than pulsed measurement schemes. Additionally, CW-ODMR is more tolerant of MW inhomogeneities \textcolor{black}{than pulsed schemes} and, when properly implemented, may yield similar sensitivities to pulsed magnetometry protocols when a larger number of sensors are interrogated with the same optical excitation power~\cite{Barry2016}.}

The shot-noise-limited sensitivity of an NV$^\text{-}$ magnetometer employing CW-ODMR is given by~\cite{Dreau2011,Barry2016}

\begin{equation}\label{eqn:CWODMR}
\eta_\text{CW}=\frac{4}{3\sqrt{3}}\frac{h}{g_e\mu_B}\frac{\Delta\nu}{C_\text{CW}\sqrt{R}},
\end{equation}
with photon detection rate $R$, linewidth $\Delta\nu$ and CW-ODMR contrast $C_\text{CW}$. The prefactor $4/(3\sqrt{3})$ originates from the steepest slope of the \textcolor{black}{resonance} lineshape when assuming a Lorentzian resonance profile \textcolor{black}{and is achieved for a detuning of $\frac{\Delta \nu}{2\sqrt{3}}$ from the linecenter~\cite{Vanier1989}.} \textcolor{black}{Operation of a CW-ODMR magnetometer can be modeled using the rate equation approach from Refs.~\cite{Dreau2011,jensen2013light}.}

However, CW-ODMR is not envisioned for many high-sensitivity applications for multiple reasons. First, CW-ODMR \textcolor{black}{precludes use of pulsed methods to improve sensitivity, such as double-quantum coherence magnetometry \textcolor{black}{(see Sec.~\ref{DQ})}, and many readout-fidelity enhancement techniques}. \textcolor{black}{In particular, the readout fidelity is quite poor compared to conventional pulsed readout schemes, as shown by the last two entries in Table~\ref{tab:parameters}.} Second, CW-ODMR methods suffer from MW and optical power broadening, degrading both $\Delta\nu$ and $C_\text{CW}$ compared to optimized Ramsey sequences. Optimal CW-ODMR sensitivity is achieved approximately when optical excitation, MW drive, and $T_2^*$ dephasing contribute roughly equally to the \textcolor{black}{resonance} linewidth~\cite{Dreau2011}. In this low-optical-intensity regime, the detected fluorescence rate per interrogated NV$^\text{-}$ center is significantly lower than for an optimized Ramsey scheme, which results in readout fidelities $\sim\!10^3$ below the spin projection limit~\cite{Barry2016}. This low optical intensity requirement becomes more stringent as $T_2^*$ increases, meaning that CW-ODMR sensitivity largely does not benefit from techniques to extend $T_2^*$.


\textcolor{black}{Overall, the combination of poor readout fidelity (and no proposed path toward improvement) combined with an inability to benefit from extended $T_2^*$ suggests that prospects are poor for further sensitivity enhancement over the best existing CW-ODMR devices~\cite{Barry2016,Schloss2018}. Moreover, \textcolor{black}{the poor readout fidelity accompanying the low required initialization intensity is} particularly deleterious to applications where volume-normalized sensitivity (i.e., the sensitivity within a unit interrogation volume) is important.}


\subsubsection{Pulsed ODMR}\label{pulsedodmr}

Pulsed ODMR is an alternative magnetometry method first demonstrated for NV$^\text{-}$ centers 
by Dr\'{e}au \textit{et al.}~in Ref.~\cite{Dreau2011}. 
Similar to Ramsey and in contrast to CW-ODMR, this technique avoids optical and MW power broadening of the spin resonances, enabling nearly $T_2^*$-limited measurements. \textcolor{black}{In contrast to Ramsey magnetometry, however, pulsed ODMR is linearly sensitive to spatial and temporal variations in MW Rabi frequency. When such variations are minimal, p}ulsed ODMR sensitivity may approach that of Ramsey magnetometry without requiring high Rabi frequency~\cite{Dreau2011}, making the method attractive when high MW field strengths are not available. 

In the pulsed ODMR protocol, depicted schematically in Fig.~\ref{fig:cwvsramsey}e, the NV$^\text{-}$ spin state is first optically initialized to $m_s =0$. Then, during the interrogation time $\tau$, a near-resonant MW $\pi$-pulse is applied with duration equal to the interrogation time, $\tau_\pi = \tau$, where the Rabi frequency $\Omega_R=\pi/\tau_\pi$. Finally, the population is read out optically. A change in the magnetic field detunes the spin resonance with respect to the MW frequency, resulting in an incomplete $\pi$-pulse and a change in the population transferred to the $m_s=\pm 1$ state prior to optical readout.

\begin{figure}[ht]
\centering
\hspace{-4pt}
\begin{overpic}[height=4 in]{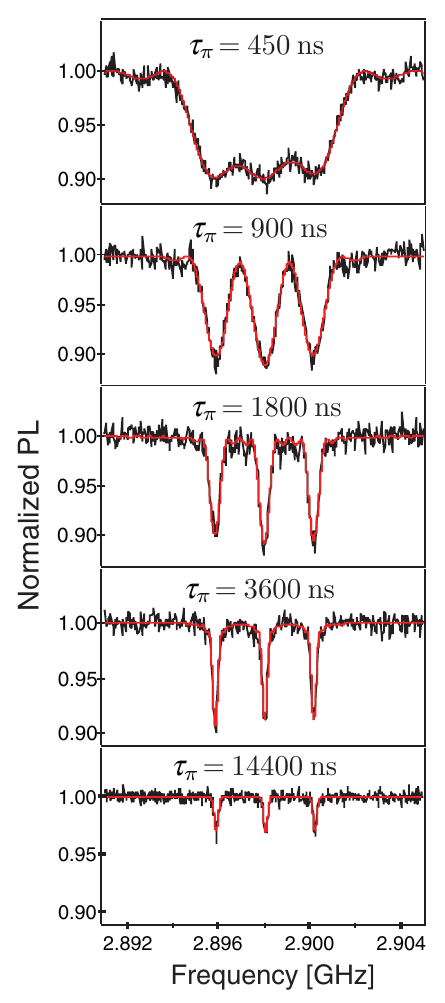}
\end{overpic}
\caption[Dreau pulsed odmr]{Pulsed ODMR spectra for various $\pi$-pulse durations $\tau_\pi$. When $\tau_\pi\ll T_2^*$, the \textcolor{black}{resonance} lineshape is Fourier-broadened beyond the \textcolor{black}{$T_2^*$-limited} linewidth. When  $\tau_\pi \gg T_2^*$ the photoluminescence (PL) contrast is diminished due to spin dephasing. Choice of $\tau_\pi\sim T_2^*$ ($\approx 3~\upmu$s here) allows nearly-$T_2^*$-limited linewidths while preserving PL contrast. From Ref.~\cite{Dreau2011}.}  \label{fig:dreau5b}
\end{figure}

For a Lorentzian \textcolor{black}{resonance} lineshape (see Appendices~\ref{linewidth} and \ref{EPR}), the expected shot-noise-limited sensitivity may be calculated starting from the shot-noise-limited CW-ODMR sensitivity given by Eqn.~\ref{eqn:CWODMR}. For pulsed ODMR, the resonance profile is given by a convolution of the $T_2^*$-limited line profile and additional broadening from the NV$^\text{-}$ spin's response to a fixed-duration, detuned MW $\pi$-pulse, as shown in Fig.~\ref{fig:dreau5b}. When the interrogation time $\tau_\pi$ is set to $\approx T_2^*$, these two broadening mechanisms contribute approximately equally to the resonance linewidth~\cite{Dreau2011}. Assuming $\tau_\pi \approx T_2^*$, we write the pulsed ODMR linewidth $\Delta \nu$ as $\Delta \nu \approx \Gamma = 1/(\pi T_2^*)$ (see Fig.~\ref{fig:cwvsramsey}f), while noting that this approximation likely underestimates the linewidth by $\lesssim 2\times$.

\begin{table*}[htbp]
  \centering
  \setlength{\tabcolsep}{3pt}
  \renewcommand{\arraystretch}{1.17}
    \begin{tabular}{rcp{40em}}
    \toprule
    \multicolumn{3}{c}{\textbf{Sensitivity optimization}} \\
    \midrule
    \multicolumn{1}{c}{\parbox{4.5em}{Parameter optimized}} & Method & \parbox{40em}{Method description and evaluation} \\
    \midrule
    \multicolumn{1}{p{4.5em}}{\multirow{18}[2]{4.5em}{Dephasing time $T_2^*$ }} & \parbox[t]{10.9em}{\raggedright Double-quantum coherence magnetometry (Sec.~\ref{DQ})} & Doubles effective gyromagnetic ratio. Removes dephasing from mechanisms inducing shifts common to the $|m_s=\pm1\rangle$ states, such as longitudinal strain and temperature. Minor additional MW hardware usually required. Generally recommended. \\[0.5ex]
          & \parbox[t]{10.9em}{\raggedright \textcolor{black}{Bias magnetic field (Sec.~\ref{efieldsuppression})}} & Suppresses dephasing from transverse electric fields and strain \textcolor{black}{at bias magnetic fields of several gauss or higher}. Generally recommended. \\[0.5ex]
          & \parbox[t]{10.9em}{\raggedright Spin bath driving (Sec.~\ref{P1driving})} & Mitigates or eliminates dephasing from paramagnetic impurities in diamond. Each impurity's spin resonance must be addressed, often with an individual RF frequency. Additional RF hardware is required. Recommended for many applications. \\[0.5ex]
          & \parbox[t]{10.9em}{\raggedright Dynamical decoupling (Sec.~\ref{DD})} & Refocuses spin dephasing using one or more MW $\pi$-pulses, extending the relevant relaxation time from $T_2^*$ to $T_2$, with fundamental limit set by $2 T_1$. Recommended for narrowband AC sensing; generally precludes DC or broadband magnetic sensing. \\[0.5ex]
          & \parbox[t]{10.9em}{\raggedright Rotary echo magnetometry (Sec.~\ref{rotaryecho})} & Extends measurement time \textcolor{black}{using a MW pulse scheme} but offers reduced sensitivity relative to Ramsey. Not recommended outside niche applications. \\[0.5ex]
          & \parbox[t]{10.9em}{\raggedright Geometric phase magnetometry (Sec.~\ref{geometricphase})} & Offers increased dynamic range\textcolor{black}{, using a MW spin manipulation method, at the cost of} reduced sensitivity relative to Ramsey. Not recommended outside niche applications. \\[0.5ex]
          & \parbox[t]{10.9em}{\raggedright Ancilla-assisted upconversion magnetometry (Sec.~\ref{upconversion})} & \textcolor{black}{Employs NV$^\text{-}$ hyperfine interaction to convert DC magnetic fields to AC fields to be sensed using dynamical decoupling. Operates} near ground-state level anticrossing ($10^3$ gauss) \textcolor{black}{and offers similar or reduced sensitivity relative to Ramsey.} Not generally recommended. \\[0.5ex]
    \midrule
    \multicolumn{1}{p{4.5em}}{\multirow{32}[4]{4.5em}{Readout fidelity $\mathcal{F}  = 1/ \sigma_R$}} & \parbox[t]{10.9em}{\raggedright Spin-to-charge conversion readout (Sec.~\ref{SCCR})} & Maps spin state to charge state of NV, increasing number of photons collected per measurement. Allows $\sigma_R \approx 3$ for single NV centers, and initial results show improvement over conventional readout for ensembles. Substantially increased readout time likely precludes application when $T_2^* \lesssim 3~\upmu$s. Requires increased laser complexity. Technique is considered promising; hence, further investigation is warranted. \\[0.5ex]
    & \parbox[t]{10.9em}{\raggedright Ancilla-assisted repetitive readout (Sec.~\ref{ancilla})} & Maps NV$^\text{-}$ electronic spin state to nuclear spin state, enabling repetitive readout and increased photon collection. Allows $\sigma_R$ to approach 1 for single NVs; no fundamental barriers to ensemble application. Substantially increased readout time likely precludes application when $T_2^* \lesssim 3~\upmu$s. Requires high magnetic field strength and homogeneity. Technique is considered promising, although further investigation is warranted. \\[0.5ex]
          & \parbox[t]{10.9em}{\raggedright Improved photon collection (Sec.~\ref{improvedcollection})} & Improves $\sigma_R$ by reducing fractional shot noise contribution, subject to unity collection and projection noise limits. Near-100\% collection efficiency is possible in principle, making this mainly an engineering endeavor. While many schemes are incompatible with wide-field imaging, the method is generally recommended for optical-based readout of single-channel bulk sensors. \\[0.5ex]
          & \parbox[t]{10.9em}{\raggedright NIR absorption readout (Sec.~\ref{NIRabsorption})} & \textcolor{black}{Probabilistically reads out} initial spin populations using optical absorption on the $^1$E$\;\leftrightarrow ^1$A$_1$ singlet transition. Demonstrated $\sigma_R$ values are on par with conventional ensemble readout, and prospects for further improvement are unknown. Technique is best used with dense ensembles and an optical cavity but is hindered by non-NV$^\text{-}$ absorption and non-radiative NV$^\text{-}$ singlet decay. Further investigation is warranted. \\[0.5ex]
          & \parbox[t]{10.9em}{\raggedright Photoelectric readout (Sec.~\ref{photoelectricreadout})} & Detects spin-dependent photoionization current. Best for small 2D ensembles; has not yet demonstrated sensitivity improvement with respect to optimized conventional readout. \\[0.5ex]
          & \parbox[t]{10.9em}{\raggedright Level-anticrossing-assisted readout (Sec.~\ref{LAC})} & Increases the number of spin-dependent photons collected per readout by operation at the excited-state level anticrossing. \textcolor{black}{Universally applicable, but} at best offers a $\sqrt{3}$ improvement in $\sigma_R$. Not recommended outside niche applications.    \\[0.5ex]
          & \parbox[t]{10.9em}{\raggedright Green absorption readout (Sec.~\ref{greenabsorption})} & \textcolor{black}{Probabilistically reads out} initial spin populations using optical absorption on the $^3$A$_2\leftrightarrow ^3$E triplet transition. Performs best with order unity optical depth. Demonstrations exhibit contrast below that of conventional readout by $3\times$ or more. Prospects are not considered promising.  \\[0.5ex]
          & \parbox[t]{10.9em}{\raggedright Laser threshold magnetometry (Sec.~\ref{laserthresholdmagnetometry})} & Probes magnetic field by measuring lasing threshold, which depends on NV$^\text{-}$ singlet state population. Moderately improved collection efficiency and contrast are predicted compared to conventional readout. Challenges include non-NV$^\text{-}$ absorption and system instability near lasing threshold. Prospects are not considered promising. \\[0.5ex]
          & \parbox[t]{10.9em}{\raggedright Entanglement-assisted magnetometry (Sec.~\ref{DRINQS})} & Harnesses strong NV$^\text{-}$ dipolar interactions to improve readout fidelity beyond the standard quantum limit. Existing proposals require 2D ensembles, impose long overhead times, and exhibit unfavorable coherence time scaling with number of entangled spins. While existing protocols are not considered promising, further investigation toward developing improved protocols is warranted. \\
    \bottomrule
    \end{tabular}%
        \caption{Summary analysis of approaches to optimize ensemble-NV-diamond magnetic sensitivity}
  \label{tab:summarysens}%
\end{table*}%

\begin{table*}[htbp]
\makebox[0pt][l]{%
  \raisebox{-13.00cm}[0pt][0pt]{%
    \vspace{5cm}\hspace{-8.16cm}\includegraphics[width=2.89 cm]{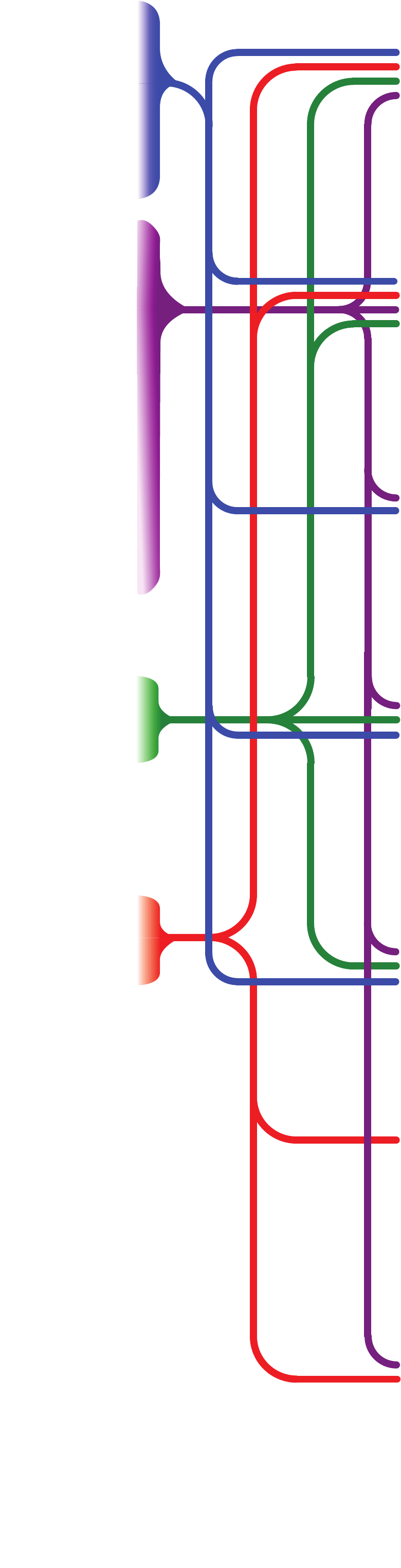}}}%
  \centering
  \setlength{\tabcolsep}{3pt}
    \begin{tabular}{lp{5.0em}lp{38.2em}}
    \toprule
    \multicolumn{4}{c}{\textbf{Diamond material optimization}} \\
    \midrule
    \multicolumn{1}{c}{\parbox{4.5em}{Parameter optimized}} & & Method & \parbox{38.2em}{Method description and evaluation} \\ [-1.2ex]
    \midrule
    \multicolumn{1}{l}{\parbox[t]{6.2em}{\raggedright \textcolor{black}{N-to-NV conversion efficiency $\chi$} (Sec.~\ref{Econvdef})}} &  & \parbox[t]{5.2em}{\raggedright CVD synthesis (Sec.~\ref{hpht})} & Common synthesis method \textcolor{black}{that} can produce high-quality ensemble-NV diamonds. Relatively easy to control dimensions and concentrations of electronic and nuclear spins. May introduce strain and unwanted impurities, which can limit achievable $\zeta$, $\chi$, and $T_2^*$. Effective for producing NV$^\text{-}$-rich-layer diamonds. \\[0.5ex]
    \multicolumn{1}{l}{\parbox[t]{6.2em}{\raggedright \textcolor{black}{NV-to-NV$^\text{-}$ charge state efficiency $\zeta$} (Sec.~\ref{Econvdef})}} &  & \parbox[t]{5.2em}{\raggedright HPHT synthesis (Sec.~\ref{hpht})} & Common synthesis method \textcolor{black}{that} can produce high-quality ensemble-NV diamonds with lower strain and fewer lattice defects than CVD. Control over doping and impurity concentration may be more difficult than in CVD. Not intrinsically amenable to creating NV$^\text{-}$-rich-layer diamonds. Ferromagnetic metals may incorporate into diamond. \\[0.5ex]
    \multicolumn{1}{l}{\parbox[t]{6.2em}{\raggedright \textcolor{black}{Paramagnetic impurities} (Sec.~\ref{chargetraps})}} &  & \parbox[t]{5.2em}{\raggedright Irradiation (Sec.~\ref{irradiation})} & Diamond treatment method that, combined with subsequent annealing, converts substitutional nitrogen to NV centers. Electrons are preferred irradiation particle. Dose should be optimized for diamond's nitrogen concentration to create high $\zeta$ without degrading $\chi$. Generally recommended with annealing for producing NV$^\text{-}$-rich diamonds. \\[0.5ex]
    \multicolumn{1}{l}{\parbox[t]{6.2em}{\raggedright \textcolor{black}{Strain~~~~~~~} (Sec.~\ref{T2*params})}} &  & \parbox[t]{5.2em}{\raggedright LPHT annealing (Sec.~\ref{LPHT})} & Low-pressure annealing that, combined with prior irradiation, converts substitutional nitrogen to NV centers. Heals some diamond lattice damage. NV$^\text{-}$ centers are created effectively at $\sim 800~^\circ$C; additional treatment at $\sim 1200~^\circ$C may eliminate some unwanted impurities. Generally recommended with irradiation for producing NV$^\text{-}$-rich diamonds. \\[0.5ex]
    \multicolumn{1}{l}{\parbox[t]{6.2em}{\raggedright \textcolor{black}{Nuclear spins} (Sec.~\ref{13ClimitT2star})}} &  & \parbox[t]{5.2em}{\raggedright HPHT treatment (Sec.~\ref{hpht})} & High-pressure annealing may reduce strain and eliminate some unwanted impurities. May enable increases in $\zeta$ and $\chi$. Recommended for diamonds with balanced aspect ratios. \\[0.5ex]
    & & \parbox[t]{5.2em}{\raggedright Isotopic enrichment (Sec.~\ref{13ClimitT2star})} & Diamond synthesis with isotopically enriched source (gas for CVD and typically solid for HPHT) allows reduction of unwanted nuclear spin concentration (e.g., $^{13}$C) and \textcolor{black}{selection} of nitrogen isotope ($^{14}$N or $^{15}$N) incorporated into NV$^\text{-}$.  CVD diamonds with [$^{13}$C]~$\approx 20$~ppm have been synthesized. Recommended for achieving long $T_2^*$. \\[0.5ex]
    & & \parbox[t]{5.2em}{\raggedright Surface treatment (Sec.~\ref{Econvdef})} & Surface termination \textcolor{black}{with favorable atomic elements} can stabilize the desired NV charge state near the surface and extend relaxation times. Generally recommended. \\
    & & \parbox[t]{5.2em}{\raggedright Preferential orientation (Sec.~\ref{preferentialorientation})} & \textcolor{black}{CVD synthesis of diamond with NV centers preferentially oriented along a single axis. At present, preferential orientation is only maintained in unirradiated diamonds, largely hindering its capability to produce NV$^\text{-}$-rich diamonds. Not generally recommended.} \\[0.5ex]
    \bottomrule
    \end{tabular}%
    \caption{Summary analysis of diamond engineering parameters and methods for high-sensitivity ensemble-NV$^\text{-}$ magnetometry. Colored lines indicate methods that may be employed to optimize each parameter. 
    }
  \label{tab:summarydiamond}%
\end{table*}%

Choosing initialization and readout times $t_I$ and $t_R$ and interrogation time $\tau_\pi = T_2^*$ reduces the time-averaged photon collection rate $R$ by the readout duty cycle $t_R/(t_I+T_2^*+t_R)$. Then, defining $\mathscr{N} = R t_R$ to be the mean number of photons collected per optical readout cycle and replacing $C_\text{CW}$ with the pulsed-ODMR contrast $C_\text{pulsed}$ yields the pulsed-ODMR sensitivity
\begin{equation}
\eta_\text{pulsed}\approx \frac{8}{3\sqrt{3}}\frac{\hbar}{g_e\mu_B}\frac{1}{C_\text{pulsed}\sqrt{\mathscr{N}}}\frac{\sqrt{t_I+T_2^*+t_R}}{T_2^*}.
\end{equation}

The value of $C_\text{pulsed}$ under optimized conditions is expected to be higher than $C_\text{CW}$ (for the same number of interrogated NV$^\text{-}$ centers and same mean photon collection rate $R$) because pulsed ODMR enables use of high optical intensities that would degrade $C_\text{CW}$~\cite{Dreau2011}. Although $C_\text{pulsed}$ may approach the Ramsey contrast $C_\text{Ramsey}$ (see Fig.~\ref{fig:cwvsramsey}a,b), $C_\text{pulsed}< C_\text{Ramsey}$ is expected in practice for several reasons: 
first, because the technique requires Rabi frequencies to be of the same order as the NV$^\text{-}$ linewidth \textcolor{black}{set by $T_2^*$}, the MW drive may be too weak to effectively address the entire inhomogeneously-broadened NV$^\text{-}$ ensemble. Second, while the high Rabi frequencies $\sim 2\pi \times 10$~MHz commonly employed in Ramsey sequences effectively drive all hyperfine-split NV$^\text{-}$ transitions of $^{14}$NV$^\text{-}$ or $^{15}$NV$^\text{-}$~\cite{Acosta2009}, the weaker $\pi$-pulses required for pulsed ODMR cannot effectively drive all hyperfine transitions with a single tone. Pulsed ODMR operation at the excited-state level anticrossing~\cite{Dreau2011} or utilizing multi-tone MW pulses~\cite{vandersypen2005nmr,Barry2016,elella2017optimised} could allow more effective driving of the entire NV$^\text{-}$ population and higher values of $C_\text{pulsed}$. However, when multi-tone pulses are employed, care should be taken to avoid degradation of $C_\text{pulsed}$ due to off-resonant MW cross-excitation, which may be especially pernicious when the \textcolor{black}{$T_2^*$-limited} linewidth (and thus MW Rabi frequency) is similar to the hyperfine splitting.  

Although pulsed ODMR may sometimes be preferable to Ramsey, the former technique ultimately provides inferior sensitivity. Several factors of order $\sqrt{2}$ (which arise from a lineshape-dependent numerical prefactor~\cite{Dreau2011}, MW Fourier broadening,  nonuniform ensemble driving, and hyperfine driving inefficiencies) combine to degrade the pulsed ODMR sensitivity with respect to that of Ramsey. Furthermore, unlike double-quantum Ramsey magnetometry (see Sec.~\ref{DQ}), pulsed ODMR has not been experimentally demonstrated to mitigate line broadening from temperature fluctuations or other dephasing mechanisms \textcolor{black}{common-mode to $|m_s \!=\! -1\rangle$ and $|m_s \!=\! +1\rangle$}. Hypothetical double-quantum analogs to pulsed ODMR~\cite{Taylor2008,Fang2013} might likely require, in addition to the sensing $\pi$-pulse, high-Rabi-frequency MW pulses to initialize the $|\pm1\rangle$ superposition states, similar to those employed for double-quantum Ramsey, which would undermine pulsed ODMR's attractive low MW Rabi frequency requirements. 

\textcolor{black}{A generalization of pulsed ODMR is Rabi beat sensing~\cite{rabi1937space,fedder2011towards}, wherein the spins are driven through multiple Rabi oscillations during the interrogation time. Under optimal conditions, Rabi beat magnetometry, like the specific case of pulsed ODMR, may exhibit sensitivity approaching that of Ramsey magnetometry. For the regime where the Rabi frequency $\Omega_R$ is large compared to the \textcolor{black}{resonance} linewidth ($\sim 1/T_2^*$), 
sensitivity is optimized when the detuning is chosen to be similar to the Rabi frequency ($\Delta \sim \Omega_R$), when the interrogation time is similar to the dephasing time ($\tau \sim T_2^*$, see Appendix~\ref{optimalprecession}), and when 
$\tau$ is chosen to ensure operation at a point of maximum slope of the Rabi magnetometry curve. However, Rabi beat magnetometry is sensitive to spatial and temporal variations in the MW Rabi frequency $\Omega_R$~\cite{ramsey1950molecular}. For high values of $\Omega_R$, MW field variations may limit the Rabi measurement's effective $T_2^*$. Hence, practical implementations of Rabi beat magnetometry on NV$^\text{-}$ ensembles likely perform best when $\Omega_R \sim 1/T_2^*$, i.e., when the scheme reduces to pulsed ODMR.} 

\subsection{Parameters limiting sensitivity}\label{physicallimits}


Examination of Eqn.~\ref{eqn:ramseyshotexact} reveals the relevant parameters limiting magnetic field sensitivity \textcolor{black}{$\eta_\text{Ramsey}^\text{ensemble}$}: (i) the dephasing time  $T_2^*$; (ii) the readout fidelity $\mathcal{F}=1/\sigma_R$; (iii) the sensor density [NV$^\text{-}$] and the interrogated diamond volume $V$, which together set the total number of sensors $N=\,[\text{NV}^\text{-}]\times V$; (iv) the measurement overhead time $t_O = t_I+t_R$; and (v) the relative precession rates of the two states comprising the interferometry measurement. Sensitivity enhancement requires improving one or more of these parameters. As we will discuss, parameters (i) and (ii) are particularly far from physical limits and therefore warrant special focus. 
{\renewcommand\labelitemi{}
\begin{itemize}
\item \textbf{(i) }\textbf{ Dephasing Time }$\boldsymbol{T_2^*}$\textbf{ |} In current realizations, dephasing times in application-focused broadband NV$^\text{-}$ ensemble magnetometers~\cite{Barry2016,Chatzidrosos2017,Kucsko2013,Clevenson2015} are typically $T_2^*\!\lesssim\! 1~\upmu$s. Considering the physical limit $T_2^*\!\leq\! 2T_1$~\cite{Levitt2008,Jarmola2012,Bauch2019,alsid2019photoluminescence}, with longitudinal relaxation time $T_1\!\approx\! 6$~ms for NV$^\text{-}$ ensembles~\cite{Jarmola2012}, a maximum $T_2^*\!\approx\! 12$~ms is theoretically achievable, corresponding to a sensitivity enhancement of $\approx\!100\times$. Although the feasibility of realizing $T_2^*$ values approaching $2T_1$ remains unknown, we consider improvement of $T_2^*$ to be an effective approach to enhancing sensitivity (see Sec.~\ref{T2*improvement}). \textcolor{black}{While the stretched exponential parameter $p$ can provide information regarding the dephasing source limiting $T_2^*$, its value (typically between 1 and 2 for ensembles) does not strongly affect achievable sensitivity~\cite{Bauch2018}.}
\item \textbf{(ii) Readout Fidelity |} Increasing readout fidelity \textcolor{black}{$\mathcal{F}=1/\sigma_R$} is another effective method to enhance sensitivity, as fractional fidelity improvements result in equal fractional improvements in sensitivity. With \textcolor{black}{conventional} $532$~nm fluorescence readout, current NV$^\text{-}$ ensemble readout fidelities $\mathcal{F}$ are a factor \textcolor{black}{$\gtrsim\!67\times$} removed from the spin projection limit \textcolor{black}{$\sigma_R=1$~\cite{LeSage2012}}, indicating large improvements might be possible. \textcolor{black}{For comparison, multiple readout methods employing single NV$^\text{-}$ centers achieve $\mathcal{F}$ within 5$\times$ of the spin projection limit, i.e., $\sigma_R < 5$~\cite{Lovchinsky2016,Shields2015,jaskula2017improved, Hopper2016, Ariyaratne2018, hopper2018spin} with Ref.~\cite{Neumann2010} achieving $\sigma_R=1.1$.}
\end{itemize}
}

In contrast, we believe prospects are modest for improving sensitivity by engineering parameters (iii), (iv), and (v).
{\renewcommand\labelitemi{}
\begin{itemize}
\item \textbf{(iii) Sensor Number, Density, or Interrogation Volume |} In theory, the number of sensors $N$ can be increased without limit. However, practical considerations \textcolor{black}{may} prevent this approach. \textcolor{black}{First, a larger value of $N$ (and an associated larger number of photons $\mathscr{N}$) can increase some types of technical noise that scale as $N$, e.g., noise from timing jitter in device electronics or from excitation-laser intensity fluctuations. As photon shot noise scales more slowly as $\sqrt{N}$, achieving a shot-noise-limited sensitivity becomes more difficult with increasing $N$.} 
Second, large values of $N$ can require impractically high laser powers, since the \textcolor{black}{number of photons needed for NV$^\text{-}$ spin initialization} scales linearly with $N$. While larger $N$ can be achieved either by increasing the NV$^\text{-}$ density or increasing the interrogation volume, both approaches \textcolor{black}{result in} distinct \textcolor{black}{technical or fundamental} difficulties. Increasing $N$ by increasing the interrogation volume with fixed [NV$^\text{-}$] \textcolor{black}{may increase the diamond cost and creates more stringent uniformity requirements for both the bias magnetic field (to avoid degrading the dephasing time $T_2^*$) and the MW field (to ensure uniform spin manipulation over the sensing volume).} Furthermore, \textcolor{black}{increasing interrogation volume} is incompatible with high-spatial-resolution sensing and imaging modalities~\cite{LeSage2013,Pham2011,Steinert2010,Glenn2015,Simpson2016,Tetienne2017,Glenn2017,Fu2014}. \textcolor{black}{On the other hand}, increasing NV$^\text{-}$ density will increase dephasing from dipolar coupling and decrease $T_2^*$ unless such effects are mitigated \textcolor{black}{ (see, e.g., Sec.~\ref{P1driving})}. Finally, because sensitivity scales as $1/\sqrt{N}$, \textcolor{black}{we expect} increasing $N$ to allow only modest enhancements \textcolor{black}{(e.g., $\lesssim 5\times$)} over standard methods. To date no demonstrated high sensitivity bulk NV-diamond magnetometer~\cite{Barry2016,Wolf2015,Chatzidrosos2017,Clevenson2015} has utilized more than a few percent of the available NV$^\text{-}$ in the diamond, suggesting limited utility for increasing sensor number $N$ in current devices. See Appendix~\ref{volumedensityconsiderations} for additional analysis.
\item \textbf{(iv) Overhead Time |} Although measurement overhead time can \textcolor{black}{likely} be decreased to $\sim 1~\upmu$s, maximum sensitivity enhancement \textcolor{black}{(in the regime where $T_2^*\sim  t_I+t_R$)} is expected to be limited to order unity, \textcolor{black}{($\lesssim 3\times$)}. See Sec.~\ref{T2*improvement} for a more detailed discussion.
\item \textbf{(v) Precession Rate |} \textcolor{black}{Use of the NV$^\text{-}$ center's full $S=1$ spin can allow $\Delta m_s = 2$ in Eqns.~\ref{eqn:ramseyshotexact} and \ref{eqn:ramseyshot}, i.e., a $2\times$ increase in the relative precession rate of the states \textcolor{black}{employed}  
compared to use of the standard $S = \nicefrac{1}{2}$\textcolor{black}{-equivalent subspace} (see Sec.~\ref{DQ})~\cite{Fang2013,Bauch2018}. However, further improvement is} unlikely, as the NV$^\text{-}$ spin dynamics are fixed.
\end{itemize}
}

We note that the derivation of Eqn.~\ref{eqn:ramseyshotexact} makes certain assumptions (\textcolor{black}{in particular, independence of the $N$ sensors}) that \textcolor{black}{do not apply to some exotic} approaches, such as exploiting strong NV$^\text{-}$-NV$^\text{-}$ interactions via Floquet techniques and harnessing entanglement for sensing (see Sec.~\ref{DRINQS})~\cite{choi2017quantum}.

\textcolor{black}{Table~\ref{tab:summarysens} summarizes our analysis of present and proposed techniques to optimize ensemble-NV$^\text{-}$ magnetic field sensitivity. Table~\ref{tab:summarydiamond} summarizes our review of engineering methods for producing optimized diamond samples for high-sensitivity ensemble-NV$^\text{-}$ magnetometry.}



%% file: sec02.tex
\section{Limits to relaxation times $T_2^*$ and $T_2$}\label{T2starlimits}

%
%


\subsection{Motivation to extend ${T_2^*}$}\label{T2*improvement}

\begin{figure}
\begin{overpic}[width = \columnwidth]{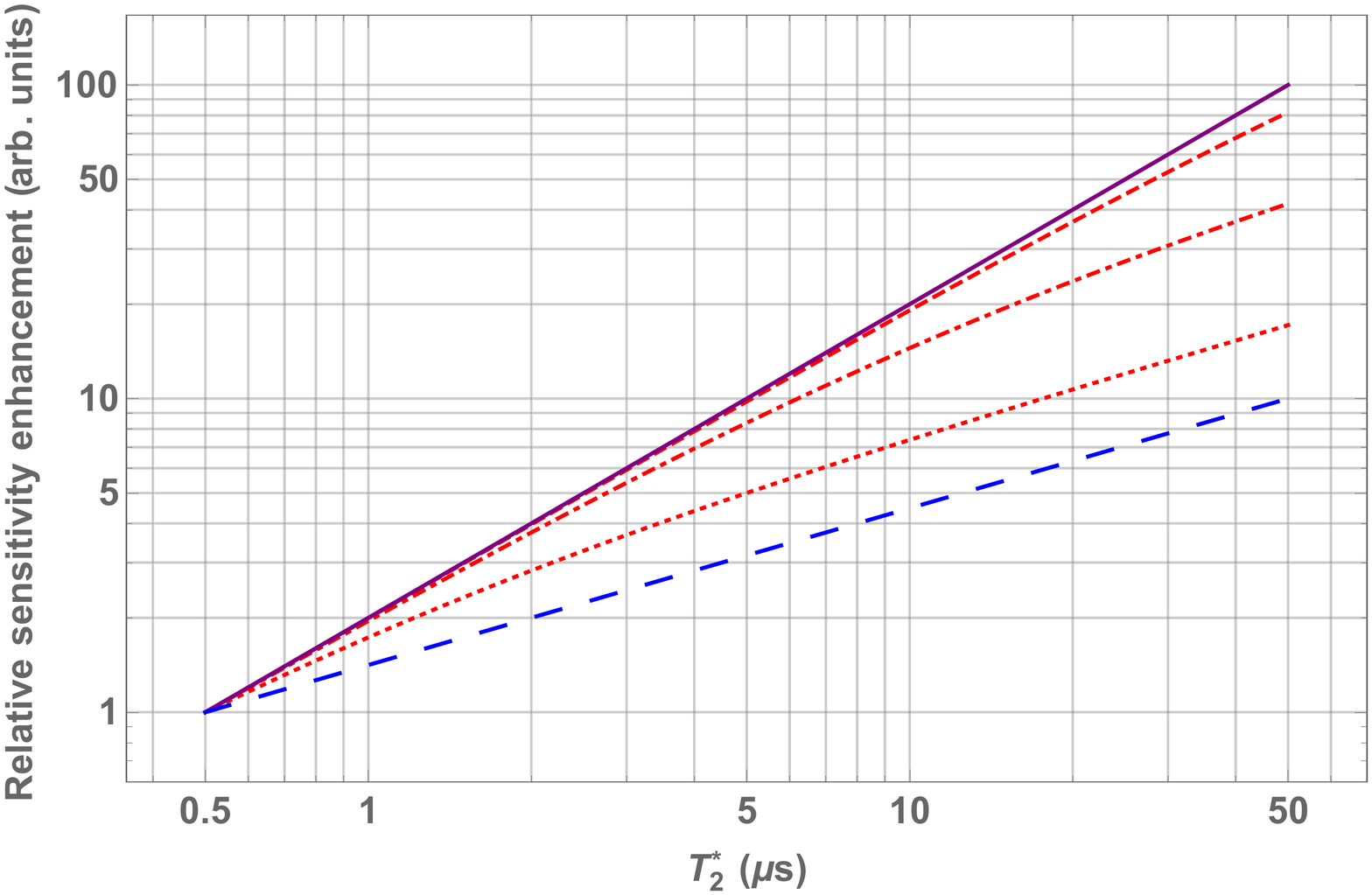}
\end{overpic}
\caption[Sensitivity scaling with $T_2^*$]{Sensitivity enhancement scaling with dephasing time $T_2^*$ for a Ramsey-type magnetometer normalized to the same device with $T_2^*=500$~ns. The different curves assume overhead times ($t_O = t_I+t_R$) of $1~\upmu$s ({\color{red}{\rule[.6mm]{.5mm}{.3mm}\;\rule[.6mm]{.5mm}{.3mm}\;\rule[.6mm]{.5mm}{.3mm}}}), $10~\upmu$s ({\color{red}{\rule[.6mm]{1mm}{.3mm}\;\rule[.6mm]{.5mm}{.3mm}\;\rule[.6mm]{1mm}{.3mm}}}), and $100~\upmu$s ({\color{red}{\rule[.6mm]{1mm}{.3mm}\;\rule[.6mm]{1mm}{.3mm}\;\rule[.6mm]{1mm}{.3mm}}}). The sensitivity enhancement is bounded \textcolor{black}{from above by} the fractional $T_2^*$ improvement ({\color{purple}{\rule[.6mm]{3mm}{.3mm}}}) \textcolor{black}{and from below by} the square root ({\color{blue}{\rule[.6mm]{1.5mm}{.3mm}\;\;\rule[.6mm]{1.5mm}{.3mm}\;\;\rule[.6mm]{1.5mm}{.3mm}}}) of the fractional $T_2^*$ improvement. For simplicity the precession time $\tau$ is set to $T_2^*$. See Appendix~\ref{optimalprecession} for details on determining the optimal precession time.}  \label{fig:sensenhancement}
\end{figure}

A promising approach to enhance DC sensitivity focuses on extending the dephasing time $T_2^*$~\cite{Bauch2018}. The effectiveness of this approach may be illustrated by close examination of Eqns.~\ref{eqn:ramseyshotexact},\textcolor{black}{~\ref{eqn:ramseyshot}}. First, optimal sensitivity is obtained when the precession time $\tau$ is similar to the dephasing time $T_2^*$ (see Appendix~\ref{optimalprecession}), so that the approximation~$\tau\sim T_2^*$ is valid for an optimized system. Therefore, for the simple arguments presented in this section, we assume that $T_2^*$ extensions translate to proportional extensions of the optimal $\tau$. When the dephasing time $T_2^*$ is similar to or shorter than the measurement overhead time ($T_2^*\lesssim t_O\equiv t_I + t_R$), which may be typical for Ramsey magnetometers employing ensembles of NV$^\text{-}$ centers in \textcolor{black}{diamonds with total nitrogen concentration} [N$^\text{T}$] = 1-20~ppm, the sensitivity enhancement may then be nearly linear in $T_2^*$, as shown in Fig.~\ref{fig:sensenhancement}.


The above outlined sensitivity scaling can be intuitively understood as follows: when the free precession time is small relative to the overhead time, i.e., $\tau \sim T_2^* \ll t_O$, doubling $T_2^*$ (thus doubling $\tau$) results in twice the phase accumulation per measurement sequence and only a slight increase in the total sequence duration; in this limit, magnetometer sensitivity is enhanced by nearly $2\times$. This favorable sensitivity scaling positions $T_2^*$ as an important parameter to optimize \textcolor{black}{when $T_2^*\lesssim t_O$.} 

Typical NV$^\text{-}$ ensemble $T_2^*$ values are $\sim 500$~ns in \textcolor{black}{$[\text{N}^\text{T}]\approx 20$\,ppm} chemical-vapor-deposition-grown diamonds from Element Six, \textcolor{black}{a popular supplier of scientific diamonds}. Even when employing extraordinarily optimistic values of $t_I = 1~\upmu$s and $t_R = 300$~ns in Ramsey sequences performed on such ensembles, only roughly one quarter of the total measurement time is allocated to free precession. In this regime, as discussed above, the sensitivity scales as $\sim 1/T_2^*$. Although values of $t_I$ and $t_R$ vary in the literature (see Table~\ref{tab:overheadtime}), the use of longer $t_I$ and $t_R$ \textcolor{black}{may be desired to} achieve better spin polarization and higher readout fidelity. Notably, initialization times are typically longer for NV$^\text{-}$ ensembles than for single NV$^\text{-}$ defects, as \textcolor{black}{higher optical excitation power is required to achieve the NV$^\text{-}$ saturation intensity over spatially-extended ensembles, and, furthermore, non-uniformity in optical intensity (e.g., from a Gaussian illumination profile)} can be compensated for by increasing the initialization time~\cite{Wolf2015}.

Longer dephasing times $T_2^*$ offer additional benefits beyond direct sensitivity improvement. For example, higher $T_2^*$ values may relax certain technical requirements \textcolor{black}{by allowing lower duty cycles for specific experimental protocol steps}. In a standard Ramsey-type experiment, the optical initialization and optical readout each occur once per measurement sequence. Assuming a fixed mean number of photons are required \textcolor{black}{for spin polarization} and \textcolor{black}{and for read out of} the NV$^\text{-}$ ensemble, the time-averaged optical power and resulting heat load are expected to scale as $1/T_2^*$. Reducing heat loads is prudent for minimizing temperature variation of the diamond, \textcolor{black}{which shifts the energy splitting between $|m_s\!=\!0\rangle$ and $|m_s\!=\! \pm 1\rangle$} and may require correction (see Sec.~\ref{DQ}). Minimizing heat load is also important for many NV-diamond sensing applications, particularly in the life sciences. Assuming a fixed overhead time $t_O$, the realization of higher values of $T_2^*$, and thus $\tau$, necessitates processing fewer photons per unit time, which may relax design requirements for the photodetector front end and associated electronics~\cite{Hobbs2011}. 

Extended $T_2^*$ times can provide similar benefits to the MW-related aspects of the measurement. A standard Ramsey-type measurement protocol employs a MW $\pi/2$-pulse before and after every free precession interval. If the length of each $\pi/2$-pulse is held fixed, the time-averaged MW power and resulting heat load will scale as $1/T_2^*$. Additionally, higher $T_2^*$ values can allow for more \textcolor{black}{sophisticated}, longer-duration MW pulse sequences, in place of simple $\pi/2$-pulses, to mitigate the effects of Rabi frequency inhomogeneities~\cite{Angerer2015, Nobauer2015,vandersypen2005nmr} or allow for other spin-manipulation protocols. Finally, higher $T_2^*$ values could make exotic readout schemes that tend to have fixed time penalties \textcolor{black}{attractive}, such as spin-to-charge conversion readout~\cite{Shields2015} (see Sec.~\ref{SCCR}) and ancilla-assisted repetitive readout~\cite{Jiang2009,Lovchinsky2016} (see Sec.~\ref{ancilla}).


%

\begin{table}[ht]
\centering 
\begin{tabular}{c c r r} 
\toprule
Reference & No.~NV$^\text{-}$ probed & \parbox[t]{3em}{$t_I$} & \parbox[t]{3em}{$t_R$} \\ [0.5ex] 
\midrule
\cite{Shields2015} & single & 150~ns & \parbox[t]{3em}{-} \\ 
\cite{DeLange2012} & single & 600~ns & 600~ns \\
\cite{Hopper2016} & single & 1~$\upmu$s & 200~ns \\
\cite{Fang2013} & single & 2~$\upmu$s & 300~ns \\
\cite{Maze2008} & single & 2~$\upmu$s & 324~ns \\
\cite{Neumann2009} & single & 3~$\upmu$s & \parbox[t]{3em}{-} \\
\cite{LeSage2012} & ensemble & 600~ns & 300~ns \\
\cite{Bauch2018} & ensemble & 20~$\upmu$s & \parbox[t]{3em}{-} \\
\cite{Wolf2015} & ensemble & 100~$\upmu$s & 10 $\upmu$s \\
\cite{Mrozek2015} & ensemble & 1~ms & \parbox[t]{3em}{-} \\
\cite{Jarmola2012} & ensemble & 1~ms & \parbox[t]{3em}{-} \\
\bottomrule 
\end{tabular}
\caption[Initialization and readout times from the literature]{Initialization and readout times in the literature used for \textcolor{black}{conventional} optical readout of NV$^\text{-}$ defects. In general, NV$^\text{-}$ ensembles require longer initialization times than single NV$^\text{-}$ defects, in part \textcolor{black}{due to} the often non-uniform optical excitation intensity applied to the ensemble~\cite{Wolf2015}. Dashed lines (-) indicate values not reported.}
\label{tab:overheadtime} 
\end{table}



\subsection{Ensemble and single-spin $T_2^*$}\label{T2*ensemble} 

As discussed \textcolor{black}{above}, the dephasing time $T_2^*$ is a critical parameter for broadband DC magnetometry. Importantly, $T_2^*$ is defined differently for a single spin than for a spin ensemble. While an ensemble's $T_2^*$ characterizes relative dephasing of the constituent spins, a single spin's $T_2^*$ characterizes dephasing of the spin with itself, i.e., the distribution of phase accumulation from repeated measurements on the spin over time~\cite{deSousa2009,Ishikawa2012}. Since this work focuses on ensemble-based sensing, single-spin dephasing times are herein denoted ${T_2^*}^\text{\{single\}}$, while the term $T_2^*$ is reserved for ensemble dephasing times.

\begin{figure}[ht]
\begin{overpic}[height=1.45 in]{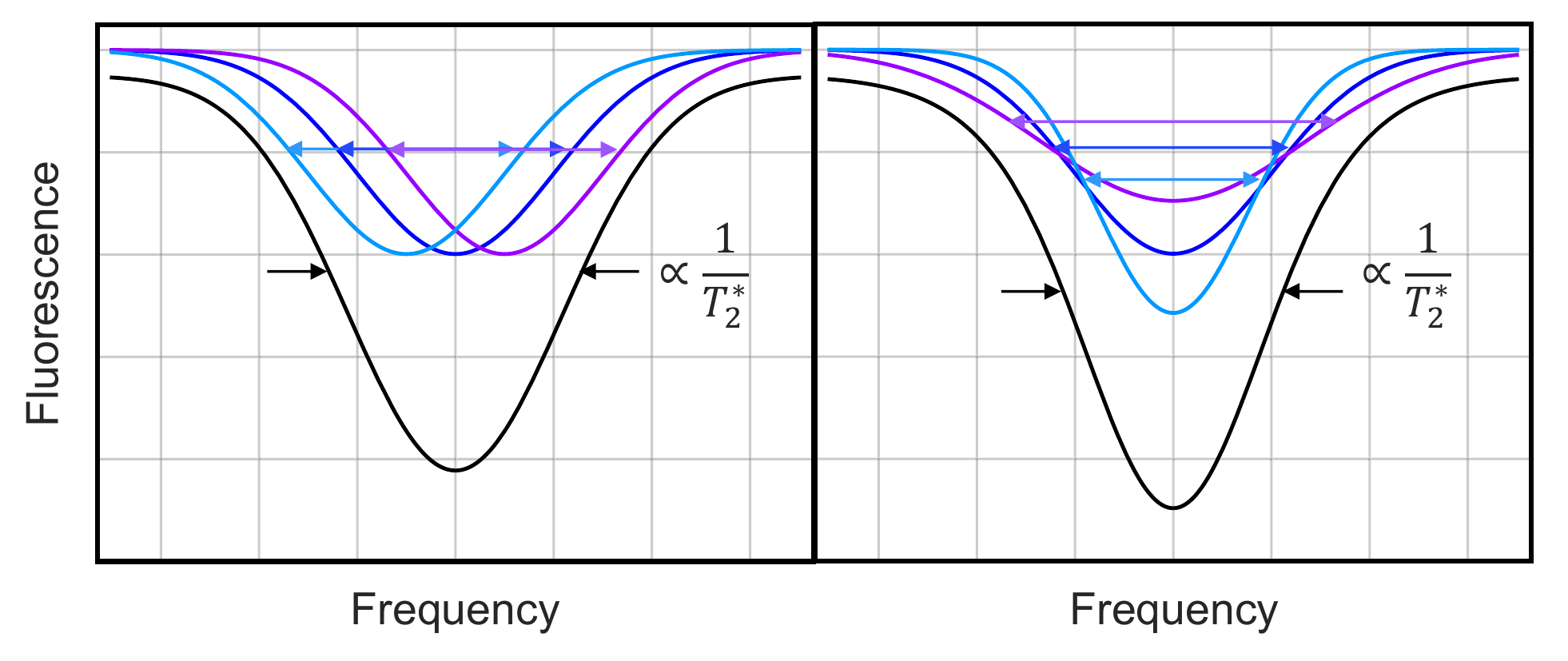}
\end{overpic}
\caption[Ensemble and single spin $T_2^*$]{Contributions of individual spin resonances to the overall spin ensemble lineshape. The ensemble resonance lineshape ({\color{black}{\rule[.5mm]{3mm}{1pt}}}) is broadened both by the distribution of line centers (left) and the distribution of linewidths (right) of the constituent spins ({\color[rgb]{0,.69,.94}{\rule[.5mm]{3mm}{1pt}}},\,{\color{blue}{\rule[.5mm]{3mm}{1pt}}},\,{\color[rgb]{.62,.33,.99}{\rule[.5mm]{3mm}{1pt}}}).}  \label{fig:t2starensemble}
\end{figure}


Values of ${T_2^*}^\text{\{single\}}$ are affected by slow magnetic, electric, strain, and temperature fluctuations. Variations in the magnetic environment may arise from dipolar interactions with an electronic or nuclear spin bath. The strength of these fluctuations can vary spatially throughout a sample due to the microscopically nonuniform distribution of bath spins. As a result, different NV$^\text{-}$ centers in the same sample display different ${T_2^*}^\text{\{single\}}$ values~\cite{Dobrovitski2008, Hanson2006, Hanson2008, Ishikawa2012}. For example, an NV$^\text{-}$ spin in close proximity to several bath spins will experience faster dephasing than an NV$^\text{-}$ spin many lattice sites away from the nearest bath spin.

Although ensemble $T_2^*$ values are also influenced by spin-bath fluctuations, as discussed in Secs.~\ref{nitrogenlimitT2*} and \ref{13ClimitT2star}, an ensemble $T_2^*$ value is not equal to the most common value of ${T_2^*}^\text{\{single\}}$ within the ensemble. 
For one, the ensemble value is limited by sources of zero-frequency noise that do not contribute to ${T_2^*}^\text{\{single\}}$, such as spatially inhomogeneous magnetic fields, electric fields, strain, or g-factors~\cite{deSousa2009}. These inhomogeneities cause a spatially-dependent distribution of the single-NV$^\text{-}$ resonance line centers, which broadens the ensemble resonance line and thus degrades $T_2^*$. Figure~\ref{fig:t2starensemble} depicts broadening contributions to $T_2^*$ from both varying single-NV$^\text{-}$ line centers and \textcolor{black}{varying} single-NV$^\text{-}$ linewidths ($\propto 1/{T_2^*}^\text{\{single\}}$). The relative contribution to an ensemble's $T_2^*$ value from these two types of broadening is expected to be sample-dependent. Although measurements in Ref.~\cite{Ishikawa2012} on a collection of single NV$^\text{-}$ centers in a sparse sample found the distribution of single-NV$^\text{-}$ line centers to be narrower than the median single-NV$^\text{-}$ linewidth, such findings are not expected to hold generally \textcolor{black}{(e.g., due to strain)}. 

\textcolor{black}{Even} in the absence of static field inhomogeneities, the spin-bath-noise-limited $T_2^*$ value of an ensemble is expected to be shorter than the most likely ${T_2^*}^\text{\{single\}}$ value, as the ensemble value is strongly influenced by the small minority of NV$^\text{-}$ centers with bath spins on nearby lattice sites~\cite{Dobrovitski2008}. In fact, theoretical calculations in Refs.~\cite{Dobrovitski2008, Hall2014} reveal that single spins and ensembles interacting with surrounding spin baths each exhibit free-induction-decay (FID) envelopes with different functional forms (see Appendices~\ref{linewidth} and \ref{FIDexponent}), a result borne out by experiments~\cite{Maze2012, Bauch2019, Bauch2018}. In general, the ensemble $T_2^*$ value cannot be predicted from ${T_2^*}^\text{\{single\}}$ of any constituent spin~\cite{Dobrovitski2008}, and application of single-spin measurements or theory to ensembles, or vice versa, should be done with great care.

\subsection{Dephasing mechanisms}\label{T2*params}

\textcolor{black}{The various contributions to an NV$^\text{-}$ ensemble's spin dephasing time $T_2^*$ can be expressed schematically} as
\begin{multline}\label{eqn:t2*depend}
\frac{1}{T_2^*} \approx \frac{1}{T_2^*\text{\{electronic spin bath\}}}+ \frac{1}{T_2^*\{\text{nuclear spin bath}\}}\\
+\frac{1}{T_2^*\{\text{strain gradients}\}}
+\frac{1}{T_2^*\text{\{electric field noise\}}} \\
+\frac{1}{T_2^*\{\text{magnetic field gradients}\}}
+\frac{1}{T_2^*\text{\{temperature variation\}}}\\
+\frac{1}{T_2^*\text{\{unknown\}}}+\frac{1}{2T_1},
\end{multline}
where the symbol notation $T_2^*$\{X\} denotes the \textcolor{black}{hypothetical} limit to $T_2^*$ solely due to mechanism X \textcolor{black}{(absent all other interactions or mechanisms)}. Equation~\ref{eqn:t2*depend} assumes that all mechanisms are independent and that associated dephasing rates add linearly. The second assumption is strictly only valid when all dephasing mechanisms lead to single-exponential free-induction-decay envelopes (i.e., Lorentzian lineshapes); see Appendices~\ref{linewidth}, \ref{EPR}, and \ref{FIDexponent}. Here we briefly discuss each of these contributions to NV$^\text{-}$ ensemble dephasing, and in later sections we examine their scalings, and how each mechanism may be mitigated. 

The electronic spin bath consists of paramagnetic impurity defects in the diamond lattice, which couple to NV$^\text{-}$ spins via magnetic dipolar interactions. The inhomogeneous spatial distribution and random instantaneous orientation of these bath spins cause dephasing of the NV$^\text{-}$ spin ensemble~\cite{Hanson2008, Dobrovitski2008,Bauch2018, Bauch2019}. Electronic spin bath dephasing can be broken down into contributions from individual constituent defect populations,
\begin{multline}\label{eqn:t2*electronic}
\frac{1}{T_2^*\{\text{electronic spin bath}\}} = \frac{1}{T_2^*\{\text{N}_\text{S}^0\}}+\frac{1}{T_2^*\{\text{NV}^\text{-}\}}\\
+\frac{1}{T_2^*\{\text{NV}^\text{0}\}} 
+ \frac{1}{T_2^*\{\text{other electronic spins}\}}.
\end{multline}
Here $T_2^*\{\text{N}_\text{S}^0\}$ denotes the $T_2^*$ limit from dephasing by paramagnetic substitutional nitrogen defects N$_\text{S}^0$ ($S=\nicefrac{1}{2}$), also called P1 centers, with concentration [N$_\text{S}^0$]~\cite{Smith1959,Cook1966,Loubser1978}. As substitutional nitrogen is a necessary ingredient for creation of NV$^\text{-}$ centers, N$_\text{S}^0$ defects typically persist at concentrations similar to or exceeding NV$^\text{-}$ (and NV$^0$) concentrations and may account for the majority of electronic spin bath dephasing~\cite{Bauch2018}. Sec.~\ref{nitrogenlimitT2*} examines $T_2^*\{\text{N}_\text{S}^0\}$ scaling with [N$_\text{S}^0$]. For NV-rich diamonds, dipolar interactions among NV$^\text{-}$ spins may also cause dephasing of the ensemble, with associated limit $T_2^*\{\text{NV}^\text{-}\}$. Sec.~\ref{NVNVlimit} examines the $T_2^*\{\text{NV}^\text{-}\}$ scaling with [NV$^\text{-}$] and other experimental parameters. In NV-rich diamonds, the neutral charge state NV$^0$  ($S=\nicefrac{1}{2})$ is also present at concentrations similar to [NV$^\text{-}$]~\cite{Hartlandthesis2014} and may also contribute to dephasing, with limit $T_2^*\{\text{NV}^\text{0}\}$. The quantity $T_2^*\{\text{other electronic spins}\}$ encompasses dephasing from the remaining defects in the electronic spin bath, such as negatively charged single vacancies~\cite{Baranov2017}, vacancy clusters~\cite{Twitchen1999b, Iakoubovskii2002} and hydrogen-containing defects~\cite{Edmonds2012}.


The quantity $T_2^*\{\text{nuclear spin bath}\}$ in Eqn.~\ref{eqn:t2*depend} describes NV$^\text{-}$ ensemble dephasing from nuclear spins in the diamond lattice. In samples with natural isotopic abundance of carbon, the dominant contributor to nuclear spin bath dephasing is the $^{13}$C isotope ($I=\nicefrac{1}{2}$), with concentration $[^{13}\text{C}] = 10700 \pm 800$~ppm~\cite{Wieser2013}, so that $T_2^*\{\text{nuclear spin bath}\} \approx T_2^*\{^{13}\text{C}\}$~\cite{Dreau2012,Balasubramanian2009,Zhao2012,Hall2014}. Other nuclear spin impurities exist at much lower concentrations and thus have a negligible effect on dephasing. The $T_2^*\{^{13}\text{C}\}$ scaling with concentration [$^{13}$C] is discussed in Sec.~\ref{13ClimitT2star} and can be minimized through isotope engineering~\cite{Balasubramanian2009,Teraji2013}. 

Another major source of NV$^\text{-}$ ensemble dephasing is non-uniform strain across the diamond lattice. Because strain shifts the NV$^\text{-}$ spin resonances~\cite{Dolde2011,Jamonneau2016,Trusheim2016}, gradients and \textcolor{black}{other} inhomogeneities in strain may dephase the ensemble, limiting $T_2^*$. Strain may vary by more than an order of magnitude within a diamond sample~\cite{Bauch2018}, and can depend on myriad diamond synthesis parameters~\cite{Gaukroger2008,Hoa2014}.  For a given NV$^\text{-}$ orientation along any of the [111] diamond crystal axes, strain couples to the NV$^\text{-}$ Hamiltonian approximately in the same way as an electric field (though with a different coupling strength)~\cite{Dolde2011,Doherty2012, Barson2017} (see Appendix~\ref{StarkZeeman} for further discussion). Thus, the quantity $T_2^*\{\text{strain gradients}\}$ may be separated into into terms accounting for strain coupling along ($\parallel$) and transverse to ($\perp$) the NV$^\text{-}$ symmetry axis,
%
\textcolor{black}{
\begin{multline}
\frac{1}{T_2^*\{\text{strain gradients}\}} = \frac{1}{T_2^*\{\text{strain}_{\parallel}\text{\,gradients}\}} \\ + \frac{1}{T_2^*\{\text{strain}_{\perp}\text{\,gradients}\}}.
\end{multline}
Application} of a sufficiently strong bias magnetic field mitigates the transverse strain contribution to dephasing~\cite{Jamonneau2016}, (see Sec.~\ref{efieldsuppression}), while the longitudinal contribution may be \textcolor{black}{mitigated} by employing double-quantum coherence magnetometry (see Sec.~\ref{DQ}).

\textcolor{black}{Inhomogeneous electric fields} also cause NV$^\text{-}$ ensemble dephasing~\cite{Jamonneau2016}, with associated limit $T_2^*$\{electric field noise\}. This dephasing source may also be broken down into components longitudinal and transverse to the NV$^\text{-}$ symmetry axis, and the contributions can be suppressed by the same methods as for strain-related dephasing.

In addition, external magnetic field gradients may cause NV$^\text{-}$ spin dephasing by introducing spatially-varying shifts in the NV$^\text{-}$ energy levels across an ensemble volume, with associated limit $T_2^*\{\text{magnetic field gradients}\}$. \textcolor{black}{Design} of uniform bias magnetic fields minimizes this contribution to NV$^\text{-}$ ensemble dephasing, \textcolor{black}{and is largely an engineering challenge given that modern NMR magnets can exhibit sub-ppb uniformities over their cm-scale sample volumes~\cite{vandersypen2005nmr}. }

Even though $T_2^*$ is considered the \textit{inhomogeneous} dephasing time, homogeneous time-varying electric and magnetic fields may \textcolor{black}{appear} as dephasing mechanisms \textcolor{black}{if these fields fluctuate over the course of multiple interrogation/readout sequences. Such a scenario could result in the unfortunate situation where the measured value of $T_2^*$ depends on the total measurement duration (see Sec.~\ref{optimalprecession}).} By the same argument, temperature fluctuations and spatial gradients can also \textcolor{black}{appear} as dephasing mechanisms and can limit the measured $T_2^*$. Temperature variations cause expansion and contraction of the diamond crystal lattice, altering the NV$^\text{-}$ center's zero-field splitting parameter $D$ ($dD/dT = -74$~kHz/K~\cite{Acosta2010}) and, \textcolor{black}{depending on experimental design,}  may also shift the bias magnetic field. 
Finally, we include a term in Eqn.~\ref{eqn:t2*depend} for as-of-yet unknown mechanisms limiting $T_2^*$, and we note that $T_2^*$ is limited to a theoretical maximum value of $2 T_1$~\cite{Levitt2008,Myers2017}.

Importantly, Eqn.~\ref{eqn:t2*depend} shows that the value of $T_2^*$ is primarily set by the dominant dephasing mechanism. Therefore, when seeking to extend $T_2^*$, one should focus on reducing whichever mechanism is dominant until another mechanism becomes limiting. \textcolor{black}{Reference~\cite{okeeffe2019hamiltonian} aptly expresses the proper strategy as a ``shoot the alligator closest to the boat'' approach}. \textcolor{black}{For example, even if the dephasing due to substitutional nitrogen is substantially decreased in a particular experiment, the improvement in $T_2^*$ may be much smaller if, say, strain inhomogeneity then becomes a limiting factor; at that point it becomes more fruitful to shift focus towards reducing strain-induced dephasing.}

\subsection{Nitrogen limit to $T_2^*$}\label{nitrogenlimitT2*}

In nitrogen-rich diamonds, the majority of electronic spins contributing to the spin bath originate from substitutional nitrogen defects, since N$_\text{S}^0$ may donate its unpaired electron to another defect X and become spinless N$_\text{S}^+$, via the process~\cite{Khan2009},
\begin{equation}
\text{N}_\text{S}^0 + \text{X}^0 \leftrightarrow \text{N}_\text{S}^+ + \text{X}^\text{-}.
\end{equation} 
In these samples, the electronic spin concentration is closely tied to the total concentration of substitutional nitrogen donors [N$_\text{S}^\text{T}$], and thus $T_2^*\{\text{electronic spin bath}\}$ is primarily set by [N$_\text{S}^\text{T}$]. In unirradiated nitrogen-rich diamonds, however, N$_\text{S}^0$ serves as the primary contributor to the electronic spin bath~\cite{Bauch2018}. The N$_\text{S}^0$ contribution to dephasing obeys
\begin{equation}\label{eqn:T2*NV-N}
\frac{1}{T_2^*\{\text{N}_\text{S}^0\}} = A_{\text{N}_\text{S}^0}\, [\text{N}_\text{S}^0]
\end{equation}
where [N$_\text{S}^0$] is the concentration of neutral substitutional nitrogen, and $A_{\text{N}_\text{S}^0}$ characterizes the magnetic dipole interaction strength between NV$^\text{-}$ spins and N$_\text{S}^0$ spins. The inverse linear scaling of $T_2^*\{\text{N}_\text{S}^0\}$ 
is supported by both theory~\cite{Abragam1983ch4, Taylor2008, Zhao2012, Wang2013,Bauch2019} and experiment~\cite{Bauch2018,Bauch2019,vanWyk1997}. However, reported values of the scaling factor $A_{\text{N}_\text{S}^0}$ from theoretical spin-bath simulations vary widely; for example, Ref.~\cite{Zhao2012} predicts \textcolor{black}{$A_{\text{N}_\text{S}^0}= 56~\text{ms}^\text{-1}\text{ppm}^\text{-1}$}, 
whereas Ref.~\cite{Wang2013} predicts \textcolor{black}{$A_{\text{N}_\text{S}^0}= 560~\text{ms}^\text{-1}\text{ppm}^\text{-1}$,} 
a $10\times$ discrepancy. The authors of Ref.~\cite{Bauch2018, Bauch2019} measure $T_2^*\{\text{N}_\text{S}^0\}$ on \textcolor{black}{five} samples in the range $[\text{N}_\text{S}^0] = 0.75 - 60$~ppm (see Fig.~\ref{fig:t2vsn}) and determine  \textcolor{black}{$A_{\text{N}_\text{S}^0} = 101 \pm 12~\text{ms}^\text{-1} \text{ppm}^\text{-1}$}, such that for a sample with [N$_\text{S}^0]= 1$~ppm,  \textcolor{black}{$T_2^*\{\text{N}_\text{S}^0\} = 9.9 \pm 1.2~\upmu$s}. 
The experimental value of $A_{\text{N}_\text{S}^0}$ is consistent with numerical simulations \textcolor{black}{in the same work~\cite{Bauch2019}. The} authors calculate the second moment of the dipolar-broadened single NV$^\text{-}$ ODMR linewidth~\cite{Abragam1983ch3, Abragam1983ch4} for $10^4$ random spin bath configurations and, by computing the ensemble average over the distribution of single-NV$^\text{-}$ linewidths~\cite{Dobrovitski2008}, \textcolor{black}{find good agreement with the experimental value $A_{\text{N}_\text{S}^0}= 101~\text{ms}^\text{-1} \text{ppm}^\text{-1}$}.

Electron paramagnetic resonance (EPR) measurements of nitrogen N$_\text{S}^0$ defects in diamond~\cite{vanWyk1997} from 63 samples also confirm the scaling $1/T_2^* \propto  [\text{N}_\text{S}^0] $ (see Appendix~\ref{EPR} and Fig.~\ref{fig:vanWykT2star}) and the approximate scaling constant $A_{\text{N}_\text{S}^0}$. With the likely assumption that the dephasing time for ensembles of substitutional nitrogen spins in a nitrogen spin bath can approximate  $T_2^*\{\text{N}_\text{S}^0\}$ for NV$^\text{-}$ ensembles~\cite{Dalethesis2015} (see Appendix~\ref{EPR}), the measurements in Ref.~\cite{vanWyk1997} suggest \textcolor{black}{$A_{\text{N}_\text{S}^0} \approx 130$~ms$^\text{-1} \text{ppm}^\text{-1}$}, which is in good agreement with the measured \textcolor{black}{$A_{\text{N}_\text{S}^0} = 101 \pm 12~\text{ms}^\text{-1} \text{ppm}^\text{-1}$} from Ref.~\cite{Bauch2019} (see Appendices~\ref{linewidth} and \ref{EPR}). 




In addition, the data in Ref.~\cite{vanWyk1997} suggest that dipolar dephasing contributions from  $^{13}$C at natural isotopic abundance [10700~ppm~\cite{Wieser2013}] and from substitutional nitrogen are equal for [N$_\text{S}^0] = 10.8$~ppm. The measured values of $A_{\text{N}_\text{S}^0}$~\cite{Bauch2018} and  $A_{^{13}\text{C}}$ (see Sec.~\ref{13ClimitT2star}) for NV$^\text{-}$ ensembles predict the two contributions to be equal at N$_\text{S}^0 = 10.3$~ppm, which is consistent to within experimental uncertainty.

In Appendix~\ref{choosingnitrogen}, we present a simple toy model~\cite{Kleinsasser2016} for the case when nitrogen-related defects dominate $T_2^*$. In this regime, under the assumption that the conversion efficiency of total nitrogen to NV$^\text{-}$, NV$^0$, and N$^+$ is independent of the total nitrogen concentration $[\text{N}^\text{T}]$, the dephasing time $T_2^*$ scales inverse-linearly with $[\text{N}^\text{T}]$, while the number of collected photons $\mathscr{N}$ scales linearly with $[\text{N}^\text{T}]$. These scalings result in a shot-noise-limited sensitivity $\eta \propto 1/\sqrt{\mathscr{N} \cdot T_2^*}$ , which is independent of $[\text{N}^\text{T}]$. However, as discussed in Sec.~\ref{physicallimits} and Appendix~\ref{volumedensityconsiderations}, technical considerations favor lower nitrogen concentrations $[\text{N}^\text{T}]$, which result in lower photon numbers $\mathscr{N}$ and longer dephasing times $T_2^*$~\cite{Kleinsasser2016}.

\subsection{Nitrogen limit to $T_2$}\label{nitrogenlimitT2}

\begin{figure}[ht]
\centering
\begin{overpic}[width = \columnwidth]{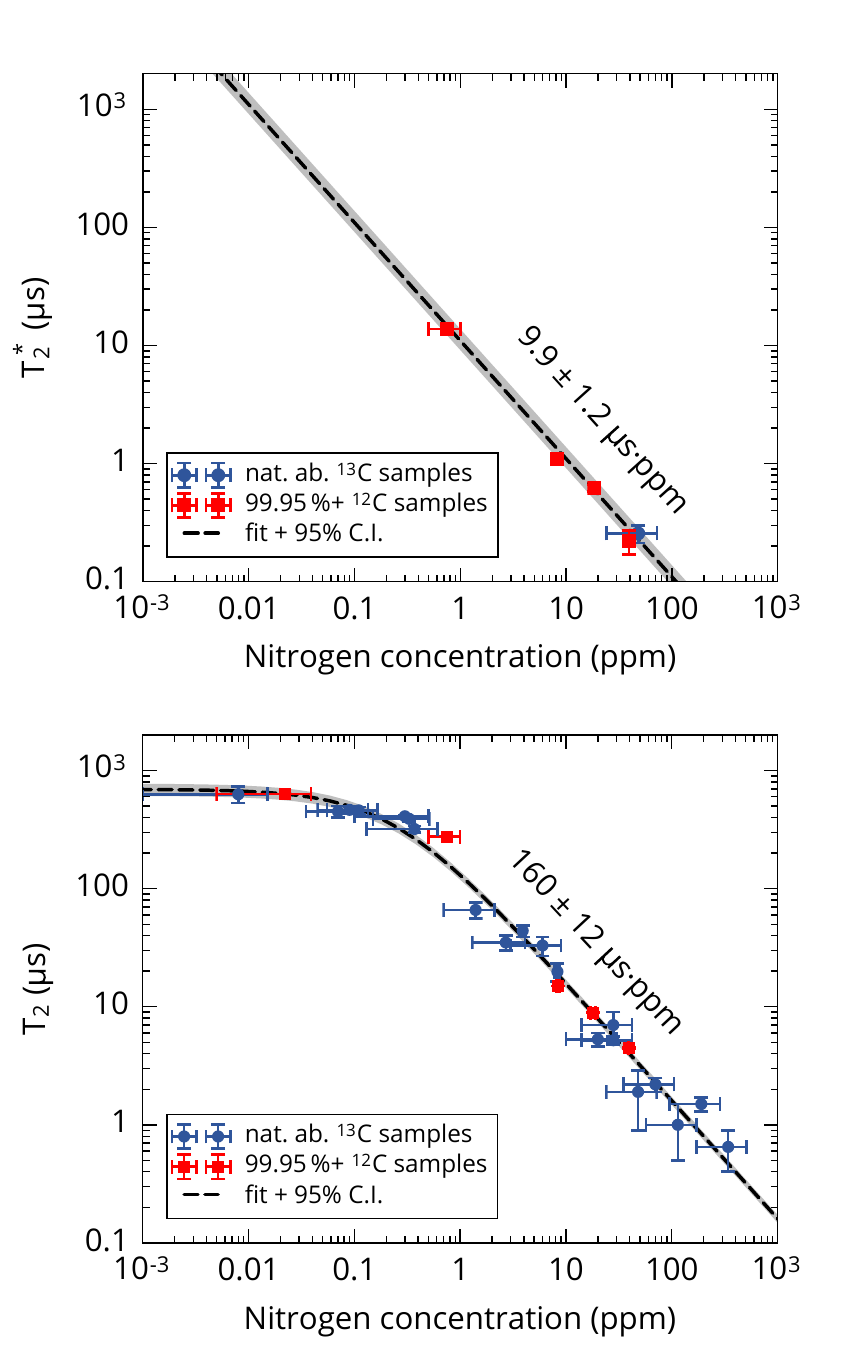}
\put (2,95.4) {\large a}
\put (2,47) {\large b}
\end{overpic}
\caption[$T_2^*$ and $T_2$ vs N]{Substitutional nitrogen spin bath contribution to ensemble-NV$^\text{-}$ dephasing time \textcolor{black}{$T_2^*$ and coherence time $T_2$. a) Measured spin-bath contribution to $T_2^*$ vs.~nitrogen concentration measured by secondary ion mass spectrometry (SIMS) 
for five diamond samples.} 
Fit yields $1/T_2^*\{\text{N}_\text{S}^0\} \!= \!A_{\text{N}_\text{S}^0} [\text{N}_\text{S}^0]$ with \textcolor{black}{$A_{\text{N}_\text{S}^0} = 101 \pm 12~\text{ms}^\text{-1} \text{ppm}^\text{-1}$.} b) Measured Hahn echo $T_2$ vs.~\textcolor{black}{nitrogen concentration} 
for 25 
diamond samples. \textcolor{black}{The linear contribution to the fit is attributed to substitutional nitrogen and} yields $1/T_2\{\text{N}_\text{S}^0\} \!= \!B_{\text{N}_\text{S}^0} [\text{N}_\text{S}^0]$ with \textcolor{black}{$B_{\text{N}_\text{S}^0} [\text{N}_\text{S}^0]\! = \!6.25 \pm 0.47~\text{ms}^\text{-1} \text{ppm}^\text{-1}$}. \textcolor{black}{The nitrogen-independent contribution to the fit is given by $T_2\{\text{other}\} = 694 \pm 82~\upmu$s.} Adapted from Ref.~\cite{Bauch2019}.
}  \label{fig:t2vsn}
\end{figure}	

Contributions to the NV$^\text{-}$ spin dephasing time $T_2^*$ from static and slowly-varying inhomogeneities are largely mitigated by employing a $\pi/2 - \pi - \pi/2$ Hahn echo pulse sequence (see Sec.~\ref{DD}). In contrast to
a $\pi/2-\pi/2$ Ramsey sequence (see Appendix~\ref{ramsey}), the added $\pi$-pulse reverses the precession direction of the sensor spins halfway through the free precession interval. As a result, any net phase accumulated by the NV$^\text{-}$ spin state due to a static magnetic field vanishes, as the accumulated phase during the first interval (before the $\pi$-pulse) cancels the accumulated phase during the second interval (after the $\pi$-pulse). Consequently, the characteristic decay time of the NV$^\text{-}$ spin state measured through Hahn echo, denoted by $T_2$ (the coherence time), is substantially longer than the inhomogeneous dephasing time $T_2^*$, typically exceeding the latter by one to two orders of magnitude~\cite{deLange2010,Bauch2019}. By design the Hahn echo sequence and its numerous extensions~\cite{Meiboom1958,Gullion1990,Wang2012} restrict sensing to AC signals, typically within a narrow bandwidth, preventing their application in DC sensing experiments. Nonetheless, the Hahn echo $T_2$ plays a crucial role in diamond sample characterization and for AC sensing protocols (Sec.~\ref{DD}) \textcolor{black}{and merits brief} discussion here. 

Like $T_2^*$, $T_2$ depends on the nitrogen concentration $[\text{N}_\text{S}^0]$, which sets both the average dipolar-coupling strength between NV$^\text{-}$ and nitrogen bath spins (i.e., $A_{\text{N}_\text{s}^0} [\text{N}_\text{s}^0]$ from Eqn.~\ref{eqn:T2*NV-N}), as well as the average coupling strength between nitrogen bath spins~\cite{deSousa2009,BarGill2012}. Furthermore, it can be shown that when nitrogen is the dominant decoherence mechanism, $T_2\{\text{N}_\text{S}^0\}$ depends inverse linearly on the nitrogen concentration $[\text{N}_\text{S}^0]$~\cite{Bauch2019}, revealing a close relationship to $T_2^*\{\text{N}_\text{S}^0\}$. The dependence of $T_2\{\text{N}_\text{S}^0\}$ \textcolor{black}{on $[\text{N}_\text{S}^0]$} was recently determined experimentally through NV$^\text{-}$ ensemble measurements on 25 
diamond samples (see Fig.~\ref{fig:t2vsn}b),  yielding~\cite{Bauch2019}
\begin{equation}\label{eqn:T2NV-N}
\frac{1}{T_2\{\text{N}_\text{S}^0\}} = B_{\text{N}_\text{S}^0} [\text{N}_\text{S}^0].
\end{equation}
Here, \textcolor{black}{$B_{\text{N}_\text{S}^0} = 6.25 \pm 0.47~\text{ms}^\text{-1} \text{ppm}^\text{-1}$, 
such that an NV$^\text{-}$ ensemble in a 1-ppm-nitrogen sample is expected to exhibit $T_2 \simeq 160 \pm 12~\upmu$s.} The scaling in Eqn.~\ref{eqn:T2NV-N} should also be compared to \textcolor{black}{that of} $T_2^*\{\text{N}_\text{S}^0\}$ (Eqn.~\ref{eqn:T2*NV-N}), with $T_2\{\text{N}_\text{S}^0\}/T_2^*\{\text{N}_\text{S}^0\} = B_{\text{N}_\text{S}^0}/A_{\text{N}_\text{S}^0} \approx 17$. A straightforward application of these results is the calibration of the total nitrogen spin concentration in diamond samples through $T_2^*$ measurements, $T_2$ measurements, or both, provided that nitrogen remains the primary source of dephasing and decoherence in such samples. Here, $T_{2}$ measurements \textcolor{black}{are} advantageous over $T_2^*$ (or linewidth) measurement schemes, \textcolor{black}{as the latter are more likely to be limited by non-nitrogen dephasing mechanisms~\cite{Bauch2018}.}


Lastly, we note that the inverse linear scaling of $T_2^*\{\text{N}_\text{S}^0\}$ and $T_2\{\text{N}_\text{S}^0\}$ with $[\text{N}_\text{S}^0]$, as well as the hierarchy $T_2 \gg T_2^*$, are consistent with earlier EPR studies of N$_\text{S}^0$ nitrogen defects in nitrogen-rich diamonds~\cite{vanWyk1997,Stepanov2016} and other comparable spin systems in silicon~\cite{Abe2010,Witzel2010}. \textcolor{black}{Predicting the values of $B_{\text{N}_\text{S}^0}$ and $A_{\text{N}_\text{S}^0}$ for NV$^\text{-}$ based on equivalent EPR scaling parameters measured with P1 centers in diamond~\cite{vanWyk1997} is expected to be crudely effective. However, accuracy at the 10$\%$ level or better likely requires accounting for various experimental specifics [e.g. the magnetic field value~\cite{Hall2014}].}

\subsection{$^{13}\text{C}$ limit to $T_2^*$}\label{13ClimitT2star}
Dipolar coupling between NV$^\text{-}$  electronic and $^{13}$C nuclear spins can also limit $T_2^*$ \cite{Dobrovitski2008,Balasubramanian2009,Zhao2012,Hall2014}. Reducing the $^{13}$C content below the natural abundance concentration $[^{13}\text{C}] = 10700 \pm 800$~ppm $\approx 1.1\%$~\cite{Wieser2013} through isotope engineering is the most direct way to mitigate this effect \cite{Balasubramanian2009,Itoh2014,Bauch2018}. In the ``dilute'' spin limit where $[^{13}\text{C}]/[^{12}\text{C}] \ll 0.01$ \cite{Kittel1953}, the $^{13}$C dephasing contribution is well-approximated by 
\begin{equation}\label{eqn:scale13c}
\frac{1}{T_2^*\{^{13}\text{C}\}} = A_{^{13}\text{C}} \,[^{13}\text{C}],
\end{equation}
where $A_{^{13}\text{C}}$ is a constant characterizing the magnetic dipole interaction strength between NV$^\text{-}$ spins and $^{13}$C nuclear spins, in accordance with theoretical predictions~\cite{Dobrovitski2008,Hall2014,Abragam1983,Kittel1953}  (see Appendix~\ref{13Cconfusion}). Although experimental measurements relating $T_2^*$ to $[^{13}$C$]$ are only available for single NV$^\text{-}$ centers \cite{Balasubramanian2009,Mizuochi2009} and not for NV$^\text{-}$ ensembles, the scaling in Eqn.~\ref{eqn:scale13c} is consistent with experimental findings in a similar ensemble spin system: EPR linewidth measurements on substitutional phosphorus spin ensembles in a $^{28}$Si crystal exhibit the same scaling for various dilute concentrations of $^{29}\text{Si}$ ~\cite{Abe2010,Morishita2011}. Figure~4b in Ref.~\cite{Abe2010} suggests that Eqn.~\ref{eqn:scale13c} is approximately valid for  $[^{29}\text{Si}]/[^{18}\text{Si}]\lesssim 0.05$, so it is \textcolor{black}{plausible} that $A_{^{13}\text{C}}$ can be inferred from measurements on diamonds with natural $^{13}$C isotopic abundance where $[^{13}\text{C}]/[^{12}\text{C}]\approx 0.0107$. We make this \textcolor{black}{assumption} henceforth. 

\begin{figure}
\centering
\begin{overpic}[width = \columnwidth]{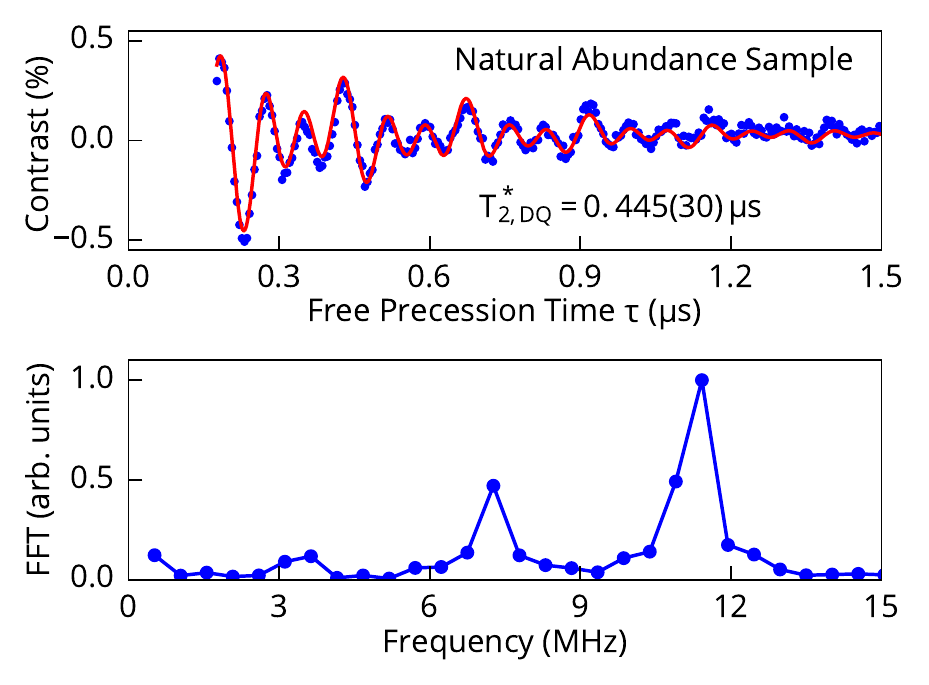} 
\put (0,70) {\large \textsf{a}}
\put (0,36) {\large \textsf{b}}
\end{overpic}
\caption[T2* vs. 13C]{$T_2^*$ measurement of a low-nitrogen-content diamond with natural abundance [$^{13}\text{C}]=10700$~ppm to assess the $^{13}\text{C}$ contribution to dephasing. a) Double-quantum Ramsey free induction decay (FID) ($\color{blue}\bullet$)  and associated fit  ({\color{red}{\rule[.5mm]{3mm}{1pt}}})  suggest $T_2^*$ is 445~ns in the double-quantum basis. This data sets a bound \textcolor{black}{$A_{^{13}\text{C}}<0.105$~ms$^\text{-1} \text{ppm}^\text{-1}$.} 
Correcting for the test diamond's approximately known $[\text{N}_\text{S}^0]\approx 0.5$~ppm content allows further refinement to \textcolor{black}{$A_{^{13}\text{C}}\approx 0.100$~ms$^\text{-1} \text{ppm}^\text{-1}$.}
b) Fourier transform of the FID shown in the top panel. \textcolor{black}{The three peaks arise from hyperfine interactions associated with the NV$^\text{-}$ center's $^{14}$N nuclear spin $I = 1$ and exhibit intra-peak spacing double that of an equivalent single-quantum Ramsey measurement. The unbalanced peak heights are attributed to nuclear spin polarization induced by the 150 gauss bias magnetic field.}}  \label{fig:wavyt2stardata}
\end{figure}

While the value of $A_{^{13}\text{C}}$ is not known precisely for NV$^\text{-}$ ensembles,  $T_2^*$ measurements in diamond with natural $^{13}$C abundance set an approximate upper 
bound on $A_{^{13}\text{C}}$, since necessarily $1/T_2^* > 1/T_2^*\{^{13}$C$\}$. Figure~\ref{fig:wavyt2stardata} shows a Ramsey FID for a diamond with natural $^{13}$C abundance \textcolor{black}{and low nitrogen concentration; these data suggest} \textcolor{black}{$A_{^{13}\text{C}}\approx 0.100$~ms$^\text{-1} \text{ppm}^\text{-1}$}. 
With this value for $A_{^{13}\text{C}}$, the expected limit for a $99.999\%$ $^{12}$C isotopically-enriched diamond is $T_2^*\{^{13}\text{C}\}\approx 1$~ms, \textcolor{black}{at which point dipolar interaction with} $^{13}$C nuclear spins \textcolor{black}{is unlikely to be the leading-order dephasing mechanism} (see Eqn.~\ref{eqn:t2*depend}). Comparing $A_{^{13}\text{C}}$ with the measured $A_{\text{N}_\text{S}^0} = 101$~ms$^\text{-1} \text{ppm}^\text{-1}$ for dephasing of NV$^\text{-}$ ensembles by substitutional nitrogen (see section~\ref{nitrogenlimitT2*}), dephasing from natural abundance $[^{13}\text{C}] = 10700$~ppm and substitutional nitrogen with concentration \textcolor{black}{$[\text{N}_\text{S}^0]= 10.6$~ppm} 
should be equivalent, in \textcolor{black}{good} agreement with Ref.~\cite{vanWyk1997}, which observes equivalence for [$\text{N}_\text{S}^0$] $\approx$ 10.8~ppm. \textcolor{black}{Conveniently, it is easy to remember that $T_2^*\{^{13}$C$\}$  is $1~\upmu$s for natural abundance $^{13}\text{C}$ diamond to better than 10$\%$. }



The bound on $A_{^{13}\text{C}}$ derived above can be crudely confirmed \textcolor{black}{using} a mix of theoretical predictions from Ref.~\cite{Hall2014} and data from Ref.~\cite{Maze2012}. The authors of Ref.~\cite{Maze2012} find the most probable $T_2^*$ for a single NV$^\text{-}$ center in natural isotopic diamond to be $T_2^{*\{\text{single,mp}\}}=1.8 \pm 0.6~\upmu$s [measured in a 20~G bias field, Fig.~4a in Ref.~\cite{Maze2012}]. From relations in Ref.~\cite{Hall2014} we estimate \textcolor{black}{$A_{^{13}\text{C}}$ in terms of the coupling constant $A_{^{13}\text{C}}^\text{single}$ for a single NV$^\text{-}$, $A_{^{13}\text{C}} \approx 2.2\, A_{^{13}\text{C}}^\text{single,mp}$,}  which yields \textcolor{black}{$A_{^{13}\text{C}}=0.11 \pm 0.04$~ms$^\text{-1} \text{ppm}^\text{-1}$.} 

Our measured value \textcolor{black}{$A_{^{13}\text{C}}\approx 0.100$~ms$^\text{-1} \text{ppm}^\text{-1}$} is also in reasonable agreement with first-principles theoretical calculations by Ref.~\cite{Hall2014}, suggesting \textcolor{black}{$A_{^{13}\text{C}} \approx 0.057$~ms$^\text{-1} \text{ppm}^\text{-1}$} 
for NV$^\text{-}$ ensembles \textcolor{black}{in natural isotopic diamond} in tens-of-gauss bias fields. Note that the experimental determination of $A_{^{13}\text{C}}$ outlined in this section represents an upper bound on the true value of  $A_{^{13}\text{C}}$ in the dilute (dipolar-broadened) limit; if substantial broadening arises from Fermi-contact contributions in addition to dipolar interactions in natural abundance $^{13}$C samples, or if [$^{13}$C] = 10700~ppm does not qualify as the dilute limit \cite {Kittel1953,Abragam1983}, the value of $A_{^{13}\text{C}}$ given here will be overestimated.

Engineering diamonds for low $^{13}$C content may be challenging \cite{Markham2011,Dwyer2013,Teraji2013}. The isotopic purity of a diamond grown by plasma-enhanced chemical vapor deposition (PE-CVD) is expected to be limited by the purity of the carbon source gas, which is most commonly methane (CH$_4$). However, diamonds grown with isotopically-enriched methane may exhibit  higher fractional $^{13}$C content than the source gas due to extraneous carbon sources in the CVD chamber \cite{Dwyer2013}. Nonetheless, Teraji \textit{et al.}~achieve $[^{12}\text{C}]=99.998\%$  as measured by secondary ion mass spectrometry (SIMS) when using isotopically-enriched methane with 99.999$\%$ $^{12}$C (i.e., $[^{13}\text{C}]\leq 10$~ppm)~\cite{Teraji2013, Teraji2015}. Although such isotopically-enriched methane is currently $10^3$ - $10^4$ times more expensive than natural abundance CH$_4$, order unity conversion of the methane's carbon content into diamond is attainable \cite{Teraji2013}.

\subsection{NV$^\text{-}$ limit to $T_2^*$}\label{NVNVlimit}

\begin{figure}[ht]
\centering
\begin{overpic}[width=\columnwidth]{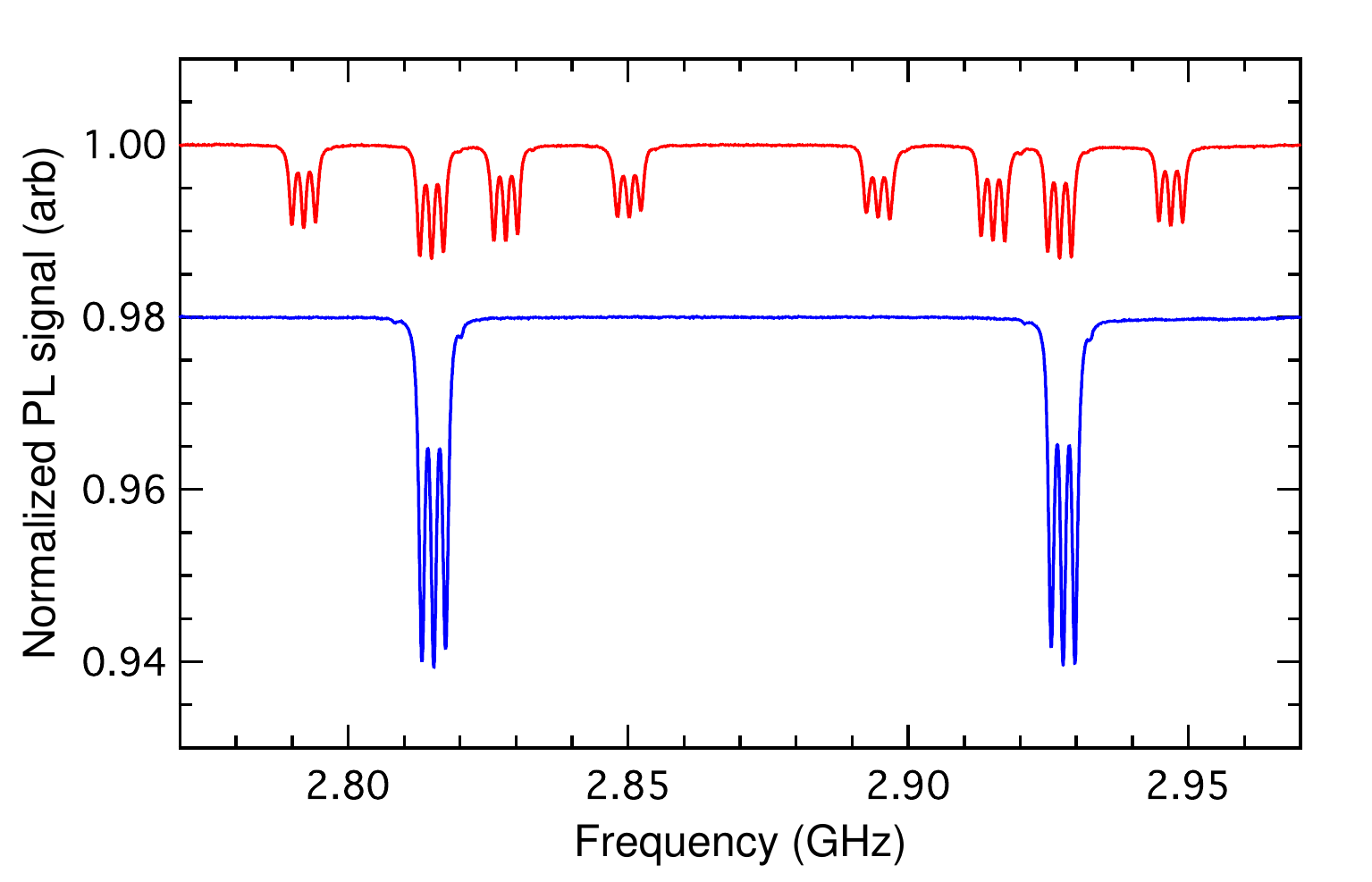}  
\end{overpic}
\caption[Experimental ODMR spectra overlapping NV$^\text{-}$ resonances]{ODMR spectra from the same NV$^\text{-}$ ensemble in different applied bias magnetic fields. A bias field with a different projection on each of the four NV$^\text{-}$ crystallographic orientations separates the $m_s =0 \leftrightarrow m_s = \pm1$ spin resonances into distinct groups ({\color{red}{\rule[.5mm]{3mm}{1pt}}}). A bias field that projects equally onto all four orientations overlaps the spin resonances ({\color{blue}{\rule[.5mm]{3mm}{1pt}}}, offset for clarity).}
\label{fig:NVgroups}
\end{figure}

Dipolar interactions among negatively-charged NV$^\text{-}$ centers may also limit the dephasing time $T_2^*$.  Dephasing from NV$^\text{-}$-NV$^\text{-}$ interactions arises from NV$^\text{-}$ spins in both the same and different groups as the NV$^\text{-}$ centers used for sensing, where groups are defined as follows: NV$^\text{-}$ centers with approximately the same spin resonance frequency are considered to be in the same group, whereas spins with different resonance frequencies are in different groups~\cite{Kucsko2016}. Depending on the strength and angle of the applied bias field, the spin resonances of the four NV$^\text{-}$ orientations may be spectrally separated, or two or more may be overlapped, changing the fraction of NV$^\text{-}$ spins in the same group (see Fig.~\ref{fig:NVgroups}). The NV$^\text{-}$-NV$^\text{-}$ dipolar contribution to $T_2^*$ is then given by




\begin{equation}
\begin{aligned}\label{eqn:NVNV}
\frac{1}{T_2^*\{\text{NV}^\text{-}\}} &= \frac{1}{T_2^*\{\text{NV}^\text{-}\}}_\parallel & \!\!\!\!\!+\,\, & \frac{1}{T_2^*\{\text{NV}^\text{-}\}}_\nparallel\\
& = \varsigma_{\parallel} \,A_{\text{NV}^\text{-}_{\parallel}}\, [\text{NV}^\text{-}_{\parallel}] & \!\!\!\!\!+\,\, & \varsigma_{\nparallel} \,A_{\text{NV}^\text{-}_{\nparallel}}\, [\text{NV}^\text{-}_{\nparallel}].
\end{aligned}
\end{equation}
Here $[\text{NV}^\text{-}_{\parallel}]$ is the concentration of NV$^\text{-}$ spins in the group being used for sensing and $[\text{NV}^\text{-}_{\nparallel}]$ is the concentration of NV$^\text{-}$ spins in other groups, with $[\text{NV}^\text{-}_{\parallel}]+ [\text{NV}^\text{-}_{\nparallel}] = [\text{NV}^\text{-}_\text{total}]$; the constants $A_{\text{NV}^\text{-}_\parallel}$ and $A_{\text{NV}^\text{-}_{\nparallel}}$ characterize the dipolar interaction strength for pairs of NV$^\text{-}$ spins in the same group and different groups respectively; and $\varsigma_{\parallel}$ and $\varsigma_{\nparallel}$ are dimensionless factors of order unity accounting for (imperfect) initialization of NV$^\text{-}$ centers~\cite{Doherty2013}. 
For example, off-resonant NV$^\text{-}$ populations polarized into the spinless $m_s = 0$ state during initialization should not contribute to dephasing of the NV$^\text{-}$ centers used for sensing, giving $\varsigma_\nparallel \simeq 0$.

Flip-flop interactions between NV$^\text{-}$ spins in different groups are off-resonant and are thus suppressed, whereas flip-flop interactions can occur resonantly between spins in the same group. The extra \textcolor{black}{resonant} interaction terms in the dipole-dipole Hamiltonian for spins in the same group result in a slightly increased dephasing rate~\cite{Abragam1983ch4,Kucsko2016}. Following Ref.~\cite{Abragam1983ch4}, it is expected that $A_{\text{NV}_\parallel} = 3/2 \,A_{\text{NV}_\nparallel}$. 

The lack of published data at present for $T_2^*\{\text{NV}^\text{-}\}$ in samples with varying NV$^\text{-}$ concentration prevents definitive determination of $A_{\text{NV}_{\parallel}}$ and $A_{\text{NV}_{\nparallel}}$. However, both terms can be estimated from the experimentally determined value of \textcolor{black}{$A_{\text{N}_\text{S}^0} = 101 \pm 12~\text{ms}^\text{-1} \text{ppm}^\text{-1}$~\cite{Bauch2019}}, which describes the scaling of NV$^\text{-}$ ensemble $T_2^*$ with substitutional nitrogen concentration (see Sec.~\ref{nitrogenlimitT2*}). Assuming an NV$^\text{-}$ electronic spin bath couples to NV$^\text{-}$ sensor spins with approximately the same strength as a substitutional nitrogen spin bath~\cite{Hanson2008}, and accounting for the higher spin multiplicity of NV$^\text{-}$ centers ($S_{\text{NV}^\text{-}}=1$) compared to substitutional nitrogen spins ($S_{\text{N}_\text{S}^0} = \nicefrac{1}{2}$), we calculate~\cite{Abragam1983ch4} 

\begin{equation}
\begin{aligned}
A_{\text{NV}^\text{-}_{\nparallel}} & \simeq \sqrt{\frac{S_{\text{NV}^\text{-}}(S_{\text{NV}^\text{-}}+1)}{S_{\text{N}_\text{S}^0}(S_{\text{N}_\text{S}^0}+1)}} A_{\text{N}_\text{S}^0}\\
& \simeq \sqrt{8/3} A_{\text{N}_\text{S}^0} \\
& \simeq \textcolor{black}{165~\text{ms}^\text{-1}\text{ppm}^\text{-1}}
\end{aligned}
\end{equation}
and find $A_{\text{NV}^\text{-}_{\parallel}} = 3/2\, A_{\text{NV}^\text{-}_{\nparallel}} \simeq$ \textcolor{black}{247~ms$^\text{-1} \text{ppm}^\text{-1}$.} 
\textcolor{black}{From this argument, although the precise} value of $T_2^*\{\text{NV}^\text{-}\}$ depends on experimental conditions including optical initialization fraction (determining $\varsigma_\parallel$ and $\varsigma_\nparallel$) and bias magnetic field orientation (setting the ratio [NV$^\text{-}_\parallel$]/[NV$^\text{-}_\nparallel$]), the value of $T_2^*\{\text{NV}^\text{-}\}$ for a given NV$^\text{-}$ concentration is well approximated (up to a factor of order unity) by the dephasing time $T_2^*\{\text{N}_\text{S}^0\}$ for \textcolor{black}{an identical} concentration of N$_\text{S}^0$ spins.






Magnetometer operation in the NV$^\text{-}$-NV$^\text{-}$ interaction limit may occur as the N-to-NV$^\text{-}$ conversion efficiency $\text{E}_\text{conv}$ approaches its theoretical limit of 50\%~\cite{Pham2011, Felton2009} (see Sec.~\ref{Econvdef}). Under these circumstances, and when other sources of dephasing can be \textcolor{black}{neglected [e.g., magnetic, electric, and strain gradients as well as $^{13}$C nuclear spins (see Sec.~\ref{13ClimitT2star}) and other paramagnetic defects (see Sec.~\ref{chargetraps})]},  the interaction among NV$^\text{-}$ spins becomes the dominant source of dephasing. However, maximal N-to-NV$^\text{-}$ conversion efficiency is not necessarily required to operate in the NV$^\text{-}$-NV$^\text{-}$ interaction limit. When $\text{E}_\text{conv} < 50\%$, dephasing due to other paramagnetic impurities may be reduced through spin bath driving techniques described in Sec.~\ref{P1driving}. Spin bath driving can also decouple the NV$^\text{-}$ centers in different groups from the NV$^\text{-}$ centers in the group used for sensing~\cite{Bauch2018}, suppressing the second term in Eqn.~\ref{eqn:NVNV} ($1/T_2^*\{\text{NV}^\text{-}\}_\nparallel$) and leaving only the first term ($1/T_2^*\{\text{NV}^\text{-}\}_\parallel$) as a fundamental limit to NV$^\text{-}$ ensemble $T_2^*$.

While this section has focused on the negatively charged NV$^\text{-}$ center, NV centers are also present in the neutral charge state NV$^0$ ($S=\nicefrac{1}{2}$) (see Sec.~\ref{Econvdef}). NV$^0$ has not been observed in its ground state in EPR, a \textcolor{black}{phenomenon} tentatively attributed to resonance line broadening from dynamic Jahn-Teller distortion~\cite{Felton2008}. Similarly, magnetic noise created by NV$^0$ may be reduced by \textcolor{black}{this} motional-narrowing-type effect to less than otherwise expected for a $S=\nicefrac12$ defect (see Sec.~\ref{P1driving}). Consequently, the contribution of NV$^0$ spins to dephasing of NV$^\text{-}$ spins may be smaller than expected. \textcolor{black}{More recent work, however, suggests an alternative hypothesis: the NV$^0$ ground state is not be visible in EPR due to large strain effects~\cite{barson2019fine}.} What little, if any, NV$^\text{+}$ is present in the sample is expected to be spinless (see Table~\ref{tab:defectspins}) and should not contribute substantially to dephasing.


Recently, several protocols have been proposed to mitigate strong NV$^\text{-}$-NV$^\text{-}$ dipolar interactions and extend $T_2^*$~\cite{okeeffe2019hamiltonian} or $T_2$~\cite{choi2017dynamical} while retaining magnetic field sensitivity. In addition, it has been proposed that under certain circumstances the NV$^\text{-}$-NV$^\text{-}$ dipolar interaction could enhance magnetometry sensitivity through enabling entanglement of multiple NV$^\text{-}$ centers~\cite{choi2017quantum} (see Sec.~\ref{DRINQS}). Harnessing entanglement could enable superior scaling of measurement SNR with number of spins addressed $N$, exceeding the standard quantum limit $\text{SNR}\propto \sqrt{N}$ and approaching the Heisenberg limit, $\text{SNR} \propto N$.
Controlled coupling of NV$^\text{-}$ spin pairs~\cite{Neumann2010b,Dolde2013,Yamamoto2013b,Jakobi2016,Bernien2013,Hensen2015} has been demonstrated; 
however, applying entanglement-enhanced techniques to larger ensembles is expected to be challenging. 



%% file: sec03.tex
\section{Methods to extend $T_2^*$ and $T_2$}\label{T2*extensionmethods}

\subsection{Dynamical decoupling for AC magnetometry}\label{DD}

\begin{figure}[ht]
\centering
\begin{overpic}[height=1.9 in]{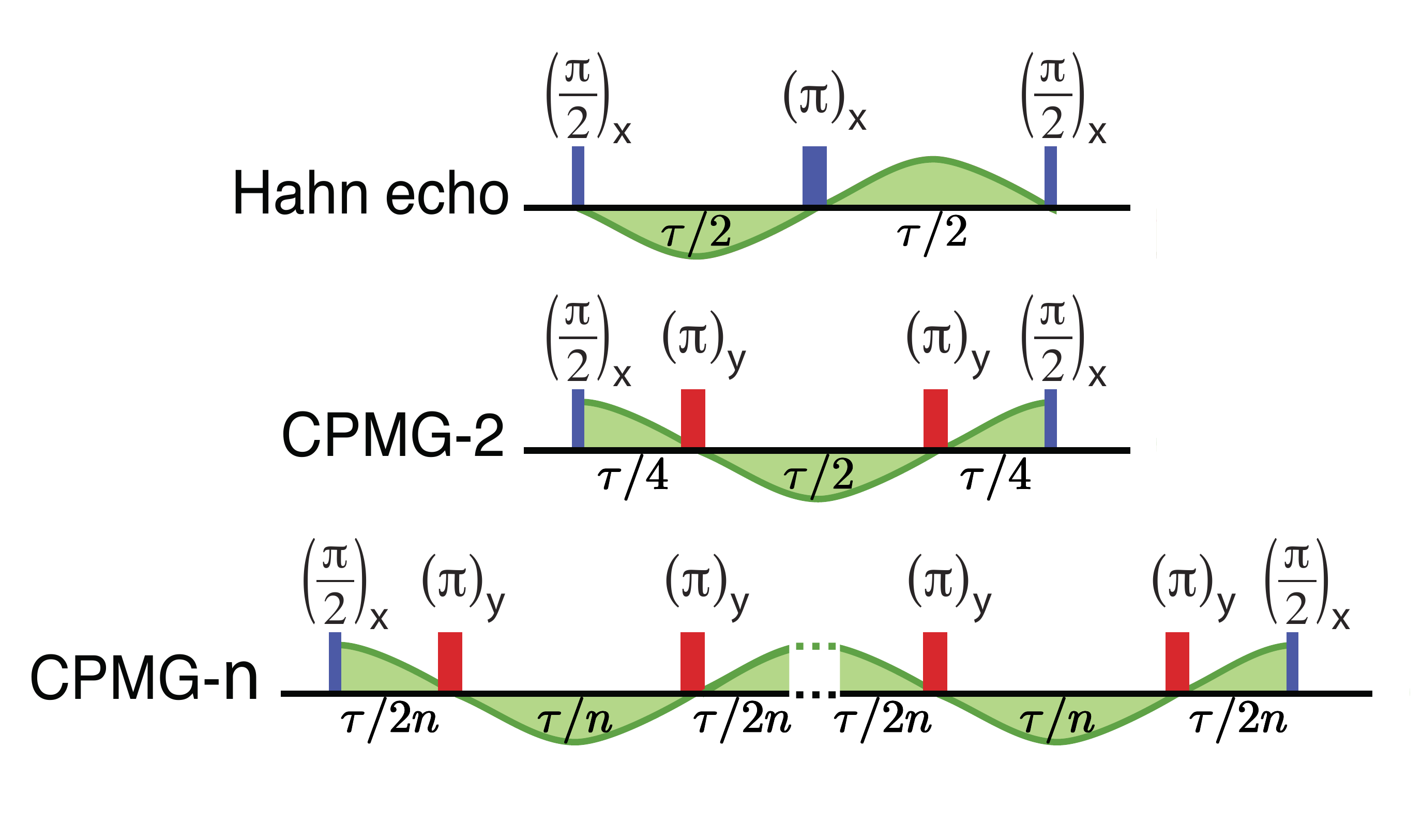}
\end{overpic}
\caption[Hahn echo and CPMG]{Select pulse sequences for AC magnetometry. The Hahn echo sequence includes a refocusing $\pi$-pulse midway through the interrogation time, allowing phase-sensitive lock-in-type measurements of AC magnetic fields (top). Hahn echo is maximally sensitive to AC fields with nodes coincident with the three MW pulses. Detection of AC fields with the quadrature phase can be achieved using the Carr-Purcell-Meiboom-Gill-2 (CPMG-2) sequence (middle). Employing additional $\pi$-pulses (CPMG-$n$) achieves more efficient decoupling of the NV$^\text{-}$ from \textcolor{black}{substitutional nitrogen and other paramagnetic defects in the diamond} and provides sensitivity to higher-frequency AC magnetic fields (bottom). From Ref.~\cite{Pham2012a}.}  \label{fig:echo}
\end{figure}

While this review primarily addresses the broadband DC sensing modality of ensemble-NV$^\text{-}$ magnetometers, many of the sensitivity-improvement techniques described herein can also be applied to detecting narrowband AC magnetic fields. Here we provide a brief overview of standard AC sensing schemes; we discuss several approaches to improving AC magnetic field sensitivity; and we highlight challenges unique to the AC sensing modality.

The Hahn echo (alternatively referred to as the spin echo) protocol, shown in Fig.~\ref{fig:echo}, builds upon the Ramsey protocol with an additional central MW $\pi$-pulse, which refocuses dephasing of the NV$^\text{-}$ spin ensemble~\cite{Hahn1950}. The decay of spin coherence measured with this pulse sequence is characterized by $T_2$, which is typically one to two orders of magnitude longer than $T_2^*$ in NV$^\text{-}$ ensemble measurements (see Secs.~~\ref{T2*params}, \ref{nitrogenlimitT2*}, and ~\ref{nitrogenlimitT2}). Furthermore, while the refocusing pulse decouples the NV$^\text{-}$ spin from DC magnetic fields, its presence makes Hahn echo measurements particularly sensitive to oscillating magnetic fields with period $T_B$ matching the spin interrogation time $\tau$ of the pulse sequence. In the ideal case where the three MW pulses are commensurate with \textcolor{black}{consecutive} nodes of the AC magnetic field, the sensitivity of a Hahn-echo-based measurement \textcolor{black}{limited by shot noise and spin projection noise} is given by 
\textcolor{black}{
\small
\begin{align}\label{eqn:hahnecho}
\eta_\text{echo} ^\text{ensemble} &\approx \\ \nonumber & \!\!\!\!\!\!\!\!\!\!\!\!\!\!\!\!\!\! \underbrace{\frac{\pi}{2} \frac{\hbar}{\Delta m_s g_e \mu_B} \frac{1}{\sqrt{N\tau}}}_{\text{Spin projection limit}} \;   \underbrace{\frac{1}{e^{-(\tau/T_2)^p}}}_{\text{Spin decoherence}} \; \underbrace{\sqrt{1\!+\!\frac{1}{C^2 n_\text{avg}}}}_{\text{Readout}} \; \underbrace{\sqrt{\frac{t_I\! +\! \tau\! +\! t_R}{\tau}}}_{\text{Overhead time}}
\end{align}
\normalsize
where $\Delta m_s$ is the spin projection quantum number difference between the two states in the interferometry measurement, $N$ is the number of interrogated spins, $\tau$ is the full field interrogation time, $p$ is a stretched exponential parameter set by ensemble averaging of the local NV$^\text{-}$ spin environments (see Appendix~\ref{FIDexponent}), $C$ is the measurement contrast prior to precession (see Sec.~\ref{magneticfieldsensitivity}), $n_\text{avg}$ is the average number of photons detected per NV$^\text{-}$ center per readout, and $t_I$ and $t_R$ are the optical initialization and readout times, respectively.} For more realistic measurements in which the pulse sequence cannot be phase-locked to the AC magnetic field, the magnetic sensitivity is degraded by $\sqrt{2}$. 


The shot-noise-limited sensitivity given by Eqn.~\ref{eqn:hahnecho} has several key differences to that of a Ramsey-based DC sensing protocol (Eqn.~\ref{eqn:ramseyshot}). 
First, since typically $T_2 \gg T_2^*$, AC \textcolor{black}{sensing}  
schemes can achieve better sensitivity than DC \textcolor{black}{sensing}  
schemes. 
Second, choice of spin interrogation time $\tau$ is more straightforward for Ramsey schemes than for echo-based schemes. For Ramsey-based sensing of DC or quasi-static fields, $\tau \sim T_2^*$ is optimal (Appendix~\ref{optimalprecession}). In contrast, while $\tau \sim T_2$ is optimal for Hahn-echo-based protocols, $\tau$ should also be matched to the period $T_B$ of the AC magnetic field to be measured. As a result the scheme is maximally sensitive to fields of period $T_B \sim T_2$, \textcolor{black}{with a detection bandwidth set by the relevant filter function~\cite{Cywinski2008}.} 
Finally, coherent interactions between the NV$^\text{-}$ spin and other spin impurities in the diamond can modulate the Hahn-echo coherence envelope. At best, these effects introduce collapses and revivals that do not affect $T_2$ and merely complicate the NV$^\text{-}$ magnetometer's ability to measure AC magnetic fields of arbitrary frequency. \textcolor{black}{When the bias magnetic field is aligned to the NV$^\text{-}$ internuclear axis in diamond samples containing a natural abundance of $^{13}$C, collapse-and-revival oscillations occur with frequency set by the $^{13}$C Larmor precession. At worst, misalignment between the bias magnetic field and the NV$^\text{-}$ internuclear axis results in anisotropic hyperfine interactions, which enhance the nuclear-spin Larmor precession rate for $^{13}$C (and $^{15}$N in $^{15}$NV-diamonds) as a function of separation between the nuclear spins and NV$^\text{-}$ centers~\cite{Childress2006, Maurer2010}}. These effects ensemble average to an effectively shorter coherence time $T_2$~\cite{Stanwix2010}, which degrades AC sensitivity.

Despite these differences, the Ramsey and spin-echo measurement schemes share many of the same components; consequently, many techniques for improving spin readout fidelity (analyzed in Sec.~\ref{fidelityimprovementmethods}) apply to both DC and AC sensing modalities. For example, ancilla-assisted repetitive readout (Sec.~\ref{ancilla}), level-anticrossing-assisted readout (Sec.~\ref{LAC}), and improved fluorescence collection methods (Sec.~\ref{improvedcollection}) increase the number of detected photons per measurement $\mathscr{N}$; preferential NV$^\text{-}$ orientation (Sec.~\ref{preferentialorientation}) enhances the measurement contrast $C$; and spin-to-charge-conversion (SCC) readout (Sec.~\ref{SCCR}) and NV$^\text{-}$ charge state optimization (Sec.~\ref{chargestate}) increase both $C$ and $\mathscr{N}$. We note that because typically $T_2\gg T_2^*$, advanced readout techniques such as repetitive readout and SCC readout presently offer greater sensitivity improvement for AC schemes than for DC schemes, as their long-readout-time requirements introduce smaller fractional overhead in AC measurements with longer interrogation times. 

Additionally, techniques to extend $T_2^*$ for DC and broadband magnetometry may also improve AC magnetic field sensitivity. For example, double-quantum (DQ) coherence magnetometry (Sec.~\ref{DQ}) is expected to improve AC sensitivity both by introducing a $2\times$ increase in the NV$^\text{-}$ spin precession rate~\cite{Fang2013,Mamin2014} and, in certain cases, by extending the NV$^\text{-}$ coherence time $T_2$~\cite{Angerer2015}. Similarly, spin bath driving (Sec.~\ref{P1driving}) and operation at a sufficiently strong bias magnetic field (Sec.~\ref{efieldsuppression}) may extend $T_2$ by suppressing magnetic and electric/strain noise, respectively.

Another technique for enhancing NV$^\text{-}$ magnetic sensitivity, unique to the AC sensing modality, is the application of multi-pulse sequences, whose timing is based on the Carr-Purcell-Meiboom-Gill (CPMG) family of pulse sequences well-known in NMR~\cite{Cywinski2008,Pham2013thesis} (see Fig.~\ref{fig:echo}). By applying additional MW $\pi$-pulses at a rate of $\frac{1}{2T_B}$, these multi-pulse sequences (i) extend the NV$^\text{-}$ coherence time $T_2$ by more effectively decoupling the NV$^\text{-}$ spins from magnetic noise and (ii) increase the time during which the NV$^\text{-}$ spins interrogate the AC magnetic field. \textcolor{black}{The coherence time has been found to scale with a power law $s$ ($T_2 \rightarrow T_2^{(k)}=T_2 k^s$) as a function of the number of pulses $k$~\cite{Pham2012a}, where $s$ is set by the noise spectrum of the decohering spin bath and is typically sub-linear.} For example, a bath of electronic spins, such as N$_\text{S}^0$ defects in diamond, exhibits a Lorentzian noise spectrum and \textcolor{black}{results in} a power-law scaling of the coherence time with $s=\nicefrac23$\textcolor{black}{, assuming} the electronic spin bath is the dominant decoherence source~\cite{deSousa2009}. The multi-pulse AC magnetic field sensitivity limited by shot noise and spin projection noise is given by\textcolor{black}{
\small
\begin{align}\label{eqn:MPsens}
\eta_\text{multi} ^\text{ensemble} &\approx \\ \nonumber & \!\!\!\!\!\!\!\!\!\!\!\!\!\!\!\!\!\! \underbrace{\frac{\pi}{2} \frac{\hbar}{\Delta m_s g_e \mu_B} \frac{1}{\sqrt{N\tau}}}_{\text{Spin projection limit}} \;   \underbrace{\frac{1}{e^{-[\tau/(k^s T_2)]^p}}}_{\text{Spin decoherence}} \; \underbrace{\sqrt{1\!+\!\frac{1}{C^2 n_\text{avg}}}}_{\text{Readout}} \; \underbrace{\sqrt{\frac{t_I\! +\! \tau\! +\! t_R}{\tau}}}_{\text{Overhead time}}
\end{align}
\normalsize
with an optimal number of pulses}
\begin{equation}\label{eqn:Kopt}
k_{\text{opt}} = \left[\frac{1}{2 p (1-s)} \left(\frac{2 T_2}{T_B}\right)^p\right]^{\frac{1}{p(1-s)}},
\end{equation}
for an AC magnetic field with period $T_B$, assuming \textcolor{black}{full} interrogation time $\tau = \frac{k}{2}T_B$ and $\pi$-pulses commensurate with the nodes of the oscillating magnetic field. As before, the sensitivity is degraded by $\sqrt{2}$ when measuring AC magnetic fields with unknown phase.

Equations~\ref{eqn:MPsens} and \ref{eqn:Kopt} illustrate that multi-pulse measurement schemes improve sensitivity to magnetic fields with periods $T_B < T_2$ and enable sensing of higher frequencies than can be accessed with Hahn-echo-based measurements. \textcolor{black}{For example, the authors of Ref.~\cite{Pham2012a}} demonstrate a 10$\times$ improvement \textcolor{black}{(compared to Hahn echo)} in ensemble AC sensitivity at 220~kHz by using a multi-pulse sequence. However, the increased number of control pulses, which are typically imperfect due to NV$^\text{-}$ hyperfine structure and inhomogeneities in the system, can result in cumulative pulse error and thus degraded AC sensitivity~\cite{Wang2012}. Compensating pulse sequences, including schemes in the XY, concatenated, and BB-$n$ families, may be employed to restore AC field sensitivity in the presence of pulse errors~\cite{Wang2012,Gullion1990,Farfurnik2015, Low2014, Rong2015}. 

\textcolor{black}{
AC sensing techniques are also pertinent to noise spectroscopy. By mapping out a diamond's spin bath spectral noise profile, tailored sensing protocols can be designed to more efficiently extract target signals. To this end, dynamical decoupling sequences, such as those in the CPMG and XY families, are employed for noise mapping~\cite{BarGill2012, BarGill2013, Romach2015, chrostoski2018electric, Bauch2019}. By varying both the total precession time and the number of refocusing pulses, noise at a variable target frequency can be isolated, ultimately allowing measurement of the entire spin bath spectral noise profile. However, such measurements are often complicated by non-idealities in certain sequences' filter functions, such as sensitivity to harmonics or the presence of sidelobes~\cite{Cywinski2008}. Recently, AC magnetometry protocols with enhanced spectral \textcolor{black}{resolution} have been demonstrated, such as the dynamic sensitivity control (DYSCO) sequence and its variants~\cite{lazariev2017dynamical, romach2019measuring}, which provide simpler, single-peaked filter functions at the cost of reduced sensitivity. \textcolor{black}{Additional dynamical decoupling sequences with increased spectral resolution or other advantages have been employed~\cite{hernandez2018noise,Glenn2018,Schmitt2017,Boss2017} or proposed~\cite{Cywinski2008, zhao2014dynamical,poggiali2018optimal}.}}

A final consideration in the application of multi-pulse sequences for enhancing AC magnetometry with NV$^\text{-}$ centers is that extension of the $T_2$ coherence time (and thus enhancement of AC magnetic field sensitivity) is eventually limited by the $T_1$ spin-lattice relaxation time, beyond which increasing the number of $\pi$-pulses is ineffective. This limitation can be overcome by reducing the magnetometer operating temperature, thereby suppressing the two-phonon Raman process that dominates NV$^\text{-}$ spin-lattice relaxation near room temperature and extending $T_1$~\cite{Jarmola2012}. Multi-pulse sequences performed at 77~K have demonstrated $> 100\times$ extensions in $T_2$ compared to room temperature measurements~\cite{BarGill2013}, and corresponding improvements to AC magnetic field sensitivity are expected.

\textcolor{black}{Although the limit $T_2$, $T_2^* \leq 2T_1$ is well established theoretically~\cite{Slichter1990,yafet1963gfactors} and observed in other spin systems~\cite{bylander2011noise}, the maximum $T_2$ values achieved in NV-diamond through dynamical decoupling protocols have historically never exceeded measured values of $T_1$, with Ref.~\cite{BarGill2013} \textcolor{black}{achieving} $T_2 = 0.53(2)T_1$ for NVs in bulk diamond and Refs.~\cite{myers2014probing,Romach2015} observing $T_2 \lesssim 0.1T_1$ for shallow NVs. While this discrepancy is not fully resolved, it is partially accounted for by the observation that the typical measurement protocol for $T_1$ \textcolor{black}{[e.g., that described in~\cite{Pham2013thesis}]} yields a $T_1$ value that does not encompass all possible decays of the full spin-1 system but rather only the decay $T_1^{(0)}$ in the \textcolor{black}{pseudo-}spin-$\nicefrac{1}{2}$ subspace of $|0\rangle$ and either $|\!+\!1\rangle$ or $|\!-\!1\rangle$~\cite{Myers2017}. The value of $T_1$ for the full $S=1$ system is typically shorter than $T_1^{(0)}$ thanks to non-negligible decay \textcolor{black}{from $|\!+\!1\rangle$ to $|\!-\!1\rangle$ and vice versa.} This spin-1 $T_1$, which can be measured \textcolor{black}{using methods} described in Ref.~\cite{Myers2017}, is the relevant relaxation time limiting $T_2$ and $T_2^*$.
}

\subsection{Double-quantum coherence magnetometry}\label{DQ}

\begin{figure*}[ht]
\centering
\vspace{-8mm}
\begin{overpic}[width = .4\textwidth, trim = 0 -5mm 0 0]{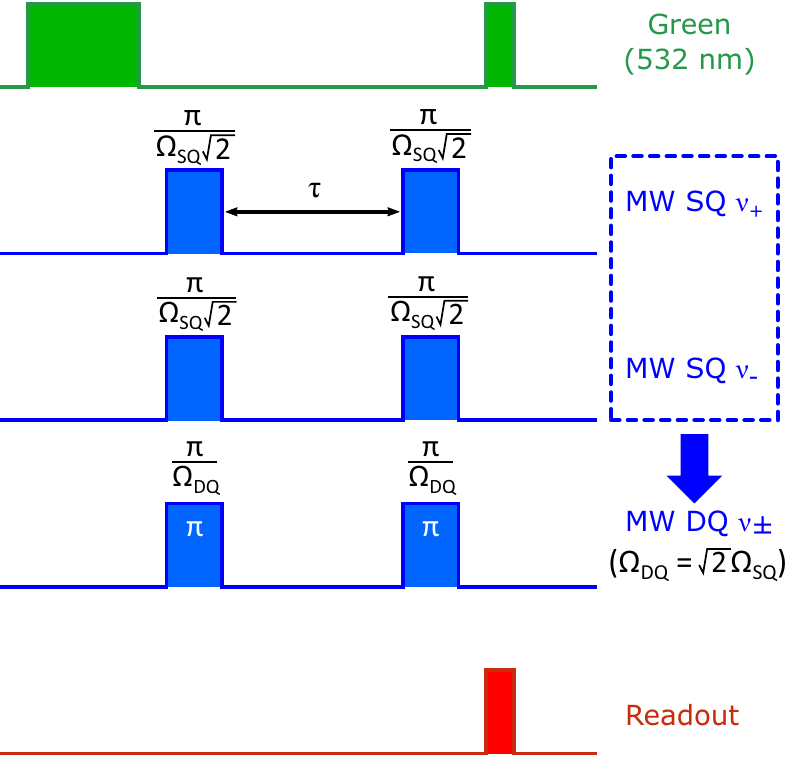}
\put (-5, 100.2) {\large a}
\end{overpic}
\hspace{8mm}
\begin{overpic}[width = .4\textwidth]{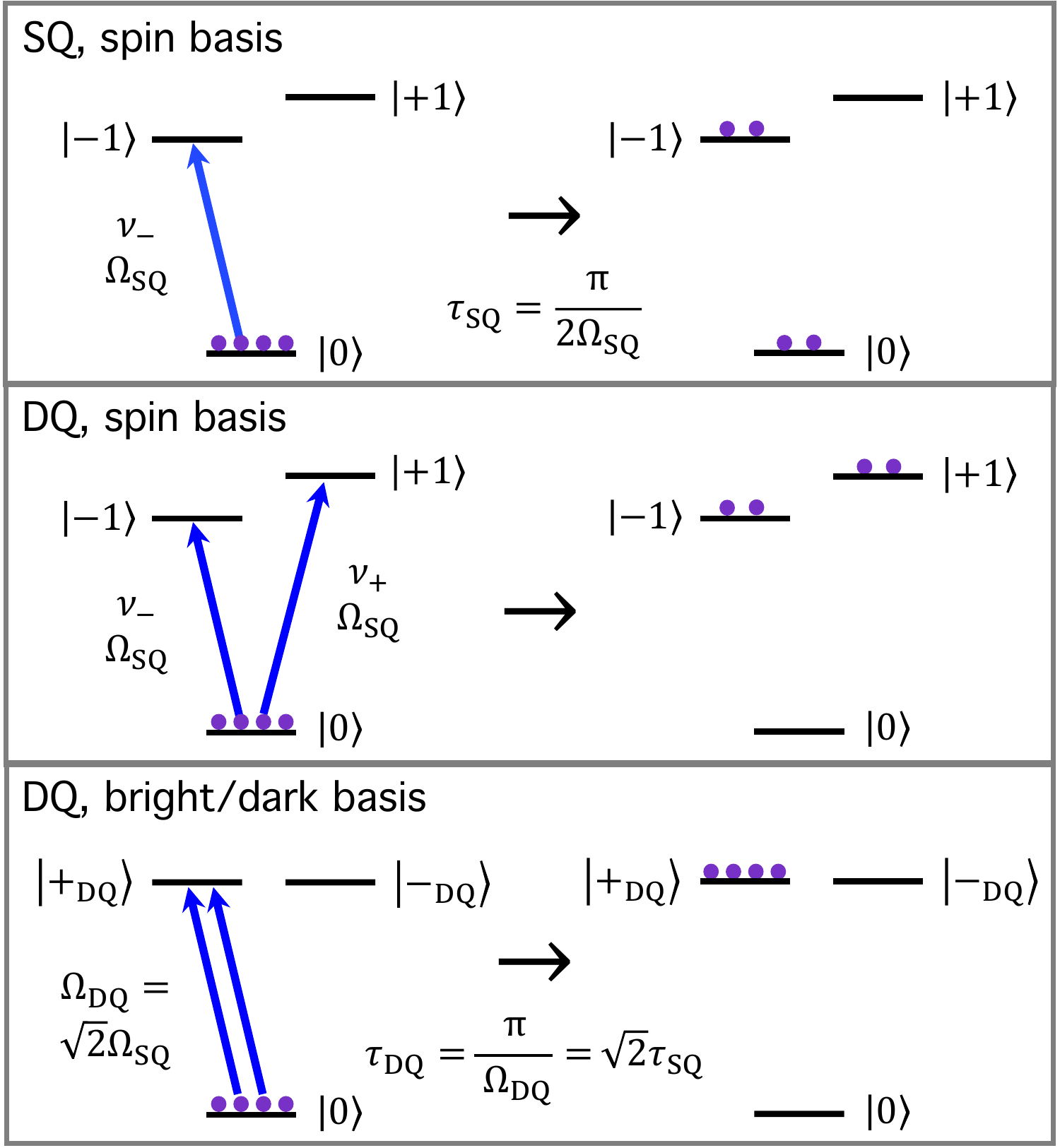}
\put (-5, 95) {\large b}
\end{overpic}
\caption[Double-Quantum Sequence Overview]{\textcolor{black}{a) Pulse sequence for Ramsey-type double-quantum coherence magnetometry, as implemented in Ref.~\cite{Bauch2018}. The single-quantum (SQ) $|0\rangle \!\leftrightarrow \!|\!+\!1\rangle$ and $|0\rangle\! \leftrightarrow \!|\!-\!1\rangle$ transitions are driven by MWs at or near respective resonance frequencies $\nu_+$ and $\nu_-$. For simultaneously applied resonant MWs with Rabi frequencies $\Omega_+ = \Omega_- = \Omega_\text{SQ}$, double-quantum (DQ) transitions occur between $|0\rangle$ and an equal superposition of $|\!+\!1\rangle$ and $|\!-\!1\rangle$ with corresponding DQ Rabi frequency $\Omega_\text{DQ} = \sqrt{2}\Omega_\text{SQ}$. A DQ Ramsey sequence requires MW pulses of duration $\tau_\text{DQ} = \pi/\Omega_\text{DQ} = \pi /(\Omega_\text{SQ}\sqrt{2})$ to prepare the state used for sensing. b) Comparison between a conventional SQ $\pi/2$-pulse on a single transition (top panel) and a DQ $\pi$-pulse used to prepare the superposition state $|+_\text{DQ}\rangle = \frac{1}{\sqrt{2}}(|\!+\!1\rangle\!+\!|\!-\!1\rangle)$ for the sequence in (a) (lower two panels). Middle panel shows the DQ state preparation pulse in the bare spin basis, while the bottom panel shows the same pulse in the basis of $|0\rangle$ and the bright and dark states $|+_\text{DQ}\rangle$ and $|-_\text{DQ}\rangle$, i.e., the orthogonal superposition states respectively coupled to and blind to the MW drive. During a DQ Ramsey free-precession interval, spin population oscillates between $|+_\text{DQ}\rangle$ and $|-_\text{DQ}\rangle$, (as these states are not energy eigenstates), at a rate proportional to the magnetic field. For additional detail see Section 5.2 in Ref.~\cite{schloss2019optimizing}. 
Adapted from Ref.~\cite{schloss2019optimizing}.}}\label{fig:doublequantumsequenceoverview}
\end{figure*}

Standard NV$^\text{-}$ magnetometry techniques, such as CW-ODMR (Sec.~\ref{cwodmr}), pulsed ODMR (Sec.~\ref{pulsedodmr}), and pulsed Ramsey or echo-type schemes (Sec.~\ref{DD}), are typically performed in the pseudo-spin-$\nicefrac{1}{2}$ \textcolor{black}{\textit{single-quantum} (SQ)} \textcolor{black}{subspace} of the NV$^{\text{-}}$ ground state, with the $|m_s=0\rangle$ and either the \textcolor{black}{$|m_s=\!+\!1\rangle$} or the \textcolor{black}{$|m_s=\!-\!1\rangle$} spin state ($\Delta m_s=1$) employed for sensing. In contrast, \textcolor{black}{\textit{double-quantum} (DQ) coherence magnetometry} ($\Delta m_s=2$) works as follows for a Ramsey-type implementation \textcolor{black}{(see Fig.~\ref{fig:doublequantumsequenceoverview})}. First, a\textcolor{black}{n equal superposition of the $|\!+\!1\rangle$ and $|\!-\!1\rangle$ states is prepared (e.g., $|+_\text{DQ}\rangle = \frac{1}{\sqrt{2}}(|\!+\!1\rangle\!+\!|\!-\!1\rangle)$).} 
Then, after a free precession interval, \textcolor{black}{the final population in $\left|+_\text{DQ}\right>$ is mapped back to $|0\rangle$, allowing for a magnetic-field-dependent population difference between $\left|0\right>$ and \textcolor{black}{$\left|-_\text{DQ}\right>$} to be read out optically.} 

Use of the full spin-1 nature of the NV$^\text{-}$ center and the double-quantum basis $\{|\!-\!1\rangle,|\!+\!1\rangle\}$ allows for several sensing advantages. First, at fixed magnetic field, an NV$^\text{-}$ spin prepared in a superposition of the $|\!+\!1\rangle$ and $|\!-\!1\rangle$ states precesses at twice the rate as in the standard \textcolor{black}{SQ \textcolor{black}{subspace} of $\{|0\rangle, |\!-\!1\rangle\}$ or $\{|0\rangle, |\!+\!1\rangle\}$}, enabling enhanced magnetometer sensitivity. Moreover, measurements in the \textcolor{black}{DQ} basis are differential, in that noise sources perturbing the \textcolor{black}{$|0\rangle \!\leftrightarrow\! |\!+\! 1\rangle$ and $|0\rangle\! \leftrightarrow \!|\!- \!1\rangle$} transitions in common-mode are effectively rejected. \textcolor{black}{Sources of common-mode noise may include} temperature fluctuations, which enter the NV$^\text{-}$ Hamiltonian via the zero-field splitting parameter $D$ ($\frac{\partial D}{\partial T} \approx -74$~kHz/K)~\cite{Acosta2010, Toyli2013, Kucsko2013}; axial strain gradients; axial electric fields; \textcolor{black}{and transverse magnetic fields}. For a detailed discussion see Ref.~\cite{Bauch2018}.


If the spin bath environment is dominated by magnetic noise, as is common for high-nitrogen and natural $^{13}$C abundance diamond samples, measurements in the \textcolor{black}{DQ} basis exhibit an increased linewidth and shortened \textcolor{black}{associated dephasing} time, as the $2\times$ enhanced sensitivity to magnetic fields causes the spin \textcolor{black}{ensemble to dephase} twice as quickly \textcolor{black}{as in the} \textcolor{black}{SQ} basis, i.e., $T_{2,\text{DQ}}^* \approx T_{2,\text{SQ}}^*/2$. This \textcolor{black}{increased dephasing and decoherence} is confirmed experimentally for single NV$^\text{-}$ centers by the authors of Ref.~\cite{Fang2013}, who observe a $2 \times$ decrease in $T_2^*$, and by the authors of Ref.~\cite{Mamin2014}, who observe an $\approx\! 2\times$ decrease in the Hahn-echo coherence time $T_2$. Similar results are reported for NV$^\text{-}$ ensembles~\cite{Kucsko2016,Bauch2018}.

\textcolor{black}{In the SQ basis, n}on-magnetic noise sources \textcolor{black}{such as temperature fluctuations, electric field noise, and inhomogeneous strain may also contribute to spin dephasing (see Sec.~\ref{T2*params}). However, \textcolor{black}{values of $T_2^*$} 
in the DQ basis are insensitive to \textcolor{black}{these common-mode} noise sources. 
When such noise 
dominates dephasing in the \textcolor{black}{SQ} basis, the \textcolor{black}{DQ} dephasing time $T_{2,\text{DQ}}$ may exceed  $T_{2,\text{SQ}}$, allowing for additional sensitivity improvement}. 
For example, \textcolor{black}{DQ} measurements reported in Ref.~\cite{Bauch2018} on NV$^\text{-}$ ensembles demonstrate a $\sim 6 \times$ increase in $T_2^*$ (narrowing of linewidth) in an isotopically purified, low-nitrogen diamond, leading to an effective $13\times$ enhancement in phase accumulation per measurement when considering the twice faster precession rate in the \textcolor{black}{DQ} basis. In Ref.~\cite{Bauch2018}, the standard \textcolor{black}{SQ} basis $T_{2}^*$ \textcolor{black}{is found to be} limited by strain inhomogeneities, whereas the $T_2^*$ value measured in the \textcolor{black}{DQ} basis is likely \textcolor{black}{primarily} limited by interactions with residual $^{13}$C nuclear spins ($\sim 100~$ppm). This $T_2^*$ limitation emphasizes the importance of isotopic purification when low-nitrogen samples are employed (see Sec.~\ref{13ClimitT2star}).

\begin{figure*}[ht]
\centering
\begin{overpic}[width = 0.9\textwidth]{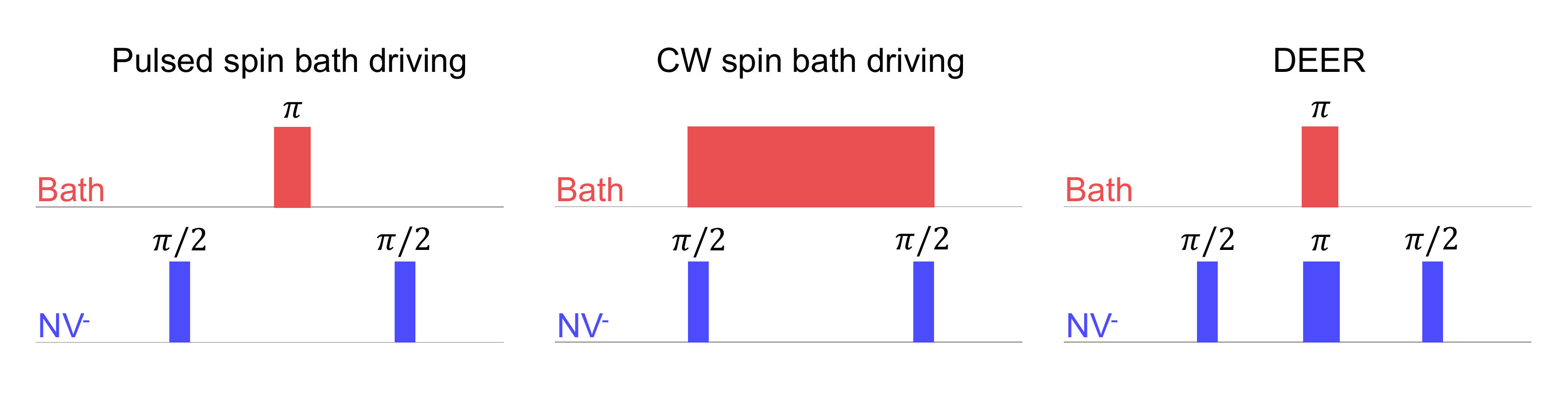}  
  \put (3,23) {\large a}
  \put (35.75,23) {\large b}
  \put (68.5,23) {\large c}
\end{overpic}
\caption[Spin bath pulse sequences]{Selected pulse sequences for concurrent manipulation of NV$^\text{-}$ spins and the surrounding paramagnetic spin bath. a) Pulsed spin bath driving protocol combining a Ramsey sequence on the NV$^\text{-}$ center(s) with a central RF $\pi$-pulse on the spin bath. b) Continuous spin bath driving protocol combining a Ramsey sequence with continuous resonant RF spin bath drive. c) Hahn echo-based double electron-electron resonance (DEER) protocol consisting of a Hahn echo sequence performed on the NV$^\text{-}$ center(s) combined with a resonant RF $\pi$-pulse performed on the spin bath. Recreated from Ref.~\cite{Bauch2018}.}
\label{fig:pulsesequences}
\end{figure*}



For AC magnetometry, dephasing due to strain inhomogeneities and temperature fluctuations can be largely alleviated by using Hahn echo or similar dynamical decoupling sequences (see Sec.~\ref{DD})~\cite{Pham2013thesis}. Nevertheless, double-quantum coherence magnetometry should still yield benefits. \textcolor{black}{First,  ensemble AC magnetometry benefits from the expected $\sqrt{2}\times$ sensitivity gain due to twice faster precession~\cite{Fang2013}. Second, sensitivity may be further enhanced if $T_{2,\text{DQ}}$ exceeds $T_{2,\text{SQ}}/2$. For example, the authors of Ref.~\cite{Angerer2015} observe $T_{2,\text{SQ}} = 1.66\pm 0.16$~ms and $T_{2,\text{DQ}}= 2.36\pm 0.09~$ms for single near-surface NV$^\text{-}$ center with  $T_{2,\text{SQ}}$ likely limited by electric field noise.}
In addition to magnetic sensing, measurements \textcolor{black}{employing the full spin-1 basis} can enhance sensitivity for temperature sensing~\cite{Toyli2013} and noise spectroscopy applications~\cite{Kim2015, Myers2017}. 

\textcolor{black}{As shown schematically in Fig.~\ref{fig:doublequantumsequenceoverview}, i}mplementation of double-quantum coherence magnetometry is a straightforward \textcolor{black}{extension of}  
standard pulsed magnetometry. The DQ technique requires applying MW pulses to drive \textcolor{black}{both the $|0\rangle\!\leftrightarrow\!|\!-\!1\rangle$ and $|0\rangle\!\leftrightarrow\!|\!+\!1\rangle$ transitions}. For sufficiently large magnetic fields, these two transitions must be addressed with separate and phase-locked MW frequencies~\cite{Mamin2014, Angerer2015}. \textcolor{black}{While equal Rabi frequencies on the two transitions are desirable, the MW pulse durations may be adjusted to compensate for unequal Rabi frequencies. MW pulses for each spin transition may  be applied simultaneously~\cite{Fang2013,Mamin2014,Bauch2018}, as depicted in Fig.~\ref{fig:doublequantumsequenceoverview}a, or sequentially~\cite{Toyli2013}.} At low magnetic field, electric field, and strain, a single MW frequency is adequate~\cite{Fang2013}. In either case, care must be taken to ensure that both the upper and lower spin transitions are addressed with adequate MW pulses to achieve an equal superposition of \textcolor{black}{$|\!+\!1\rangle$ and $|\!-\!1\rangle$}~\cite{Mamin2014,Bauch2018}. Due to the minimal increase in experimental complexity, the ability to suppress common-mode noise sources, and the increased spin precession rate, we expect \textcolor{black}{DQ} coherence magnetometry to become standard for \textcolor{black}{high-performance pulsed-measurement DC magnetometers employing NV$^\text{-}$ ensembles}.

\subsection{Spin bath driving}\label{P1driving}


Residual paramagnetic impurity spins in diamond contribute to NV$^\text{-}$ dephasing, thereby reducing $T_2^*$. This effect can be mitigated by directly driving the impurity spins, which is particularly useful when dynamical decoupling (see Sec.~\ref{DD}) of the NV$^\text{-}$ sensor spins is not applicable, such as in DC sensing protocols. \textcolor{black}{This technique, termed spin bath driving,} has been successfully demonstrated with substitutional nitrogen spins N$_\text{S}^0$ ($S=\nicefrac{1}{2}$)~\cite{DeLange2012,Knowles2014,Bauch2018}. Due to the high typical concentrations of N$_\text{S}^0$ spins in NV-rich diamonds, we focus our discussion on this implementation.

In pulsed spin bath driving (see Fig.~\ref{fig:pulsesequences}a), a resonant $\pi$-pulse is applied to the N$_\text{S}^0$ spins halfway through the NV$^\text{-}$ Ramsey sequence, decoupling the N$_\text{S}^0$ spins from the NV$^\text{-}$ spins in analogy with a refocusing $\pi$-pulse in a \textcolor{black}{Hahn} echo sequence (see Fig.~\ref{fig:echo})~\cite{DeLange2012,Bauch2018}. Alternatively, the spin bath can be driven continuously (see Fig.~\ref{fig:pulsesequences}b)~\cite{DeLange2012,Knowles2014,Bauch2018}. In the latter case, the driving Rabi frequency $\Omega_\text{N}$ must significantly exceed the NV$^\text{-}$-N$_\text{S}^0$ coupling rate $\gamma_\text{N}$ \textcolor{black}{(i.e.,~satisfy $\Omega_\text{N}/\gamma_\text{N} \gg 1$)} to achieve effective decoupling. \textcolor{black}{[Note that $\gamma_\text{N} \sim 2\pi \times (0.01 - 10)~$MHz for nitrogen concentrations in the $1 - 1000~$ppm range (see Sec.~\ref{nitrogenlimitT2*}).]} Under this condition, the nitrogen spins undergo many Rabi oscillations during the characteristic dipolar interaction time $1/\gamma_\text{N}$. As a result, the NV$^\text{-}$ ensemble is decoupled from the nitrogen spin bath and the NV$^\text{-}$ dephasing time is enhanced. This phenomenon is similar to motional narrowing observed in many NMR and ESR systems, such as \textcolor{black}{rotation- and diffusion-induced time-averaging of} magnetic field imhomogeneities ~\cite{Abragam1983, Slichter1990}.

\begin{figure}[t]
\hspace{-6pt}\begin{overpic}[width = \columnwidth]{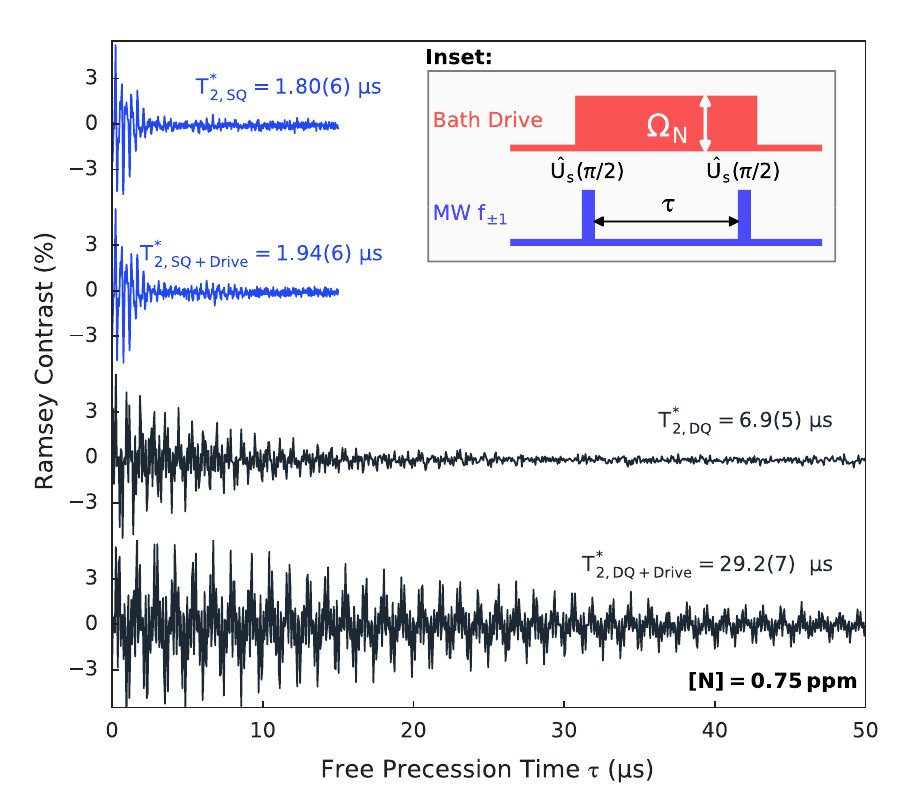}
\end{overpic}
\caption[$T_2^*$ extension using P1DQ]{Ensemble free induction decay envelopes as measured using \textcolor{black}{SQ} and \textcolor{black}{DQ} Ramsey magnetometry, with and without continuous spin bath driving. Measurements are shown for the following: in the \textcolor{black}{SQ} basis without spin bath driving (\textcolor{blue}{\rule[.5mm]{3mm}{1pt}}, first from top); in the \textcolor{black}{SQ} basis with spin bath driving (\textcolor{blue}{\rule[.5mm]{3mm}{1pt}}, second from top); in the \textcolor{black}{DQ} basis without spin bath driving (\textcolor{black}{\rule[.5mm]{3mm}{1pt}}, third from top); in the \textcolor{black}{DQ} basis with spin bath driving (\textcolor{black}{\rule[.5mm]{3mm}{1pt}}, fourth from top). \textcolor{black}{Measurements in the \textcolor{black}{DQ} basis mitigate strain-induced dephasing, while spin bath driving mitigates dipolar dephasing from the paramagnetic substitutional nitrogen in the diamond. The data illustrate the synergistic effect of combining \textcolor{black}{DQ} coherence magnetometry and spin bath driving; the aggregate approach vastly outperforms either technique employed independently.} Even with twice faster precession, $T_2^*$ is extended from 1.8~$\upmu$s to 29~$\upmu$s. \textcolor{black}{The DQ protocol with spin-bath driving is depicted in the inset.} From Ref.~\cite{Bauch2018}.}  \label{fig:bauch2018fig3}
\end{figure}

The authors of Ref.~\cite{DeLange2012} perform pulsed spin bath driving in a diamond with $[\text{N}^\text{T}] \lesssim 200~$ppm and increase $T_2^*$ for a single NV$^\text{-}$ $1.6\times$, from 278~ns to 450~ns. Similarly, in Ref.~\cite{Knowles2014}, $T_2^*$ for an individual NV$^\text{-}$ is extended from 0.44~$\upmu$s to 1.27~$\upmu$s, a 2.9$\times$ improvement, using continuous spin bath driving in nanodiamonds with $[\mathrm N] \lesssim 36~$ppm. An NV$^\text{-}$ ensemble study in Ref.~\cite{Bauch2018} finds that if another mechanism, such as lattice strain or magnetic field gradients, is the dominant source of dephasing, spin bath driving becomes less effective, as shown in Fig.~\ref{fig:bauch2018fig3} (see also Sec.~\ref{T2*params}). Nonetheless, at high nitrogen concentrations ($[\text{N}_\text{S}^\text{T}] \gtrsim 1~$ppm), NV$^\text{-}$ ensemble dephasing due to dipolar interaction with nitrogen spins can be greatly reduced by spin bath driving~\cite{Bauch2018}, \textcolor{black}{as also demonstrated in single-NV$^\text{-}$ experiments~\cite{DeLange2012, Knowles2014}.}

\begin{figure*}[bth]
\centering
\begin{overpic}[height=3.5 in]{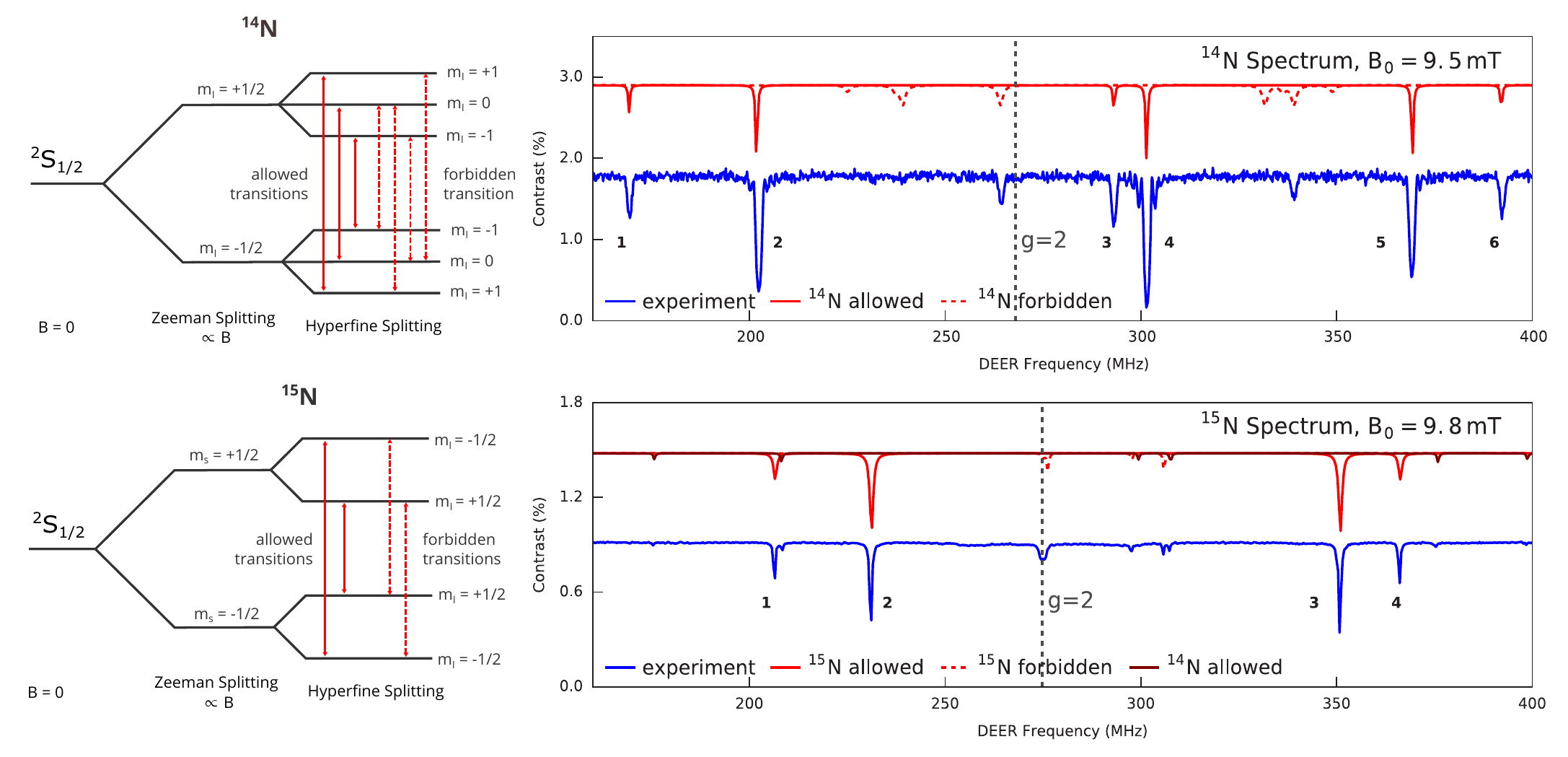}
\end{overpic}
\caption[$^{14}$N and $^{15}$N transitions and DEER]{Substitutional nitrogen N$_\text{S}^0$ spin energy levels (left panes) and associated double electron-electron resonance (DEER) spectra (right panes), for $^{14}$N (top) and $^{15}$N (bottom). Simulated spectra depict allowed-transition resonances ($\Delta m_I = 0$) of the primary nitrogen isotope ({\color{red}{\rule[.5mm]{3mm}{1pt}}}), forbidden-transition resonances ($\Delta m_I \neq 0$) of the primary nitrogen isotope ({\color{red}{\rule[.5mm]{1mm}{1pt}\;\rule[.5mm]{1mm}{1pt}\;\rule[.5mm]{1mm}{1pt}}}), and spurious features associated with allowed transitions of impurity isotopes ({\color{Maroon}{\rule[.5mm]{3mm}{1pt}}}). The simulated data resonance linewidths and amplitudes are chosen to approximately match the experimental data ({\color{blue}{\rule[.5mm]{3mm}{1pt}}}). Spectra are simulated for and experimentally measured in an external magnetic field aligned along the diamond crystallographic [111] axis. Adapted from Ref.~\cite{Bauch2018}.
}
\label{fig:deerspectra}
\end{figure*}

To effectively suppress NV$^\text{-}$ dephasing, all nitrogen spin transitions must typically be driven. \textcolor{black}{Elemental nitrogen occurs in two stable isotopes, $^{14}$N with 99.6$\%$ natural isotopic abundance, and $^{15}$N with 0.4$\%$ natural isotopic abundance. Diamonds may \textcolor{black}{contain predominantly} 
$^{14}$N, where the 99.6$\%$ natural abundance purity is typically deemed sufficient, or $^{15}$N, which requires isotopic purification. $^{14}$N exhibits nuclear spin $I=1$ while $^{15}$N exhibits nuclear spin $I=\nicefrac{1}{2}$, resulting in 3 and 2 magnetic-dipole-allowed transitions for each isotope, respectively~\cite{Smith1959, Cook1966, Loubser1978}.} 
Like NV$^\text{-}$ centers, substitutional nitrogen defects possess a trigonal symmetry as a result of a Jahn-Teller distortion~\cite{Davies1979, Davies1981, Ammerlaan1981}. The Jahn-Teller distortion defines a symmetry axis along any of the 4 
crystallographic $[111]$ axes, leading to 4 
groups of N$_\text{S}^0$ spins. \textcolor{black}{For an axial bias magnetic field $B_0$ satisfying $g_e\mu_B / \hbar B_0 \gg A_\text{HF}$ where $A_\text{HF}\sim 100$ MHz is the substitutional nitrogen hyperfine interaction}, $m_s$ and $m_I$ are good quantum numbers, and the $^{14}$N spectrum consequently exhibits up to 12 distinct resonances, \textcolor{black}{each of which needs} to be driven~\cite{DeLange2012, Belthangady2013}. If $B_0$ is aligned with any of the diamond [111] axes, the 12 resonances reduce to 6 partially-degenerate groups with \textcolor{black}{multiplicity} 1:3:1:3:3:1 (see Fig.~\ref{fig:deerspectra}a). Similarly, the $^{15}$N spectrum shows up to 8 distinct resonances, which reduce to 4 partially-degenerate groups with \textcolor{black}{multiplicity} 1:3:3:1 \textcolor{black}{for} $B_0$ \textcolor{black}{aligned to an NV internuclear axis} (see Fig.~\ref{fig:deerspectra}b). 

\textcolor{black}{Spin-bath driving} is expected to be easiest to execute when the bias magnetic field $B_0$ and hyperfine coupling $A_\text{HF}$ are not of the same order. When $\frac{g \mu_B}{h} B_0 \sim A_\text{HF}$, additional nuclear-spin-non-conserving transitions arise, \textcolor{black}{resulting in reduced oscillator strength for the nuclear-spin-conserving transitions. Thus,} given fixed RF power, the drive efficiency for each addressed transition decreases. Although spin bath driving has to date only been demonstrated in the regime $\frac{g \mu_B}{h} B_0\gtrsim A_\text{HF}$~\cite{DeLange2012, Knowles2014, Bauch2018}, \textcolor{black}{we expect} driving in the $\frac{g \mu_B}{h} B_0\ll A_\text{HF}$ regime to \textcolor{black}{also} be effective.

The N$_\text{S}^0$ electron spin resonance spectra for $^{14}$N and $^{15}$N are readily observed in EPR experiments [see for example Ref.~\cite{Smith1959} and~\cite{Drake2016}]. Alternatively, the \textcolor{black}{nitrogen} resonance spectra in a diamond can be characterized with NV$^\text{-}$ centers using a \textcolor{black}{Hahn-echo-based} double electron-electron resonance (DEER) technique~\cite{DeLange2012,Bauch2018}. In this \textcolor{black}{case}, the NV$^\text{-}$ electronic spin is made sensitive to decoherence from N$_\text{S}^0$ target \textcolor{black}{impurity} spins via application of frequency-selective $\pi$-pulses at the \textcolor{black}{targeted spins'} resonance frequency. A schematic of the DEER pulse sequence is shown in Fig.~\ref{fig:pulsesequences}c, and the resulting DEER spectra for both nitrogen isotopes are compared in Fig.~\ref{fig:deerspectra}. Extra resonance features associated with substitutional-nitrogen-related dipole-forbidden transitions and additional paramagnetic spins are also commonly observed and may reveal additional sources of dephasing. 

\begin{figure}[ht]
\centering
\begin{overpic}[height=2.7 in]{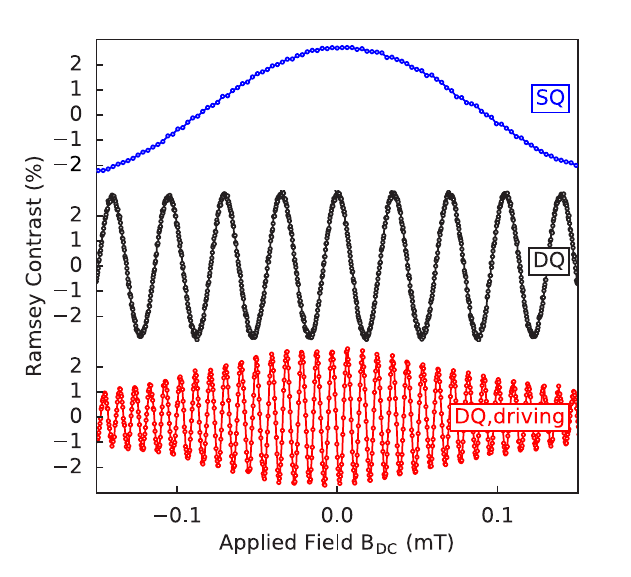}
\end{overpic}
\caption[P1DQ magnetometry fringes]{Ramsey interference fringes versus an applied test magnetic field, measured in the \textcolor{black}{SQ} basis (\textcolor{blue}{\rule[.5mm]{3mm}{1pt}}, top), \textcolor{black}{DQ} basis (\textcolor{black}{\rule[.5mm]{3mm}{1pt}}, middle), and \textcolor{black}{DQ} basis with N$_\text{S}^0$ spin bath driving (\textcolor{red}{\rule[.5mm]{3mm}{1pt}}, bottom). The longer dephasing times achieved when combining \textcolor{black}{DQ} coherence magnetometry and spin bath driving allow for denser Ramsey fringes and enhanced sensitivity. \textcolor{black}{The decreased contrast for magnetic fields $>\!0.05$ mT in the bottom plot results from magnetic-field-induced detuning of the nitrogen spin resonances with respect to the RF drive frequencies.} From Ref.~\cite{Bauch2018}.}  \label{fig:ramseyinterferencefringesp1dq}
\end{figure}

The experimental requirements for effective spin bath driving depend on the substitutional nitrogen concentration. \textcolor{black}{At lower impurity concentrations, reduced} spin bath drive strength (i.e., RF power) is needed to mitigate nitrogen-induced dephasing. However, dephasing mechanisms \textcolor{black}{unrelated to nitrogen} may exhibit larger relative contributions to $T_2^*$ in this regime, limiting the achievable $T_2^*$ increase from nitrogen spin bath driving. 
In particular, in samples with nitrogen content $[\mathrm N_\text{S}^0] \lesssim 1~$ppm, lattice-strain gradients may dominate the ensemble dephasing time, as is found in Ref.~\cite{Bauch2018}. In this instance, strain-insensitive measurement techniques, such as double-quantum coherence magnetometry (see Sec.~\ref{DQ}) must be employed for spin bath driving to extend $T_2^*$. In the intermediate regime ($[\mathrm N_\text{S}^0] \sim 1~$ppm), where strain gradients and NV$^\text{-}$ dipolar interactions with the nitrogen spin bath are of similar magnitude, neither spin bath driving nor \textcolor{black}{DQ} coherence magnetometry alone can achieve significant enhancement of the dephasing time. However, Ref.~\cite{Bauch2018} demonstrates a $\sim 16\times$ improvement in $T_2^*$ (effectively a $\sim 32\times$ improvement when considering the twice faster precession rate in the \textcolor{black}{DQ} basis) for a $[\text{N}_\text{S}^0] \simeq~0.75~$ppm diamond when both techniques are combined, as shown in Fig.~\ref{fig:ramseyinterferencefringesp1dq}. \textcolor{black}{In contrast, employing spin bath driving alone improves the dephasing time only by $\sim\!1.1\times$ (see Fig.~\ref{fig:bauch2018fig3}), as strain-induced dephasing is left unmitigated.}

In nitrogen-rich diamonds ($[\text{N}_\text{S}^0] \gtrsim 1~$ppm), achieving the motional narrowing condition $\Omega_\text{N}/\gamma_\text{N} \gg 1$ may be technically difficult; increases in [$\text{N}_\text{S}^0$] necessitate linear proportional increases in $\Omega_\text{N}$, which correspond to quadratic increases in the RF power required ~\cite{Bauch2018}. 
We  expect both pulsed and continuous spin bath driving in nitrogen-rich samples to be ultimately limited by parasitic effects. These effects include induced AC Zeeman shifts, strain gradients, and sample heating due to the strong applied RF fields~\cite{Knowles2014}.

We note that spin bath driving should be applicable to any paramagnetic spin species in diamond, such as \textcolor{black}{dark electron spins (i.e., non-optically-active paramagnetic defects)}, NV$^\text{-}$ centers, or even nuclear spins. The effectiveness of the driving for dilute bath spins \textcolor{black}{(fractional concentration $\ll0.01$)} is expected to depend on the target spin's concentration but not its gyromagnetic ratio, as both $\Omega_\text{N}$ and $\gamma_\text{N}$ vary linearly with the target spin's gyromagnetic ratio. \textcolor{black}{In other words, species with small gyromagnetic ratios are difficult to drive but also do not contribute much to dephasing for a given concentration~\cite{Bauch2018}.}
\textcolor{black}{Given its relatively high concentration}, spin bath driving of the $^{13}$C in a natural abundance diamond ([$^{13}$C]$=10700$~ppm) is expected to be quite challenging~\cite{Bauch2019}. \textcolor{black}{Lastly, as the nitrogen spin bath contributes to $T_2$ decoherence~\cite{Pham2012a,BarGill2012}, nitrogen spin bath driving would be expected to extend the Hahn echo $T_2$ and, to a lesser extent, coherence times achieved with dynamical decoupling sequences~\cite{Bauch2018}, although neither application has been demonstrated at present.}

\begin{figure}[ht]
\begin{overpic}[width=0.46\textwidth]{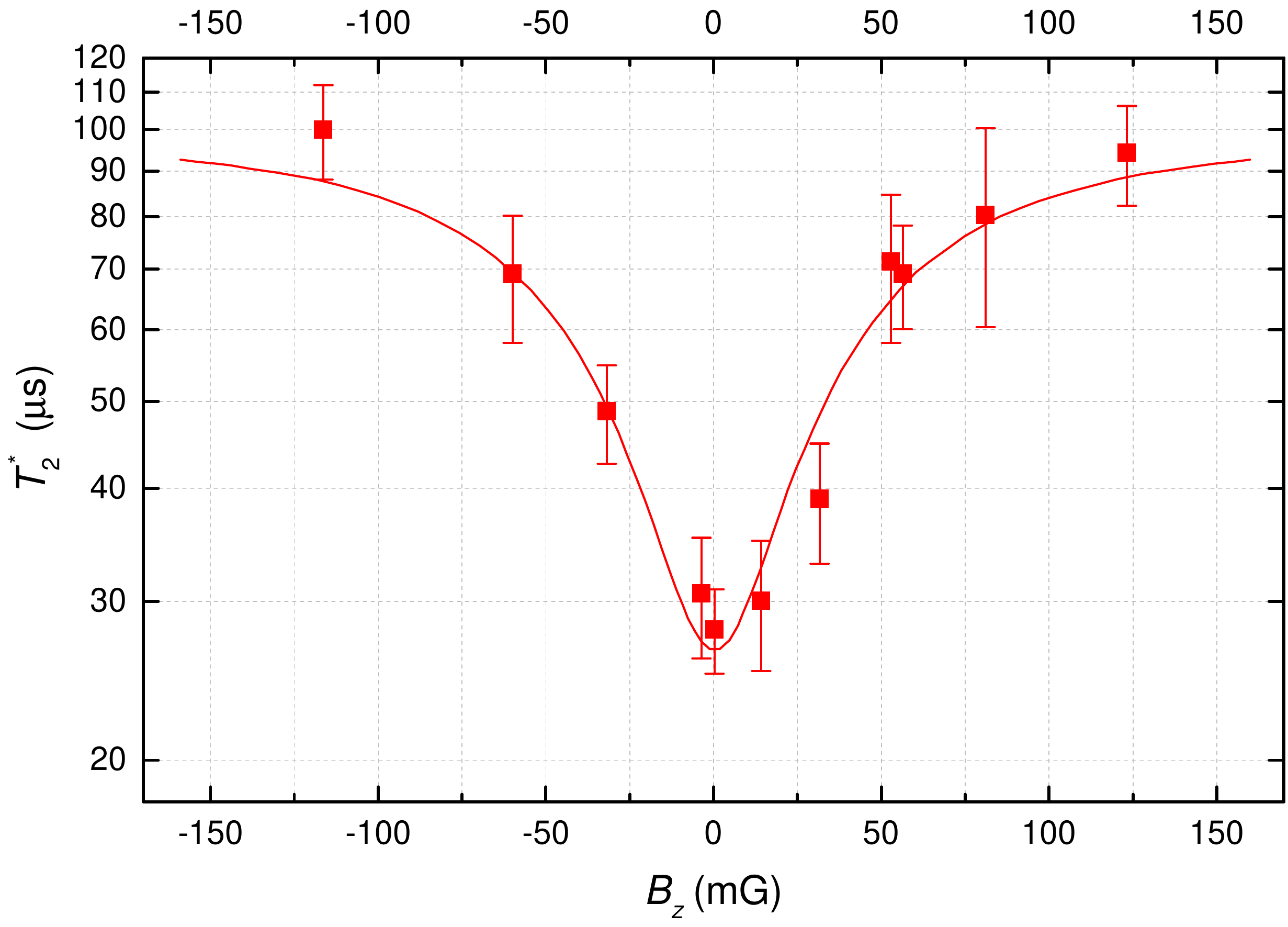}
\end{overpic}
\caption[Electric field and strain noise suppression]{Suppression of dephasing from transverse electric fields and strain. For the single NV$^\text{-}$ center measured in Ref.~\cite{Jamonneau2016}, the dephasing time ${T_2^*}^\text{\{single\}}$ at zero magnetic field is limited by electric field fluctuations transverse to the NV$^\text{-}$ symmetry axis, $E_x$ and $E_y$. An applied axial magnetic field suppresses this source of dephasing by decoupling the NV$^\text{-}$ center from transverse electric fields and strain. For magnetic fields larger than $\sim 100$~mG, the value of ${T_2^*}^\text{\{single\}}$ is limited by magnetic noise, reaching $\sim 100~\upmu$s in this isotopically enriched, [$^{13}$C] = 20~ppm sample. For NV$^\text{-}$ ensembles at zero magnetic field, \textcolor{black}{in addition to temporal fluctuations in $E_x$ and $E_y$ that limit ${T_2^*}^\text{\{single\}}$, spatial variations in $E_x$ and $E_y$ and in the transverse spin-strain coupling terms $\mathscr{M}_x$ and $\mathscr{M}_y$ may also limit $T_2^*$ for ensembles}. 
Recreated from Ref.~\cite{Jamonneau2016}.}
\label{fig:electricfieldnoise}
\end{figure}

\subsection{Transverse strain and electric field mitigation}\label{efieldsuppression}

Spatial and temporal variations in electric fields or in diamond crystal strain can degrade $T_2^*$, as described in Secs.~\ref{T2*params} and \ref{DQ}. Measurements performed in the NV$^\text{-}$ spin's double-quantum basis are insensitive to variations in the \textit{axial} components of the electric field $E_{z}$ and spin-strain coupling $\mathscr{M}_{z}$, as these terms cause common-mode shifts in the \textcolor{black}{energies of the $|+1\rangle$ and $|-1\rangle$ states}~\cite{Barson2017,Glenn2017}. In contrast, broadening due to \textit{transverse} electric fields $E_{x}$, $E_{y}$ and transverse spin-strain couplings $\mathscr{M}_x$ and $\mathscr{M}_y$ may remain in \textcolor{black}{DQ} measurements~\cite{Barson2017,udvarhelyi2018spinstrain}. However, by operating at a sufficiently strong axial bias magnetic field $B_{0,z}$, the resonance line broadening from \textcolor{black}{inhomogeneities in} $E_{x}$, $E_{y}$, $\mathscr{M}_x$, and $\mathscr{M}_y$ can be mitigated~\cite{Jamonneau2016,Schloss2018}, as illustrated in Fig.~\ref{fig:electricfieldnoise} and discussed further in Appendix~\ref{StarkZeeman}.

\textcolor{black}{The frequency shifts of the NV$^\text{-}$ ground state spin resonances due to transverse strain and electric fields at zero magnetic field are given by}
\begin{equation}
\pm \xi_\perp = \pm \sqrt{\left(d_\perp E_x + \mathscr{M}_x\right)^2 + \left(d_\perp E_y + \mathscr{M}_y\right)^2},
\end{equation}
where $d_\perp = 0.17$ Hz/(V/m)~\cite{VanOort1990, Dolde2011,michl2019robust} is the transverse electric dipole moment of the NV$^\text{-}$ ground state spin. Application of an external axial magnetic field $B_{0,z}$ \textcolor{black}{introduces additional magnetic-field-dependent shifts and suppresses the effect of $\pm \xi_\perp$ on the spin resonances. When} $ \beta_z  \equiv (g_e \mu_B/h)B_{0,z} \gg \xi_\perp$, contributions to $T_2^*$ from temporal fluctuations or spatial variations in $\xi_\perp$~\cite{Jamonneau2016, Fang2013} are diminished (see Appendix.~\ref{StarkZeeman}). For the nanodiamonds characterized in Ref.~\cite{Jamonneau2016}, with $\xi_\perp = 7$~MHz, $B_{0,z} \sim 30$~G is required to \textcolor{black}{suppress} the contribution to $T_2^*$ from transverse electric fields and strain. For the lower-strain bulk diamonds used in Refs.~\cite{Fang2013, Jamonneau2016}, with $\xi_\perp \sim 10$ kHz, $B_{0,z} \lesssim 100$~mG is sufficient.



%% file: sec04.tex
\section{Methods to increase readout fidelity}\label{fidelityimprovementmethods}

\subsection{Spin-to-charge conversion readout}\label{SCCR}

Spin-to-charge conversion (SCC) readout is an alternative to conventional fluorescence-based readout \textcolor{black}{of the NV$^\text{-}$ spin state.} The technique has been demonstrated for single NVs~\cite{Shields2015,Hopper2016,jaskula2017improved,Ariyaratne2018} and for small ensembles in nanodiamonds~\cite{hopper2018amplified} and bulk diamond~\cite{Jayakumar2018}. In SCC readout, the NV$^\text{-}$ center's spin state is mapped optically onto the NV's neutral and negative charge states (NV$^0$ and NV$^\text{-}$). The charge state, and thus the original NV$^\text{-}$ spin information, can then be accurately read out by exploiting differences in the NV$^0$ and NV$^\text{-}$ wavelength-dependent excitation and associated fluorescence~\cite{Aslam2013,Waldherr2011}. Key advantages \textcolor{black}{of SCC readout} over conventional spin-state-dependent fluorescence readout are: (i), a slightly increased spin contrast~\cite{jaskula2017improved}; and (ii), the ability to read out the charge state for extended durations and thus collect more photons per readout, leading to \textcolor{black}{high-fidelity charge-state determination.} Larger photon-numbers per readout reduce the relative contribution of shot noise to the measurement, allowing for readout fidelities within order unity of the spin-projection limit $\sigma_R = 1$ (see definition in Sec.~\ref{magneticfieldsensitivity}).

\begin{figure}[ht]
\centering
\hspace{1.3cm}\begin{overpic}[width=0.60\columnwidth]{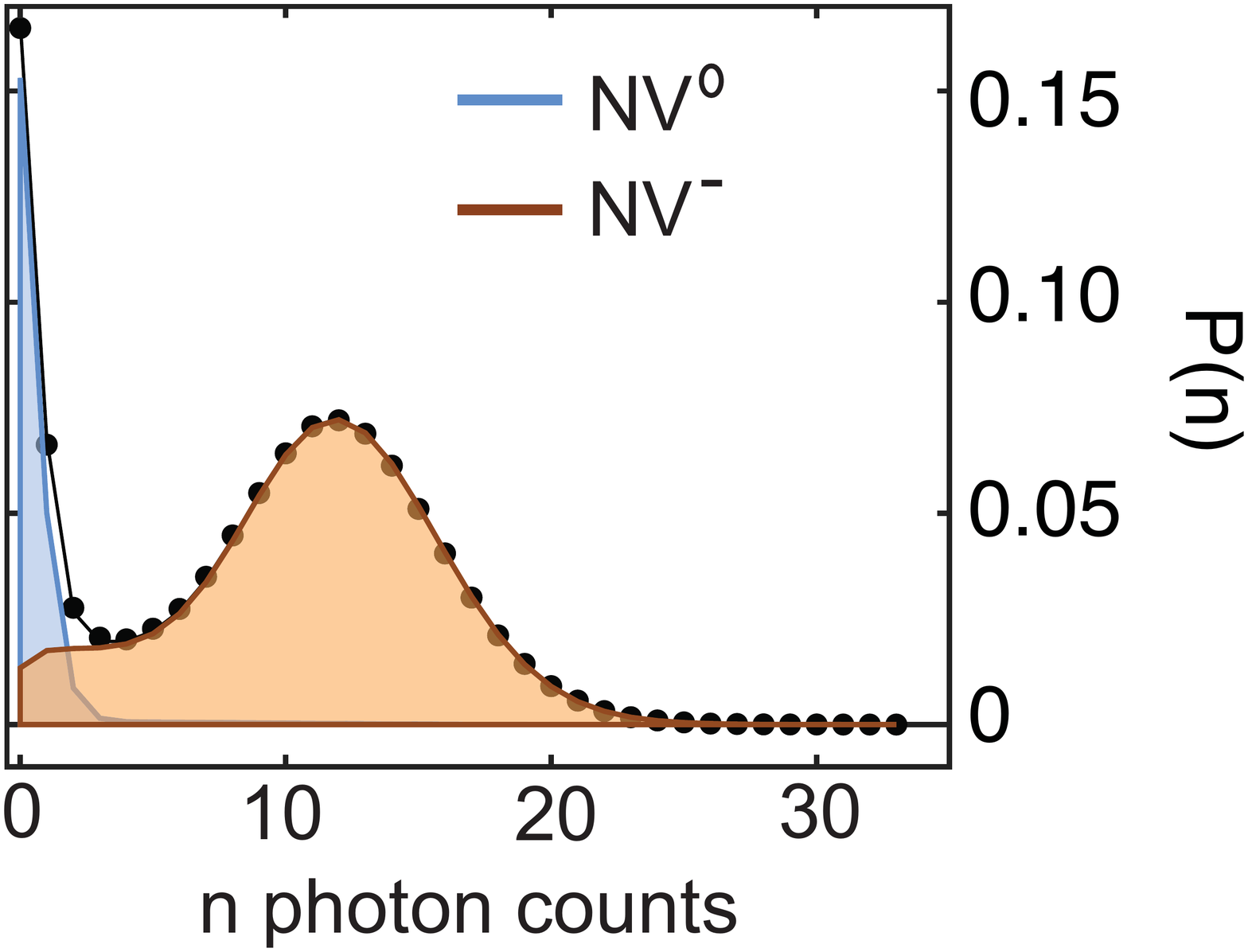}  
\put (-40, 65) {\large a}
\end{overpic}
\centering
\begin{overpic}[width=0.90\columnwidth]{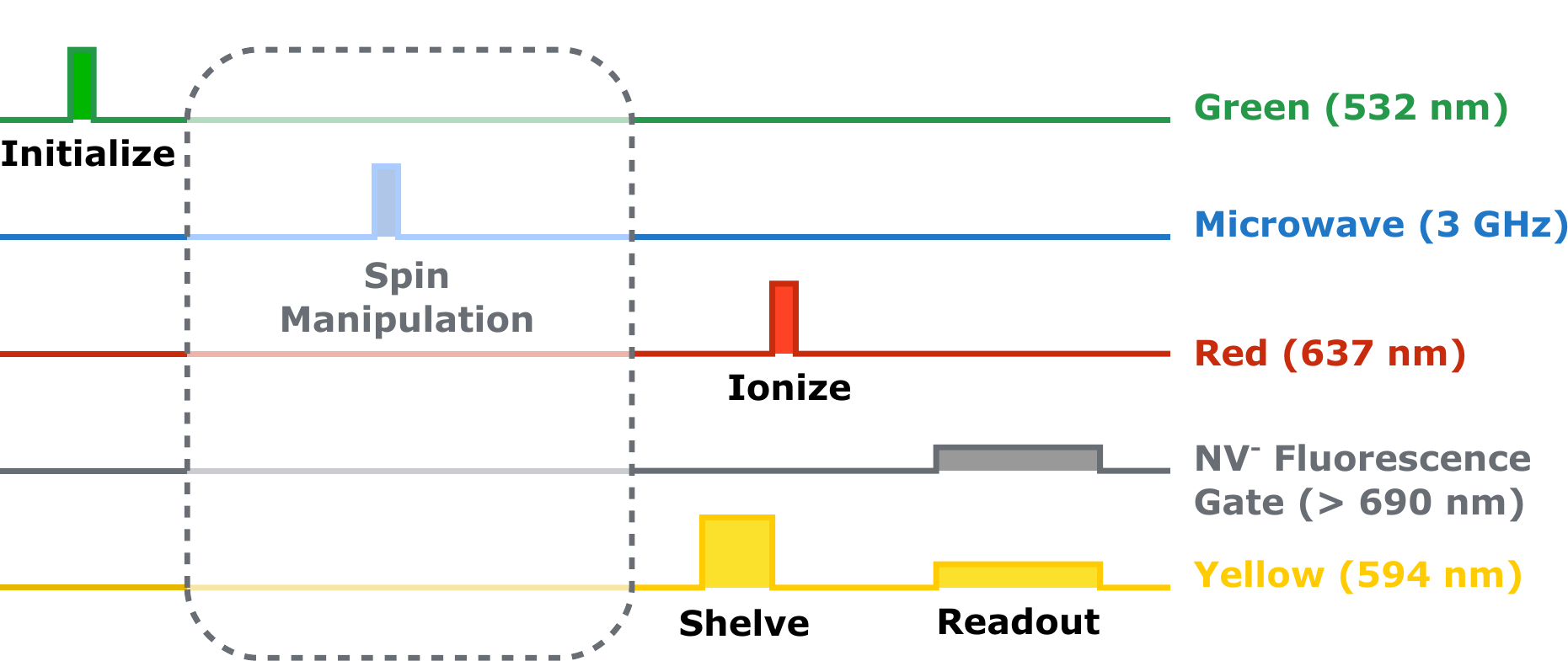}  
\put (-2, 44) {\large b}
\end{overpic}
\caption[SCC readout]{
a) \textcolor{black}{Probability histogram depicting photon emission from NV$^0$ and NV$^\text{-}$ under weak yellow excitation. The striking difference in photon emission rate between NV$^\text{0}$ and NV$^\text{-}$ allows the NV charge state to be determined with fidelity $\gtrsim 99\%$. Adapted from Ref.~\cite{bluvstein2019identifying}.}
b) Schematic of the SCC readout protocol used by Shields \textit{et al.}~\cite{Shields2015}. Adapted from Ref.~\cite{jaskula2017improved}.}\label{fig:SCC}
\end{figure}

Successful spin-to-charge conversion requires control of the NV charge state. Characterization of charge dynamics under optical excitation~\cite{Aslam2013, Beha2012,Hacquebard2018, Manson2018} indicate power- and wavelength-selective photo-ionization processes, which allow for controlled switching between NV$^\text{-}$ and NV$^0$. For example, green $\sim\!532~$nm light transfers single NV centers preferentially to NV$^\text{-}$ with $70-75\%$ probability~\cite{Waldherr2011,Beha2012,Aslam2013}; strong yellow $\sim\!589~$nm~\cite{hopper2018amplified} or red $\sim \!637~$nm~\cite{Shields2015,jaskula2017improved} light can selectively ionize NV$^\text{-}$ to NV$^0$ via absorption of two photons by an electron in the triplet ground state; and near-infrared $\sim \!900\>\text{-}1000$~nm~\cite{Hopper2016} can similarly ionize NV$^{\text{-}}$ via absorption of two photons by an electron in the singlet metastable state. Readout of the NV$^\text{-}$ charge state is commonly performed by applying weak yellow laser light at $\sim \!594~$nm. At \textcolor{black}{intensities} well below the NV$^\text{-}$ saturation intensity $I_\text{sat} \sim \!1\>$-$\>3$~mW/$\upmu\text{m}^2$~\cite{Wee2007}, yellow light efficiently excites the NV$^\text{-}$ electronic spin transition with zero phonon line (ZPL) at 637~nm without \textcolor{black}{inducing ionization,} while hardly exciting the NV$^0$ transition (with ZPL at 575~nm)~\cite{Waldherr2011,Beha2012,Aslam2013}. \textcolor{black}{Through introduction of a photon-detection threshold combined with appropriate spectral filtering, NV$^\text{-}$ (which fluoresces under the yellow excitation) may thus be distinguished from NV$^\text{0}$ (which produces little if any fluorescence).} Figure~\ref{fig:SCC}a displays a photon-count histogram characteristic of single-NV charge readout reproduced from Ref.~\cite{bluvstein2019identifying}. The clear separation of photon distributions from NV$^0$ and NV$^\text{-}$ at low excitation powers \textcolor{black}{allows charge-state determination with fidelity $> 99\%$~\cite{hopper2018spin}.}

The original work by Ref.~\cite{Shields2015} demonstrates SCC readout on a single NV center in Type IIa diamond nanobeams (see Appendix \ref{diamondtypeclassification} for overview of diamond types). First, utilizing green laser light (see Fig.~\ref{fig:SCC}b) and appropriate MWs, the NV center is prepared in the $m_s=0$ or one of the $m_s=\pm1$ spin states of the NV$^\text{-}$ triplet electronic ground state. A moderate power, 594~nm yellow ``shelving'' pulse ($145~\upmu$W, $\sim \!0.9~$mW/$\upmu\text{m}^2$) then excites the spin population to the triplet excited state. Due to the spin-dependent intersystem crossing from the triplet excited state, the $m_s = 0$ population is more likely to decay back to the ground state, whereas the $m_s = \pm 1$ population is more likely to be shelved into the metastable singlet states. The spin-to-charge conversion is then realized with a $\sim \!10~$ns high intensity resonant 637-nm pulse ($22.5~$mW, $\sim \!140~$mW/$\upmu\text{m}^2$), \textcolor{black}{which ionizes (i.e., converts NV$^\text{-}$ to NV$^0$)} the triplet ground state population, corresponding to $m_s=0$, but leaves the shelved population corresponding to $m_s=\pm 1$ unaffected. \textcolor{black}{Last, the NV charge state is read out by applying weak $\sim\!594$~nm light. The $\sim\!594$~nm light with lower energy than the NV$^\text{0}$ ZPL at 575~nm, ensures that only NV$^\text{-}$ is excited while the weak intensity ($\sim \!1 - 10~\upmu$W, $\sim \!6 - 60~\upmu$W/$\upmu\text{m}^2$, Fig.~\ref{fig:SCC}a) ensures that NV$^\text{-}$ is not ionized during readout.} 


The single-NV SCC result by Ref.~\cite{Shields2015} achieves \textcolor{black}{a factor over spin projection noise} $\sigma_R = 2.76$ ($\mathcal F = 1/\sigma_R=0.36$, see comparison in Table~\ref{tab:parameters}). As the fidelity of the charge readout process itself approaches unity ($\mathcal F^\text{CR} = 0.975$), the dominant inefficiency is attributed to the imperfect spin-to-charge conversion step ($\mathcal F^\text{SCC} = 0.37$). Several alternative SCC readout variants 
have been \textcolor{black}{demonstrated, providing similar sensitivity gains while offering reduced experimental complexity ~\cite{hopper2018amplified}, or utilizing the singlet state for ionization~\cite{Hopper2016}. For all SCC readout implementations, however, the improved values of $\sigma_R$} come at the cost of substantially prolonged spin readout times $t_R$, which increase the sequence's overhead time and diminish the overall sensitivity improvement (see Sec.~\ref{magneticfieldsensitivity}). For example, the best reported readout fidelity ($\mathcal F = 0.36$)~\cite{Shields2015} is achieved for readout times $t_R=700~\upmu$s, which \textcolor{black}{exceed} conventional fluorescence-based readout times ($t_R \sim \!300$~ns) by $\sim\!1000 \times$. SCC readout is therefore most advantageous for measurement modalities with long sensing intervals (e.g., $T_1$ relaxometry and AC field sensing), where the penalty due to additional readout overhead is less severe. To date, the best SCC readout demonstrations improve field sensitivity only when interrogation times exceed $\sim \!10~\upmu$s~\cite{Shields2015,hopper2018amplified}, which further motivates improvement of spin ensemble properties to achieve sufficiently long \textcolor{black}{dephasing} times (see Sec.~\ref{T2*improvement}).


\textcolor{black}{Given the clear success of SCC readout with single NVs, application to NV$^\text{-}$-rich ensembles is a logical progression, especially given the low conventional readout fidelities achieved for NV$^\text{-}$-rich ensembles ($\mathcal{F}\lesssim 0.015$, see Table~\ref{tab:parameters}). However, the prospect for SCC readout to substantially improve $\mathcal{F}$ in NV$^\text{-}$ ensembles likely hinges on whether the additional complex charge dynamics present in NV-rich diamonds can be mitigated~\cite{hopper2018amplified}. Promising SCC readout results on small NV ensembles in Type Ib nanodiamonds demonstrate $\sigma_R\!=\!20$, compared to $\sigma_R\!=\!70$ with conventional readout in the same setup, allowing the authors to observe improved sensing performance for interrogations times > 6~$\upmu$s~\cite{hopper2018amplified}. However, this and other studies~\cite{choi2017depolarization,Manson2018} report intricate NV$^\text{-}$ and NV$^0$ charge dynamics absent in single NV experiments. The effectiveness of SCC readout in the complex charge environment inherent to NV-rich ensembles (e.g., due to ionization and charge dynamics of substitutional nitrogen and other impurity defects) warrants further investigation  (see Sec.~\ref{chargestate}). Nevertheless, SCC readout overcomes one sensing disadvantage specific to ensembles, namely that NV$^\text{-}$ orientations not being used for sensing can be preferentially transferred to NV$^0$ during the ionization step. This results in reduced background fluorescence and potentially allows for an additional $\sim\! 2\times$ sensitivity improvement relative to conventional NV$^\text{-}$ readout. Overall, beyond the long overhead times already discussed, SCC readout's demanding power requirements are expected to further hamper ensemble-based implementation. In particular, high required optical intensities ($\gtrsim 150~$mW/$\upmu\text{m}^2$)~\cite{Shields2015,Hopper2016,hopper2018amplified} suggest scaling of SCC readout to larger bulk sample sizes ($\gtrsim 100 \times 100~\upmu\text{m}^2$) will be challenging.}

\subsection{Photoelectric readout}\label{photoelectricreadout}



\begin{figure}
\begin{overpic}[height=2.2 in]{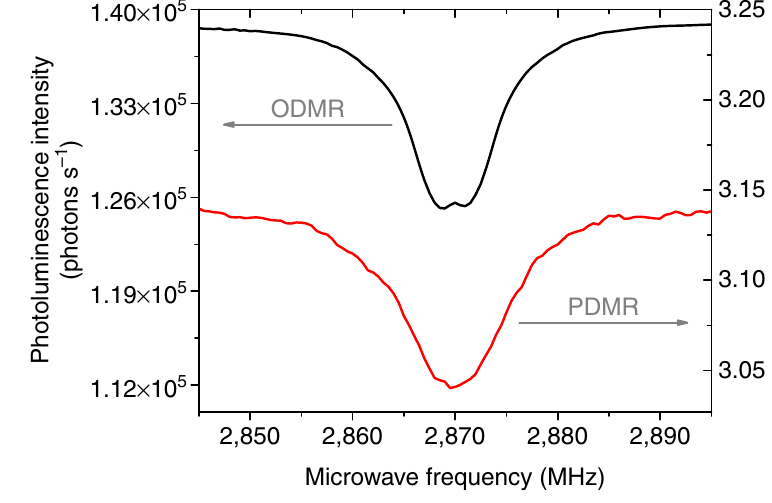}  
\end{overpic}
\caption[PDMR vs ODMR]{Photoelectrically detected magnetic resonance (PDMR) of NV$^\text{-}$ centers. Spectra are simultaneously measured by ODMR ({\color{black}{\rule[.5mm]{3mm}{1pt}}}) and PDMR ({\color{red}{\rule[.5mm]{3mm}{1pt}}}) in the absence of an external magnetic field. From Ref.~\cite{Bourgeois2015}.
}  \label{fig:bourgeois}
\end{figure}

Another method to interrogate the NV$^\text{-}$ spin state is photoelectric (PE) readout, which relies on measuring a current of charge carriers resulting from NV$^\text{-}$ photo-ionization~\cite{Bourgeois2015,Gulka2017,Bourgeois2017,Hrubesch2017,siyushev2019photoelectrical}. Since NV$^\text{-}$ photo-ionization is spin-state dependent, (see Secs.~\ref{SCCR} and \ref{chargestate})~\cite{Shields2015}, the spin state can be inferred from the photocurrent signal in analogy to fluorescence-based readout. Figure~\ref{fig:bourgeois} shows a photoelectrically detected magnetic resonance (PDMR) spectrum measured simultaneously with an ODMR spectrum, from Ref.~\cite{Bourgeois2015}.
One promised benefit of PE readout is that the photoelectron collection efficiency can approach unity~\cite{Bourgeois2015}.

In PE readout, a bias voltage is applied across electrodes fabricated on the diamond surface. An excitation laser induces NV$^\text{-}$ photo-ionization, and the ejected electrons generate a current, which is collected at the electrodes. NV$^\text{-}$ photo-ionization may occur via single- or two-photon excitation. Single-photon ionization of the NV$^\text{-}$ $^3$A$_2$ electronic ground state requires photon energies of $2.7\pm 0.1$ eV or higher (wavelength $\lesssim$ 460~nm)~\cite{Aslam2013, Bourgeois2017,Londero2017,Deak2014}. PE readout implementations employing lower photon energies, such as from 532~nm (2.33 eV) light common for spin-state initialization, ionize the NV$^\text{-}$ centers via a two-photon process, namely $^3$A$_2\! \rightarrow\; ^3$E$\;\rightarrow$\,conduction band (see Fig.~\ref{fig:Hopper})~\cite{Heremans2009,Bourgeois2015,Gulka2017,Bourgeois2017}. Whereas the rate of single-photon ionization scales linearly with optical intensity~\cite{Hacquebard2018}, two-photon ionization depends quadratically on intensity~\cite{Aslam2013}.

Optically illuminating the diamond for PE readout may also induce background photocurrent from ionization of other defects present in the sample. \textcolor{black}{Most unfortunately}, 532~nm green light ionizes substitutional nitrogen N$_\text{S}^0$ defects in a single-photon process~\cite{Heremans2009}. The background N$_\text{S}^0$ photocurrent may exceed the signal NV$^\text{-}$ photocurrent, resulting in poor NV$^\text{-}$ measurement contrast.  This problem is exacerbated for excitation intensities well below the NV$^\text{-}$ saturation intensity, where two-photon NV$^\text{-}$ ionization may be weak compared to single-photon ionization of N$_\text{S}^0$, and at elevated nitrogen concentrations [N$_\text{S}^0$]\,$\gg$\,[NV$^\text{-}$]~\cite{Bourgeois2017,Londero2017}.  

Multiple approaches can partially mitigate the unwanted photocurrent associated with N$_\text{S}^0$ ionization. For example, lock-in techniques can remove the DC background from the nitrogen photocurrent~\cite{Gulka2017}. Additionally, a shorter-wavelength laser can be employed to induce single-photon ionization from the NV$^\text{-}$ $^3$A$_2$ state, thereby improving the NV$^\text{-}$ ionization rate relative to that of N$_\text{S}^0$. However, the authors of Ref.~\cite{Bourgeois2017} observe that under optimized experimental conditions, single-photon ionization using 450~nm light provides no contrast improvement compared to two-photon ionization with 532~nm light.


A variety of challenges accompany implementation of PE readout, \textcolor{black}{not only for single NV$^\text{-}$ centers and} small NV$^\text{-}$ ensembles~\cite{Gulka2017, Bourgeois2015, Bourgeois2017, Hrubesch2017,siyushev2019photoelectrical} \textcolor{black}{but also} for envisioned extensions to larger detection volumes $\gtrsim (100~\upmu\text{m})^3$ using NV-rich diamonds. In addition to background photocurrent from ionization of nitrogen and other defects, another expected obstacle to PE readout is electrical cross-talk between MW-delivery electrodes (used to manipulate the NV$^\text{-}$ spin states) and photocurrent-detection electrodes~\cite{Gulka2017,siyushev2019photoelectrical}. Fluctuations in the applied electric field could also add additional measurement noise by coupling to fluctuations in photoelectric collection efficiency. 

Scaling PE readout implementations to larger NV$^\text{-}$ ensembles may introduce additional challenges. Because the electrodes reside on the diamond surface, collecting photocurrent from NV$^\text{-}$ centers located $\gtrsim 100~\upmu$m deep may prove difficult~\cite{Bourgeois2015}. Achieving the necessary bias electric field strength and uniformity over $\gtrsim (100~\upmu\text{m})^3$ volumes may also be challenging; bias electric field gradients across large detection volumes could reduce NV$^\text{-}$ ensemble $T_2^*$ values. Moreover, the presence of charge traps in NV-rich diamonds might hinder photoelectric collection efficiency (see Sec.~\ref{chargetraps}), especially from deeper NV$^\text{-}$ centers. In addition, Johnson noise in the readout electrodes may induce magnetic field fluctuations that could limit the achievable sensitivity~\cite{Kolkowitz2015}. 

In certain PE readout implementations, the detected signal amplitude may be increased by photoelectric gain, an intrinsic charge-carrier amplification arising from the diamond's charge dynamics and the electrode boundary conditions~\cite{Bourgeois2015,Hrubesch2017,Rose1963}. However, photoelectric gain is expected to be diminished in NV-rich diamonds due to charge traps, non-uniform electric fields, and space-charge limitations~\cite{Bourgeois2015, Bube1960, Rose1963}. The applicability of photoelectric gain to improving PE readout fidelity in ensemble-based extensions remains to be shown. \textcolor{black}{Although PE readout shows promise for nanoscale sensing and integrated quantum devices~\cite{Morishita2018}, and may prove beneficial when combined with PIN structures~\cite{Kato2013}, this technique's utility for ensemble magnetometry in NV-rich diamonds remains unknown.}

\subsection{Ancilla-assisted repetitive readout}\label{ancilla}

\begin{figure}

\begin{minipage}{.5\textwidth}
  \centering
 \includegraphics[width=4.2cm]{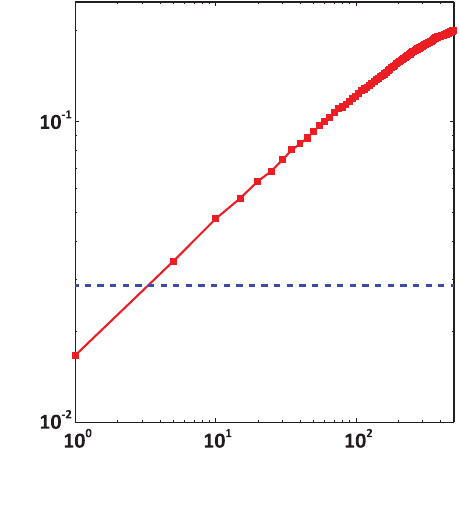}
\hspace{.25cm}
 \includegraphics[width=4cm]{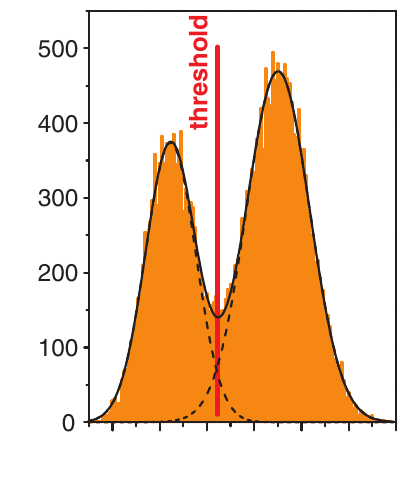}
\end{minipage}%

\begin{overpic}[height=1.76 in]{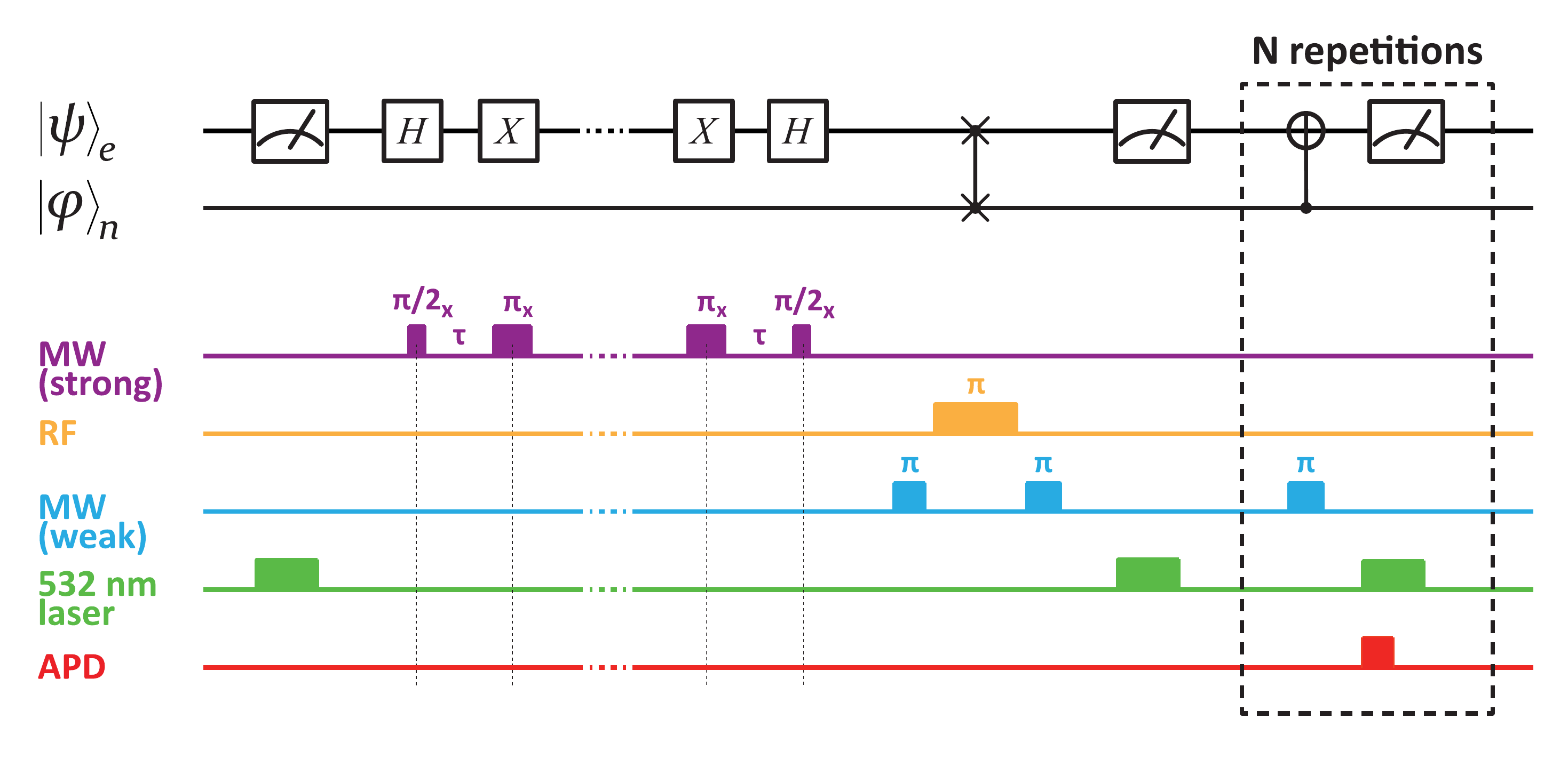}
\put(0,100){\textbf{a}}
\put(55,100){\textbf{b}}
\put(0,45){\textbf{c}}

\put(11,50){Repetitive readout cycles}
\put(2,65){\rotatebox{90}{Readout fidelity}}
\put(23,93){\textcolor{black}{Repetitive}}
\put(23,89.5){\textcolor{black}{readout}}

\put(23,68){\textcolor{blue}{Single}}
\put(23,64.5){\textcolor{blue}{conventional}}
\put(23,61){\textcolor{blue}{readout}}

\put(60,50){Number photons / 5~ms}
\put(54,66){\rotatebox{90}{Number of events}}

\end{overpic}

\caption[Ancilla-assisted repetitive readout circuit diagram]{Overview of ancilla-assisted repetitive readout. a) Readout fidelity $\mathcal{F}$ is improved with the number of repetitive readout cycles. Fidelity for repetitive readout (red) is plotted relative to a single conventional readout (blue, dashed). From Ref.~\cite{Lovchinsky2016}. b) The clear difference in total number of collected photons associated with the initial $m_s$ states allows determination of $m_s$ with fidelity approaching 1 in some implementations. Here $\mathcal{F} \approx 0.92$ in Ref.~\cite{Neumann2010}. From Ref.~\cite{Neumann2010}. c) Quantum circuit diagram and magnetometry pulse sequence with detection via ancilla-assisted repetitive readout. Application of an RF $\pi$-pulse between two weak MW $\pi$-pulses maps the NV$^\text{-}$ electronic spin superposition onto the ancilla nuclear spin. Subsequently the superposition state may be repeatedly mapped back onto the electronic spin via a weak MW $\pi$-pulse and optically read out without destroying the ancilla spin's quantum state. Adapted from Ref.~\cite{Lovchinsky2016}.
}  \label{fig:lovchinsky}
\end{figure}
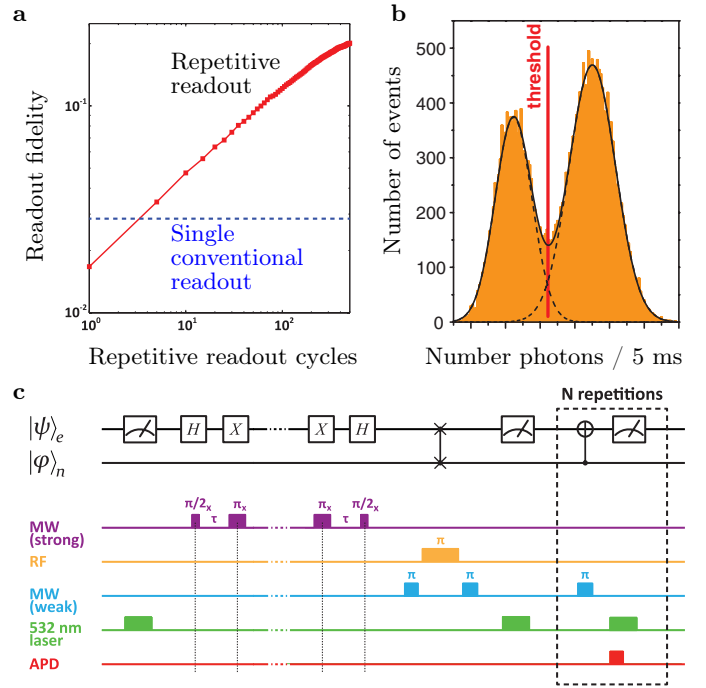

\textcolor{black}{In conventional readout, the fast $\sim\!500$~ns repolarization of the NV$^\text{-}$ electronic spin limits the number of photons an NV$^\text{-}$ emits before all initial spin state information is lost (see Fig.~\ref{fig:spindependentfluorescencecontrast}). Even when implementing conventional readout with the best present collection efficiencies, the average number of collected photons per NV$^\text{-}$ center $n_\text{avg}$ is less than 1, and for many implementations $n_\text{avg} \ll 1$, making photon shot noise the dominant contributor to the parameter $\sigma_R$ (see Eqn.~\ref{eqn:readnoise2}, Table~\ref{tab:parameters}, Sec.~\ref{improvedcollection}). An alternative method to increase the readout fidelity $\mathcal{F}=1/\sigma_R$ circumvents this problem by instead first mapping the initial NV$^\text{-}$ electronic spin state information onto an ancilla nuclear spin. In the second step, the ancilla nuclear spin state is mapped back onto the electron spin, which is then detected using conventional fluorescence-based readout. This second step may be repeated many times with each marginal readout improving the aggregate readout fidelity, as shown in Fig.~\ref{fig:lovchinsky}a,b. While first demonstrated with a nearby $^{13}$C nuclear spin as the ancilla~\cite{Jiang2009}, the technique was later realized using the NV$^\text{-}$ center's $^{14}$N~\cite{Neumann2010} and $^{15}$N nuclear spin~\cite{Lovchinsky2016}. In the $^{13}\text{C}$ realization~\cite{Jiang2009,Maurer2012}, the coupling to the ancilla spin depends on the distance between the NV$^\text{-}$ defect and the nearby $^{13}\text{C}$ atom, making the technique difficult to implement for NV$^\text{-}$ ensembles where this distance varies. This discussion instead focuses on the more scalable realization using the NV$^\text{-}$ nitrogen nuclear spin as the ancilla, which ensures the electron spin to ancilla spin coupling remains fixed over the NV$^\text{-}$ ensemble.}

Figure~\ref{fig:lovchinsky}c shows a quantum circuit diagram from Ref.~\cite{Lovchinsky2016} depicting the repetitive readout scheme. After the final MW pulse in an NV$^\text{-}$ sensing protocol, the NV$^\text{-}$ electronic spin state (denoted by subscript $e$) is mapped onto the nitrogen nuclear spin state (subscript $n$). In Ref.~\cite{Lovchinsky2016}, this mapping is achieved using a SWAP gate ($\text{CNOT}_{e|n} -\text{CNOT}_{n|e} - \text{CNOT}_{e|n}$), where CNOT denotes a controlled NOT gate. The SWAP gate consists of a MW $\pi$-pulse, then an RF $\pi$-pulse, then another MW $\pi$-pulse, where the MW pulses flip the electronic spin and the RF pulse flips the nuclear spin. \textcolor{black}{This procedure swaps the electronic and nuclear spin states, importantly, storing the electronic spin state information in the ancilla nuclear spin. Then an optical pulse re-polarizes the electronic spin to $m_s = 0$.  Next, a set of repetitive readouts is performed. In each readout, the nuclear spin state is copied back onto the electronic spin with a MW pulse, (a $\text{CNOT}_{e|n}$ gate), and then the electronic spin is optically read out without affecting the nuclear spin \textcolor{black}{state}. This process can be repeated many times ($\gtrsim 10^2$), and is limited in principle by the nuclear spin lifetime $T_{1,n}$. In Ref.~\cite{Lovchinsky2016}, while the initial RF pulse used in the SWAP gate requires $\sim$~50-60~$\upmu$s, each readout cycle requires only $\sim\!1$~$\upmu$s. The large number of readouts allow the aggregate readout fidelity $\mathcal{F}=1/\sigma_R$ to approach 1; notably, Ref.~\cite{Neumann2010} achieves $\mathcal{F} = 0.92 \;(\sigma_R = 1.1)$ as depicted in Fig.~\ref{fig:lovchinsky}b.}

\textcolor{black}{Extending ancilla-assisted repetitive readout to ensembles is expected to be fruitful but necessitates overcoming further challenges. The scheme requires a large magnetic field to minimize coupling between the NV$^\text{-}$ nuclear and electronic spins, with Refs.~\cite{Neumann2010,Lovchinsky2016} employing fields of 2500 gauss and 6500 gauss respectively. Further, the bias magnetic field must be precisely aligned along a single NV$^\text{-}$ symmetry axis, presently precluding its use for sensing from more than one NV$^\text{-}$ orientation~\cite{Schloss2018}. Even slight angular misalignments introduce measurement back action on the nuclear spin $I_z$, which spoils $T_{1,n}$~\cite{Neumann2010}. The reduction in $T_{1,n}$ limits the available readout duration. Ensemble implementations would therefore require highly uniform bias magnetic fields over ensemble sensing volumes, conceivably on the $\sim (100~\upmu \text{m})^3$ scale. Engineering such fields is within current technical capability but difficult nevertheless (see Sec.~\ref{T2*params} and Ref.~\cite{vandersypen2005nmr}). Additionally, the MW and RF control pulses would ideally manipulate the entire ensemble uniformly; spatial inhomogeneities of the control pulses are likely to result in reduced readout fidelity unless mitigated~\cite{vandersypen2005nmr}. Assuming sufficiently strong and homogeneous $B_0$ fields and MW driving can be realized, and that the additional overhead time is acceptable, repetitive readout appears to be a promising but technically demanding method to improve $\mathcal{F}$ for ensembles.}




\subsection{Level-anticrossing-assisted readout}\label{LAC}

\begin{figure*}[ht]
\centering
\begin{overpic}[height=1.94 in]{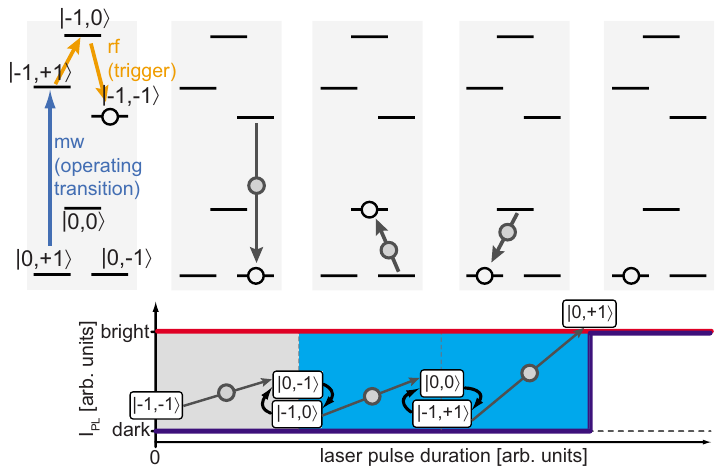}
\put (-2, 65) {\large a}
\end{overpic}
\begin{overpic}[height=1.87 in]{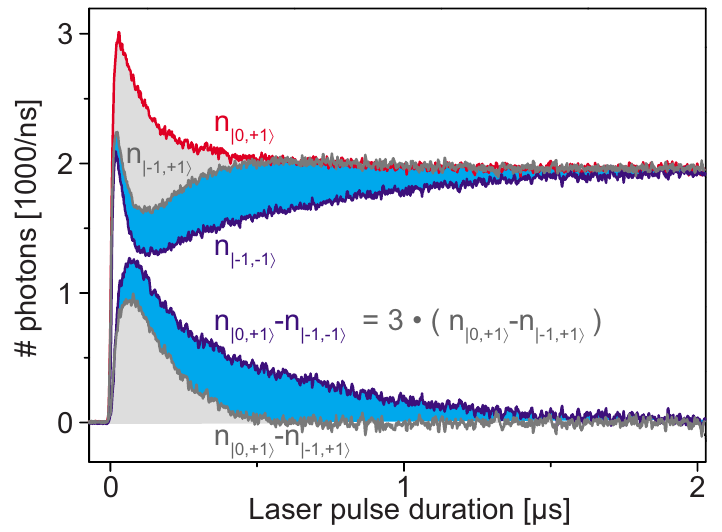}
\put (2, 76) {\large b}
\end{overpic}
\caption[LAC-assisted readout]{Level-anticrossing-assisted readout as demonstrated in Ref.~\cite{Steiner2010}. At the excited-state level anticrossing  near $B=500$~G, green optical excitation polarizes NV$^\text{-}$ into the spin state $|m_s=0,m_I=+1\rangle$. a) Upon completion of a sensing sequence, two RF pulses transfer population in the electronic spin state $|m_s=-1\rangle$ from the nuclear spin state $|m_I=+1\rangle$ to $|m_I=-1\rangle$ without affecting the $|m_s=0\rangle$ state. During optical readout, this population passes three times through the singlet states before being repolarized to $|m_s=0,m_I=+1\rangle$, increasing the time over which the state-dependent fluorescence contrast persists. b) Time-resolved photon detection comparing conventional readout (gray) and LAC-assisted readout (blue). The optimal readout duration is extended by $3\times$ and the difference in detected photon number between the two spin states is increased by $3\times$. From Ref.~\cite{Steiner2010}.}\label{fig:lacrfig}
\end{figure*}

In conventional readout~\cite{Doherty2013}, the readout fidelity $\mathcal{F} = 1/\sigma_R$ depends on the number of photons $n_\text{avg}$ collected per measurement sequence (see Eqn.~\ref{eqn:readnoise2}). The value of $n_\text{avg}$ is limited by the time the spin population originally in the $m_s = \pm 1$ states spends shelved in the singlet state before decaying to the triplet $m_s = 0$ state. 
Steiner \textit{et al.}~engineer the NV$^\text{-}$ spin to pass through the singlet state multiple times before repolarization, extending the readout duration per sequence to increase  $n_\text{avg}$~\cite{Steiner2010}, as depicted in Fig.~\ref{fig:lacrfig}. Using NV$^\text{-}$ centers with  $^{14}$N, which has nuclear spin $I=1$, three cycles through the singlet state occur during readout, yielding a  $\sim 3 \times$ increase in $n_\text{avg}$ and thus a $\sim \sqrt{3}\times$ improvement in the fidelity $\mathcal{F}$. For NV$^\text{-}$ centers with $^{15}\text{N}$ with $I=\sfrac{1}{2}$, the spin only passes twice through the singlet state before repolarization, yielding only a $\sim \sqrt{2}\times$ improvement in $\mathcal{F}$.


The technique is implemented as follows:~the bias field $B_0$ is tuned to the excited-state level anticrossing at $B_\text{LAC} \approx \, 500 ~$G~\cite{Fuchs2008, Neumann2010} to allow resonant flip-flops between the NV$^\text{-}$ center's electronic spin and its $^{14}$N nuclear spin ($I=1$). Operation at the level anticrossing polarizes the nuclear spin into the state $|m_I\!=\!+1\rangle$~\cite{Jacques2009}. At completion of a sensing sequence, immediately prior to readout, the NV$^\text{-}$ electronic spin occupies a superposition of the states $|m_s\!=\!0,\,m_I\!=\!+1\rangle$ and $|m_s\!=\!-1,\,m_I\!=\!+1\rangle$. Before the NV$^\text{-}$ electronic spin state is read out using a conventional green laser pulse, two sequential RF $\pi$-pulses flip the nuclear spin into the $m_I=-1$ state, conditional on the electronic spin occupying the $m_s=-1$ state. This CNOT gate relies on the RF drive being resonant with the nuclear transitions between the $m_I$ states for population in the $m_s = -1$ state and off-resonant for population in the $m_s = 0$ state. 
During readout, the population in $|m_s\!=\!-1, m_I\!=\!-1\rangle$ cycles through the long-lived singlet state three times before the information stored in the NV$^\text{-}$ electronic spin is lost, allowing more signal photons to be collected. After the first and second pass through the singlet to the $m_s = 0$ state, an electron-nuclear spin flip-flop returns the electronic spin state to $m_s = -1$, as shown in Fig.~3a of Ref.~\cite{Steiner2010}, enabling another cycle through the singlet state. The third pass repolarizes the NV$^\text{-}$ spin into the stable $|m_s\!=\!0, m_I\!=\!+1\rangle$ state.

This technique's utility for magnetic sensing depends on whether the $\leq \sqrt{3} \times$ increase in fidelity $\mathcal{F}$ outweighs the cost of additional overhead time (see Eqn.~\ref{eqn:duty}) introduced by the RF pulses. Although the authors of Ref.~\cite{Steiner2010} assert that microsecond-scale RF nuclear spin $\pi$-pulses are attainable, 
achieving such nuclear Rabi frequencies over large ensemble volumes $\sim (100~\upmu \text{m})^3$ may prove very difficult, making this method impractical for NV$^\text{-}$ ensembles with $T_2^* \lesssim 1~\upmu$s. 
Additional challenges for implementation with NV$^\text{-}$ ensembles include realizing the requisite uniformity in the MW/RF fields and in the 500~G bias magnetic field over ensemble volumes. \textcolor{black}{Finally, the scheme presently precludes sensing from more than one NV$^\text{-}$ orientation~\cite{Schloss2018}.}


\subsection{Improved photon collection methods}\label{improvedcollection}

\textcolor{black}{In the limit of low contrast, the readout fidelity $\mathcal{F}$ is proportional to the square root of the average number of photons collected per NV$^\text{-}$ per measurement, i.e., $\mathcal{F} \propto \sqrt{\mathscr{N}/N}=\sqrt{n_\text{avg}}$ (see Eqn.~\ref{eqn:readnoise2}).}  Under these conditions, sensitivity can be enhanced by increasing the geometric collection efficiency $\eta_\text{geo}$, defined as $\mathscr{N}/\mathscr{N}_\text{max}$, where $\mathscr{N}$ and $\mathscr{N}_\text{max}$ are the number of photons collected and emitted respectively by the NV$^\text{-}$ ensemble per measurement.

\begin{figure}[ht]
\centering
\begin{overpic}[width=\columnwidth]{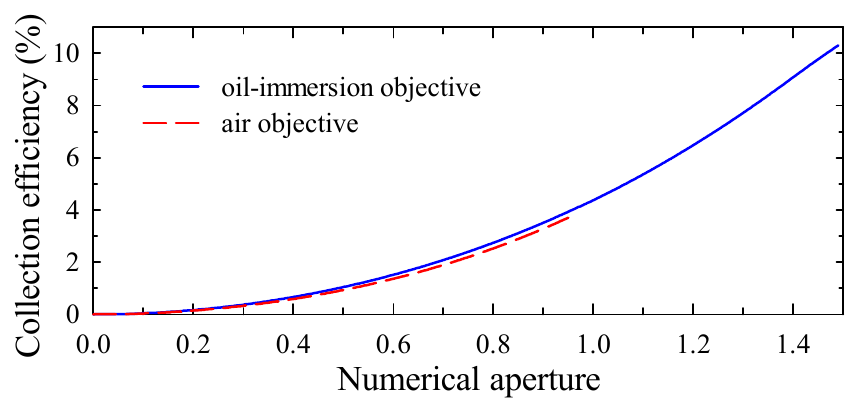}  
\end{overpic}
\caption[Collection efficiency vs. NA]{Calculated collection efficiencies of NV$^\text{-}$ fluorescence by oil-immersion or air microscope objectives through the \textcolor{black}{\{100\}} surface of a diamond chip, as a function of numerical aperture. Figure and caption from Ref.~\cite{LeSage2012}.}\label{fig:collectionlesage}
\end{figure}

Efficient photon collection in diamond is hindered by total-internal-reflection confinement resulting from diamond's high refractive index of approximately 2.41. For example, air and oil immersion objectives, with numerical apertures of 0.95 and 1.49 respectively, provide calculated collection efficiencies of only $3.7\%$ and $10.4\%$ respectively for photons emitted directly through the \textcolor{black}{\{100\}} diamond surface~\cite{LeSage2012}, as depicted in Fig.~\ref{fig:collectionlesage}. Although anti-reflection coatings can allow for higher collection efficiencies, present implementations demonstrate only  modest improvement~\cite{Yeung2012}. While great effort has resulted in high values of $\eta_\text{geo}$ for single NV$^\text{-}$ centers through use of various nano-fabrication approaches~\cite{Hadden2010,Babinec2010,Marseglia2011,Choy2011,Schröder2011,Yeung2012,Choy2013,Riedel2014,Neu2014,Momenzadeh2015,Li2015,Shields2015,Haberle2017}, such methods do not easily translate to large ensembles.



\begin{figure}[ht]
\centering
\begin{overpic}[width=\columnwidth]{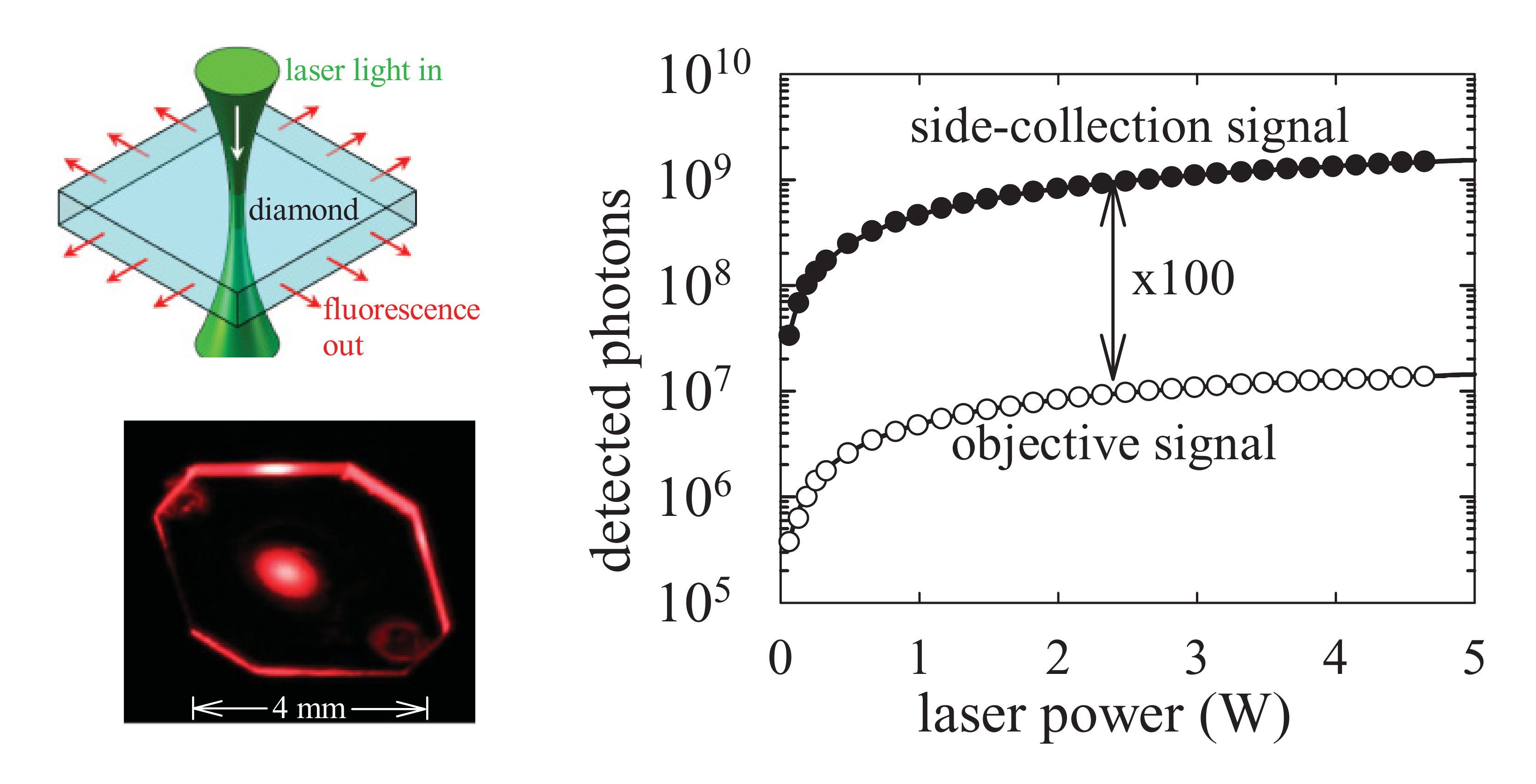}
\put (2.5, 48) {\large a}
\put (2.5, 24) {\large b}
\put (37, 48) {\large c}
\end{overpic}
\caption[Side Collection]{Fluorescence side-collection method~\cite{LeSage2012}. a) Green optical excitation is applied normal to the large face of the diamond chip, and red fluorescence is collected from the sides. b) Red fluorescence from actual diamond chip. c) The depicted implementation results in a 100$\times$ increase in detected photons relative to a 0.4 numerical aperture air objective. From Ref.~\cite{LeSage2012}.}  \label{fig:sidecollection}
\end{figure}

Successful approaches for bulk ensemble magnetometers have so far focused on collecting NV$^\text{-}$ fluorescence that has undergone total internal reflection in the diamond~\cite{LeSage2012,Wolf2015}.  
\textcolor{black}{While absorption of NV$^\text{-}$ fluorescence by various defects may limit $\eta_\text{geo}$~\cite{Khan2013,Khan2009} for some diamonds, nitrogen~\cite{DeWeerdt2008} and NV$^\text{-}$ centers~\cite{Fraczek2017} are expected to hardly absorb in the NV$^\text{-}$ PL band $\sim\!600\>\text{-}\>850$~nm.} A collection efficiency of 39$\%$ is demonstrated in Ref.~\cite{LeSage2012} by detecting fluorescence from the four sides of a rectangular diamond chip surrounded by four photodiodes (see Fig.~\ref{fig:sidecollection}). However, the increased experimental complexity associated with employing four detectors in contact with the diamond may be problematic for certain applications. In another approach, Wolf \textit{et al.}~employ a trapezoidally-cut diamond chip and a parabolic concentrator to improve collection efficiency~\cite{Wolf2015}. Although the authors calculate $\eta_\text{geo}$ to be between $60\%$ and $65\%$, this result is not confirmed experimentally. Ma \textit{et al.}~\cite{Ma2018} demonstrate a collection efficiency of 40$\%$ by eliminating all air interfaces between the diamond and detector, in conjunction with coupling prisms which direct light exiting the diamond's four side faces to the detector. 


In the future, collection efficiency in bulk NV-diamond magnetometers is expected to improve to near 100$\%$, limited only by losses due to absorption. For example, light lost from the top and sides of the diamond in Ref.~\cite{Wolf2015} could be redirected to the detector by coating these sides of the diamond with a metallic~\cite{Choy2011} or dielectric reflector~\cite{LMpatentWO2016118791}. The authors of Ref.~\cite{Wolf2015} might also see an improvement in collection efficiency by designing an optimized parabolic concentrator rather than using a commercially available part. Hypothetical geometries for light collection using parabolic or ellipsoidal reflectors are discussed in Ref.~\cite{LMpatentWO2016118791}. Whereas multiple-reflection methods are suitable for bulk magnetometers, increasing $\eta_\text{geo}$ by collecting light undergoing multiple reflections in the diamond may substantially complicate accurate image reconstruction for NV$^\text{-}$ magnetic imaging microscopes~\cite{Steiner2010,Pham2011,LeSage2013}.

\textcolor{black}{Is is also natural to consider exploiting the Purcell effect to improve readout fidelity. By engineering physical structures around a chosen NV$^\text{-}$ center or ensemble, several works have increased the triplet excited state's radiative decay rate~\cite{kaupp2016purcell,Choy2011,riedel2017deterministic}. The increased radiative decay allows for more PL photons to be collected from population originally in $|m_s \!= 0\rangle$ while population originally in $|m_s \!=\!\pm 1\rangle$ is shelved in the singlet states.  Although theoretical investigations suggest that Purcell enhancement might improve readout fidelity~\cite{wolf2015purcell} or leave fidelity unchanged~\cite{babinec2012design}, the only reported experimental demonstration of Purcell-enhanced NV$^\text{-}$ spin readout to date finds reduced fidelity~\cite{bogdanov2017electron}. The authors surmise that charge effects related to dense NV$^\text{-}$ ensembles may contribute to the observed fidelity decrease.  Achieving high Purcell factors for NV$^\text{-}$ ensemble applications may also impose undesirable geometric constraints. For a clear and detailed discussion of radiative lifetime engineering for NV$^\text{-}$ spin readout, we recommend Ref.~\cite{hopper2018spin}. Along similar lines, a recent proposal suggests that NIR fluorescence-based readout could exhibit improved fidelity over conventional readout when combined with Purcell enhancement~\cite{meirzada2019enhanced}. While this scheme requires high IR excitation intensities likely incompatible with large NV$^\text{-}$ ensembles, it may show utility for small ensembles within a confocal volume.}

\subsection{Near-infrared absorption readout}\label{NIRabsorption}


The sensitivity of conventional fluorescence-based readout is limited by shot noise on the collected photons due to low fluorescence contrast $C$ (see Eqn.~\ref{eqn:readnoise}). As an alternative to fluorescence-based readout, population in one or both NV$^\text{-}$ singlet states may be directly probed via absorption, giving a probabilistic measure of the initial $m_s$ spin state prior to readout. \textcolor{black}{While the upper singlet state $^1$A$_1$ lifetime $\lesssim 1$~ns at room temperature~\cite{Acosta2010b,ulbricht2018excited} is likely too short for such an approach to be effective}, the lower singlet state $^1$E lifetime  $\sim 140-220$~ns at room temperature~\cite{Gupta2016,Robledo2011,Acosta2010b} makes measuring the $^1$E population via absorption on the $^1$E$\;\leftrightarrow ^1\!\!\text{A}_1$ transition at 1042~nm viable. 


Near-infrared (NIR) absorption has attractive benefits for certain applications: a) Contrast (and therefore sensitivity) is not reduced by background fluorescence from non-NV$^\text{-}$ defects (such as NV$^0$). b) The directional nature of the 1042~nm probe light allows maximal collection efficiency (ignoring absorptive losses) to be obtained more easily than in a fluorescence-based measurement; for example, this benefit was exploited in the first demonstration of microwave-free magnetometry with NV$^\text{-}$ centers~\cite{Wickenbrock2016}. c) \textcolor{black}{Owing to the upper singlet $^1$A$_1$ lifetime of $\lesssim 1$~ns~\cite{Acosta2010b,ulbricht2018excited}, the saturation intensity} of the $^1$E$\;\leftrightarrow ^1\!\!\text{A}_1$ transition is unusually large ($I^\text{sat}_{1042}\sim\!50$ megawatts/cm$^2$~\cite{Dumeige2013}), allowing high intensity 1042~nm probe radiation to be used, so that fractional shot noise is reduced to well below that of an equivalent fluorescence-based measurement~\cite{AcostaThesis2011}. d) NIR absorption readout is nondestructive, allowing for a single NV$^\text{-}$ center in the $^1$E singlet state to absorb multiple 1042~nm photons before eventual decay to the $^3$A$_2$ triplet. In principle such absorption by a single NV$^\text{-}$ center can allow readout fidelity near the spin-projection limit, even in the presence of non-negligible optical losses. 



\begin{figure}
\centering
\begin{overpic}[width=\columnwidth]{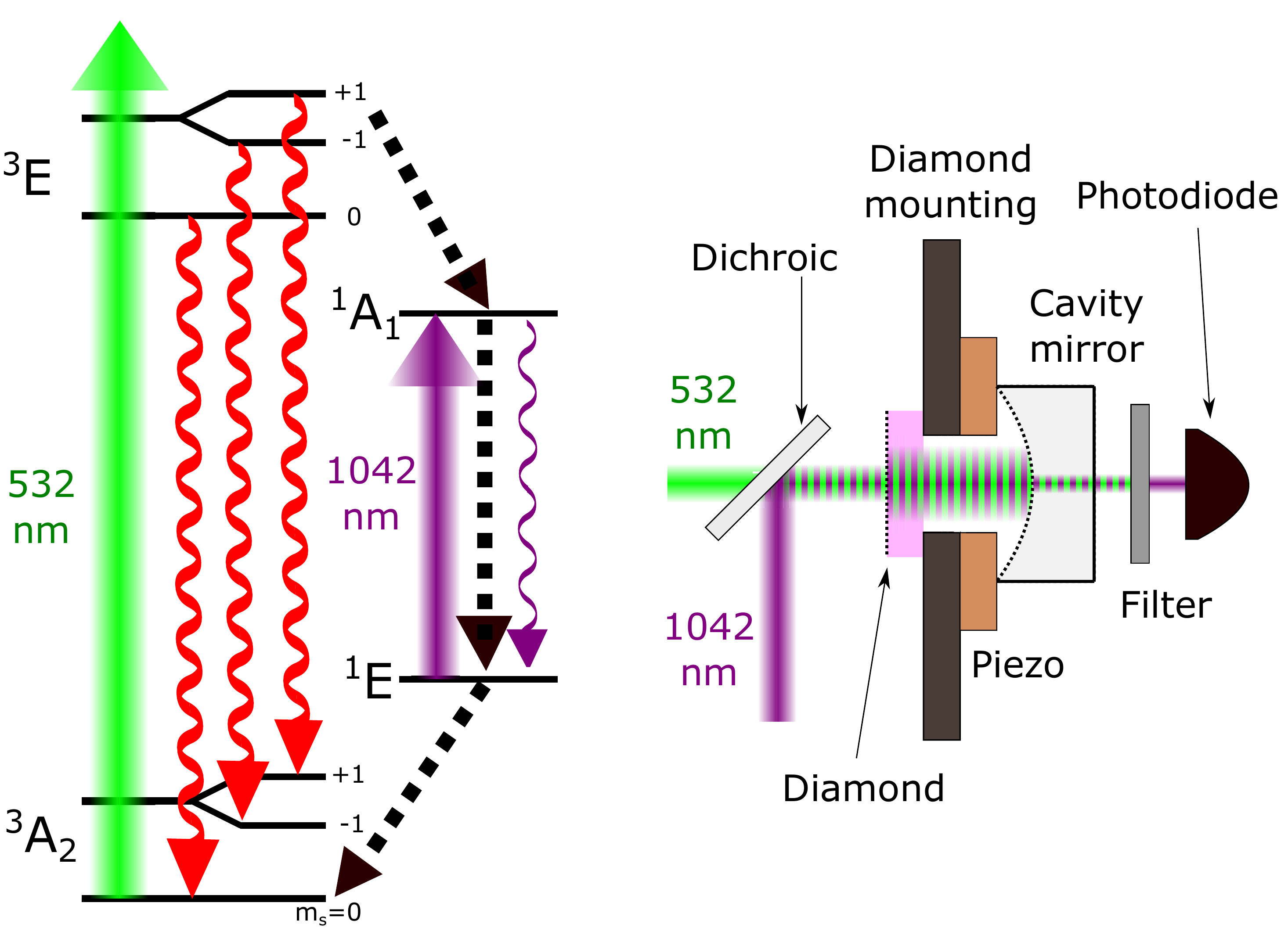}  
\put (-0.5, 65) {\large a}
\put (50, 65) {\large b}
\end{overpic}
\caption[Near-infrared absorption]{Near-infrared (NIR) absorption readout. a) Level diagram for the NV$^\text{-}$ center. Population accumulating in the $^1$E singlet state is probed via absorption of 1042~nm radiation resonantly addressing the $^1$E$\;\leftrightarrow ^1\!\!\!\text{A}_1$ transition. b) Miniature NIR cavity-enhanced diamond magnetometer as described in Ref.~\cite{Chatzidrosos2017}. Dashed black lines depict surfaces forming the dual-wavelength optical cavity. Components pertaining to MW delivery are omitted for clarity.}  \label{fig:NIRabsorption}
\end{figure}

NIR absorption readout has been successfully implemented in several proof-of-principle magnetometers. In the first demonstration~\cite{Acosta2010c}, a diamond containing $ [\text{NV}^\text{-}] \sim\!16$~ppm is continuously illuminated with 532~nm radiation (driving the $^3$A$_2\! \leftrightarrow ^3$E transition to optically polarize the NV$^\text{-}$ spin state) and 1042~nm NIR radiation (resonantly addressing the $^1$E $\! \leftrightarrow ^1\!\!\text{A}_1$ transition), as shown in Fig.~\ref{fig:NIRabsorption}a. MW radiation transfers population between the ground state Zeeman sublevels. In this first demonstration~\cite{Acosta2010c}, a single pass of the 1042~nm radiation through the diamond sample resulted in a peak-to-peak contrast of $\sim\!0.003$ at room temperature.



The contrast can be enhanced by increasing the number of NV$^\text{-}$ defects interacting with each 1042~nm photon, such as by employing a higher NV$^\text{-}$ density or a larger diamond. Alternatively, for a fixed number of NV$^\text{-}$ centers, the 1042~nm radiation can be recirculated through the diamond. An NIR absorption magnetometer employing an optical cavity to increase the 1042~nm interaction length is analyzed in Ref.~\cite{Dumeige2013} and experimentally realizes a peak-to-peak contrast of $7.1\%$ in Ref.~\cite{Jensen2014} when the bias magnetic field is directed along the [100] crystallographic direction (making the magnetic resonances of all four NV$^\text{-}$ orientations degenerate). For this demonstration, the diamond is anti-reflection coated~\cite{Yeung2012} and placed in the center of a dual-wavelength optical cavity, which allows recirculation of both 1042~nm and 532~nm radiation. The more efficient use of the 532~nm light enabled by the cavity allows both a larger NV$^\text{-}$ ensemble to be addressed and \textcolor{black}{a higher degree of} spin initialization into the $m_s=0$ state. Ultimately the device in Ref.~\cite{Jensen2014} achieves a sensitivity of 2.5 nT/$\sqrt{\text{Hz}}$, well above the shot-noise limit of 70 pT/$\sqrt{\text{Hz}}$.



A notable recent implementation of NIR absorption~\cite{Chatzidrosos2017} is depicted in Fig.~\ref{fig:NIRabsorption}b. One diamond face forms a reflector while the addition of a dual-wavelength concave mirror results in an optical cavity with a finesse of 160 and a cylindrical sensing volume of $\sim\!76~\upmu$m diameter and  $\sim\!390~\upmu$m length~\cite{Chatzidrosos2017}. With 500 mW of 532~nm radiation and 80~mW of 1042~nm radiation, a DC magnetic field sensitivity of 28 pT/$\sqrt{\text{Hz}}$ is achieved with this compact setup, with a bandwidth of about 530 Hz. The shot-noise-limited sensitivity is 22 pT/$\sqrt{\text{Hz}}$ and the spin-projection-noise-limited sensitivity is 0.43 pT/$\sqrt{\text{Hz}}$.


The NIR absorption approach is hindered, however, by several non-idealities, which so far limit readout fidelity to values far from the spin projection noise limit, similar to conventional optical readout (i.e., $\sigma_R = 65$ for the NIR absorption approach in Ref.~\cite{Chatzidrosos2017} \textcolor{black}{versus} 
$\sigma_R\approx 67$ for conventional readout in Ref.~\cite{LeSage2012}). First, the predominantly non-radiative decay of the $^1$A$_1$ singlet greatly reduces the absorption cross section $\sigma_\text{1042}$ of the $^1$E$\;\leftrightarrow ^1\!\!\text{A}_1$ transition compared to a radiative-decay-only transition~\cite{Rogers2008,Acosta2010b}. Estimates suggest $\sigma_\text{1042} = 3^{+3}_{-1} \times 10^{-18}$~cm$^2$~\cite{Kehayias2013,Dumeige2013,Jensen2014}, whereas the purely radiative $^3$A$_2\!\leftrightarrow ^3$E transition is measured to have a much \textcolor{black}{larger absorption cross section $\sigma_\text{532} \approx 3 \times 10^{-17}$~cm$^2$ for 532~nm excitation, as shown in Table~\ref{tab:532nmabsorptioncrosssectioncompiled}.} Realizing the full potential of this method requires 1042~nm laser intensities of order $I^\text{sat}_{1042} \sim 50$ MW/cm$^2$~\cite{Jensen2014}. This saturation intensity appears to limit interrogation cross sections to $\lesssim\!\!100$~$\upmu$m$^2$ for $\sim\!\!100$ mW-scale 1042~nm radiation powers, assuming a cavity finesse of $\sim\!\!160$. Laser intensities of this magnitude may lead to undesirable ionization dynamics (see Sec.~\ref{chargestate}). \textcolor{black}{We note that many absorption cross section measurements neglect ionization/recombination dynamics, which may skew reported values~\cite{meirzada2018negative,meirzada2019enhanced}.} Second,  as described in Ref.~\cite{Dumeige2013}, non-resonant losses for 1042~nm radiation compromise sensitivity by reducing the effective achievable collection efficiency. For example, in Ref.~\cite{Chatzidrosos2017}, 80 mW of 1042~nm radiation input to the dual-wavelength cavity results in 4.2 mW transmitted to the detector. Third, the NIR absorption has only been demonstrated for dense ensembles with $[\text{NV}^\text{T}] \sim 10$~ppm to ensure appreciable 1042~nm absorption; the performance of this method for diamonds with more dilute NV$^\text{-}$ concentrations and longer $T_2^*$ values remains unknown, and will likely depend on the scaling of cavity finesse with [N$^\text{T}$] or [NV$^\text{T}$] density. 




While NIR absorption readout is effective and may find preference for certain applications~\cite{Wickenbrock2016,Chatzidrosos2017}, without further advances  enabling readout fidelity enhancement, (e.g., reduced 1042~nm non-resonant absorption or reduced non-radiative $^1$A$_1$ singlet decay rate), this method will remain approximately on par with conventional fluorescence readout while requiring the non-trivial overhead of an NIR single frequency laser and an optical cavity.








\subsection{Green absorption readout}\label{greenabsorption}



\begin{figure}
\begin{overpic}[width=\columnwidth]{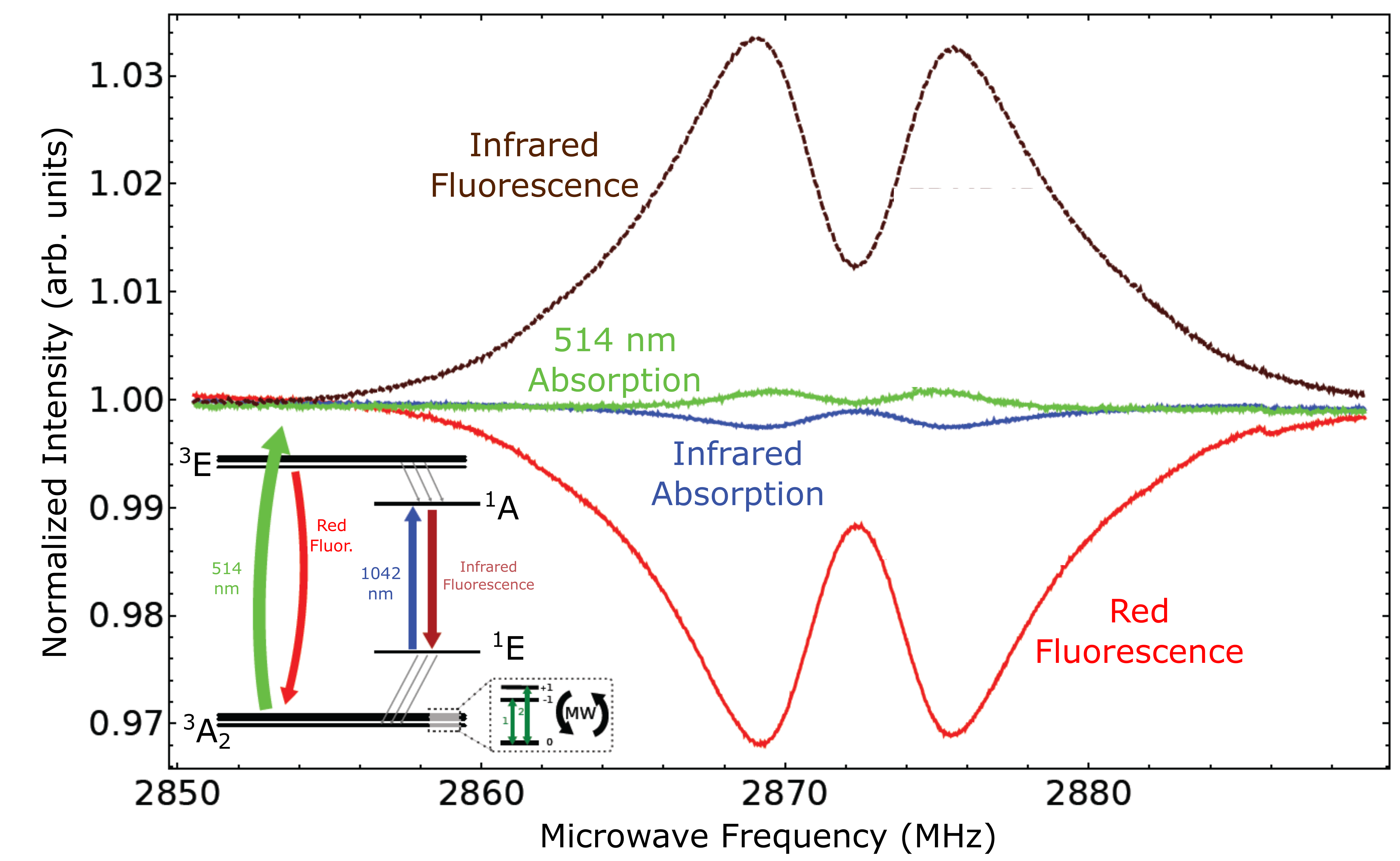}  
\end{overpic}
\caption[Green absorption and PL spectrum]{Simultaneous measurement of absorption and fluorescence on both the triplet and singlet NV$^\text{-}$ electronic transitions (see inset lower left). For both transitions, the absorption and fluorescence features have opposite signs and mirror one another up to a scaling factor. Adapted from Ref.~\cite{BauchThesis2010}.}  \label{fig:GreenAbsorptionDataErik}
\end{figure}

Alternatively, NV$^\text{-}$ readout may be achieved by monitoring absorption of green probe laser radiation, which off-resonantly drives the triplet $^3$A$_2\! \leftrightarrow \! ^3$E transition~\cite{BauchThesis2010,Walsworth2012}. When resonant MWs drive the $m_s\!=\!0\!\leftrightarrow\!m_s\!=\!-1$ or $m_s\!=\!0\!\leftrightarrow\!m_s\!=\!+1$ ground state spin transitions and facilitate population transfer to the NV$^\text{-}$ singlet states, it is expected that the $^3$A$_2 $ state will be depleted, resulting in \textit{increased} green probe transmission and \textit{decreased} red fluorescence. For absorption measurements (both NIR and green), the change in transmitted light upon resonant MW drive is expected to mirror the change in fluorescence light up to a scaling constant, since transmission is minimal when fluorescence is maximal and vice versa~\cite{BauchThesis2010}. Data consistent with this understanding is shown in Fig.~\ref{fig:GreenAbsorptionDataErik} for NV$^\text{-}$ centers \textcolor{black}{illuminated} 
with 514~nm light. 

\textcolor{black}{The absorption contrast, denoted $C_\text{absorb}$, may differ substantially in magnitude from the fluorescence contrast $C_\text{fluor}$ (see Fig.~\ref{fig:GreenAbsorptionDataErik}). Because absorption measurements monitor transmitted light, the detected signal (and thus $C_\text{absorb}$) depends on the optical depth of the absorbing material. For example, even for the idealized case where $C_\text{fluor} = 1$, if only a small fraction of the incident light is absorbed in the absence of MWs, the change in transmission upon application of resonant MWs will necessarily also be small, yielding a low absorption contrast $C_\text{absorb}$. Additionally, the absorption contrast may be further decreased due to the presence of non-radiative decay pathways.}

 



\begin{figure}
\centering
\begin{overpic}[height=2.6 in]{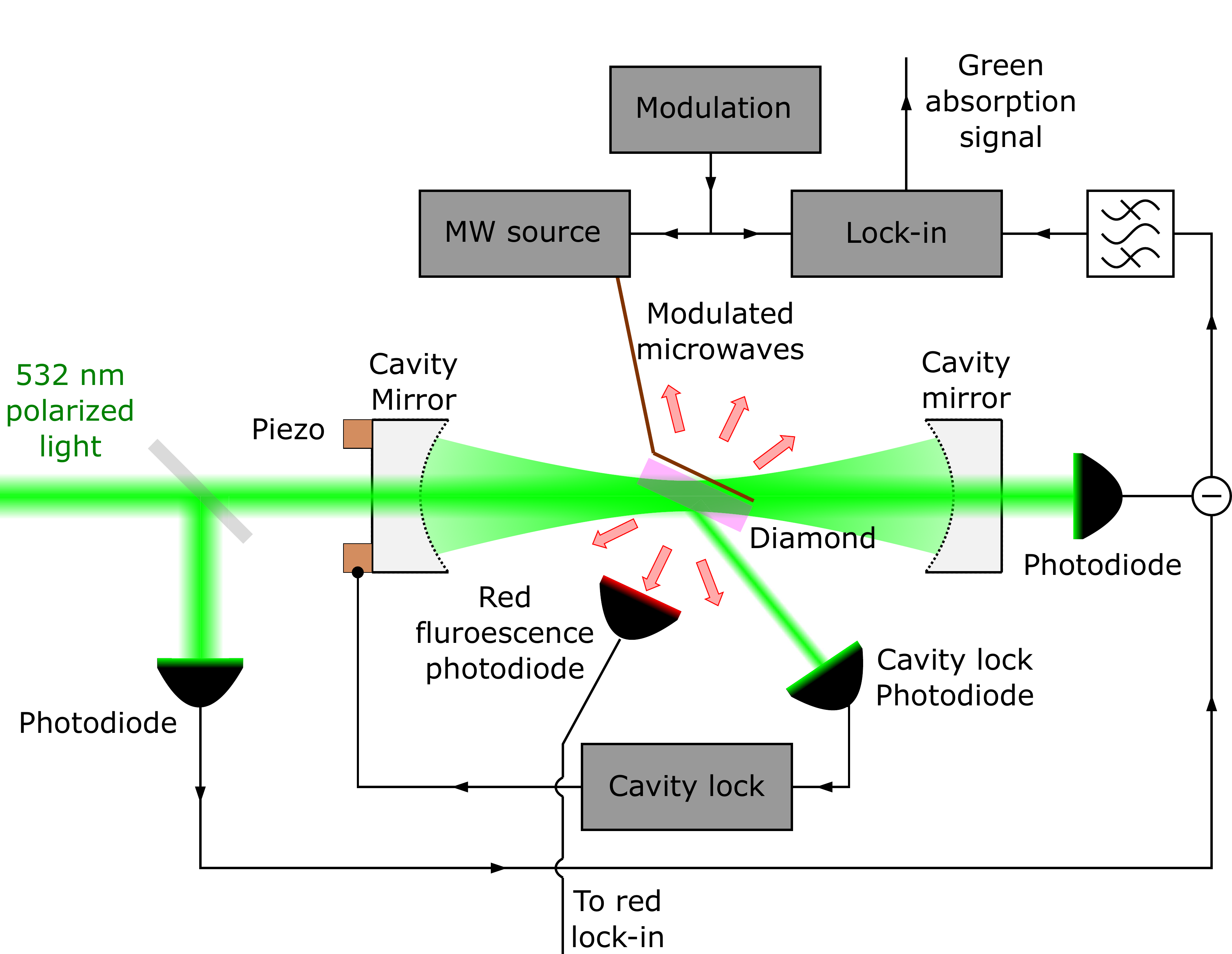}  
\end{overpic}
\caption[Cavity-enhanced green absorption magnetometry]{Cavity-enhanced magnetometry based on green absorption as demonstrated in Refs.~\cite{Ahmadi2018,Ahmadi2018b}. A power build-up cavity allows green excitation light to pass through the diamond sample multiple times, increasing the effective path length. The red fluorescence is measured simultaneously along with the green absorption. Adapted from Ref.~\cite{Ahmadi2018}. }  \label{fig:Greenabsorbtion}
\end{figure}

Observed magnitudes of $C_\text{absorb}$ in the literature are lower than $C_\text{fluor}$ by $\sim\! 3\times$~\cite{BauchThesis2010,LeSageKeigoPrivate} or more. For example, the authors of Refs.~\cite{Ahmadi2017,Ahmadi2018,Ahmadi2018b} use a CW-ODMR-based magnetometer employing a resonant optical cavity to recycle the green excitation light through the diamond multiple times, and observe $C_\text{fluor} \!\!\sim\!\! 0.01$, (which is typical), while measuring $C_\text{absorb} \!\!\sim\!\! 10^{-6}$. In Ref.~\cite{Ahmadi2018b} the same experimental setup performs magnetometry simultaneously using both green absorption and red fluorescence, as shown in Fig.~\ref{fig:Greenabsorbtion}. The green absorption yields $\sim\!\! 100 \text{ nT}/\sqrt{\text{Hz}}$ sensitivity while the conventional readout based on red fluorescence reaches  $\sim 400\!\! \text{ pT}/\sqrt{\text{Hz}}$, about 250$\times$ better. As with NIR absorption readout (see Sec.~\ref{NIRabsorption}), recycling the green excitation light via a resonant optical cavity can increase the absorption signal by (i) addressing a larger NV$^\text{-}$ population, (ii) improving initialization into the $m_s=0$ state, or (iii) enhancing $C_\text{absorb}$. Although effectively implemented absorption readout may achieve higher optical collection efficiency than fluorescence detection, the low realized absorption contrasts are a current major drawback.



Furthermore, absorption behavior for 532~nm probe radiation can result in \textit{increased} probe laser transmission under resonant MW application~\cite{BauchThesis2010,LeSageKeigoPrivate}, leading to an anomalous inversion of the green absorption signal. This deviation from expected behavior has been independently observed in multiple research groups~\cite{BauchThesis2010,LeSageKeigoPrivate}. The anomalous $C_\text{absorb}$ reveals a strong wavelength and power dependence~\cite{BauchThesis2010}, which suggests that green absorption readout is hindered by an unknown effect competing with and sometimes dominating otherwise expected behavior. The wavelength and power dependence of this effect suggests NV$^0$/NV$^\text{-}$ charge dynamics could play a role. Further investigation of this behavior might reveal presently unknown NV dynamics. Overall, given the low absorption contrast $C_\text{absorb}$, and yet unknown mechanism of anomalous absorption behavior, the utility of green absorption readout remains questionable. 



\subsection{Laser threshold magnetometry}\label{laserthresholdmagnetometry}

Another approach for bulk NV$^\text{-}$ magnetometry is the creation of a NV-diamond-based laser threshold magnetometer, as suggested by Ref.~\cite{Jeske2016}. \textcolor{black}{Lasing is induced on the NV$^\text{-}$ $^3\text{E}\,(v=0)\leftrightarrow ^3\text{A}_2\,(v'\geq 1)$ transition; then, when a magnetic-field-dependent population accumulates in the singlet state, the lasing threshold increases, and the laser's output power is reduced.} 
As theoretically outlined~\cite{Jeske2017,Savitski2017}, \textcolor{black}{the laser threshold approach} has a number of benefits relative to generic CW-ODMR methods (Sec.~\ref{cwodmr}): (i), effective contrast is enhanced near the lasing threshold due to competition between stimulated and spontaneous emission, (ii) collection efficiency is substantially improved by virtue of the lasing process. Although the emission cross sections for NV$^\text{-}$ and NV$^0$ have been measured~\cite{Fraczek2017}, and stimulated emission from NV$^\text{-}$ was recently demonstrated~\cite{Jeske2017}, substantial work remains to address potential problems. For example, absorption by substitutional nitrogen or other defects may obstruct the lasing process~\cite{Dodson2011}, and it will need to be shown that other sources of noise affecting the lasing threshold or output power can be either controlled or normalized out~\cite{Jeske2016}. More concerning, however, is that both theory~\cite{DeGiorgio1970} and experiment~\cite{Lim2002} find large laser field fluctuations in the vicinity of the lasing threshold.

%% file: sec05.tex
\section{Diamond material engineering}\label{sampleengineering}





\subsection{Conversion efficiency}\label{Econvdef}

In an idealized case in which all other parameters are held constant, increasing the NV$^\text{-}$ density in a fixed detection volume will result in enhanced sensitivity. Since the NV$^\text{-}$ density is limited by the density of nitrogen introduced into the diamond, the N-to-NV$^\text{-}$ conversion efficiency 
\begin{equation}
\text{E}_\text{conv} \equiv \frac{[\text{NV}^\text{-}]}{[\text{N}^\text{T}]}
\end{equation}
must be increased in order to achieve a high density of NV$^\text{-}$ spins while minimizing the concentration of residual paramagnetic substitutional nitrogen. Converting a substitutional nitrogen N$_\text{S}$ into a NV$^\text{-}$ defect requires both introducing a vacancy to a lattice site adjacent to a substitutional nitrogen (to create the NV), and capturing an electron (to change the NV center's charge state to NV$^\text{-}$). We denote the efficiency with which nitrogen atoms in the diamond are converted to NVs as
\begin{equation}
\chi = \frac{[\text{NV}^\text{T}]}{\text{[N}^\text{T}\text{]}}, 
\end{equation}
where $[\text{N}^\text{T}] = [\text{N}_\text{S}^0] + [\text{N}_\text{S}^+] + [\text{NV}^\text{-}]+[\text{NV}^0] +[\text{NV}^+] +[\text{N}^\text{other}]$ accounts for the concentration of substitutional nitrogen N$_\text{S}$ in the neutral and ionized charge states, NV centers in all three charge states, and other nitrogen-containing defects in the diamond
, such as NVH (see Sec.~\ref{chargetraps}). We define the fraction of NV centers residing in the negative charge state as the charge state efficiency $\zeta$,
\begin{equation}\label{eqn:chargestateefficiency}
\zeta =\frac{[\text{NV}^\text{-}]}{[\text{NV}^\text{T}]}= \frac{[\text{NV}^\text{-}]}{[\text{NV}^\text{-}]+[\text{NV}^0]+[\text{NV}^+]},
\end{equation}
so that $\text{E}_\text{conv} = \chi \cdot \zeta$. Although Refs.~\cite{Pfender2017,Hauf2014} show evidence for NV$^+$, this state has so far required application of external voltages for observation. The rest of this section therefore assumes [NV$^+$] is negligible and can be ignored.

As the N-to-NV$^\text{T}$ conversion efficiency $\chi$ is determined by the physical location of nitrogen and vacancies in the diamond lattice, the value of $\chi$ is expected to be invariant under ambient conditions. Modification of $\chi$  requires conditions severe enough to rearrange atoms within the diamond lattice, such as irradiation, implantation, high temperature, or high pressure. With suitable electron irradiation and subsequent annealing,  N-to-NV$^\text{T}$ conversion efficiencies approaching 1 can be achieved, although such high values are not necessarily desirable (see Secs.~\ref{irradiation} and \ref{LPHT}).  

In contrast, the charge state efficiency $\zeta$ depends on local conditions in the diamond and can be affected by external fields and optical illumination. Increasing  $\zeta$ benefits sensitivity in two ways: first, by increasing the NV$^\text{-}$ concentration and thus the number of collected photons $\mathscr{N}$ from the NV$^\text{-}$ ensemble; and second, by decreasing the concentration of  NV$^\text{0}$ and the associated background fluorescence, which improves measurement contrast. In the following section we discuss factors contributing to the charge state efficiency and methods to optimize it for sensing.

\subsection{NV charge state efficiency}\label{chargestate}




The charge state efficiency $\zeta$ from Eqn.~\ref{eqn:chargestateefficiency}, 
depends on many factors both internal and external to the diamond.  For both native NVs~\cite{Iakoubovskii2000} and NVs created by irradiation and annealing of nitrogen-rich diamonds~\cite{Mita1996}, the NV$^\text{-}$ and NV$^0$ charge states can coexist in a single sample. In general, for a given sample and experimental procedure, the steady-state charge state efficiency is difficult to predict. Contributing factors include the concentration of substitutional nitrogen and other defects serving as charge donors or acceptors~\cite{Groot-Berning2014} and their microscopic distributions~\cite{Collins2002, Doi2016}; the wavelength, intensity, and duty cycle of optical illumination~\cite{Doi2016, Manson2005,Aslam2013,Ji2016}; the application of a bias voltage~\cite{Grotz2012, Kato2013, Schreyvogel2014, Doi2014}; and, for near-surface NVs, the diamond surface termination~\cite{Rondin2010, Fu2010, Hauf2011, Cui2013, Groot-Berning2014, Chu2014,Yamano2017,Kageura2017,Osterkamp2015,Newell2016,Santori2009}. \textcolor{black}{The charge state efficiency is likely affected by the conditions of diamond growth, as well as the irradiation dose~\cite{Mita1996} (see Sec.~\ref{irradiation})}, the annealing duration and temperature, and possibly the operation temperature~\cite{Manson2005}. Moreover,  the value of the charge state efficiency $\zeta$ during an NV$^\text{-}$ sensing procedure can be difficult to measure. NVs may be reversibly converted between NV$^\text{-}$ and NV$^0$ by various optical and non-optical processes~\cite{Khan2009,Aslam2013,Bourgeois2017}. Because $\zeta$ is strongly affected by the illumination laser intensity and wavelength~\cite{Aslam2013,Bourgeois2017},  characterization of $\zeta$ by methods such as Fourier-transform infrared spectroscopy (FTIR), ultraviolet-visible (UV-Vis) spectroscopy, and electron paramagnetic resonance (EPR) may be misrepresentative of NV charge state behavior under the optical illumination employed in most NV-diamond sensing devices. 

\subsubsection{Non-optical effects on NV charge state efficiency}

Here we discuss the charge state efficiency $\zeta$ in nitrogen-rich diamond in the absence of optical illumination. For shallow NVs, the charge state is strongly affected by the surface chemical termination~\cite{Rondin2010, Fu2010, Hauf2011, Cui2013, Groot-Berning2014}. Based on the work in Ref.~\cite{Groot-Berning2014}, surface termination should provide enhanced charge state stability to a depth of at least 60~nm and possibly farther~\cite{Santori2009,Malinauskas2008}. The charge state efficiency $\zeta$ can also be controlled electrically~\cite{Grotz2012, Kato2013,Schreyvogel2014,Doi2014,Schreyvogel2015,Hauf2014,Murai2018,Forneris2017,Karaveli2016}. Because diamond is an approximately 5.47 eV wide band gap insulator~\cite{Wort2008}, Ref.~\cite{Collins2002} contends that an NV center's charge state depends less on the position of the Fermi level and more on the distance to the nearest charge donor. In nitrogen-rich diamonds, these donors are mainly substitutional nitrogen defects N$_\text{S}$, and the charge state efficiency $\zeta$ is seen to increase with the concentration $[\text{N}_\text{S}]$~\cite{Collins2002,Manson2005}. Other defects in the diamond lattice can alter $\zeta$ as well; for example, in Ref.~\cite{Groot-Berning2014}, the NVs in separate implanted regions containing phosphorus (an electron donor) and boron (an electron acceptor), were seen to have increased, and respectively decreased, NV charge state efficiencies. 

Introduction of electron donors other than nitrogen into diamond might appear to be a promising avenue for increasing the NV charge state efficiency. For example, phosphorus~\cite{Groot-Berning2014,Doi2016,Murai2018}, with donor level 0.6~eV below the conduction band~\cite{Katagiri2004}, is a shallower donor than nitrogen, which lies 1.7~eV below the conduction band~\cite{Farrer1969,Wort2008}. \textcolor{black}{However, creating n-doped diamond through introduction of phosphorus has proven difficult~\cite{Kalish1999}, as phosphorus doping is correlated with the introduction of a deep acceptor tentatively identified as the phosphorus vacancy (PV)~\cite{Jones1996}. Moreover, irradiation and annealing to create NV centers may further convert desirable substitutional phosphorus into PVs~\cite{Jones1996}. PVs in diamond will compete with NVs for electrons, undermining the benefit of phosphor doping to the charge state efficiency. Additionally, PL emission at wavelengths overlapping the NV$^\text{-}$ PL spectrum was observed in phosphorus-doped diamond~\cite{Cao1995}, further complicating the use of phosphorus in NV-diamond sensing. For additional discussion see Ref.~\cite{doherty2016towards}.}


The irradiation and annealing procedures applied to increase the N-to-NV$^\text{T}$ conversion efficiency $\chi$ can also affect the charge state efficiency $\zeta$. In Type Ib diamonds grown by high-pressure-high-temperature (HPHT) synthesis (see Sec.~\ref{hpht}), with $[\text{N}_\text{S}] \gtrsim 50~$ppm, $\zeta$ approaching 1 is seen after low- and moderate-dose irradiation and annealing~\cite{Manson2005, Mita1996}. As discussed in Sec.~\ref{irradiation}, at higher irradiation doses, the NV$^0$ concentration is seen to abruptly increase~\cite{Mita1996}, which can be attributed to the combination of an insufficient concentration of nitrogen defects N$_\text{S}$ available to donate electrons to the increasing overall NV population, and an increase in deep acceptor states such as multi-vacancy defects~\cite{Twitchen1999b,Pu2001}. 



\subsubsection{Optical effects on NV charge state efficiency}

Optical illumination of diamond may also change the NV charge state efficiency $\zeta$ through ionization of NV$^\text{-}$ to NV$^0$ and also recombination of NV$^0$ back to NV$^\text{-}$~\cite{Manson2005, Aslam2013, Waldherr2011, Hopper2016, hopper2018amplified}. 
The steady-state value of $\zeta$ is seen to depend on the illumination intensity and wavelength, although most of the reported measurements have been taken on single NV centers~\cite{Aslam2013,Waldherr2011,Hopper2016}. For example, an excitation wavelength band from 510~nm to 540~nm was found to produce the most favorable single-NV charge state efficiency in steady state compared to longer and shorter wavelengths~\cite{Aslam2013}. \textcolor{black}{In particular, when single NVs were illuminated by 532~nm light at intensities typical for pulsed sensing protocols~\cite{Hopper2016, Waldherr2011} or similar wavelength light at lower intensities~\cite{Aslam2013},} a charge state efficiency $\zeta \sim 70\,\text{-}\,75\%$ was observed. However, the value of $\zeta$ under these conditions is likely to differ for dense NV ensembles~\cite{Manson2005,meirzada2018negative}. For example, measurements in Ref.~\cite{Manson2005} on an NV ensemble in a diamond with $[\text{N}_\text{S}^\text{T}] \sim 70$~ppm and $[\text{NV}^\text{T}] \sim 1$~ppm show the charge state efficiency dropping to $\sim 50\%$ as the 532~nm power approaches the saturation power of the NV$^\text{-}$ optical transition. More study is required to determine the relative contributions to the NV ensemble $\zeta$ of optical charge-state switching, the presence of nearby charge donors/acceptors, and other effects.



\begin{figure}[ht]
\centering
\begin{overpic}[width = \columnwidth]{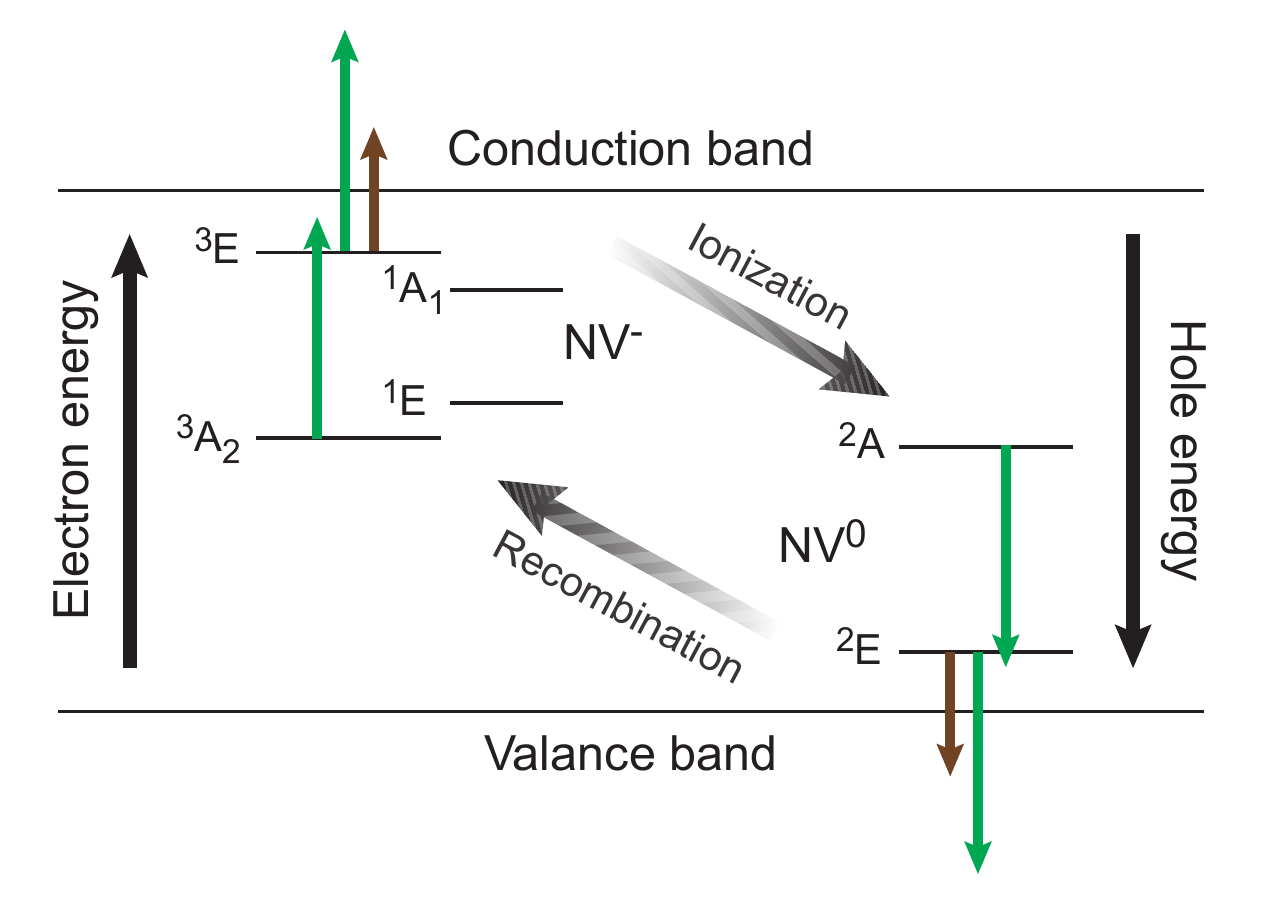}  
\end{overpic}
\caption[NVEnergyLevelDiagram]{Energy level diagrams for NV$^\text{-}$ and NV$^0$, representing optical ionization and recombination processes \textcolor{black}{through either absorption of two 532~nm photons [green arrows (\textcolor{igorgreen}{$\rightarrow$})] or a 532~nm photon and an NIR photon [brown arrows (\textcolor{brown}{$\rightarrow$})]}
}  \label{fig:Hopper}
\end{figure}

Recently, several studies on single NV centers have shown improved optical initialization to NV$^\text{-}$ by applying near-infrared radiation (NIR) in combination with the 532~nm green excitation light~\cite{Hopper2016, Ji2016, Chen2017}. This enhanced charge-state initialization has been \textcolor{black}{demonstrated with 780~nm CW radiation~\cite{Chen2017}, 1064~nm CW radiation~\cite{Ji2016}, and 900~nm\,-\,1000~nm pulsed radiation, achieving in the third case $\zeta >90\%$~\cite{Hopper2016}.} The effect is theoretically explained as follows: after absorption of a green photon to enter the electronically excited state, an NV$^0$ absorbs an NIR photon, which promotes a hole to the valence band and forms NV$^\text{-}$~\cite{Ji2016, Hopper2016}. The mechanism, visualized schematically in Fig.~\ref{fig:Hopper}, is the same as the two-photon ionization and recombination of NV$^\text{-}$ and NV$^0$ by 532~nm radiation, but with the second absorbed photon being an NIR photon. In Ref.~\cite{Hopper2016} the NV$^0$-to-NV$^\text{-}$ recombination process is found to occur with up to a $\sim 7\times$ higher likelihood than the analogous ionization process converting NV$^\text{-}$ to NV$^0$, wherein the excited-state NV$^\text{-}$ absorbs an NIR photon, promoting an electron to the conduction band. 

NIR-enhancement of charge state efficiency is expected to be compatible with pulsed initialization and readout. However, when employing 532~nm intensities $I \approx I_\text{sat} \approx 2.7$~mW/$\upmu$m$^2$~\cite{Wee2007} typical for pulsed experiments, Ref.~\cite{Hopper2016} finds enhancements in $\zeta$ to be lessened compared to \textcolor{black}{operation at} lower green intensity.  Furthermore, if the charge switching rate under green-plus-NIR illumination approaches \textcolor{black}{or exceeds} the optical \textcolor{black}{spin polarization} rate, spin readout fidelity could be degraded by the increased photoionization during the readout pulse. Refs.~\cite{Ji2016} and~\cite{Chen2017} report charge switching rates near $\sim 1~\upmu\text{s}^{\text{-}1}$, approaching the singlet state decay rate of $4 ~\upmu\text{s}^{\text{-}1}$~\cite{Acosta2010b}. Nonetheless, Hopper \textit{et al.}~achieve enhanced charge state initialization with much lower charge switching rates of $\sim 10$~ms$^{\text{-}1}$. 

Further work is required to determine if 
\textcolor{black}{the NIR-plus-green illumination technique can be extended to increase the charge state efficiency $\zeta$ in NV ensembles. While the technique has shown success for near-surface NV centers, Ref.~\cite{meirzada2018negative} observes no enhancement in the NV$^\text{-}$ PL from NIR-plus-green illumination compared to green-only excitation for NV centers in bulk diamond.} Moreover, even if NIR-plus-green illumination can enhance the ensemble value of $\zeta$, the power requirements may limit the technique's application to large ensembles. Although the required NIR power for confocal setups addressing single NV$^\text{-}$ centers or small ensembles is modest ($\sim$\,mW), the NIR intensity is $\gtrsim 10 \times$ higher than the typical 532~nm intensities used for NV$^\text{-}$ spin initialization ($I_\text{NIR} \approx 23 \,I_\text{sat}^{532~\text{nm}}$ in Ref.~\cite{Hopper2016}). Thus, when applying the technique to macroscopic ensemble volumes, the maximum addressable ensemble size will quickly be limited by the available laser power. For example, a $(50~\upmu\text{m})^2$ spot would require $\gtrsim 100$~W of NIR~\cite{Wee2007}. At present, NIR-enhancement of charge state efficiency appears unlikely to yield substantial improvements to ensemble-NV$^\text{-}$ magnetometer sensitivities.


\subsection{Diamond synthesis and high pressure high temperature treatment}\label{hpht}



Fabricated bulk diamonds are commonly synthesized using one of two methods. In high pressure high temperature (HPHT) synthesis, a process mimicking natural diamond formation, a carbon source material is mechanically compressed (pressure > 5 GPa) and heated (temperatures $\gtrsim 1250$ $^\circ$C) to create conditions where diamond is the thermodynamically favored carbon allotrope.  Dissolving the carbon source (typically graphite) in a metal "solvent-catalyst" can increase the growth rate, decrease the required temperature and pressure, and allow for better composition control. Consequently, solvent-catalysts are nearly always employed. A small seed diamond facilitates the growth; the dissolved carbon precipitates out of the metal catalyst solution and crystallizes onto the seed diamond, growing the size. Nitrogen easily incorporates into the diamond lattice, and is historically the primary impurity element in HPHT diamonds. However, nitrogen content in HPHT-synthesized diamonds can be reduced by varying the atomic composition of the metal solvent catalyst to ``getter'' the nitrogen, and recent advances in getter technology have allowed direct creation of electronic grade HPHT diamond with [N$_\text{S}^0$] $\lesssim$ 5 ppb~\cite{
DHaenensJohansson2015,Tallaire2017}. References ~\cite{Kanda2000,Dobrinets2013,Palyanov2015} discuss HPHT synthesis in detail. 

Plasma-enhanced chemical vapor deposition (PE-CVD) diamond synthesis ~\cite{Angus1968} is a popular alternative to HPHT synthesis, and can leverage established semiconductor fabrication techniques. In the most widespread variant of this method, employing homo-epitaxial growth, a diamond seed is exposed to a hydrocarbon plasma consisting of \textcolor{black}{approximately 95 - 99\%} hydrogen, with the balance composed of carbon and possibly other species such as oxygen or argon. Methane is the most popular carbon source. Radicalized carbon atoms bond with the growth surface, forming a mixture of sp$^2$ and sp$^3$ bonded orbitals. Although  hydrogen etches both sp$^2$ and sp$^3$ bonded carbon, the etch rate for sp$^2$ bonded carbon is much greater~\cite{Schwander2011} and, if the hydrogen etching and carbon deposition rates are carefully tuned, diamond synthesis can be achieved~\cite{Angus1968}. Unlike HPHT synthesis, PE-CVD (alternatively simply called CVD) synthesis can easily allow the production of thin or delta doped layers from nanometer to micron scale~\cite{Ohno2012,McLellan2016}, masked synthesis of diamond structures~\cite{Zhang2016}, or layered epitaxial growth required for PIN~\cite{Kato2013} or NIN structures~\cite{Murai2018}.



In the past 15 years, the majority of NV-diamond literature has employed diamonds grown by PE-CVD. First, much early work focused on single NV$^\text{-}$ centers; and most HPHT-synthesized diamonds were not available at that time with the requisite low nitrogen concentration ($\lesssim$ 100 ppb), as HPHT impurity control can be challenging~\cite{Martineau2009,Gaukroger2008}. Second, the layered deposition inherent to CVD allows straightforward growth of epitaxial layers (as would be required for magnetic imaging devices) and the application of semi-conductor techniques to control diamond composition. Third, the PE-CVD method was historically more popular with commercial collaborators (such as Element Six and Apollo Diamond)  responsible for producing the majority of scientific diamonds containing NV$^\text{-}$ centers. 





\textcolor{black}{In addition, several challenges accompany direct HPHT synthesis of high-quality NV-diamonds. For one, solvent-catalyst incorporation into the diamond lattice may result in metal inclusions with size visible to the naked eye. Such inclusions could be particularly problematic for magnetic sensing applications, since the common materials employed in the solvent catalyst alloys are the ferromagnetic elements Fe, Co, and Ni~\cite{Palyanov2015}. The purity of HPHT-synthesized diamonds may be limited by the solid precursor materials, which may not be available with as high chemical or isotopic purity as the gas-phase precursor elements employed for CVD synthesis. Finally, the HPHT process is not intrinsically amenable to fabrication of NV$^\text{-}$-rich layers, as are needed for imaging applications. In spite of these challenges, HPHT-fabricated diamonds with good characteristics for ensemble-NV$^\text{-}$ DC magnetometry - including long $T_2^*$ ($\gtrsim 2~\upmu$s), high $\text{E}_\text{conv}$ ($\sim 30\%$), and [N$^\text{T}]\sim$ 1-4~ppm - have been recently reported in the literature~\cite{Wolf2015,Grezes2015,sturner2019compact} (see Table~\ref{tab:diamondsurvey}).}

\begin{table*}[ht]
\centering 
\begin{tabular}{c r r r r r r r c} 
\toprule
Reference & \parbox[t]{3em}{$T_2^*$} & \parbox[t]{3em}{$T_2$} & \parbox[t]{3em}{E$_\text{conv}$} & \parbox[t]{3em}{[NV$^\text{-}$]} & \parbox[t]{3em}{[NV$^0$]} & \parbox[t]{3em}{[N$^\text{T}$]} & \parbox[t]{3em}{$[^{13}\text{C}]$} & Synthesis \\ [0.5ex] 
\midrule 
\cite{Grezes2015} & $\sim 2.6~\upmu$s & 84~$\upmu$s & 29$\%$ & 0.4~ppm & 0.4~ppm  & 1.4~ppm & 300~ppm & HPHT \\ 
\cite{Wolf2015} & \parbox[t]{3em}{-}  & $\sim 50$~$\upmu$s & 30$\%$ & 0.9~ppm & \parbox[t]{3em}{-}  & 3~ppm & \parbox[t]{3em}{-}  & HPHT\\
\cite{zheng2018zero}
& $\geq1.4~\upmu$s & \parbox[t]{3em}{-}  & \parbox[t]{3em}{-}  & $\sim 0.9$~ppm & \parbox[t]{3em}{-}  & $ > 2.9$~ppm & 300~ppm
& HPHT \\
\cite{Hartlandthesis2014} & \parbox[t]{3em}{-}  & \parbox[t]{3em}{-}  & 28$\%$ & $1.2$~ppm & 0.7~ppm & 4.1~ppm & 10700~ppm & CVD+HPHT \\
\cite{schloss2019optimizing}
& $1.55~\upmu$s & $15.7~\upmu$s & $\sim 30 \%$ & $\sim 3$~ppm & \parbox[t]{3em}{-}  & $\sim 10$~ppm & $100$~ppm 
& CVD \\
\cite{Barry2016} 
& 580~ns & 5.1~$\upmu$s & 6.3$\%$ & $\sim 1.7$~ppm & \parbox[t]{3em}{-}  & 27~ppm & 10~ppm & CVD \\
\cite{Schloss2018} 
& 450~ns & $7~\upmu$s & $\sim 14 \%$ & $3.8$~ppm & $2.0$~ppm & $\sim 28$~ppm & 10700~ppm & CVD \\
%
\bottomrule 
\end{tabular}
\caption[TableSurveyLiteratureDiamondsAndT2star]{Partial literature survey of diamonds with properties well-suited to ensemble-NV$^\text{-}$ magnetometry. Diamonds with long $T_2^*$, high N-to-NV$^\text{-}$ conversion efficiency E$_\text{conv}$, and [NV$^\text{-}]\gtrsim 1$~ppm, are expected to be particularly favorable \textcolor{black}{for high sensitivity magnetometry applications}. Dashed lines (-) indicate values not reported or unknown.}
\label{tab:diamondsurvey} 
\end{table*}

\textcolor{black}{While the exact motivation for HPHT diamond synthesis is not always explicitly stated~\cite{Teraji2013}, HPHT synthesis may circumvent undesired characteristics inherent to CVD-synthesized diamonds~\cite{Hartlandthesis2014,Charles2004}. A serious disadvantage of CVD synthesis is the incorporation of unwanted impurities and charge traps into the lattice (see Sec.~\ref{chargetraps}). In addition, CVD-grown diamonds may} display undesirable strain non-uniformities or contain a high dislocation density. 
For example, CVD-grown diamonds sometimes exhibit a brown coloration, which is attributed to vacancy cluster incorporation during synthesis~\cite{Hounsome2006,Khan2013}.  As vacancy clusters, chains, and rings are typically paramagnetic~\cite{Iakoubovskii2002,Yamamoto2013,Lomer1973,Baker2007}, these clusters can increase NV$^\text{-}$ ensemble dephasing, reducing $T_2^*$. Additionally, since such vacancy chains and clusters are deeper \textcolor{black}{electron} acceptors than NV$^\text{-}$~\cite{Khan2009,Edmonds2012}, their presence may decrease measurement contrast~\cite{Tallaire2017b}. \textcolor{black}{Naturally occurring diamond that has not undergone irradiation rarely contains} vacancies~\cite{Mainwood1999}, \textcolor{black}{suggesting that} vacancies and vacancy clusters \textcolor{black}{should be} uncommon in well-synthesized HPHT diamond.
\textcolor{black}{As p}oint defects, dislocations, and other extended defects are believed to be the dominant sources of strain in Type IIa diamonds~\cite{Fisher2006}, \textcolor{black}{HPHT-synthesized diamonds may also exhibit lower strain than their CVD-grown counterparts}. While dislocation densities of $\approx 10^4\text{ - }10^6$~cm$^{\text{-}2}$ are typical in CVD-grown diamonds~\cite{Achard2014}, certain HPHT-synthesized diamonds can demonstrate dislocation densities of $\approx 100\text{ - }1000$~cm$^{\text{-}2}$~\cite{Martineau2009,Tallaire2017} and substantially lower strain~\cite{DHaenensJohansson2014,DHaenensJohansson2015}.

Although more research is needed, it is observed that the high quantity of hydrogen present during CVD growth can result in hydrogen incorporation into the diamond lattice~\cite{Charles2004,Goss2014}, (see Sec.~\ref{chargetraps}). In contrast, diamonds synthesized directly by HPHT are unlikely to have hydrogen defects, as only one hydrogen-related defect has been found to  incorporate into HPHT-synthesized diamond~\cite{Hartlandthesis2014}.

Alternatively, mixed-synthesis approaches can combine the strengths of CVD and HPHT. One popular method is HPHT treatment, where an existing CVD diamond is heated and subjected to high pressure, resulting in atomic-scale reconfigurations of atoms in the lattice while leaving the macro-scale diamond largely unchanged ~\cite{Dobrinets2013}. HPHT treatment effectively removes single vacancies~\cite{Collins2000,Dobrinets2013} and causes vacancy clusters to dissociate~\cite{Collins2000,Dobrinets2013} or aggregate~\cite{Bangert2009}. Thus, this method is effective to treat CVD-grown diamonds, which can exhibit vacancies and vacancy clusters~\cite{Charles2004,Hartlandthesis2014,Khan2013}. The approach of applying HPHT treatment to CVD diamonds was proposed in Ref.~\cite{E6patent2010WO149775} and realized by the author of Ref.~\cite{Hartlandthesis2014}, wherein a CVD-grown diamond was HPHT treated after synthesis but prior to irradiation and subsequent annealing (see Secs.~\ref{irradiation} and \ref{LPHT}). The diamond produced in Ref.~\cite{Hartlandthesis2014} exhibits a notably high 30$\%$ conversion efficiency $\text{E}_\text{conv}$= [NV$^\text{-}$]/[N$^\text{T}$] as shown in Table~\ref{tab:hartlanddiamond}. A similar process pioneered by Lucent Diamonds employs HPHT treatment of diamonds prior to irradiation and annealing~\cite{Vinspatent20070053823}. This process results in a final material with an intense red hue and photoluminescence dominated by NV$^\text{-}$ emission~\cite{Wang2005,Dobrinets2013}, suggesting that HPHT treatment can be effective to increase the charge state efficiency $\zeta$, likely by eliminating charge traps. 

However, HPHT treatment cannot address all diamond deficiencies, CVD-related or otherwise. For example, should a CVD-synthesized diamond incorporate high concentrations of hydrogen or other elemental impurities into the diamond lattice during growth, HPHT treatment is ineffective to remove these impurities~\cite{Charles2004}. Such treatment is also limited to diamonds with balanced aspect ratios, as thin plates or rods will likely crack under the high applied pressure.

In addition to HPHT treatment of existing diamonds, other mixed-synthesis approaches have also been pursued. For example, utilizing type IIa HPHT seeds for CVD growth rather than CVD-grown seeds can yield material with lower strain and reduced densities of dislocations and other unwanted defects~\cite{Martineau2009,Tallaire2017,Gaukroger2008,Hoa2014}. Another mixed-synthesis method exploits the fine composition control and high chemical purities available with CVD synthesis to create the carbon precursor for HPHT synthesis~\cite{Teraji2013}. \textcolor{black}{The diamond composition can thus be carefully controlled, and HPHT synthesis} can take advantage of high-purity or isotopically enriched gaseous sources (e.g.,~methane or  $^{15}$N$_2$). 






Given the prominent role lattice defects and elemental impurities play in \textcolor{black}{determining} the charge state efficiency and coherence times for NV$^\text{-}$, additional research focused on synthesizing sensing-optimized diamonds is warranted.

\subsection{Electron irradiation}\label{irradiation}





\begin{table*}[ht]
\centering 
\begin{tabular}{c c c c} 
\toprule
$\text{E}_\text{conv}$ & $[$N$^\text{T}]$ & Growth location & Reference \\ [0.5ex] 
\midrule
$0.0007-0.005$ & $0.3 - 30$~ppm & Element Six & \cite{Edmonds2012}  \\ 
$0.0006-0.03$ & $0.35 - 2.4$~ppm & Apollo Diamond Inc. & \cite{Edmonds2012}  \\
$0.02-0.03$ &$4$~ppm & Warwick University &  \cite{Hartlandthesis2014}\\
\bottomrule 
\end{tabular}
\caption[TableNtoNVConversionEfficincy]{Native N-to-NV$^\text{-}$ conversion efficiencies $\text{E}_\text{conv}$ and total nitrogen concentrations [N$^\text{T}$] in unmodified bulk CVD diamond}
\label{tab:nativeconversion} 
\end{table*}

\begin{table}[ht]
\centering 
\setlength{\extrarowheight}{1.5pt}
\begin{tabular}{r c} 
\toprule 
$[\text{V}^0+\text{V}^\text{-}]$  & Reference  \\ [1ex] 
\midrule 
\midrule 
$60$~ppb & \cite{Rutledge1998}  \\ 
$\lesssim$ $20$~ppb & \cite{E6patent2010WO149775}  \\
$\lesssim$ $0.03$~ppb & \cite{Mainwood1999}  \\
\bottomrule 
\end{tabular}
\caption[TableNativeFractionalMonovacancies]{Native monovacancy concentrations in unmodified bulk CVD diamond} 
\label{tab:nativevacancydensity}
\end{table}


For unmodified as-grown CVD diamond, realized conversion efficiency values $\text{E}_\text{conv}$ can be far less than unity, as shown in Table~\ref{tab:nativeconversion}, where the majority of substitutional nitrogen is not converted to NV$^\text{-}$~\cite{Edmonds2012,Hartlandthesis2014}. In fact, for some CVD diamonds (see Table~\ref{tab:nativevacancydensity}) the concentration of grown-in monovacancies is insufficient to achieve good $\text{E}_\text{conv}$ for total nitrogen concentration $[\text{N}^\text{T}] \gtrsim$ 1~ppm regardless of  location; even if every monovacancy were adjacent to a substitutional nitrogen, the conversion efficiency $\text{E}_\text{conv}$ would still be low~\cite{Mainwood1999,Deak2014}. However, the monovacancy concentration can be augmented after growth by irradiating the diamond with energetic particles. The high-energy irradiating particles knock carbon atoms out of the diamond lattice, creating both interstitial carbon atoms and monovacancies~\cite{Twitchen1999,Newton2002}. Although theoretical calculations have not yet completely converged with experimental observations~\cite{Deak2014,Zaitsev2017}, the widely accepted model posits that upon subsequent annealing (discussed in Sec.~\ref{LPHT}), diffusing vacancies are captured by substitutional nitrogen atoms, forming NV centers~\cite{Acosta2009}. Primary considerations in the irradiation process are the particle type, energy, and dose. 


The irradiation of diamond has been performed using a variety of particles: protons, ionized deuterium atoms, neutrons, and electrons~\cite{Ashbaugh1988}. Gamma ray irradiation from $^{60}$Co has also been used~\cite{Ashbaugh1988,Campbell2000}. Many of these particles are suboptimal for NV creation, however, where only single monovacancies V$^0$ are desired, and other created defects are likely deleterious.  A particular problem for certain irradiation methods is the production of ``knock-on-atoms''~\cite{Campbell2000,Davies2001}, where the irradiating particle has sufficient energy not only to displace an initial carbon atom from the lattice, but to impart enough kinetic energy to that carbon that it displaces additional carbon atoms, resulting in localized lattice damage~\cite{Buchan2015}. Although annealing  (see Sec.~\ref{LPHT})  can partially alleviate such damage,  the lattice damage can never be completely repaired~\cite{Balmer2009,E6patent2010WO149775,Lobaev2017,Oliveira2016,Oliveira2017} and may result in unwanted paramagnetic defects or charge traps. For irradiation with protons, neutrons, or ionized deuterium atoms, damage from such knock-on-atoms can be severe. Similar lattice damage occurs from ion implantation of various species \textcolor{black}{such as nitrogen~\cite{Oliveira2017,Yamamoto2013,Naydenov2010}, carbon~\cite{Naydenov2010,Schwartz2011}, and helium nuclei~\cite{Waldermann2007,Schwartz2011,Himics2014,McCloskey2014,Kleinsasser2016}.} Electrons, with their lower mass, transfer less kinetic energy to the carbon atoms and are therefore better suited to creating isolated monovacancies. Electron irradiation is favored over gamma ray irradiation because the former can be accomplished in hours whereas the latter, when implemented using $^{60}$Co, can take weeks~\cite{Collins2007}. In summary, electron irradiation is preferred to create NV$^\text{-}$ ensembles optimized for sensing applications~\cite{Uedono1999,Campbell2002,E6patent2010WO149775}, as this method allows for evenly distributed monovacancies to be created throughout the diamond in a timely manner, with less lattice damage than alternative methods. 

\textcolor{black}{Theoretical calculations predict monovacancy creation requires electron energies $\gtrsim$ 165 keV~\cite{Campbell2002}, roughly consistent with experiments observing vacancy creation down to 145 keV~\cite{McLellan2016,eichhorn2019optimizing}. Older, less reliable experiments find vacancy creation for electron irradiation along the [100] direction at $180~\text{keV}$ but not $170~\text{keV}$~\cite{Koike1992}. Crude estimates suggest electron irradiation energies lower than $\sim~0.8$ MeV will create mainly single vacancies~\cite{Loubser1978,Mitchell1965} and avoid producing multi-vacancy complexes.} While this estimate is consistent with Ref.~\cite{Dannefaer1992} where divacancies are detected after irradiation with 3.5 MeV electrons,  Ref.~\cite{E6patent2010WO149775}, however, finds no evidence of vacancy pairs after irradiation with 4.6 MeV electrons, suggesting that several-MeV irradiation energies may be safe. The optimal irradiation energy may also depend on sample geometries; thicker diamonds should require higher energies to ensure vacancies are created uniformly through the entire thickness~\cite{E6patent2010WO149775,Campbell2000}. For small ensembles close to the diamond surface, an electron microscope can provide the needed irradiation~\cite{Kim2012,Farfurnik2017,Schwartz2012,McLellan2016}. More study is required to resolve remaining discrepancies between experimental data and detailed simulations of the electron irradiation process~\cite{Campbell2000,Campbell2002}. For example, recent measurements of monovacancy density profiles versus depth, as judged by GR1 intensities in 1 MeV electron irradiated diamonds~\cite{Zaitsev2017}, are inconsistent with Monte Carlo simulations in Refs.~\cite{Campbell2000,Campbell2002}.

The irradiation dose should also be approximately matched to the diamond's total nitrogen concentration $[$N$^\text{T}]$ as suggested in Sec.~\ref{Econvdef}; if too many vacancies are created, then >~50$\%$ of N$_\text{s}$ will be converted to NV$^0$, and the number of electrons donated by the remaining N$_\text{s}$ will be insufficient to convert every NV$^0$ to NV$^\text{-}$. Figure 2 in Ref.~\cite{Mita1996} illustrates the importance of matching the irradiation dose to [N$^\text{T}$] to achieve maximal $\text{E}_\text{conv}$. When determining irradiation dose, in-situ recombination between a vacancy and an interstitial carbon should be accounted for~\cite{Davies2001,Campbell2000}. Current estimates suggest approximately 30$\%$~\cite{Campbell2000} to 50$\%$~\cite{Davies2001} of initially created vacancies are immediately lost to spontaneous recombination. For example, using 1~MeV electrons (generating $\sim2\times 10^{-4}$ vacancies/electron/$\upmu$m \textcolor{black}{according to Ref.~\cite{Campbell2000}}, and assuming 40$\%$ of vacancies recombine immediately and two nitrogens are required to make a single NV$^\text{-}$ center, we expect a sample with $[\text{N}^\text{T}] \sim 1$~ppm to require a dose of $7.3\times 10^{16}$ cm$^{\text{-}2}$. However, fine-tuning of the irradiation dose is often done empirically, suggesting either the presence of dynamics more complicated than those included in the simple model presented here (i.e.,~the presence of other vacancy traps, the formulation of divacancies, loss at surfaces, etc.) or errors in the measured electron flux or substrate temperature~\cite{Campbell2000}. For example,  while the production rate of neutral monovacancies from irradiation with 2~MeV electrons is found to be temperature-independent from room temperature to  $\sim 300~^\circ$C, the rate decreases notably for higher temperatures~\cite{Newton2002}.

\subsection{Low pressure high temperature annealing}\label{LPHT}




For the successful creation of NV$^\text{-}$ centers, substitutional nitrogen and monovacancies must be relocated to occupy adjacent sites in the diamond lattice. This process can be accomplished via diffusion at elevated temperature, i.e., annealing. Since monovacancies migrate in the neutral charge state $\text{V}^0$~\cite{Breuer1995} with an activation energy of $E_a = 2.3 \pm 0.3$ eV~\cite{Davies1992,Mainwood1999}, compared to measured values of $E_a=4.8-6.2$ eV  for substitutional nitrogen~\cite{Jones2015,Deak2014,Dobrinets2013}, neutral monovacancies diffuse throughout the lattice during annealing until they reach the more immobile nitrogens. The negatively charged monovacancy's higher activation energy~\cite{Breuer1995} ensures that monovacancy diffusion occurs predominantly in the neutral charge state~\cite{Breuer1995}, although a negative monovacancy can convert to a neutral monovacancy in a reversible charge transfer process~\cite{Davies1992}. The diffusion constant $D$ of the neutral monovacancy is~\cite{Orwa2012,Hu2002}
\begin{equation}\label{eqn:diffusionactivation}
D = D_0 e^{-E_a/k_B T},
\end{equation}
where $k_B$ is the Boltzmann constant, $T$ is the temperature, and $D_0$ is a diffusion prefactor (see Appendix~\ref{exampleannealing}). \textcolor{black}{Measurements of the diffusion constant $D$ have yielded $\sim 1.1$~nm$^2$/s at 750~$^\circ$C~\cite{Martin1999} and 1.8~nm$^2$/s at 850~$^\circ$C~\cite{alsid2019photoluminescence}, suggesting values for $D_0$ in agreement with an independently measured upper bound ~\cite{Acosta2009}, and in moderate agreement with first principles theoretical calculations~\cite{Fletcher1953} (see Appendix~\ref{exampleannealing}).} Other sources, however find or employ different values for $D_0$ or $E_a$~\cite{Orwa2012,Hu2002,Onoda2017}, suggesting that further measurements are warranted. Once an NV center is formed, the deeper binding energy of the nitrogen-vacancy bond relative to the neutral vacancy ensures that the bound vacancy does not diffuse away~\cite{Goss2005,Hartlandthesis2014}.







The procedure described here is commonly termed low pressure high temperature (LPHT) annealing to distinguish it from high pressure high temperature (HPHT) annealing (discussed in Sec.~\ref{hpht}). Given the role of diffusion in LPHT treatment,  the  annealing temperature and annealing duration are important control parameters. A  temperature of $\sim\!\!800$ $^\circ$C is usually employed~\cite{Botsoa2011}, given that monovacancies become mobile around $600~ ^\circ$C~\cite{Davies1992,davies1976optical,Uedono1999,Kiflawi2007}, and annealing times of several hours are typical, e.g., 2 hours in Ref.~\cite{Acosta2009}, 4 hours in Ref.~\cite{Lawson1998}, 8 hours in Ref.~\cite{E6patent2010WO149775}, 12 hours in Ref.~\cite{Barry2016}, and 16 hours in Ref.~\cite{Fraczek2017}. Diamonds with lower values of $[\text{N}^\text{T}]$ are expected to require longer annealing times due to the greater initial distances between vacancies and substitutional nitrogens. A study by Element Six found no observable deleterious changes in diamond properties between samples that were annealed at $\sim\!800$ $^\circ$C for $\sim 8$ hours and samples that were annealed at the same temperature for longer periods ~\cite{E6patent2010WO149775}. This $\sim 800$ $^\circ$C annealing step is typically performed under vacuum or in a non-oxidizing, inert gaseous environment to avoid graphitization~\cite{Dobrinets2013}. \textcolor{black}{Under vacuum, present understanding is that diamond graphitization begins roughly around $1500~ ^\circ$C~\cite{Davies1972}.}







Although the 800 $^\circ$C LPHT treatment is effective to create NVs, unwanted defects may form as well. For example, diffusing monovacancies can combine to form divacancies~\cite{Twitchen1999b}, which are immobile at $800~^\circ$C. As deeper electron acceptors than NVs~\cite{Deak2014,Miyazaki2014}, the presence of divacancies reduces $\text{E}_\text{conv}$. To mitigate divacancy formation, electron irradiation with in-situ (i.e.,~simultaneous) annealing has been proposed~\cite{Nobauer2013}.  Under such conditions, single vacancies are continuously created in an environment consisting primarily of substitutional nitrogen (and, as the process progresses, NVs), thereby reducing divacancy formation. Although preliminary work in Ref.~\cite{Nobauer2013} finds electron irradiation with in-situ annealing increases $T_2^*$, no increase in $\text{E}_\text{conv}$ is observed, and further investigation is warranted. 
 


Following NV formation, further LPHT annealing above 800~$^\circ$C may reduce strain or paramagnetic impurities resulting from lattice damage. For example, divacancies can combine into other defects at $\sim\!900$ $^\circ$C~\cite{Twitchen1999b}. Reduction of a given defect species may be effected by  consolidation into other larger defect species, which may be paramagnetic~\cite{Yamamoto2013,Hartlandthesis2014,Lomer1973,Baker2007}. Annealing to temperatures of 1000 $^{\circ}$C to 1200 $^{\circ}$C is shown to extend the $T_2$ of both single NV$^\text{-}$ centers~\cite{Naydenov2010,Yamamoto2013} and ensembles~\cite{Tetienne2018} created by ion implantation. As this increase is attributed to a reduction in paramagnetic multi-vacancy defects~\cite{Yamamoto2013,Tetienne2018}, improvement in $T_2^*$ is expected as well, though this expectation has not been systematically confirmed \textcolor{black}{in experiment.} Practically, this additional LPHT treatment is limited by the temperature at which NVs anneal out, which is typically around 1400~$^{\circ}$C to 1500~$^{\circ}$C~\cite{Zaitsev2001,Hartlandthesis2014,Pinto2012} and can vary depending on the presence of other defect species within the diamond~\cite{Zaitsev2001}. While a systematic study of annealing temperatures and durations is warranted for engineering optimal samples for ensemble-NV$^\text{-}$ sensing, a standard recipe for samples is at least several hours at $\sim\!\!800~^{\circ}$C followed by several more hours at $\sim\!\!1200~^\circ$C~\cite{Fraczek2017,Chu2014,breeze2018continuous}. Some example calculations for annealing are detailed in Appendix~\ref{exampleannealing}.










\begin{table}
\begin{center}
    \begin{tabular}{ l  l }
    \toprule
    \parbox[t]{8em}{\raggedright Diamond defect} & \parbox[t]{8em}{\raggedright Ground state spin}  \\ \midrule
    N$_\text{S}^0$ & $S=\nicefrac{1}{2}$    \\ 
    N$_\text{S}^+$ & $S=0$ \\ 
    NV$^\text{-}$ & $S=1$ \\ 
    NV$^0$ & $S=\nicefrac{1}{2}$ \\ 
    NV$^+$ & $S=0$ \\ 
   NVH$^\text{-}$ & $S=\nicefrac{1}{2}$  \\ 
   NVH$^{0}$ & $S=0$  \\ 
   N$_2^0$ & $S=0$~\cite{Tucker1994} \\ 
   N$_2^+$ & $S=\nicefrac{1}{2}$ \\ 
   N$_2$V$^\text{-}$ & $S=\nicefrac{1}{2}$ \\ 
   N$_2$V$^0$ & $S = 0$ \\ 
   N$_3$V$^0$ & $S=\nicefrac{1}{2}$ \\ 
   N$_2$VH$^0$ & $S=\nicefrac{1}{2}$ \\ 
   VH$^\text{-}$ & $S=1$ \\ 
   VH$^0$ & $S=\nicefrac{1}{2}$ \\ 
   V$_n$H$^\text{-}$ & $S=1$ \\ 
   V$^\text{-}$ & $S=\nicefrac{3}{2}$~\cite{Baranov2017} \\ 
   V$^0$ & $S=0$~\cite{Baranov2017} \\ 
   V$^+$ & $S=\nicefrac{1}{2}$~\cite{Baranov2017} \\ 
   VV$^\text{-}$ & $S=\nicefrac{3}{2}$~\cite{Kirui2013} \\ 
   VV$^0$ & \parbox[t]{13.5em}{\raggedright $S=1$~\cite{Twitchen1999b}} \\ \bottomrule 
    \end{tabular}
\end{center}
\caption{Common defects in diamond and their ground state electronic spin}\label{tab:defectspins}
\end{table}

\subsection{Other common impurities in synthetic or treated single crystal diamond}\label{chargetraps}

Unwanted species in the diamond lattice can degrade magnetometer performance by decreasing the NV charge state efficiency $\zeta = [\text{NV}^\text{-}]$/[NV$^\text{T}$], creating local magnetic noise, or reducing the fraction of substitutional nitrogen N$_\text{S}$ converted to NV$^\text{-}$. This section restricts detailed discussion to multivacancy clusters and NVH~\cite{Khan2013}, species present in diamond at sufficient concentrations to likely affect NV spin and charge dynamics. Extended discussion of other defects can be found in Refs.~\cite{Newton2007,Deak2014}; see also Table~\ref{tab:defectspins} for relevant defects commonly found in diamond. 

Multivacancy clusters are common in some diamonds grown by chemical vapor deposition (CVD)~\cite{Pu2000, Hounsome2006}, and are believed to cause the brown coloration in CVD-grown diamond~\cite{Hounsome2006,Fujita2009}. During CVD synthesis, the diamond surface can become rough and stepped. When these steps are rapidly covered with additional deposited material, small voids, i.e.,~clusters of vacancies, can be left  in the diamond~\cite{Hounsome2006,Khan2013}. Multivacancy cluster incorporation has been observed to increase at high growth rates~\cite{Hounsome2006}, and may be correlated with nitrogen content~\cite{Pu2000}.  Using positron annihilation, the authors of Ref.~\cite{Dannefaer1993} found the density of multivacancy clusters was found  to be roughly $10^{17}-10^{18}$ cm$^{\text{-}3}$ for their growth conditions. Such vacancy clusters can  trap electrons~\cite{Campbell2002,Jones2007,Edmonds2012,Deak2014}, reducing the ratio of NV$^\text{-}$ to NV$^0$ and also generating magnetic noise resulting from their trapped unpolarized electron spins. The neutral divacancy V$_2^0$~\cite{Lea-wilsonf1995,Twitchen1999b,Deak2014,Slepetz2014} and neutral multivacancy chains (V$_n^0$, $n \geq 3$) are paramagnetic~\cite{Iakoubovskii2002,Lomer1973,Baker2007}, and increase environmental magnetic noise. Irradiation or implantation followed by annealing can also produce such defects ~\cite{Naydenov2010,Yamamoto2013}. Low pressure high temperature annealing is effective to remove certain multivacancy clusters. However, as the removal of multivacancy clusters is effected by aggregating these species together or combining them with other defects, the reduction of smaller multivacancy defects may be accompanied by an increase in larger multivacancy clusters or other defects. High pressure high temperature (HPHT) treatment effectively removes single vacancies~\cite{Dobrinets2013} and causes some vacancy clusters to dissociate~\cite{Dobrinets2013}, which may aggregate to form different multivacancy clusters~\cite{Bangert2009}. See Sec.~\ref{hpht}.




Another common impurity in diamond is hydrogen, which gives rise to many defects~\cite{Dobrinets2013,Zaitsev2001,Goss2014}. For typical CVD diamond growth, the plasma is composed predominantly of hydrogen ($\gtrsim 95\%$)~\cite{Tokuda2015}, which can incorporate into single crystal diamond at concentrations as high as 1000~ppm~\cite{Sakaguchi1999}. The hydrogen incorporation rate into the lattice is partially dependent upon the diamond growth recipe~\cite{Tang2004}, and further investigation into the hydrogen quantity incorporated and methods to mitigate hydrogen incorporation \textcolor{black}{is warranted. Hydrogen-related defects may influence the NV charge state~\cite{lyons2016surprising,Hauf2011}. Additionally, at high enough concentrations the nuclear spin of hydrogen may result in non-negligible dephasing or decoherence.} At present we are unaware of any published method to effectively remove hydrogen from the bulk diamond lattice~\cite{Charles2004,Hartlandthesis2014}.

The presence of hydrogen in the diamond lattice can enable formation of the NVH defect~\cite{Glover2003}, wherein the hydrogen occupies the vacancy of an NV. In as-grown nitrogen-enriched CVD diamond, the ratio of ($[$N$_\text{S}^+]$+$[$N$_\text{S}^0]$):$[$NVH$^\text{-}]$:$[$NV$^\text{-}]$ was found to be approximately 300:30:1 in Ref.~\cite{Edmonds2012} and 52:7:1 in Ref.~\cite{Hartlandthesis2014}. The NVH species is undesirable because: (i) it lowers the conversion efficiency of incorporated nitrogen to NV centers; (ii) it reduces the concentration of substitutional nitrogen N$_\text{S}$ available to donate electrons to turn NV$^0$ defects into NV$^\text{-}$; (iii) NVH competes with NV as an electron acceptor; \textcolor{black}{(iv) NVH$^\text{-}$ is paramagnetic, causing magnetic noise; and (v) the hydrogen in NVH may rapidly tunnel among the three adjacent carbon atoms at GHz frequencies, resulting in high-frequency magnetic or electric noise~\cite{Edmondsthesis2008}.} 

No known treatment can transform existing NVH defects into NV defects. The NVH complex is stable against annealing up to approximately 1600~$^\circ$C but anneals out completely by 1800 $^\circ$C~\cite{Khan2013,Hartlandthesis2014}. However, removal of NVH via annealing is not associated with increased NV concentration; rather, further isochronal annealing to 2000 $^\circ$C and 2200 $^\circ$C is accompanied by increases in N$_2$VH$^0$ and N$_3$VH$^0$ species~\cite{Hartlandthesis2014}, suggesting the NVH concentration is reduced via aggregation of NVH with one or more nitrogen atoms. \textcolor{black}{NVH$^0$ exhibits absorption at 3123 cm$^{\text{-}1}$~\cite{Cannthesis2009} but is otherwise not known to be optically active.}




%

Diamonds subject to temperatures at which substitutional nitrogen or interstitial nitrogen become mobile may exhibit defects consisting of aggregated nitrogen, such as N$_2$~\cite{davies1976thea,Boyd1994,Tucker1994}, N$_2$V~\cite{Green2015}, N$_2$VH~\cite{Hartlandthesis2014}, N$_3$V~\cite{Green2017}, N$_3$VH~\cite{Ligginsthesis2010,Hartlandthesis2014}, N$_4$V~\cite{Bursill1985}, or other aggregated nitrogen defects~\cite{Goss2004}. The presence of aggregated nitrogen defects reduces the quantity of nitrogen available to form NV centers or donate electrons to NV$^0$ to form NV$^\text{-}$, and can cause additional paramagnetic noise. Other defects such as VH~\cite{Glover2004,Gloverthesis2003}, V$_2$H~\cite{Shaw2005,Cruddacethesis2007}, and OV~\cite{Cannthesis2009,Hartlandthesis2014} have been identified in synthetic diamond and may act as charge acceptors or create additional paramagnetic noise. However most defects discussed in this paragraph are observed at concentrations  low enough to be neglected for diamonds fabricated for NV$^\text{-}$ magnetometry, as shown in Table~\ref{tab:hartlanddiamond}, reproduced from Ref.~\cite{Hartlandthesis2014}. Additional defect species are inferred to exist from charge conservation arguments but have not been directly observed~\cite{Khan2009}. More research is needed to better  understand defects in synthetic diamond grown for magnetometry applications. 


\begin{table*}[ht]
\centering 
\begin{tabular}{l c c c c} 
\toprule 
 \parbox[t]{6em}{\raggedright Defect} & \parbox[t]{8em}{As-grown} & \parbox[t]{8em}{1500~$^\circ$C anneal} & \parbox[t]{8em}{Irradiation} & \parbox[t]{8em}{800~$^\circ$C anneal}  \\ [0.5ex] 
\midrule 
$[$N$_\text{S}^0]$ (ppb) & 1620 (160) & 1100 (100) & 200 (20) & 120 (15) \\
$[$N$_\text{S}^+]$ (ppb) & 1500 (150) & 2200 (250) & 3000 (300) & 1000 (100) \\
$[$NV$^0]$ (ppb) & $\leq$ 10 & $\leq 10$ & $\leq 10$ & 695 (70) \\
$[$NV$^\text{-}]$ (ppb) & 60 (5) & 40 (5) & 35 (5) & 1160 (120) \\
$[$NVH$^0]$ (ppb) & 500 (50) & 310 (30) & 380 (40) & 290 (30) \\
$[$NVH$^\text{-}]$ (ppb) & 405 (40) & 200 (20) & obscured & 20 (5) \\
$[$N$_2$VH$^0]$ (ppb) & $<0.1$ & 22 (3) & obscured & 24 (5) \\
$[$V$_n$H$^\text{-}]$ (ppb) & 3.1 (1) & $\leq 0.1$ & 25 (3) & 41 (4) \\
\bottomrule 
\end{tabular}
\caption[TableHartlandDiamond]{Concentrations of quantifiable defects in sample GG1 in the as-grown state and after each treatment stage. From Ref.~\cite{Hartlandthesis2014}.}
\label{tab:hartlanddiamond} 
\end{table*}

\subsection{Preferential orientation}\label{preferentialorientation}

In naturally occurring and many fabricated diamonds, NV$^\text{-}$ centers are distributed evenly among all four crystallographic orientations. However, under certain circumstances, CVD-grown diamond can exhibit preferential orientation of NV$^\text{-}$ centers along certain crystallographic axes~\cite{Edmonds2012, Pham2012b}. Several research groups have achieved almost perfect alignment of all NV$^\text{-}$ centers along the a single [111] axis. Michl \textit{et al.}~demonstrated 94$\%$ alignment~\cite{Michl2014}, Lesik \textit{et al.}~demonstrated 97$\%$ alignment~\cite{Lesik2014}, and Fukui \textit{et al.}~demonstrated 99$\%$ alignment~\cite{Fukui2014}. The mechanism for preferential orientation is explained in Ref.~\cite{Miyazaki2014}.

An ensemble-NV$^\text{-}$ magnetometer utilizing a single NV$^\text{-}$ orientation in a diamond with no preferential orientation suffers from reduced measurement contrast due to unwanted PL from NV $^\text{-}$ centers of other orientations. A diamond with 100\% preferential orientation may allow a $4 \times$ increase in contrast. In practice, though, the enhancement is typically somewhat less than $4\times$, since polarized excitation light can already be used to selectively address particular NV$^\text{-}$ orientations~\cite{Lesik2014}, and high bias fields can suppress fluorescence from off-axis NV$^\text{-}$ centers~\cite{Epstein2005, Tetienne2012}.

Diamonds grown with preferential orientation have at least two main drawbacks. First, NV$^\text{-}$ concentrations for preferentially grown diamonds in the literature are currently relatively low~\cite{Michl2014,Lesik2014,Fukui2014}, typically around $10^{12}$ cm$^{\text{-}3}$ although concentrations up to $10^{15}$ cm$^{\text{-}3}$ have been achieved~\cite{Tahara2015}.  Second, it appears that the N-to-NV$^\text{-}$ conversion efficiency cannot be increased through irradiation and subsequent annealing without destroying the preferential alignment, although conflicting evidence on this topic has been reported~\cite{Fukui2014}. Since electron irradiation followed by annealing can increase the N-to-NV conversion efficiency by $\sim 10\times$ to $100\times$, preferential orientation is not currently believed to be a viable method to achieve better ensemble magnetometry sensitivity. However it is possible that future technical advances or treatment could alter this understanding. Additionally, preferential orientation precludes the implementation of vector magnetometry~\cite{Schloss2018}.


%% file: sec06.tex
\section{Miscellaneous sensing techniques}\label{miscellaneoustechniques}

\subsection{Rotary echo magnetometry}
\label{rotaryecho}
Broadband magnetometry can also be performed using a MW pulse scheme called rotary echo~\cite{Aiello2013,Mkhitaryan2014,Mkhitaryan2015}. In this technique pioneered by Aiello \textit{et al.}~\cite{Aiello2013}, rotary echoes are produced by periodic reversals of the driving field. The simplest protocol inverts the phase of the driving field to reverse the sign of the Rabi oscillations. The rotary echo technique may have utility for certain niche applications such as event detection~\cite{Aiello2013}, but the method so far yields worse sensitivity than a Ramsey protocol. Like other dynamical-decoupling-type methods, rotary echo can be tailored to reject noise at certain frequencies and also has applications for certain narrowband AC sensing, such as detection of individual nuclear spins~\cite{Mkhitaryan2015}. 

\subsection{Geometric phase magnetometry}
\label{geometricphase}

\begin{figure}
\begin{overpic}[height=2.2 in]{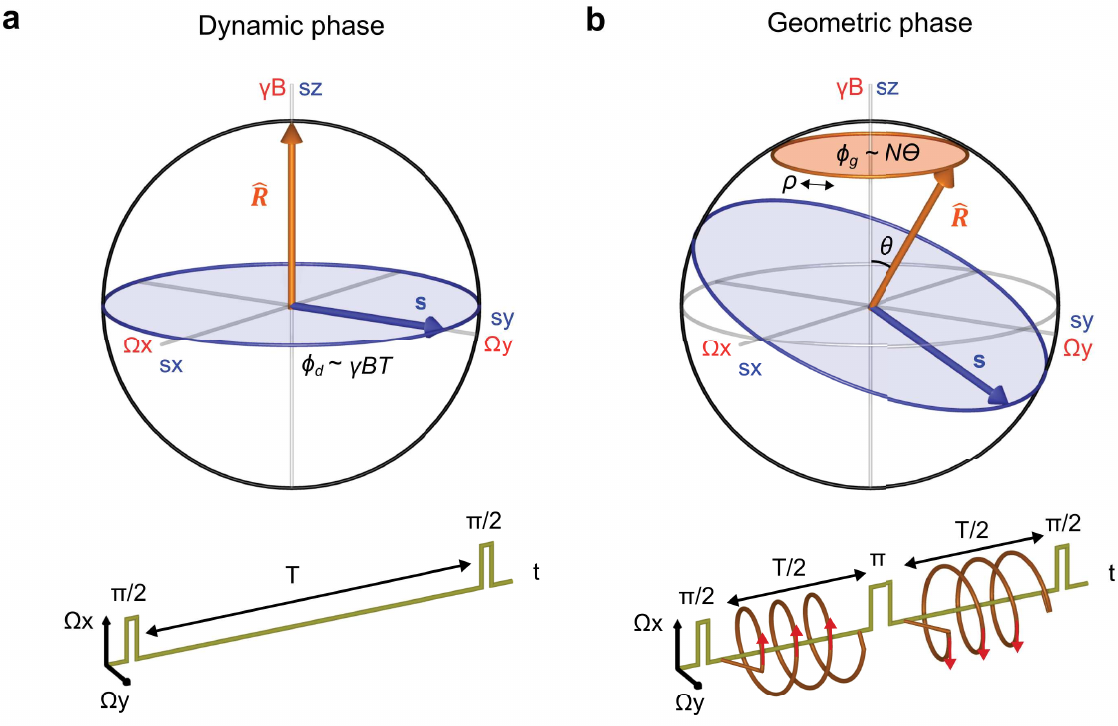}  
\end{overpic}
\caption[Dynamic versus geometric phase magnetometry]{Comparison of dynamic and geometric phase magnetometry. For dynamic phase magnetometry (i.e., Ramsey), the Bloch vector (blue arrow), is optically prepared and then rotated by a $\pi/2$-pulse to the equator. The Bloch vector then precesses about the fixed Larmor vector [orange arrow (\textcolor{orange}{$\rightarrow$})] before being mapped into a population difference by a second $\pi$/2-pulse and read out optically. b) For geometric-phase magnetometry, the Bloch vector is optically prepared and then rotated to the equator. Additional off-resonant driving then rotates the Larmor vector about the z-axis. As the spins precess, a geometric phase proportional to the product of the solid angle (orange disk) and the number of Larmor vector rotations is acquired in addition to the dynamic phase. To cancel the dynamic phase while continuing geometric phase accrual, a $\pi$-pulse and a reversal of the off-resonant drive are inserted at the sequence midpoint. Lastly, the Bloch vector is mapped onto a population difference by a second $\pi$/2-pulse and read out optically.   From Ref.~\cite{arai2018geometric}.}  \label{fig:geometricphasefigure}
\end{figure}

In the presence of particular DC and RF magnetic fields, an NV$^\text{-}$ spin may accumulate a measurable geometric phase~\cite{berry1984quantal} in addition to a dynamical phase. Following demonstrations of control and readout of an NV$^\text{-}$ center's geometric phase~\cite{maclaurin2012measurable,zu2014experimental,arroyocamejo2014room,yale2016optical,zhou2017holonomic}, the authors of Ref.~\cite{arai2018geometric} implemented geometric phase measurements for DC magnetometry. In their protocol, \textcolor{black}{depicted in Fig.~\ref{fig:geometricphasefigure},} the phase of a MW Rabi drive is swept adiabatically around a closed phase-space loop during two intervals separated by a central $\pi$-pulse. Whereas the $\pi$-pulse cancels the dynamic phase accumulated during the sequences, the acquired geometric phase depends on the strength of the DC magnetic field. While this technique enables wide-dynamic-range field sensing by avoiding a $2\pi$ phase ambiguity inherent to Ramsey magnetometry, it is unlikely to enhance sensitivity with respect to optimized Ramsey. 

\subsection{Ancilla-assisted upconversion magnetometry}\label{upconversion}

A clever and novel magnetometry scheme pioneered by Ajoy \textit{et al.}~in Ref.~\cite{liu2019nanoscale} utilizes frequency upconversion via an ancilla nuclear spin to make broadband measurements of an external magnetic field. The method works as follows: A large magnetic field is aligned along the NV$^\text{-}$ internuclear axis and tuned to near the ground state level anti-crossing (GSLAC) at $\approx$ 1024 gauss, allowing the relative strengths of the Zeeman term and the hyperfine coupling of the NV$^\text{-}$ electronic spin to the ancilla nuclear spin to be precisely tuned. In this regime, the NV$^\text{-}$ electronic spin is first-order insensitive to magnetic fields perpendicular to the NV$^\text{-}$ symmetry axis. However, an applied transverse magnetic field $B_\perp$ modulates the strength of the hyperfine interaction, resulting in amplitude modulation of the electronic spin energy level at the nuclear spin precession frequency. The modulation deviation is proportional to $B_\perp$. Thus, by performing standard AC magnetometry at the nuclear spin precession frequency, the magnitude of the perpendicular magnetic field $B_\perp$ can be detected.

The technique is intriguing because (i), it allows the effective gyromagnetic ratio of the sensor to be tuned and (ii), it enables the use of AC magnetometry techniques including dynamical-decoupling protocols to sense DC fields for durations on the order of $T_2$ or longer (see Sec.~\ref{DD}). However, the method is expected (and observed) to upmodulate both magnetic signals and magnetic noise, including spin bath noise, to the AC measurement band. Further, the improved dephasing times are achieved primarily by decreasing the effective gyromagnetic ratio (i.e.,~the ratio relating $B_\perp$ to an energy level shift) relative to the native NV$^\text{-}$ electronic gyromagnetic ratio. Although the scheme enables vector sensing from a single NV$^\text{-}$ center and may be compatible with NV$^\text{-}$ spin ensembles, \textcolor{black}{the method presently precludes sensing from multiple NV$^\text{-}$ orientations}. So far there has been no experimental demonstration of improved sensitivity using this method relative to that of an optimized Ramsey-type equivalent. The requirement for $\approx 1000$ gauss axial fields is also disadvantageous and likely prevents utilization of off-axis NV$^\text{-}$ centers for sensing. 

\subsection{Techniques for the strong NV$^\text{-}$-NV$^\text{-}$ interaction regime}\label{DRINQS}

Dipolar interactions among NV$^\text{-}$ spins contribute to ensemble-NV$^\text{-}$ dephasing, as described in Sec.~\ref{NVNVlimit}. When NV$^\text{-}$ centers comprise the majority of spin defects in diamond, or when a different majority spin species is decoupled from the NV$^\text{-}$ centers via spin bath driving, NV$^\text{-}$-NV$^\text{-}$ interactions may degrade relaxation times $T_2^*$, $T_2$, and $T_1$~\cite{choi2017depolarization}, limiting the sensitivity of both DC and AC magnetometers. Measurement protocols that decouple or leverage these like-spin interactions while retaining sensitivity to magnetic signals offer an avenue to surpass this sensitivity limit.
 
Proposed techniques to improve sensitivity in the limit of strong NV$^\text{-}$-NV$^\text{-}$ interactions may be separated into two categories. Protocols in the first category mitigate dipolar interactions between like spins to extend either the dephasing time $T_2^*$~\cite{okeeffe2019hamiltonian} for DC sensing or the coherence time $T_2$~\cite{choi2017dynamical} for AC sensing. However, these techniques partially decouple the spins from the fields to be sensed, which may counteract the sensitivity enhancement from $T_2^*$ or $T_2$ extension. Protocols in the second category harness like-spin interactions\textcolor{black}{~\cite{raghunandan2018high} and may} generate entangled many-body states\textcolor{black}{~\cite{choi2017quantum}}. Measurements of an entangled spin state comprising $N$ spins can beat the standard quantum limit for spin projection noise ($\eta \propto 1/\sqrt{N}$, see Eqn.~\ref{eqn:spinprojlim}), and may approach the Heisenberg limit ($\eta \propto 1/N$) \textcolor{black}{~\cite{choi2017quantum, gammelmark2011phase}}.

\begin{figure}[ht]
\centering
\begin{overpic}[width =\columnwidth]{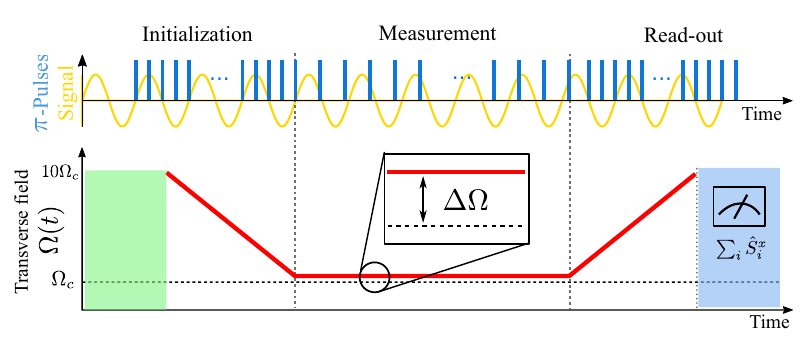}
\end{overpic}
\caption[DRINQS]{Schematic diagram of entanglement-enhanced sensing protocol proposed in Ref.~\cite{choi2017quantum}. During the initialization, measurement, and readout steps, the amplitude of a transverse magnetic field $\Omega$ and the repetition frequency of \textcolor{black}{additional transverse-magnetic-field} $\pi$-pulses are tuned. In the initialization stage, a correlated many-body spin state is generated as $\Omega$ is decreased toward a quantum critical point at $\Omega_\text{C}$. At the end of the measurement period, an axial AC magnetic field signal is mapped onto the total magnetization of the ensemble, which for NV$^\text{-}$ centers can be detected using conventional readout. From Ref.~\cite{choi2017quantum}.}  \label{fig:DRINQS}
\end{figure}

To illuminate the promise and challenges associated with entanglement-enhanced techniques, we focus on the specific protocol proposed in Ref.~\cite{choi2017quantum}. The technique, which is expected to be applicable to NV$^\text{-}$ centers, utilizes strong like-spin interactions to create quantum correlated states sensitive to AC magnetic fields. The proposed scheme, outlined schematically in Fig.~\ref{fig:DRINQS} generates entanglement within a 2D array of spins by first polarizing the individual spins along a transverse magnetic field \textcolor{black}{(which for NV$^\text{-}$ centers may be a MW-frequency field)} and then adiabatically decreasing the field toward a quantum critical point. For the measurement to be compatible with global NV$^\text{-}$ ensemble readout, the system approaches the quantum critical point, generating entanglement, without crossing over the quantum phase transition to a Greenberger-Horne-Zeilinger (GHZ) state. In the measurement step, periodic \textcolor{black}{transverse-magnetic-field} $\pi$-pulses are applied to the ensemble, allowing axial AC magnetic fields to excite the many-body system. The number of excitations detected after the transverse field is increased to its original value provides a measure of the strength of the AC magnetic field.

Importantly, the entangled state's coherence time, denoted $\overline{T}_2$, is no longer limited by like-spin interactions, but by external noise. That is, if the coherence time $T_2$ is separated into contributions from NV$^\text{-}$-NV$^\text{-}$ dipolar interactions and from other noise (including spin-lattice relaxation) as
\begin{equation}
\frac{1}{T_2} = \frac{1}{T_2\{\text{NV}^\text{-}\text{-}\text{NV}^\text{-}\}} + \frac{1}{T_2\{\text{other}\}},
\end{equation} 
then the entangled state's coherence time $\overline{T}_2$ is only a function of $T_2\{\text{other}\}$. Therefore, when NV$^\text{-}$-NV$^\text{-}$ interactions dominate, $\overline{T}_2$ may be comparable to or exceed $T_2$, yielding improved AC magnetic field sensitivity via both the increased coherence time and reduced readout noise. 

However, when the noise on each spin in the entangled ensemble is independent, $\overline{T}_2$ is expected to diminish linearly with the number of entangled spins $N$ (i.e., $\overline{T}_2 \propto 1/N$), which at best cancels the sensitivity enhancement obtained from the $1/\sqrt{N}$ reduction in spin projection noise compared to the standard quantum limit. Even without improved AC magnetic sensitivity, the scheme is expected to provide an increased measurement bandwidth by enabling faster field sampling than conventional sensing. When the dominant noise limiting the NV$^\text{-}$ ensemble's spin coherence time is instead set by spatially-correlated noise, such as dipolar interactions with nearby magnetic dipoles of a different species~\cite{choi2017quantum}, enhanced AC magnetic field sensitivity from reduced spin projection noise may again be possible. Although the protocol may also be compatible with broadband DC magnetometry, the scaling of the correlated ensemble's effective $T_2^*$ with entangled number of spins $N$ remains unclear.  Further investigation is required to determine if this protocol could yield a sensitivity improvement over conventional DC magnetometry.

While the approach proposed in Ref.~\cite{choi2017quantum} represents an important milestone towards magnetometry enhanced by NV$^\text{-}$-NV$^\text{-}$ dipolar interactions, the protocol is expected to be challenging to execute. First, the mean NV$^\text{-}$-NV$^\text{-}$ separation \textcolor{black}{distance} $\langle r_{\text{NV}^\text{-},\text{NV}^\text{-}} \rangle$ must be small compared to the average distance to the nearest paramagnetic defect $\langle r_{\text{NV}^\text{-},\text{other}} \rangle$, but large compared to the thickness $L$ of any (quasi) two-dimensional NV$^\text{-}$ layer, i.e., $\langle r_{\text{NV}^\text{-},\text{other}} \rangle $\textgreater $ \langle r_{\text{NV}^\text{-},\text{NV}^\text{-}} \rangle$\textgreater $L$. This hierarchy indicates that for a typical NV$^\text{-}$-rich diamond with $\langle r_{\text{NV}^\text{-},\text{NV}^\text{-}} \rangle \! \sim \! 10~$nm, the NV$^\text{-}$ layer thickness $L$ should be less than a few nanometers. Shallow nitrogen implantation into diamond ~\cite{Pham2011, Glenn2017} or nitrogen delta-doping during CVD growth~\cite{Ohno2012,Osterkamp2015} can yield layers that approach the appropriate thickness. Crucially, this 2D requirement restricts the practical NV$^\text{-}$ ensemble size, which may limit achievable sensitivity (see Sec.~\ref{physicallimits} and Appendix~\ref{volumedensityconsiderations}) when considering wide-field imaging and bulk magnetometry applications. Second, since the preparation and readout steps require a slow adiabatic field ramp, the practical requirement that these steps occur within time $\overline{T}_2$ limits the degree of achievable entanglement. 
Consequently, the protocol will mostly likely entangle sub-ensembles much smaller than the total ensemble size. Disorder (i.e., static field inhomogeneity) in the ensemble, e.g., from the random positioning of NV$^\text{-}$ centers, also restricts the maximum entangled sub-ensemble size. Both of these mechanisms are expected to increase the measurement's spin projection noise above the Heisenberg limit, further restricting the parameter regime where sensitivity enhancements are possible. In spite of the serious challenges and limitations, the proposed technique remains a promising first step toward practical schemes harnessing the full quantum nature of NV$^\text{-}$ ensembles for sensitivity enhancement.

\textcolor{black}{Additionally, with entangled and individually addressable spins, a wealth of proposed quantum error correction (QEC) sensing schemes may become more experimentally feasible~\cite{kessler2014quantum,unden2016quantum, layden2018spatial, layden2019ancilla, waldherr2014quantum,bonato2015optimized,cramer2016repeated}. QEC protocols mitigate certain relaxation mechanisms by encoding quantum information into redundant degrees of freedom~\cite{Degen2017}. Certain QEC implementations have already shown extension of NV$^\text{-}$ spin coherence~\cite{unden2016quantum}, although all demonstrations so far have employed at most a handful of spins. To date, most QEC studies concentrate on methods to correct noise along a different axis from the signal, (e.g., errors that cause spin flips rather than dephasing), limiting their use for extending $T_2^*$ or $T_2$. Recently, however, QEC schemes to correct dephasing-type errors have been proposed~\cite{layden2018spatial, layden2019ancilla}.}

\textcolor{black}{In the near future, ensemble-NV$^\text{-}$ sensing at the standard quantum limit is likely to outperform entanglement-enhanced schemes, as argued in Ref.~\cite{braun2018quantum}. Nonetheless, further development of these techniques remains an important endeavor toward enabling long-term sensitivity improvements approaching fundamental limits.}






%% file: sec07.tex
\section{Conclusion and outlook}\label{conclusion}

\textcolor{black}{The recent excitement accompanying quantum sensing with nitrogen-vacancy centers in diamond is well motivated, as NV-diamond sensors promise many advantages over alternative sensing technologies. NV$^\text{-}$ centers provide precision and repeatability similar to atomic systems in a robust solid-state package with less experimental complexity. Furthermore, NV$^\text{-}$-based devices can operate under ambient conditions and record spatial variations at length scales inaccessible to most other quantum sensors.} 

\textcolor{black}{Efforts to optimize performance of ensemble-NV$^\text{-}$ sensors are particularly warranted, as these devices at present have greater potential for improvement than other NV$^\text{-}$ sensing platforms. Historically, the NV-diamond community has focused on optimizing few- or single-NV$^\text{-}$ sensors, while the best demonstrated ensemble-NV$^\text{-}$ devices exhibit sensitivities orders of magnitude away from theoretical limits~\cite{Taylor2008}.}

\textcolor{black}{Consequently, this work provides a comprehensive survey of methods for optimizing broadband magnetometry from DC to $\sim 100$~kHz using ensembles of NV$^\text{-}$ centers. We explore strategies to enhance sensitivity toward physical limits, both through highlighting key parameters (Sec.~\ref{physicallimits}) and through evaluating proposed methods to improve those parameters. After identifying Ramsey magnetometry as the most promising sensing protocol (Secs.~\ref{magneticfieldsensitivity} and \ref{cwpulsed}), we focus on understanding and improving the spin dephasing time $T_2^*$ (Secs.~\ref{T2starlimits} and  \ref{T2*extensionmethods}), the spin readout fidelity $\mathcal{F} \equiv 1/\sigma_R$ (Sec.~\ref{fidelityimprovementmethods}), and the host diamond material properties (Sec.~\ref{sampleengineering}).} 
Below, we summarize our analyses within these broad categories, and we recommend areas where future study could lead to improvements in magnetometer sensitivity and performance.

Measurements employing Ramsey-type protocols with NV$^\text{-}$ ensembles are limited by $T_2^*$, which presently remains orders of magnitude shorter than the physical limit of $2T_1$. The magnetic field sensitivity improves nearly linearly with $T_2^*$ extension when the measurement overhead time is significant ($t_O\gtrsim T_2^*$), as is common for present-day ensemble-NV$^\text{-}$ magnetometers. Therefore, this work focuses on understanding limitations to $T_2^*$ and methods to extend $T_2^*$ in NV-rich diamonds. Among the factors limiting $T_2^*$ are magnetic-field, electric-field, and strain gradients. External bias-magnetic-field gradients may be mitigated through experimental design. Whereas internal strain and electric-field gradients can be more difficult to eliminate outright, the NV$^\text{-}$ ensemble can be made insensitive to such gradients through operation at sufficiently strong bias magnetic fields (Sec.~\ref{efieldsuppression}) and employment of double-quantum coherence magnetometry (Sec.~\ref{DQ}). Ensemble-NV$^\text{-}$ $T_2^*$ values may also be limited by dipolar interactions with the diamond's inhomogeneous paramagnetic spin bath. We determine the individual contributions to $T_2^*$ from substitutional nitrogen N$_\text{S}^0$ electronic spins (Sec.~\ref{nitrogenlimitT2*}), $^{13}$C nuclear spins (Sec.~\ref{13ClimitT2star}), and NV$^\text{-}$ spins (Sec.~\ref{NVNVlimit}). Recent experiments determine $T_2^\text{ }$- and $T_2^*$-dependencies on nitrogen concentration to better than $10\%$~\cite{Bauch2018,Bauch2019}. We suggest reducing the unwanted bath-spin concentrations through (i) diamond growth using isotopically-purified $^{12}$C (Sec.~\ref{13ClimitT2star}), and (ii) diamond treatment via optimized electron irradiation and annealing procedures (Sec.~\ref{sampleengineering}). We also identify spin bath driving using strong, resonant RF fields as an effective measure to decouple N$_\text{S}^0$ and other impurity spins from the NV$^\text{-}$ ensemble (Sec.~\ref{P1driving}). Recent work \textcolor{black}{implementing} spin bath driving combined with double-quantum coherence magnetometry in NV$^\text{-}$ ensembles \textcolor{black}{demonstrates} $T_2^*$ extension by more than 16$\times$~\cite{Bauch2018}. We expect continued progress on this front; one avenue opened up when $T_2^*$ is increased to the NV$^\text{-}$-NV$^\text{-}$ dipolar interaction limit is the exploration of enhanced sensing techniques harnessing quantum entanglement~\cite{choi2017quantum} (Sec.~\ref{DRINQS}). 

In Sec.~\ref{fidelityimprovementmethods} we survey existing techniques to improve ensemble-NV$^\text{-}$ readout fidelity $\mathcal{F} = 1/\sigma_R$, which, for conventional fluorescence-based readout, is currently limited to $\sim\!0.015$ (see Table~\ref{tab:parameters}). We analyze methods that allow readout fidelities for single NV$^\text{-}$ centers and small ensembles in nanodiamonds to approach the spin projection limit, including spin-to-charge conversion readout (Sec.~\ref{SCCR}) and ancilla-assisted repetitive readout (Sec.~\ref{ancilla}). However, no demonstrated method has substantially outperformed conventional fluorescence-based readout for large NV$^\text{-}$ ensembles (Table~\ref{tab:parameters}). Nonetheless, we anticipate that with careful experimental design and advances in diamond-sample engineering, fidelity-enhancement methods so far limited to single spins or small ensembles may be extended to large NV$^\text{-}$ ensembles. Additionally, given that any method employing optical readout benefits from increased collection efficiency, such optimizations (Sec.~\ref{improvedcollection}) remain worthwhile for improving magnetometer sensitivity.


As optimal sensing techniques require co-development with diamond samples tailored to these techniques, this work reviews diamond fabrication and relevant material properties in Sec.~\ref{sampleengineering}. In particular, we focus on methods to engineer lab-grown diamond samples optimized for ensemble-NV$^\text{-}$ magnetometry. We analyze growth via chemical vapor deposition, high-pressure-high-temperature synthesis, and mixed-synthesis methods (Sec.~\ref{hpht}). We examine how diamond synthesis and treatment can be used to engineer high N-to-NV$^\text{-}$ conversion efficiencies E$_\text{conv}$, and we investigate methods to improve and stabilize the charge state efficiency $\zeta = [$NV$^\text{-}$]/[NV$^\text{T}]$ (Sec.~\ref{chargestate}). We also investigate undesired defects commonly found in NV-rich diamond samples (Sec.~\ref{chargetraps}). These defects, including multi-vacancy clusters and hydrogen-related impurities, may both trap charges in the diamond and contribute to the dipolar spin bath, reducing both E$_\text{conv}$ and $T_2^*$.

Although present understanding of diamond synthesis, treatment, and characterization is extensive and spans multiple decades, further work is needed to reproducibly create NV$^\text{-}$-rich diamond samples with low strain, low concentrations of unwanted impurities, and high NV$^\text{-}$ concentrations. In particular, advancing diamond materials science to enable longer native $T_2^*$ values is a worthwhile pursuit; e.g., although the NV$^\text{-}$ center's sensitivity to strain can be reduced (Secs.~\ref{DQ} and \ref{efieldsuppression}), employing low-strain host diamonds is preferable regardless. Importantly, a robust and optimized protocol for diamond irradiation and annealing that takes nitrogen concentration into account should be established \textcolor{black}{(Secs.~\ref{irradiation} and \ref{LPHT})}. Furthermore, widespread access to high-quality \textcolor{black}{scientific} diamonds is imperative and would greatly accelerate advances in NV-diamond-related research. Presently, diamonds with natural carbon isotopic abundance, suboptimal nitrogen concentrations, and undesired strain and surface characteristics are widely employed by the community solely because most research groups lack access to optimized diamond samples.





In addition, many aspects of NV physics, and charge dynamics for ensembles in particular, remain poorly understood and warrant further investigation.  We anticipate that additional knowledge could be harnessed to improve sensor performance, similar to how the study of NV$^\text{-}$ and NV$^0$ ionization characteristics under low optical intensity by Aslam \textit{et al.}~\cite{Aslam2013} prompted the development of spin-to-charge conversion readout. Further examination of charge dynamics under magnetometer operating conditions (e.g., high optical intensity) is expected to yield fruitful insights. For NV-rich diamonds, systematic studies of (i) NV$^\text{-}$ ionization (both from the singlet and triplet excited states), and (ii) recombination from the NV$^0$ excited state versus optical wavelength and intensity, would be particularly useful. Such studies would address present knowledge gaps and could inform diamond-engineering protocols to better stabilize the NV$^\text{-}$ charge state in ensemble-based devices. These investigations could also lay the groundwork for new sensitivity-enhancement techniques tailored to ensembles. In addition, continued basic research into the NV$^\text{-}$ center is warranted. For example, while four electronic states of NV$^\text{-}$ have been observed, two additional predicted states have not yet been experimentally confirmed~\cite{Jensen2017}. 


We also expect unanticipated creative ideas to emerge that further enhance readout fidelity, dephasing time $T_2^*$, and overall magnetic field sensitivity. Ensemble-NV$^\text{-}$ magnetometers are already relevant in wide-varying sensing applications, thanks to key advances made over the past decade, which we have summarized here. Moreover, NV-diamond quantum sensing is a quickly developing platform, well positioned to continue improving, with significant advancements possible before fundamental limits are reached. By combining the knowledge collected here with likely future advances, we expect further expansion of applications of quantum sensors based on NV$^\text{-}$ ensembles in diamond.

\subsection*{Acknowledgments}

\textcolor{black}{The authors thank Eisuke Abe, Victor Acosta, Scott Alsid, Ashok Ajoy, Keigo Arai, Nithya Arunkumar, Ania Bleszynski-Jayich, Dolev Bluvstein, Danielle Braje, Dominik Bucher, Alexei Bylinskii, Paola Cappellaro, Soonwon Choi, Alexandre Cooper, Colin Connolly, Andrew Edmonds, Michael Geis, David Glenn, Ben Green, Kohei Itoh, Jean-Christophe Jaskula, Pauli Kehayias, Mark Ku, Junghyun Lee, David LeSage, Igor Lovchinsky, Matthew Markham, Claire McLellan, Idan Meirzada, Gavin Morley, Mark Newton, Michael O'Keeffe, David Phillips, Emma Rosenfeld, Brendan Shields, Swati Singh, Matthew Steinecker, and Daniel Twitchen for helpful comments and discussions. This material is based upon work supported by, or in part by, the United States Army Research Laboratory and the United States Army Research Office under Grant No.~W911NF-15-1-0548; the National Science Foundation Electronics, Photonics and Magnetic Devices (EPMD), Physics of Living Systems (PoLS), and Integrated NSF Support Promoting Interdisciplinary Research and Education (INSPIRE) programs under Grants No.~ECCS-1408075, No.~PHY-1504610, and No.~EAR-1647504, respectively; Air Force Office of Scientific Research award FA9550-17-1-0371; Defense Advanced Research Projects Agency Quantum Assisted Sensing and Readout (DARPA QuASAR) program under Contract No.~HR0011-11-C-0073; and Lockheed Martin under Contract No.~A32198.  J.~M.~S.~was partially supported by a Fannie and John Hertz Foundation Graduate Fellowship and a NSF Graduate Research Fellowship under Grant No.~1122374.}

%% file: sec08.tex
\appendix
\section{}







\subsection{Derivations}\label{derivations}

\subsubsection{Ramsey DC magnetic field measurement}\label{ramsey}

The following is a derivation of a Ramsey-type pulsed magnetometry sequence (see Fig.~\ref{fig:cwvsramsey}) using a magnetic dipole moment. Here the magnetic moment is taken to be an NV$^\text{-}$ center's ground-state electronic spin, although this discussion applies to any two-level system sensitive to magnetic fields, including atomic vapors and other solid state defects. Although the NV$^\text{-}$ ground state spin is a triplet with $S=1$, a bias magnetic field $B_0$ can be applied along the NV$^\text{-}$ symmetry axis to split the $m_s=+1$ and $m_s=-1$ energy levels so that resonant MWs may selectively drive the $m_s=0$ to $m_s=+1$ (or $m_s=0$ to $m_s=-1$) transition. Any off-axis magnetic field component $B_\perp$ can be ignored so long as $(\gamma_e B_\perp)^2/[(2\pi D)^2 \pm (\gamma_e B_0)^2] \ll 1$, where $D=2.87$~GHz is the zero-field splitting and $\gamma_e = g_e \mu_B/\hbar $ is the gyromagnetic ratio of the NV$^\text{-}$ electronic spin. Here the NV$^\text{-}$ center's nuclear spin is also ignored, as well as static electric fields or strain. We describe this two-level subspace as a pseudo-spin-$\nicefrac{1}{2}$ system with $|m_s=+1\rangle = |\!\!\downarrow\rangle$, $|m_s=0\rangle = |\!\!\uparrow\rangle$, and Hamiltonian
\begin{align}
\begin{split}
H&= (2\pi D + \gamma_e B) S_z\\
&=\frac{\hbar}{2} 
\begin{pmatrix}
   2 \pi D + \gamma_e B  & 0 \\
    0 & -2 \pi D - \gamma_e B
  \end{pmatrix},
  \end{split}
  \label{H}
\end{align}
\textcolor{black}{expressed in the $\{|\!\!\downarrow\rangle, |\!\!\uparrow\rangle\}$ basis}, where \textcolor{black}{here we take $S_z$ to be the spin-\nicefrac{1}{2} z-projection operator with units of $\hbar/2$}; and $B =B_0 + B_{\text{sense}}$ is the total magnetic field projection along the NV$^\text{-}$ symmetry axis (the z-axis), which is the sum of the applied bias field and an unknown DC field to be sensed. Here terms in the Hamiltonian proportional to the identity matrix have been dropped, as they introduce only a global phase to the states' time evolution. In the bias field $B_0$ the spin resonance frequency is $\omega_0 =  2 \pi D + \gamma_e B_0$. Spin operators are expressed in the $S_z$ basis in terms of the Pauli matrices $\vec{S} = \frac{\hbar}{2} \vec{\sigma}$, yielding
\begin{equation}
H=\frac{\hbar \omega_0}{2}\sigma_z + \frac{\hbar}{2}\gamma_e B_\text{sense}\sigma_z.
\end{equation}

As described herein, a Ramsey sequence consists of two $\pi/2$-pulses of an oscillating magnetic field resonant with the transition between $|\!\!\uparrow\rangle$ and $|\!\!\downarrow\rangle$, which are separated by a free precession time $\tau$. The sequence begins at time $t=0$, with the spin polarized to $|\psi(0)\rangle=|\!\!\uparrow\rangle$. An oscillating magnetic field oriented perpendicular to the NV$^\text{-}$ symmetry axis $\vec{B}_1(t) = B_1 \cos(\omega t) \hat{y}$ with angular frequency $\omega \approx \omega_0$ is turned on abruptly. Without loss of generality $\vec{B}_1$ is assumed to be polarized along the y-axis. For $B_1\gg B_\text{sense}$, the second term in $H$ can be dropped, thereby ignoring effects of the unknown DC sensing field while the oscillating field is on. The Hamiltonian for the system driven by this oscillating field, denoted $H_\text{driv}$, becomes 
\begin{equation}
H_\text{driv}=\frac{\hbar \omega_0}{2}\sigma_z +\frac{\hbar}{2} \gamma_e B_1 \cos(\omega t) \sigma_y.
\end{equation}
We proceed in the interaction picture, with $H_0=\frac{\hbar \omega_0}{2}\sigma_z$ and $H_{1} = \frac{\hbar}{2} \gamma_e B_1 \cos(\omega t)\sigma_y$. This step is equivalent to transforming into a rotating frame with angular frequency $\omega_0$. The interaction-picture state vector $|\tilde{\psi}(t)\rangle$ is defined in terms of the Schr\"{o}dinger-picture state vector $|\psi(t)\rangle$ as $|\tilde{\psi}(t)\rangle=U_0^\dag(t)|\psi(t)\rangle$ with $U_0(t)=e^{-iH_0t/\hbar}$. This state evolves according to $|\tilde{\psi}(t)\rangle = \tilde{U}_1(t)|\tilde{\psi}(0)\rangle$ where $\tilde{U}_1(t)=e^{-i\tilde{H_1}t/\hbar}$, with
\small
\begin{align}
\begin{split}
&\tilde{H_1} 
= U_0^{\dag}(t) H_1 U_0(t)\\
&=\!\frac{\hbar}{4}\ \gamma_e B_1\!
\begin{pmatrix}
   0  & -i (e^{i (\omega_0\!+\!\omega)t}\!+\!e^{i (\omega_0\!-\!\omega)t}) \\
    i(e^{-i (\omega_0\!-\!\omega)t}\!+\!e^{-i (\omega_0\!+\!\omega)t}) & 0
  \end{pmatrix}.
  \end{split}
\end{align}
\normalsize
The transformed interaction Hamiltonian $\tilde{H_1}$ is simplified by assuming resonant driving of the spin with $\omega = \omega_0$ and by making the rotating wave approximation, dropping off-resonant terms rotating at $2\omega_0$, to yield
\begin{align}
\tilde{H_1} &\approx \frac{\hbar}{4} \gamma_e B_1 \sigma_y.
\end{align}
\textcolor{black}{This Hamiltonian causes the spin system to undergo Rabi oscillations at angular frequency $\Omega = \gamma_e B_1/2$}. The oscillating field $\vec{B}_1(t)$ is turned off abruptly after a duration \textcolor{black}{$\tau_{\frac{\pi}{2}}= \frac{\pi}{2\Omega}=\frac{\pi}{\gamma_e B_1}$}, so that
\begin{align}
\begin{split}
|\tilde{\psi}(\tau_{\frac{\pi}{2}})\rangle 
&= \exp\left(-i \frac{\gamma_e B_1 \sigma_y \tau_{\frac{\pi}{2}}}{4}\right) |\tilde{\psi}(0)\rangle \\
&=\exp\left(-i \frac{\pi}{4}\sigma_y\right) |\!\uparrow\rangle \\
&=\frac{1}{\sqrt{2}}
\begin{pmatrix}
   1  & -1 \\
    1 & 1
  \end{pmatrix}
  \begin{pmatrix}
   0 \\
   1
  \end{pmatrix}\\
 &= \frac{1}{\sqrt{2}}\left(-|\!\downarrow\rangle+|\!\uparrow\rangle\right),
 \end{split}
\end{align}
which uses the identity \textcolor{black}{$e^{-i\theta \hat{n}\cdot \vec{\sigma}} = \cos\left(\theta \right)I - i \sin\left(\theta \right)(\hat{n}\cdot \vec{\sigma})$} where $\hat{n}$ is a unit vector on the Bloch sphere. This constitutes a $\pi/2$-pulse on the spin. 

Next, the magnetic moment undergoes free precession in the absence of $\vec{B_1}(t)$ for a sensing time $\tau$. During this time the system Hamiltonian returns to $H$ from Eqn.~\ref{H}. We continue to use the interaction picture with $H_0=\frac{\hbar \omega_0}{2}\sigma_z$,  and with new interaction Hamiltonian $H_1'$ determined by $\vec{B}_{\text{sense}}=B_{\text{sense}}\hat{z}$ as

\begin{equation}
H_1' = \frac{\hbar}{2}\gamma_e B_{\text{sense}} \sigma_z.
\end{equation}

Recognizing that $H_1'$ commutes with $H_0$, the transformed interaction Hamiltonian $\tilde{H_1'} \equiv U_0^\dag(t) H_1' U_0(t) = H_1'$, and thus the interaction-picture state vector $|\tilde{\psi}(t)\rangle$ evolves under $H_1'$ into
\vspace{.5mm}
\begin{align}
\begin{split}
|\tilde{\psi}(\tau_{\pi/2}+\tau)\rangle &= e^{-i H_1' \tau/\hbar}|\tilde{\psi}(\tau_{\pi/2})\rangle\\
&=\frac{1}{\sqrt{2}}(-e^{-i\phi/2}|\!\downarrow\rangle + e^{i\phi/2}|\!\uparrow\rangle),
\end{split}
\end{align}
where 
\begin{equation}
\phi  = \gamma_e B_{\text{sense}} \tau
\end{equation}
is the phase accumulated due to $B_\text{sense}$ in the interaction picture. (If $B_{\text{sense}} = 0$, the state vector $|\tilde{\psi}(t)\rangle$ accumulates no phase, as $H_1'$ vanishes and the entire Hamiltonian $H = H_0$.)

To complete the sequence, a second oscillating field \textcolor{black}{$\vec{B_2}(t) = \vec{B_2}\cos(\omega t)$, again with $\omega = \omega_0$,} is applied for a $\pi/2$-pulse. As with the first $\pi/2$-pulse, $B_{\text{sense}}\ll B_2$ is assumed so that additional spin state evolution due to $B_{\text{sense}}$ can be ignored. The polarization of $\vec{B_2}(t)$ is chosen to be along $\hat{n}$ in the x-y plane at an angle $\vartheta$ with respect  $\hat{y}$, the polarization direction of the first $\pi/2$-pulse $\vec{B_1}(t)$. After again making the rotating wave approximation, the transformed interaction Hamiltonian, $\tilde{H}_1''$ is given by 
\begin{equation}
\tilde{H}_1'' \approx \frac{\hbar}{4} \gamma_e B_2 \left(\cos\left(\vartheta\right)\sigma_y -\sin\left(\vartheta\right)\sigma_x\right)
\end{equation}
and 
\begin{align}
\begin{split}
&|\tilde{\psi}(\tau_{\frac{\pi}{2}}+\tau+\tau_{\frac{\pi}{2}})\rangle\\ 
&=e^{-i \tilde{H}_1'' \tau_{\frac{\pi}{2}}/\hbar} |\tilde{\psi}(\tau_{\frac{\pi}{2}}+\tau)\rangle \\
&=\frac{1}{\sqrt{2}}
\begin{pmatrix}
   1  & -e^{-i\vartheta} \\
    e^{i\vartheta} & 1
  \end{pmatrix} \cdot \frac{1}{\sqrt{2}}
  \begin{pmatrix}
   -e^{-i\phi/2} \\
   e^{i\phi/2}
  \end{pmatrix},
  \end{split}
\end{align}
which, up to a global phase, is equal to 
\begin{align}
|\tilde{\psi}\rangle = \cos\left( \frac{\phi-\vartheta}{2}\right)|\!\downarrow\rangle +ie^{i\vartheta}\sin \left( \frac{\phi-\vartheta}{2}\right)|\!\uparrow\rangle. 
\label{eqn:finalstate}
\end{align}

The phase accumulated during $\tau$ is thus mapped on to a population difference between the $|\!\!\downarrow\rangle$ and $|\!\!\uparrow\rangle$ states. The population difference is detected by measuring \textcolor{black}{the rotating-frame observable $\tilde{S}_z$, which is equal to the fixed-frame spin projection operator $S_z$, as $S_z$ commutes with $H_0$}. The value of $B_{\text{sense}}$ is determined by relating this measured observable to $\phi$:
\begin{align}\label{eqn:sz}
\begin{split}
\langle S_z \rangle 
&=  \frac{\hbar}{2}  \langle \tilde{\psi} | \sigma_z | \tilde{\psi} \rangle  \\ 
&=   \frac{\hbar}{2}\left( \cos^2 \left(\frac{\phi-\vartheta}{2}\right) -   \sin^2 \left(\frac{\phi-\vartheta}{2}\right)\right)  \\ 
&=  \frac{\hbar}{2}\cos (\phi-\vartheta)\\
&= \frac{\hbar}{2}\cos (\gamma_e B_{\text{sense}} \tau-\vartheta).
\end{split}
\end{align}

The cosinusoidal fluctuations in $\langle S_z \rangle$ are termed Ramsey fringes. Common choices of $\vartheta$ are 0 and $\pi/2$. The case where $\vartheta = 0$ (respectively, $\vartheta = \pi/2$) is commonly called cosine (sine) magnetometry, as the observable $\langle S_z \rangle$ varies as the cosine (sine) of $B_{\text{sense}}$ for fixed $\tau$. \textcolor{black}{For ensembles of NV$^\text{-}$ centers, $\langle S_z \rangle$ is measured by reading out the spin-state-dependent fluorescence over a predetermined readout window of several hundred nanoseconds (see Fig.~\ref{fig:spindependentfluorescencecontrast}), as discussed in Sec.~\ref{magneticfieldsensitivity} and later in Appendix~\ref{shot}.}

For small $B_{\text{sense}}$ such that $\phi\ll 2\pi$, Eqn.~\ref{eqn:sz} can be linearized about $\phi = 0$ for any value of $\vartheta$ except $\vartheta = 0$.  The values of $B_{\text{sense}}$ and $\phi$ can then be related to a small change in the observable $\delta \langle S_z \rangle = \langle S_z \rangle|_{\phi} - \langle S_z \rangle |_{0}$ as follows:
\begin{align}
\begin{split}
B_{\text{sense}} =  \frac{\phi}{\gamma_e \tau} &\approx   \frac{1}{\gamma_e \tau} \frac{\delta \langle S_z \rangle}{\frac{d\langle S_z \rangle }{d\phi} |_0} \\
& \approx   \frac{1}{\gamma_e \tau} \frac{\frac{2}{\hbar} \langle S_z \rangle|_\phi - \cos(\vartheta)}{\sin (\vartheta)}.
\end{split}
\label{eqn:linearize}
\end{align}
For $\vartheta = \pi/2$, the slope of the Ramsey fringe is maximized, and Eqn.~\ref{eqn:linearize} reduces to
\begin{equation}
B_{\text{sense}} \approx  \frac{2}{\hbar \gamma_e \tau} \langle S_z \rangle.
\end{equation}
For $\vartheta = 0$, \textcolor{black}{where the linearization method fails}, the linear term $\delta \langle S_z \rangle$ vanishes, as the slope of the Ramsey fringe goes to zero; a small $B_{\text{sense}}$ produces to lowest order a quadratic change in $\langle \tilde{S}_z \rangle$. 

\subsubsection{Spin-projection-noise-limited sensitivity}\label{spinproj}


The spin-projection-noise-limited magnetic field sensitivity is defined as the field $\delta B$ at which the size of the signal $\delta \langle S_z \rangle $ due to $\delta B$ is equal to the uncertainty in the signal, i.e., when $\delta \langle S_z \rangle = \Delta S_z$, where $ \Delta S_z = \sqrt{ \langle S_z^2 \rangle - \langle S_z \rangle^2} $ is the standard deviation of a series of identical measurements of $\delta B$. 
\textcolor{black}{From Eqn.~\ref{eqn:linearize} assuming} precession time $\tau$, this minimum field is
\begin{equation}
\delta B_\text{sp}  =  \frac{1}{\gamma_e \tau} \frac{\Delta S_z }{|\frac{d\langle S_z \rangle }{d\phi}|}.
\end{equation}

When $M$ uncorrelated consecutive measurements are taken, each with precession time $\tau$ over a total measurement time $t_\text{meas}$, the minimum field is modified by the factor $\sqrt{1/M} = \sqrt{\tau/t_\text{meas}}$, yielding 
\begin{equation}
\delta B_\text{sp}  =  \frac{1}{\gamma_e }\frac{1}{\sqrt{\tau t_\text{meas}}} \frac{\Delta S_z }{|\frac{d\langle S_z \rangle }{d\phi}|} .
\end{equation}
The spin-projection-noise-limited sensitivity of a Ramsey magnetometry measurement is then
\begin{equation}\label{eqn:sensspinproj}
\eta_\text{sp} = \delta B_\text{sp} \sqrt{t_\text{meas}} = \frac{1}{\gamma_e \sqrt{\tau}} \frac{\Delta S_z }{|\frac{d\langle S_z \rangle }{d\phi}|}. 
\end{equation}

The quotient $\frac{\Delta S_z }{|\frac{d\langle S_z \rangle }{d\phi}|}$ is calculated: 
\begin{align}
\begin{split}
\langle S_z \rangle &= \frac{\hbar}{2} \cos (\phi-\vartheta),
\end{split}\\
\begin{split}\\
\frac{d\langle S_z \rangle}{d\phi} &= -\frac{\hbar}{2} \sin( \phi - \vartheta),
\label{eqn:spinproj}
\end{split}\\
\begin{split}\\
\langle S_z^2 \rangle &= \frac{\hbar^2}{4} \langle \tilde{\psi} | \sigma_z^2 | \tilde{\psi} \rangle \\
&=  \frac{\hbar^2}{4}\left( \cos^2 \left(\frac{\phi - \vartheta}{2}\right) + \sin^2 \left(\frac{\phi - \vartheta}{2}\right) \right)\\&= \frac{\hbar^2}{4},
\end{split}\\
\begin{split}\\
\Delta S_z &=\sqrt{ \langle S_z^2 \rangle - \langle S_z \rangle^2} \\  
&=   \sqrt{\frac{\hbar^2}{4} \left(1-\cos^2 (\phi - \vartheta)\right)}  \\ &=  \frac{\hbar}{2} | \sin (\phi - \vartheta)|,
\end{split}\\
\begin{split}\\
\frac{\Delta S_z }{|\frac{d\langle S_z \rangle }{d\phi}|} &= 1.
\end{split}\label{eqn:spsnr}
\end{align}


\textcolor{black}{Equation~\ref{eqn:spsnr} illustrates that the signal-to-noise ratio of a spin-projection-noise-limited measurement is independent of the value of $\phi$ or $\vartheta$. The projection noise $\Delta S_z$ is exactly equal to the slope of the Ramsey fringe $|\frac{d\langle S_z \rangle }{d\phi}|$.  As a result, a magnetometer limited by spin projection noise has the same sensitivity regardless of where on the Ramsey fringe the measurement is taken, which is given by}
\begin{equation}
\eta_\text{sp} = \delta B_{sp} \sqrt{t_\text{meas}} = \frac{1}{\gamma_e \sqrt{\tau}}.
\end{equation}
For sensing with an ensemble of $N$ independent spins, the sensitivity $\eta_\text{sp}^\text{ensemble} = \eta_\text{sp}/\sqrt{N}$ such that
\begin{equation}
\eta_\text{sp}^\text{ensemble} = \frac{1}{\gamma_e \sqrt{N \tau}}.
\end{equation}

\subsubsection{Photon-shot-noise-limited sensitivity}\label{shot}

The above discussion considered a direct measurement of $S_z$. The measurement technique for NV$^\text{-}$ spins - optical readout - instead indirectly probes the spin through measuring the spin-state-dependent fluorescence. Shot noise in the collected fluorescence must be incorporated into the measurement uncertainty and sensitivity. 

To phenomenologically introduce Poisson fluctuations from the fluorescence photons into the sensitivity, the optical readout procedure is treated as a mapping of the spin eigenstates onto two light field modes: $|m_s=+1\rangle =|\!\!\downarrow\rangle \rightarrow |\beta\rangle$ and $|m_s=0\rangle =|\!\!\uparrow\rangle \rightarrow|\alpha\rangle$, where $|\alpha\rangle$ and $|\beta\rangle$ are coherent states defined by $\hat{a}|\alpha\rangle = \alpha |\alpha\rangle$ and $\hat{b}|\beta\rangle=\beta|\beta\rangle$. 
We define $a = |\alpha|^2$ as the mean number of photons in $ |\alpha\rangle$ and $b = |\beta|^2$ as the mean number of photons in $ |\beta\rangle$. Since the $|m_s=0\rangle$ state produces more fluorescent photons during readout than the $|m_s=+1\rangle$ state, $a > b$.  The final spin state $|\tilde{\psi} \rangle$ from Eqn.~\ref{eqn:finalstate} is mapped onto the photon field state 
 \begin{equation}
|\psi_\text{ph} \rangle = \cos{ \left(\frac{\phi-\vartheta}{2}\right)} | \beta \rangle +ie^{i\vartheta} \sin \left(\frac{\phi-\vartheta}{2}\right) | \alpha \rangle .
 \end{equation}
A measurement of the spin state has become a measurement of the number of photons collected from the two light fields $\hat{N} = \hat{a}^\dag\hat{a}+\hat{b}^\dag \hat{b}$. Defining $\varphi = \phi - \vartheta$,
\begin{align}
\begin{split}
\langle \hat{N} \rangle &= \langle\psi_\text{ph}|(\hat{a}^\dag\hat{a}+\hat{b}^\dag \hat{b})|\psi_\text{ph}\rangle = b \cos^2\left(\frac{\varphi}{2}\right) + a \sin^2\left(\frac{\varphi}{2}\right) \\
&= b \left( \frac{1+\cos(\varphi)}{2}\right) + a \left( \frac{1-\cos(\varphi)}{2}\right).
\end{split}
\end{align}
where the two light fields are assumed to be noninterfering so that $\hat{a}|\beta\rangle = \hat{b}|\alpha\rangle = 0$ and $\langle \alpha | \beta \rangle = \langle \beta | \alpha \rangle = 0$. 

The sensitivity of a magnetometer employing optical readout is written in the same way as the spin-projection-noise-limited sensitivity given in Eqn.~\ref{eqn:sensspinproj}, but with the observable $S_z$ replaced by $\hat{N}$:
\begin{equation}\label{eqn:sensoptreadout}
\eta_\text{opt} = \delta B_\text{opt} \sqrt{t_\text{meas}}  =  \frac{1}{\gamma_e\sqrt{\tau}} \frac{\Delta \hat{N} }{|\frac{d\langle \hat{N} \rangle }{d\phi}|},
\end{equation}where $ \Delta \hat{N}  = \sqrt{ \langle \hat{N}^2 \rangle - \langle \hat{N} \rangle^2} $. The derivative of $\langle \hat{N} \rangle$ with respect to $\phi$ is
\begin{equation} \label{eqn:dndphi}
\frac{d\langle \hat{N} \rangle}{d \phi} = \frac{d\langle \hat{N} \rangle}{d \varphi} =\frac{(a-b)}{2} \sin (\varphi).
\end{equation}
Recalling the operator commutation relation $[\hat{a},\hat{a}^\dag] = 1$, $\Delta \hat{N} $ is calculated:
\begin{align} \label{eqn:n2avg}
\begin{split}
\langle \hat{N^2} \rangle 
&= \langle (\hat{a}^\dag\hat{a}+\hat{b}^\dag \hat{b}) (\hat{a}^\dag\hat{a}+\hat{b}^\dag \hat{b}) \rangle\\
& =  \langle\psi_\text{ph}|
(\hat{a}^\dag\hat{a}\hat{a}^\dag\hat{a}+\hat{b}^\dag\hat{b}\hat{b}^\dag\hat{b})
|\psi_\text{ph}\rangle \\
&= \langle\psi_\text{ph}|
(\hat{a}^\dag(\hat{a}^\dag\hat{a}+1)\hat{a}+\hat{b}^\dag(\hat{b}^\dag\hat{b}+1)\hat{b})
|\psi_\text{ph}\rangle \\
&= b(b+1) \left( \frac{1+\cos(\varphi)}{2}\right) + a(a+1) \left( \frac{1-\cos(\varphi)}{2}\right),
\end{split}
\\
\begin{split}\\
 \langle \hat{N} \rangle^2 &= b^2\!\left(\!\frac{\nicefrac{1}{2}\!+\!\cos(\varphi)}{2}\!\right)\! +  \!a^2\!\left(\!\frac{\nicefrac{1}{2}\!-\!\cos(\varphi)}{2}\!\right)\! \\ &\thickspace\thickspace\thickspace\thickspace\thickspace +\! \left(\! \frac{b^2}{4}\! + \!\frac{a^2}{4} \!\right)\! \cos^2(\varphi) + \frac{ba}{2} \sin^2 (\varphi),
\end{split}
\label{eqn:navg2}
\\
\small
\begin{split}\\
\Delta \hat{N}  &= \sqrt{\! \langle \hat{N}^2 \rangle - \langle \hat{N} \rangle^2} \\ &= \small{\sqrt{ \left(\!\frac{b^2}{4}\! -\!\frac{ba}{2}\! +\! \frac{a^2}{4}\! \right)\! \sin^2(\varphi)\! +\!b\! \left(\! \frac{1\!\!+\!\!\cos(\varphi)}{2}\!\right) \!+\! a\! \left(\! \frac{1\!\!-\!\!\cos(\varphi)}{2}\!\right)}} \\
& = \sqrt{ \frac{(a-b)^2}{4} \sin^2(\varphi) + b \cos^2\left(\frac{\varphi}{2}\right) + a \sin^2\left(\frac{\varphi}{2}\right) }.
\end{split}
\label{eqn:deltan}
\end{align}
\normalsize

Using Eqns.~\ref{eqn:deltan} and \ref{eqn:dndphi}, the sensitivity reduces to 
\begin{equation}\label{eqn:noisen}
\frac{\Delta \hat{N} }{|\frac{d\langle N\rangle}{d\phi}|} = \sqrt{\frac{ \frac{(a-b)^2}{4} \sin^2(\varphi) + b \cos^2\left(\frac{\varphi}{2}\right) + a \sin^2\left(\frac{\varphi}{2}\right)}{\frac{(a-b)^2}{4} \sin^2 (\varphi)} }.
\end{equation}

For the case $\varphi = \pi/2$, the sensitivity is optimized, yielding 
\begin{equation}
\frac{\Delta \hat{N} }{|\frac{d\langle \hat{N}\rangle}{d\phi}|} = \sqrt{\frac{ \frac{(a-b)^2}{4} + \frac{a+b}{2}}{\frac{(a-b)^2}{4}} } =\sqrt{1+\frac{2(a+b)}{(a-b)^2}}.
\end{equation}

We identify $C = \frac{a-b}{a+b}$ as the measurement contrast, (i.e., the fringe visibility),  and $n_\text{avg} = \frac{a+b}{2}$ as the average number of photons collected per measurement (per spin, if the measurement is on an ensemble). The contrast $C$ depends on the degree of initial polarization of the spin state and the readout duration. Measurement contrast also diminishes with increased free precession time due to spin dephasing and docoherence, parameterized by $T_2^*$, as shown in Eqn.~\ref{eqn:dephasing}. However, since this degradation affects both the shot-noise and spin-projection-noise terms in the measurement sensitivity $\eta_\text{opt}$, it is included explicitly rather than incorporated into $C$. Thus, the sensitivity \textcolor{black}{(neglecting overhead time)} for a Ramsey measurement on a single spin with both photon shot noise and spin-projection noise is given by

\begin{equation}
\eta_\text{opt} = \delta B_\text{opt} \sqrt{t_\text{meas}}  =  \frac{1}{\gamma_e e^{-(\tau/T_2^*)^p} \sqrt{\tau}}\sqrt{1+\frac{1}{C^2 n_\text{avg}}}. 
\end{equation}
(See Appendix~\ref{FIDexponent} for discussion of the stretched exponential parameter $p$.)
When sensing with an ensemble of $N$ independent spins, the sensitivity is given by
\begin{equation}
\eta_\text{opt}^\text{ensemble} = \frac{\eta_\text{opt}}{\sqrt{N}}.
\end{equation}

In conventional NV$^\text{-}$ optical readout, measurement contrast is low $(\lesssim 15 \%)$, and the number of photons $n_\text{avg}$ collected per spin at most order unity (see Table~\ref{tab:parameters}), and is often much less due to imperfect collection efficiency. Thus, $C^2 n_\text{avg} \ll 1$, and shot noise becomes the dominant contribution to the magnetic field sensitivity, \textcolor{black}{which in the absence of overhead time is given by}

\begin{equation}
\eta_\text{shot} \approx  \frac{1}{\gamma_e}\frac{1}{C e^{-(\tau/T_2^*)^p} \sqrt{n_\text{avg} \tau}}.
\end{equation}
and 
\begin{equation}
\eta_\text{shot}^\text{ensemble}  \approx  \frac{1}{\gamma_e}\frac{1}{C e^{-(\tau/T_2^*)^p} \sqrt{N n_\text{avg} \tau}}.
\end{equation}

\subsubsection{Overhead time}\label{overhead}

The sensitivity equations above have neglected any optical initialization or readout time, as well as the finite duration of the two $\pi/2$-pulses of the Ramsey sequence. Grouping all of these factors into an experimental dead time $t_O$, we find the sensitivity factor for $M$ measurements each with sensing time $\tau$ over a total time $t_\text{meas}$ is $\sqrt{1/M} = \sqrt{(\tau+t_O) / t_\text{meas}}$, yielding a sensitivity limited by shot noise and spin-projection noise of 
\begin{equation}
\eta_\text{opt} = \delta B_\text{opt} \sqrt{t_\text{meas}}  =  \frac{1}{\gamma_e e^{-(\tau/T_2^*)^p}} \frac{\sqrt{\tau+t_O}}{\tau} \sqrt{1+\frac{1}{C^2 n_\text{avg}}}.
\end{equation}

\subsection{Optimal precession time}\label{optimalprecession}

\begin{figure}
\begin{overpic}[height=3.3 in]{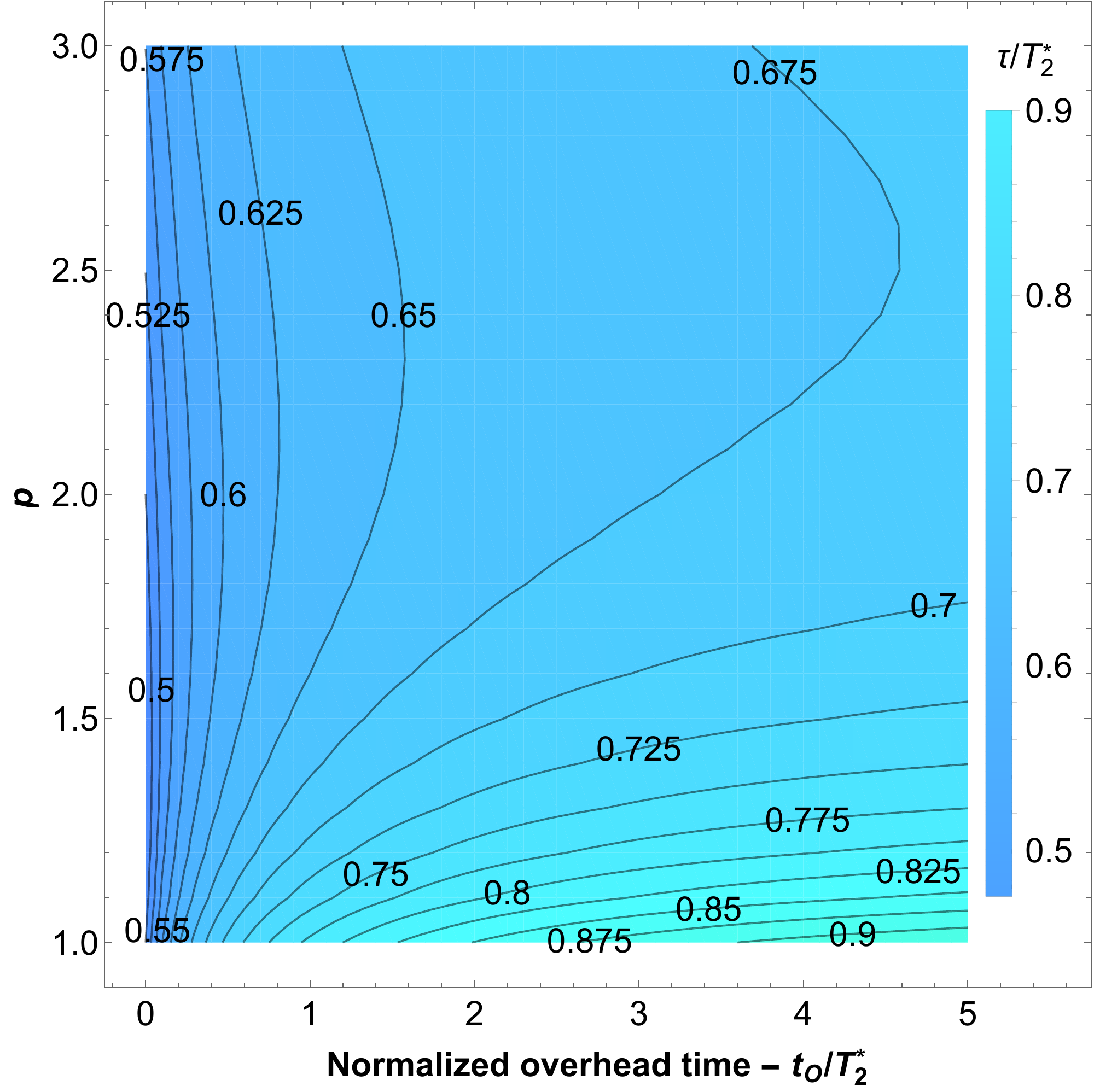}  
\end{overpic}
\caption[OptimalPrecessionTime]{Optimal precession time $\tau$ for a pulsed magnetometry protocol. Contour plot shows precession time $\tau$ to achieved optimal sensitivity, in units of $T_2^*$, for different stretched exponential parameters $p$ and different overhead times $t_O$.}  \label{fig:optimalprecessiontime}
\end{figure}

The optimal precession time $\tau$ to achieve best Ramsey magnetometry sensitivity (Eqn.~\ref{eqn:ramseyshot}) depends on the value of the stretched exponential parameter $p$, the initialization time $t_I$, and the readout time $t_R$. By defining the overhead time per measurement as $t_O = t_I+t_R$, Eqn.~\ref{eqn:ramseyshot} reduces to
\begin{equation}\label{eqn:optimalprecession}
\eta \propto \frac{1}{e^{-(\tau/T_2^*)^p}}\frac{\sqrt{\tau+t_O}}{\tau}.
\end{equation}
For $t_O \ll T_2^*$, sensitivity is optimized when $\tau \approx T_2^*/2$ for $1 \leq p \leq 2$. Particularly, for $t_O=0$ and $p=1$ or $p=2$ (see Appendix~\ref{FIDexponent}), sensitivity is exactly optimized for $\tau=T_2^*/2$.  As $t_O$ increases from zero, the optimal precession time increases as well, asymptotically approaching $\tau = T_2^*$ when $t_O\!\gg\!T_2^*$ for $p=1$.   Figure~\ref{fig:optimalprecessiontime} shows the optimal precession time $\tau$ for various combinations of $p$ and $t_O$. For clarity the optimal precession time is normalized to the dephasing time in the employed measurement basis (DQ or SQ). \textcolor{black}{Equation~\ref{eqn:optimalprecession}, and thus Fig.~\ref{fig:optimalprecessiontime}, also apply for Hahn echo (Eqn.~\ref{eqn:hahnecho}) with $T_2^*$ replaced by $T_2$ (see Section~\ref{DD}).}

In practice, additional experimental factors warrant consideration when choosing the Ramsey free precession time $\tau$. For example, because time-varying electric and magnetic fields and temperature may mask as dephasing mechanisms, the measured value of $T_2^*$ depends on the measurement duration. Thus, if the time required to measure the value of $T_2^*$ is significantly longer than the duration of a magnetic field measurement, field fluctuations may artificially reduce the measured value of $T_2^*$ compared to the value relevant for sensing. This spoiled $T_2^*$ measurement could lead to a suboptimal choice of  $\tau$~\cite{Bauch2018}. Therefore, care should be taken when choosing the appropriate free precession time $\tau$ for a magnetometry experiment. 



\subsection{Considerations for increasing sensor number}\label{volumedensityconsiderations}


Increasing the number $N$ of interrogated NV$^\text{-}$ centers by increasing either the interrogation volume or the NV$^\text{-}$ density may be partially effective to improve magnetic field sensitivity. In this instance, the number of photons detected per measurement $\mathscr{N}$ increases with the number of sensors $N$. However, a series of practical factors may hinder this strategy. First, sensitivity enhancement exhibits sublinear scaling with $N$ and the associated number of photons detected per measurement $\mathscr{N}$, i.e., $\eta \propto \frac{1}{\sqrt{\mathscr{N}}}$, making significant sensitivity improvements from increasing $N$ difficult.

Additional technical difficulties may arise when increasing $N$, such as the photon number requirement for optical initialization. Assuming that each interrogated NV$^\text{-}$ center requires $m$ photons for optical initialization, each measurement is expected to require an energy of 
\begin{equation}
    E_\text{init}=N m \frac{h c}{\lambda},
\end{equation} 
where $h$ is Planck's constant, $c$ is the speed of light and $\lambda$ is the excitation wavelength. If measurements are performed every $T_2^*$, the required mean power is 
\begin{equation}\label{eqn:initpowerrequirement}
    P_\text{init} = \frac{N m}{T_2^*} \frac{h c}{\lambda}. 
\end{equation}
For example, initialization of all $1.76\times 10^{14}$ NV$^\text{-}$ centers in a 1~mm$^3$ diamond with 1~ppm [NV$^\text{-}$] would require $E_\text{init}= 200~\upmu$J, using a crude guess of $m=3$ (see Table~\ref{tab:branchingratios}). Assuming $T_2^*=1~\upmu$s, the required power is $P_\text{init} = 200$ W. Eqn.~\ref{eqn:initpowerrequirement} illustrates that achieving a sensitivity improvement by increasing the NV$^\text{-}$ ensemble size will increase $P_\text{init}$ unless $T_2^*$ is increased as well. For experimental approaches employing an acousto-optic modulator to gate a CW laser, the required CW laser power will be higher as many photons are wasted. 

Another difficulty encountered when increasing the number of interrogated NV$^\text{-}$ centers $N$ (and thus detected photon number $\mathscr{N}$) is that reaching the shot noise limit can become challenging for large values of $\mathscr{N}$. For example, the absolute noise contributed by some systematic (not stochastic) noise sources scales linearly with the number of photons detected, i.e., $\propto k_1\mathscr{N}$, where $k_1 \ll 1$. In comparably proportional units, shot noise scales as $\propto\sqrt{\mathscr{N}}$. For $\mathscr{N}>\frac{1}{k_1^2}$, the systematic noise will be larger than shot noise.  Primary examples of such noise sources include laser intensity noise in all implementations, timing jitter in the readout pulse length for Ramsey and pulsed ODMR, and MW amplitude noise in CW-ODMR. 

Lastly, increases in interrogation volume or NV$^\text{-}$ density are both accompanied by unique challenges independent of those associated with increases in sensor number $N$. For example, larger sensing volumes require better engineering to ensure the bias magnetic field and MW field are uniform over the sensing volume~\cite{eisuke2018tutorial,eisenach2018broadband}. Alternatively, increasing NV$^\text{-}$ density necessarily positions NV$^\text{-}$ spins (and all other nitrogen-related paramagnetic spins) closer together, which results in increased dipolar dephasing and shorter associated $T_2^*$ values, canceling the sensitivity improvement from addressing more NV$^\text{-}$ spins.




\subsection{Choosing nitrogen concentration in diamond samples}\label{choosingnitrogen}

The following discussion parallels the clear analysis presented in Ref.~\cite{Kleinsasser2016}, which the reader is encouraged to review. Equation~\ref{eqn:t2*depend} can be simplified by grouping all non-nitrogen-related broadening mechanisms together, yielding

\begin{equation}\label{eqn:both}
\frac{1}{T_2^*} = \frac{1}{T_2^*\{\text{N}_\text{S}^0\}} +\frac{1}{T_2^*\{\text{NV}^\text{-}\}} +\frac{1}{T_2^*\{\text{NV}^0\}} + \frac{1}{T_2^*\{\text{other}\}}, 
\end{equation}
where we have ignored typically less common defects in fully treated diamond (i.e.,~irradiated, annealed, etc.) such as NVH$^\text{-}$, N$_2$V$^\text{-}$, etc.~\cite{Hartlandthesis2014}; and $T_2^*\{\text{other}\}$ denotes the $T_2^*$ limit from all non-nitrogen-related dephasing mechanisms. The above equation can be rewritten as
\begin{align}\label{eqn:expandedallnitrogen}
\frac{1}{T_2^*} &= A_{\text{N}_S^0} [\text{N}^\text{T}]\lbrack 1\!-\!\text{E}_\text{conv}\!-\!\text{E}_\text{conv}^0\!-\!\text{E}^{\text{N}_\text{S}^+}_\text{conv}\rbrack+A_{\text{NV}^\text{-}}[\text{N}^\text{T}]\lbrack \text{E}_\text{conv}\rbrack \\ \nonumber & \;\;\;\;\;\;\;\;\;\;\;\; +A_{\text{NV}^0}[\text{N}^\text{T}]\lbrack \text{E}_\text{conv}^0\rbrack +\frac{1}{T_2^*\{\text{other}\}}
\end{align}
where E$_\text{conv}\equiv [\text{NV}^\text{-}]/[\text{N}^\text{T}],\,\text{E}_\text{conv}^0 \equiv [\text{NV}^\text{0}]/[\text{N}^\text{T}]$, and E$^{\text{N}_\text{S}^+}_\text{conv} \equiv [\text{N}_\text{S}^\text{+}]/[\text{N}^\text{T}]$ are the conversion efficiencies from the total nitrogen concentration $[\text{N}^\text{T}]$ to $[\text{NV}^\text{-}]$, $[\text{NV}^0]$, and $[\text{N}_\text{S}^+]$  respectively. \textcolor{black}{The $A_\text{X}$ coefficients characterize the magnetic dipole interaction strength between $\text{NV}^\text{-}$ spins and spin species $\text{X}$. The value of $A_{\text{N}_S^0} $ is defined in Eqn.~\ref{eqn:T2*NV-N}, the value of $A_{\text{NV}^\text{-}}$ is defined in Sec.~\ref{NVNVlimit}, and in this section for reasons of compactness we do not differentiate between $A_{\text{NV}^\text{-}_\parallel}$ and $ A_{\text{NV}^\text{-}_\nparallel}$. The value of $A_{\text{NV}^0}$ is defined so that the $\text{NV}^\text{-}$ dephasing from $\text{NV}^0$ satisfies $\frac{1}{T_2^*\{\text{NV}^0\}}= A_{\text{NV}^0} [\text{NV}^0]$.} Under the assumption that E$_\text{conv},\,\text{E}_\text{conv}^0,\text{E}^{\text{N}_\text{S}^+}_\text{conv}$ are independent of $[\text{N}^\text{T}]$, consolidation yields
\begin{equation}\label{eqn:expandedallnitrogen2}
\frac{1}{T_2^*} = \kappa [\text{N}^\text{T}]+ \frac{1}{T_2^*\{\text{other}\}},
\end{equation}
where $\kappa = A_{\text{N}_\text{S}^0}\lbrack1-\text{E}_\text{conv}-\text{E}_\text{conv}^0-\text{E}^{\text{N}_\text{S}^+}_\text{conv}\rbrack+ A_{\text{NV}^\text{-}}\lbrack \text{E}_\text{conv}\rbrack+A_{\text{NV}^0}\lbrack \text{E}_\text{conv}^0\rbrack$. The detected number of PL photons per measurement is $\mathscr{N} \propto [\text{N}^\text{T}]\text{E}_\text{conv} V n_\text{avg}$ where $V$ is the interrogation volume. For simplicity we consider the limit where initialization and readout times $t_I$ and $t_R$ are negligible, so that sensitivity is 
\begin{equation}
\eta \propto \sqrt{\frac{1}{\mathscr{N} T_2^*}} = \sqrt{\frac{1}{\text{E}_\text{conv} V n_\text{avg}}}\times \sqrt{\kappa+\frac{1}{[\text{N}^\text{T}]T_2^*\{\text{other}\}}},
\end{equation} 
which suggests that for $[\text{N}^\text{T}]\gg \frac{1}{\kappa \;T_2^*\{\text{other}\}}$, sensitivity is independent of $[\text{N}^\text{T}]$. Qualitatively, this can be interpreted as follows: when $T_2^*$ is limited by nitrogen-related dephasing mechanisms (i.e., NV$^\text{-}$, NV$^0$, N$_\text{S}^0$), and again assuming $\text{E}_\text{conv},\,\text{E}_\text{conv}^0$, and E$^{\text{N}_\text{S}^+}_\text{conv}$ are independent of $[\text{N}^\text{T}]$, decreasing $[\text{N}^\text{T}]$ increases $T_2^*$ by the same fractional quantity that the NV$^\text{-}$ ensemble photoluminescence $\mathscr{N}$ is decreased. However, when $T_2^*$ is limited by other broadening mechanisms unrelated to nitrogen, decreasing $[\text{N}^\text{T}]$ decreases  the collected fluorescence $\mathscr{N}$ without any corresponding $T_2^*$ increase. The implications here are significant: this analysis suggests that while there is not a unique value of $[\text{N}^\text{T}]$ for maximal sensitivity, there is a minimum value. \textcolor{black}{In other words, if nitrogen-related broadening is a small contributor to $T_2^*$, the nitrogen content should be increased; the increased resulting PL will favorably offset the increase in $T_2^*$, resulting in overall enhanced sensitivity.}

A few points are in order regarding the above analysis. Experimental considerations can also set an upper bound on the most desirable total nitrogen concentration $[\text{N}^\text{T}]$. For example, the larger detected photon number $\mathscr{N}$ associated with higher values of $[\text{N}^\text{T}]$ can present technical challenges (see Appendix~\ref{volumedensityconsiderations}). Moreover, the above analysis considers the simple limit where the initialization and readout times are negligible; accounting for this fixed overhead time (see Eqns.~\ref{eqn:duty}, \ref{eqn:ramseyshotexact}, and \ref{eqn:ramseyshot}) favors trading off nitrogen concentration density for longer values of $T_2^*$, in order to reduce the fractional overhead time devoted to initialization and readout. \textcolor{black}{Overall, combined experimental and theoretical considerations suggest that for best sensitivity nitrogen content should be decreased until nitrogen-related broadening is similar to broadening unrelated to nitrogen, i.e., $\kappa [\text{N}^\text{T}] \approx \frac{1}{T_2^*\{\text{other}\}}$.}





\subsection{Spin resonance linewidth and $T_2^*$}\label{linewidth}

The quantity $T_2^*$, which characterizes the time scale of the free induction decay (FID), is inversely proportional to the natural spin resonance linewidth in the absence of power broadening. Exact conversion between $T_2^*$ and linewidth requires knowledge of the functional form of the FID or the resonance lineshape~\cite{Abragam1983, Kwan1979}. 
Ramsey fringes decaying with an FID envelope $\propto e^{-t/T_2^*}$ indicate a Lorentzian spin resonance profile with full width at half maximum (FWHM) $\Gamma = \frac{1}{\pi T_2^*}$, as shown by the Fourier transform pair: 

\begin{align}
\begin{split}
\mathcal{F}_t[e^{2\pi i f_0 t} e^{- t/T_2^*}](f) &= \frac{1}{\pi}\frac{\frac{1}{2\pi T_2^*}}{\left(\frac{1}{2\pi T_2^*}\right)^2+(f-f_0)^2} \\
&= \frac{1}{\pi}\frac{\Gamma/2}{(\Gamma/2)^2+(f-f_0)^2},
\label{eqn:FTlor}
\end{split}
\end{align}
valid for $t\geq 0$, where $f_0$ is the Ramsey fringe frequency. 

A Gaussian decay envelope $\propto e^{-(t/T_2^*)^2}$ corresponds to a resonance with a Gaussian profile, with standard deviation $\sigma = \frac{1}{\sqrt{2}\pi T_2^*}$ as shown by the Fourier transform pair:

\begin{align}
\begin{split}
\mathcal{F}_t[e^{2\pi i f_0 t} e^{- (t/T_2^*)^2}](f) &= \sqrt{\pi} T_2^* e^{- (\pi T_2^* (f-f_0))^2}\\ 
&= \frac{1}{\sigma \sqrt{2 \pi}}e^{- (f-f_0)^2/(2\sigma^2)}.
\label{eqn:FTgau}
\end{split}
\end{align}

\subsection{Estimating $T_2^*$ from spin resonance linewidths of N$_\textnormal{S}^0$}\label{EPR}

\begin{figure*}[ht]
\centering
\begin{overpic}[height=2.9 in]{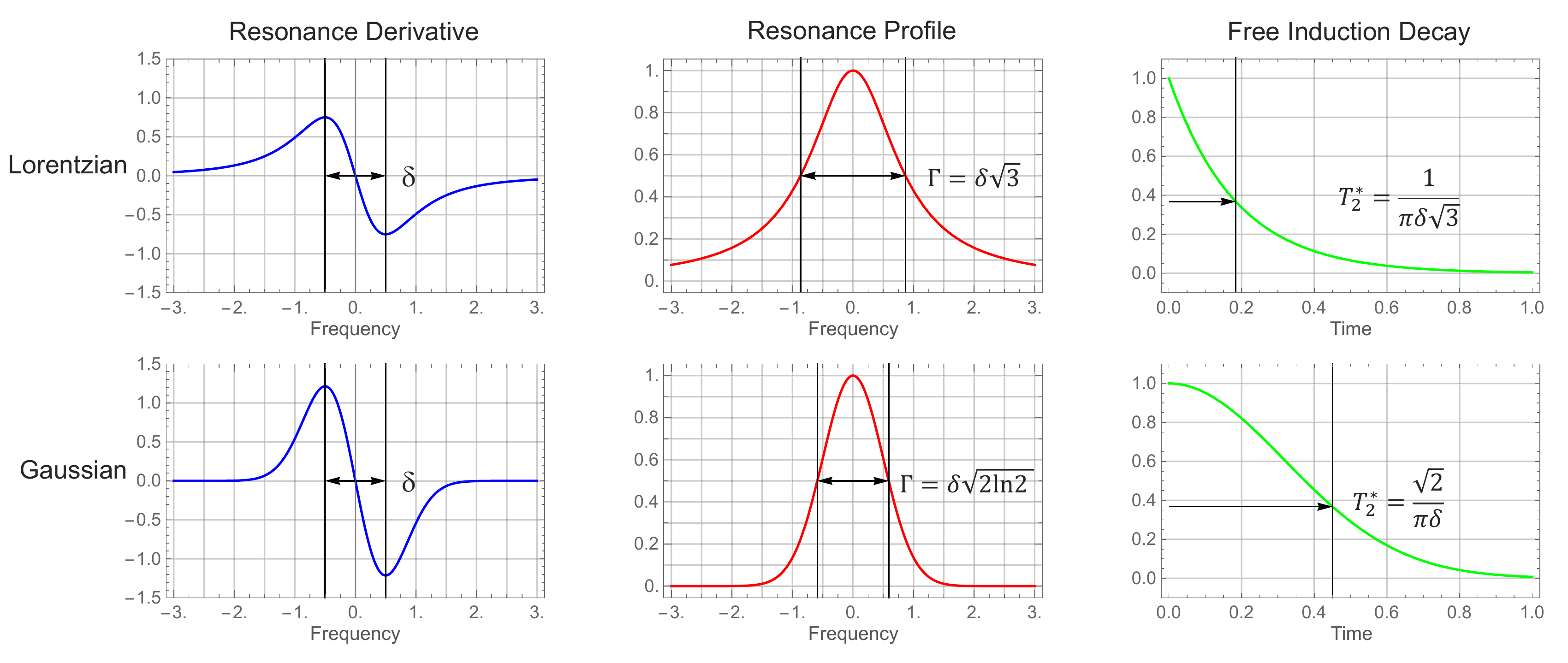} 
\end{overpic}
\caption[ResonanceDerivativeProfilesandFIDenvelopes]{Resonance derivatives ({\textcolor{blue}{\rule[.5mm]{3mm}{1pt}}}), resonance profiles ({\textcolor{red}{\rule[.5mm]{3mm}{1pt}}}), and free induction decay (FID) envelopes ({\textcolor{green}{\rule[.5mm]{3mm}{1pt}}}) for Lorentzian and Gaussian lineshape profiles with the same peak-to-peak widths $\delta$. Full-width-at-half-max linewidths $\Gamma$ and FID decay envelope times $T_2^*$ are indicated and expressed in terms of the peak-to-peak width $\delta$, a commonly reported parameter characterizing linewidth in electron paramagnetic resonance (EPR) data.}  \label{fig:lineshape}
\end{figure*}

\begin{figure}[ht]
\centering
\begin{overpic}[height=3.1 in]{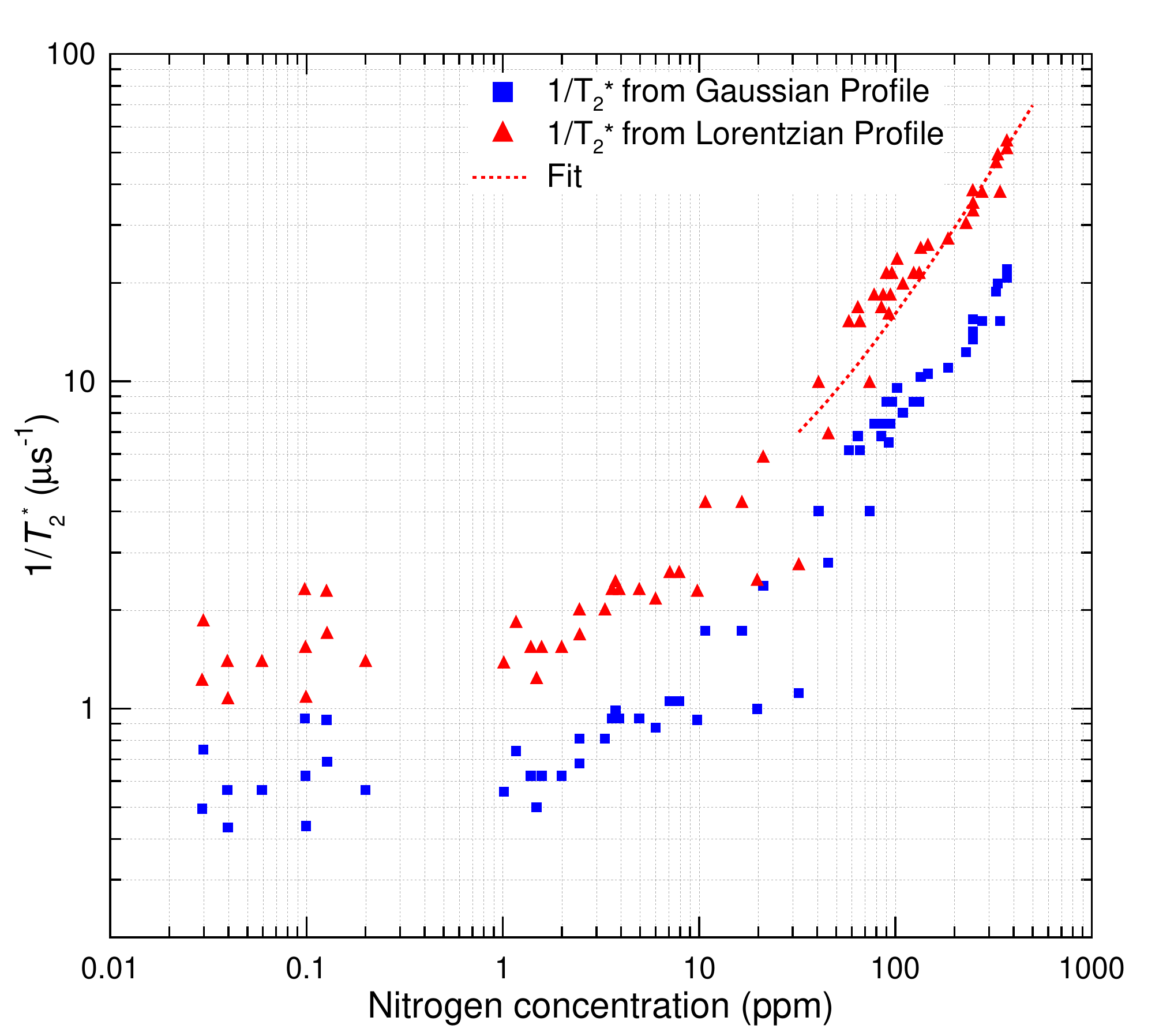} 
\end{overpic}
\caption[Bounds on T2star from Van Wyk]{Bounds on $1/T_2^*$ associated with EPR linewidth measurements of N$_\text{S}^0$ defects (P1 centers) from Ref.~\cite{vanWyk1997} in diamonds with a range of nitrogen impurity concentrations, calculated assuming Gaussian ({\textcolor{blue}{\rule[.25mm]{1mm}{1mm}}}) and Lorentzian (\raisebox{.4mm}{\tiny \textcolor{red}{$\blacktriangle$}}) EPR lineshapes. A fit to the function $1/T_2^* = A_{\text{N}_\text{S}^0}[\text{N}_\text{S}^0] + b$ assuming a Lorentzian lineshape (\textcolor{red}{\rule[.5mm]{.3mm}{1pt}\,\rule[.5mm]{.3mm}{1pt}\,\rule[.5mm]{.3mm}{1pt}\,\rule[.5mm]{.3mm}{1pt}}) yields \textcolor{black}{$A_{\text{N}_\text{S}^0} \approx 130$~ms$^\text{-1} \text{ppm}^\text{-1}$} for nitrogen spins in a nitrogen spin bath (see main text).}  \label{fig:vanWykT2star}
\end{figure}

Although sensor performance is dictated by $T_2^*$ of the NV$^\text{-}$ ensemble, $T_2^*$ values of other paramagnetic defects within the diamond, such as substitutional nitrogen defects, can provide useful information on sources of NV$^\text{-}$ spin dephasing. Such $T_2^*$ values can be extracted from linewidth measurements, for example from electron paramagnetic resonance (EPR). Accurate conversion from EPR linewidth to paramagnetic-defect $T_2^*$ enables leveraging of existing diamond EPR data~\cite{vanWyk1997} to better understand the contributions of different noise sources to NV$^\text{-}$ ensemble $T_2^*$ values. 

EPR linewidths are commonly tabulated by their peak-to-peak widths $\Delta B$, where $\Delta B$ denotes the magnetic field spacing between extrema of the resonance line first derivative~\cite{Poole1996}. In (linear) frequency units, this peak-to-peak width is $\delta = \frac{g\mu_B}{h} \Delta B$. Accurately relating $\delta$ and $T_2^*$ requires the resonance lineshape to be known~\cite{Kwan1979}. For example, a Lorentzian profile with full width at half maximum (FWHM) $\Gamma$, expressed in frequency units, has $\delta = \Gamma/\sqrt{3}$ and $\Gamma = \frac{1}{\pi T_2^*}$ (see Appendix~\ref{linewidth}). Combining these relations yields ${T_2^*}_\text{Lor} = \frac{1}{\sqrt{3} \pi\delta}$. A Gaussian lineshape with the same measured peak-to-peak linewidth $\delta$ has standard deviation $\sigma = \sfrac{\delta}{2}$ and $\sigma = \frac{1}{\sqrt{2}\pi T_2^*}$ (see Appendix~\ref{linewidth}). Thus, ${T_2^*}_\text{Gau} = \frac{\sqrt{2}}{\pi\delta}$, which is $\sqrt{6}\times$ longer than ${T_2^*}_\text{Lor}$. A visual comparison of these relationships is displayed in Fig.~\ref{fig:lineshape}.

Diamond EPR literature results may report values of $\delta$ without giving the associated resonance lineshape, preventing accurate determination of $T_2^*$ from $\delta$.  For example, linewidth measurements in Ref.~\cite{vanWyk1997} on substitutional nitrogen defects N$_\text{S}^0$ in diamond indicate a scaling $1/T_2^*\{\text{N}_\text{S}^0\} = A_{\text{N}_\text{S}^0}\,[\text{N}_\text{S}^0]$ with varying nitrogen concentration $[\text{N}_\text{S}^0]$, but the scaling factor $A_{\text{N}_\text{S}^0}$ cannot be accurately determined without knowledge of the lineshape. 

Theoretical and experimental results on dipolar-coupled spin systems suggest a Lorentzian resonance lineshape when spin-bath interactions are the dominant source of line-broadening~\cite{Kittel1953,Abragam1983,Dobrovitski2008,Hall2014}. Furthermore, Ramsey measurements with NV$^\text{-}$ spin ensembles show FID envelopes well fit by $e^{-(t/T_2^*)^p}$ with $p \sim 1$, corresponding to a Lorentzian lineshape (see Appendix~\ref{FIDexponent}, Fig.~\ref{fig:lineshape} and Ref.~\cite{Bauch2018}) when $T_2^*$ is expected to be spin-bath limited. 

Assuming a Lorentzian profile when converting $\delta$ values from Ref.~\cite{vanWyk1997} to $T_2^*$ values yields \textcolor{black}{$A_{\text{N}_\text{S}^0} \approx 130$~ms$^\text{-1} \text{ppm}^\text{-1}$} for nitrogen spins in a nitrogen spin bath (see Fig.~\ref{fig:vanWykT2star}). This calculated scaling factor is considered to be an upper bound because (i), a Gaussian or Voigt profile would result in a smaller value of $A_{\text{N}_\text{S}^0}$ than that calculated by assuming a Lorentzian profile, as $1/{T_2^*}_\text{Gau} =1/(\sqrt{6}\times {T_2^*}_\text{Lor})$; and (ii), other sources of broadening may contribute to the EPR linewidths observed in Ref.~\cite{vanWyk1997}. In the latter case the true contribution to dephasing from dipolar interactions between N$_\text{S}^0$ spins would be smaller than that estimated from the measured $\delta$. Nitrogen-spin-bath induced dephasing of N$_\text{S}^0$ and of NV$^\text{-}$ are expected to be similar, as the dipolar coupling between two N$_\text{S}^0$ spins is similar to the dipolar coupling between a N$_\text{S}^0$ and an NV$^\text{-}$ for equivalent separation~\cite{Hanson2008}. Thus, the spin-bath-limited linewidth of nitrogen defects in diamond measured via EPR can serve as a proxy for the spin-bath limited linewidth of NV$^\text{-}$ centers. The value of  \textcolor{black}{$A_{\text{N}_\text{S}^0} \approx 130$~ms$^\text{-1} \text{ppm}^\text{-1}$} for N$_\text{S}^0$ from the data in Ref.~\cite{vanWyk1997} serves as an independent estimate of $A_{\text{N}_\text{S}^0}$ for NV$^\text{-}$ centers in a nitrogen spin bath. This value is in reasonable agreement with the measured  \textcolor{black}{$A_{\text{N}_\text{S}^0} \approx 101$~ms$^\text{-1} \text{ppm}^\text{-1}$} for NV$^\text{-}$ ensembles from Ref.~\cite{Bauch2018}.

\subsection{Stretched exponential parameter}\label{FIDexponent}

\begin{table*}[ht]
\centering 
\setlength{\tabcolsep}{4pt}
\begin{tabular}{c c c c} 
\toprule 
Ramsey $T_2^*$ decay & $p$  & Reference (experiment) & Reference (theory) \\ [0.2ex] 
\midrule 
Single NV$^\text{-}$ & 2 & \cite{Maze2012} & \cite{Dobrovitski2008,deSousa2009,Hall2014}\\
NV$^\text{-}$ ensemble &  1 & \cite{MacQuarrie2015,Bauch2018} &\cite{Dobrovitski2008,Hall2014}\\
\bottomrule 
\end{tabular}
\caption[TableStretchedExponentialParameters]{Stretched exponential parameters $p$ associated with free induction decay envelopes for single NV$^\text{-}$ centers and NV$^\text{-}$ ensembles in dipolar-coupled spin baths}
\label{tab:FIDp} 
\end{table*}

Equations~\ref{eqn:FTlor} and \ref{eqn:FTgau} show that the spin resonance lineshape can be parameterized by the stretched exponential parameter $p$ of the free induction decay (FID) envelope $e^{-(t/T_2^*)^p}$. We note that for the idealized case of a purely Lorentzian lineshape, $p=1$, and for a purely Gaussian lineshape, $p=2$. The exact ODMR lineshape and value of $p$ are well characterized for single spins under a variety of environmental conditions~\cite{deSousa2009, Hanson2008, Dobrovitski2008, Maze2012,Hall2014}. For example, a single spin experiencing dipolar coupling to a surrounding bath of spins displays an FID envelope with stretched exponential parameter $p=2$~\cite{deSousa2009,Dobrovitski2008, Maze2012,Hall2014} (Gaussian ODMR lineshape, see Table~\ref{tab:FIDp} and Fig.~\ref{fig:lineshape}). Meanwhile, NV$^\text{-}$ ensembles with linewidth limited by dipolar coupling to a spin bath are predicted~\cite{Dobrovitski2008,Hall2014} and measured~\cite{MacQuarrie2015, Bauch2018} to exhibit FID envelopes with $p=1$ (Lorentzian ODMR lineshape, see Table~\ref{tab:FIDp} and Fig.~\ref{fig:lineshape}). However, experimental Ramsey measurements on NV$^\text{-}$ ensembles may sometimes exhibit decay envelopes with $p\neq 1$, suggesting the presence of other broadening mechanisms such as strain gradients, magnetic field gradients, or temperature fluctuations~\cite{Bauch2018}. A noninteger $p$ for an ensemble 
may also indicate the presence of more complex dephasing and decoherence dynamics, including spatial inhomogeneity, than can be encompassed by a single decay time constant. In some cases the decay may be better described by a sum~\cite{Cao1994} or a product of multiple decay curves with different values of $T_2^*$ and $p$. For example, a product of two FID decays, one with $p=1$ and one with and $p=2$, corresponds to a Voigt profile lineshape. Allowing $p$ to vary when fitting FID envelopes crudely accounts for these sorts of lineshape variations while only requiring a single additional fit parameter. Therefore, Ramsey FID measurements exhibiting $1\leq p\leq 2$ for some NV$^\text{-}$ ensembles may suggest contributions to the ODMR lines from both Lorentzian and Gaussian broadening mechanisms~\cite{Bauch2018}. 

Hahn echo $T_2$ decays of single NV$^\text{-}$ spins have been predicted~\cite{deSousa2009} and measured~\cite{deLange2010} to exhibit a stretched exponential parameter $p=3$ when $T_2$ is limited by spin-bath noise. In contrast, Hahn echo decay envelopes for ensembles of NV$^\text{-}$ spins have been seen to exhibit $p$ varying from $\sim 0.5$ to 3, depending on the dominant contributors to the spin bath and the bias magnetic field angle~\cite{Stanwix2010,Bauch2019}.

\subsection{Isotopic purity confusion in the literature}\label{13Cconfusion}

In Sec.~\ref{13ClimitT2star}, we discussed the $T_2^*$ limit imposed by $[^{13}\text{C}]$, which is described by an inverse linear scaling in Eqn.~\ref{eqn:scale13c}, reproduced below,
\begin{equation}\label{eqn:scale13cappendix}
\textcolor{black}{\frac{1}{T_2^*\{^{13}\text{C}\}} = A_{^{13}\text{C}}\,[^{13}\text{C}],}
\end{equation}
where \textcolor{black}{$A_{^{13}\text{C}}\approx 0.100$~ms$^\text{-1} \text{ppm}^\text{-1}$}. Although such inverse linear scaling with $[^{13}\text{C}]$ is predicted by several theoretical calculations (see Refs.~\cite{Dobrovitski2008,Hall2014,Kittel1953,Abragam1983}), some experiments based on single NV$^\text{-}$ centers (incorrectly we believe) suggest an inverse square root scaling~\cite{Mizuochi2009,Balasubramanian2009}, i.e.,~
\begin{equation}\label{eqn:scale13csinglewrong}
\textcolor{black}{\frac{1}{T_2^{*\{\text{single}\}}\{^{13}\text{C}\}} = A^{\{\text{single}\}}_{^{13}\text{C}}\,\sqrt{[^{13}\text{C}]}.}
\end{equation}
for single NV$^\text{-}$ centers in the dilute limit \textcolor{black}{($[^{13}\text{C}]/[^{12}\text{C}]\ll 0.01$)}. In Ref.~\cite{Mizuochi2009}, the data were derived from mean $T_2^*$ values taken from many single NV$^\text{-}$ defects in the diamond. However, Eqn.~\ref{eqn:scale13csinglewrong} conflicts with theoretical calculations by both Dobrovitski \textit{et al.}~\cite{Dobrovitski2008} and Hall \textit{et al.}~\cite{Hall2014} explicitly for single NV$^\text{-}$ centers. Both sources instead suggest that for the mean single NV$^\text{-}$ center,
\begin{equation}\label{eqn:scale13csingleright}
\textcolor{black}{\frac{1}{T_2^{*\{\text{single}\}}\{^{13}\text{C}\}} = A^{\{\text{single}\}}_{^{13}\text{C}}\,[^{13}\text{C}],}
\end{equation}
similar to Eqn.~\ref{eqn:scale13c}.

We hypothesize that the origin of this discrepancy is omission of nitrogen broadening in the study from Ref.~\cite{Mizuochi2009}, as summarized in Table~\ref{tab:3diamonds}. \textcolor{black}{Using the relation $1/T_2^{*\{\text{single}\}}\{\text{N}_\text{S}^0\} = A_{\text{N}_\text{S}^0}^\text{\{single\}}\,[\text{N}_\text{S}^0]$ (see Sec.~\ref{nitrogenlimitT2*}), we roughly estimate $A_{\text{N}_\text{S}^0}^\text{\{single\}} =56$~ms$^{-1} \text{ppm}^\text{-1}$} from Ref.~\cite{Zhao2012}. For the lowest $^{13}\text{C}$ sample in the data from Ref.~\cite{Mizuochi2009}, which has \textcolor{black}{$[\text{N}_\text{S}^0]\sim 1$~ppm}, this estimate predicts a nitrogen-limited  $T_2^{*\{\text{single}\}}\{\text{N}_\text{S}^0\}$ of $\sim\!18~\upmu$s, close to the actual reported $T_2^*$ measurement. Neglecting the additional nitrogen contribution to  $T_2^{*\{\text{single}\}}$ for the lowest $^{13}\text{C}$ sample likely caused Mizuochi \textit{et al.}~to overestimate the contribution of $^{13}\text{C}$ to $T_2^*$ and draw incorrect conclusions on the scaling of $T_2^*$ with $[^{13}\text{C}]$.

\textcolor{black}{Reference~\cite{Balasubramanian2009} report linewidths of 210~kHz and 55~kHz for diamonds with 1.1$\%$ and $0.3\%$ $^{13}$C respectively, data which is clearly consistent with Eqn.~\ref{eqn:scale13c}. However, the authors of Ref.~\cite{Balasubramanian2009} interpret their data using formalism appropriate for $^{13}$C $\gtrsim 10\%$~\cite{Abragam1983}, which results in them employing Eqn.~\ref{eqn:scale13csinglewrong}.}

As discussed in Sec.~\ref{13ClimitT2star}, Eqn.~\ref{eqn:scale13c} has been experimentally verified in the dilute limit in a similar system~\cite{Abe2010}. Given that the mean single-NV$^\text{-}$ FID time is longer than the ensemble FID time by $\sim\!2\times$ in diamond with natural abundance $^{13}\text{C}$ (see Section~\ref{13ClimitT2star} and Ref.~\cite{Maze2012}), if Eqn.~\ref{eqn:scale13csinglewrong} were also correct, then at sufficiently low $^{13}\text{C}$ concentration, an NV$^\text{-}$ ensemble would dephase more slowly than its constituent spins. This prediction conflicts with the present understanding that $T_2^{*\{\text{single}\}}\{^{13}\text{C}\} > T_2^*\{^{13}\text{C}\}$ regardless of concentration (see Sec.~\ref{T2*ensemble}).

\begin{table*}[ht]
\centering 
\setlength{\tabcolsep}{4.25pt}
\begin{tabular}{c c c c c c} 
\toprule 
$[^{13}$C$]$~(ppm) & Measured $T_2^{*\{\text{single}\}}$~($\upmu$s)  & Synthesis & [N$_\text{S}$]~(ppm) & Calculated $T_2^{*\{\text{single}\}}\{^{13}$C$\}$~($\upmu$s) & Calculated $T_2^{*\{\text{single}\}}\{$N$_\text{S}^0\}$~($\upmu$s)  \\ [0.2ex] 
 & \cite{Mizuochi2009} &  &  & (This work) & (This work)\\ 
\midrule 
10700 & 3.3 & CVD & $< 0.001$ & $2.3$& $\gtrsim 18000$ \\ 
3500 & 6.2 & CVD & $< 0.001$ & $7$& $\gtrsim 18000$ \\
300 & 18 & HPHT & $\sim 1$ & $8$2 & $\sim 18$  \\
\bottomrule 
\end{tabular}
\caption[3DiamondsMizouchi]{The three diamonds used in Ref.~\cite{Mizuochi2009}. The calculated value of $T_2^{*{\{\text{single}\}}}\{^{13}$C$\}$  is derived using the mean value of $T_2^*= 2.3~\upmu$s for single NV$^\text{-}$ centers in a natural abundance $^{13}\text{C}$ sample measured with a bias field of 20~G from Ref.~\cite{Maze2012}, so that
$T_2^{*{\{\text{single}\}}}\{^{13}$C$\} = 2.3~\upmu\text{s} \times \frac{0.0107}{[^{13}\text{C}]}$. The calculated value of  $T_2^{*{\{\text{single}\}}}\{$N$_\text{S}^0\}$ is estimated using the simulation in Fig.~1 of Ref.~\cite{Zhao2012} which predicts $T_2^{*{\{\text{single}\}}}\{$N$_\text{S}^0\}=18\pm1 \text{ ps} /[$N$_\text{S}^0]$.  }
\label{tab:3diamonds} 
\end{table*}

\subsection{Linear Stark and Zeeman regimes}\label{StarkZeeman}

Here we describe coupling of electric fields, strain, and magnetic fields to the NV$^\text{-}$ spin resonances in the regimes of both low and high axial bias magnetic field $B_{0,z}$. This treatment draws heavily on equations and analysis in Ref.~\cite{Jamonneau2016}. While understanding of strain's effect on the NV$^\text{-}$ spin continues to evolve~\cite{Doherty2013,Barson2017,udvarhelyi2018spinstrain,barfuss2018spinstress}, we take the NV$^\text{-}$ ground state spin Hamiltonian in the presence of a bias magnetic field $\vec{B}_0$, an electric field $\vec{E}$, and intrinsic crystal strain to be~\cite{udvarhelyi2018spinstrain,Doherty2013}
\begin{align}
    \begin{split}
    H/h =& \left(D + \mathscr{M}_z+d_\parallel E_z\right)S_z^2 \\ +&\frac{g_e\mu_B}{h}\left(B_{0,z}S_z + B_{0,x}S_x + B_{0,y}S_y\right) \\
    +&\left(d_\perp E_x +\mathscr{M}_x\right)\left(S_y^2-S_x^2\right)\\
    +&\left(d_\perp E_y + \mathscr{M}_y\right)\left(S_x S_y+S_y S_x\right)  \\ 
    +&\mathscr{N}_x\left(S_x S_z+S_z S_x\right)+\mathscr{N}_y\left(S_y S_z+S_z S_y\right).
    \end{split}
\end{align}
Here $S_i$ with $i = x,y,z$ are the dimensionless spin-1 projection operators; $D$ is the NV$^\text{-}$ zero field splitting ($\approx 2.87$~GHz at room temperature);  $d_\parallel = 3.5 \times 10^{-3}$ Hz/(V/m) and $d_\perp = 0.17$ Hz/(V/m) are the axial and transverse electric dipole moments~\cite{VanOort1990, Dolde2011,michl2019robust}, see Table~\ref{tab:hamconstants}; and $\mathscr{M}_z$, $\mathscr{M}_x$, $\mathscr{M}_y$, $\mathscr{N}_x$, and $\mathscr{N}_y$ are spin-strain coupling parameters.

The Hamiltonian can be simplified when $D$ is large compared to all other coupling terms, i.e., in the regime of low magnetic field, electric field, and strain. In particular, energy level shifts associated with transverse magnetic field components $B_{0,x}$ and $B_{0,y}$~\cite{Jamonneau2016}, and with spin-strain coupling parameters $\mathscr{N}_x$ and $\mathscr{N}_y$, are suppressed by $D$ and thus may be neglected from the Hamiltonian~\cite{kehayias2019diamond}. This low-field Hamiltonian $H_\text{LF}$ is given by
\begin{align}
    \begin{split}
    H_\text{LF}/h =& \left(D + \mathscr{M}_z+d_\parallel E_z\right)S_z^2 + \frac{g_e\mu_B}{h}B_{0,z}S_z \\
    +&\left(d_\perp E_x +\mathscr{M}_x\right)\left(S_y^2-S_x^2\right)\\
    +&\left(d_\perp E_y + \mathscr{M}_y\right)\left(S_x S_y+S_y S_x\right).
    \end{split}
\end{align}
We focus on the interplay between different terms in $H_\text{LF}$ that shift the NV$^\text{-}$ spin resonance frequencies in opposite directions, including $B_{0,z}$, $E_x$, $E_y$, $\mathscr{M}_x$, and $\mathscr{M}_y$. In contrast, dephasing associated with variations in terms that shift the resonance frequencies in common-mode ($D$, $E_z$, and $\mathscr{M}_z$), can be mitigated by employing double-quantum coherence magnetometry (see Sec.~\ref{DQ}) and are ignored herein. 
%
In addition to shifting the spin resonance frequencies, transverse electric fields, $E_{x}$ and $E_{y}$, and transverse spin-strain coupling terms, $\mathscr{M}_{x}$ and $\mathscr{M}_{y}$, mix the $m_s = \pm 1$ spin states into
\begin{align}
|+\rangle &= \cos\left(\frac{\theta}{2}\right)|\!+\!1\rangle+e^{i\phi}\sin\left(\frac{\theta}{2}\right)|\!-\!1\rangle, \\
|-\rangle &=\sin\left(\frac{\theta}{2}\right)|\!+\!1\rangle-e^{i\phi}\cos\left(\frac{\theta}{2}\right)|\!-\!1\rangle,
\end{align}
where $\tan(\phi) = (d_\perp E_y +\mathscr{M}_y)/(d_\perp E_x +\mathscr{M}_x)$ and $\tan(\theta) = \xi_\perp/\beta_z$. Here $\beta_z = (g_e \mu_B/h) B_{0,z}$ represents the magnetic field coupling to the NV$^\text{-}$ spin and $\xi_\perp = \sqrt{(d_\perp E_x +\mathscr{M}_x)^2+(d_\perp E_y +\mathscr{M}_y)^2}$ combines the effects of transverse strain and electric fields. The transition frequencies $|0\rangle \leftrightarrow |+\rangle$ and $|0\rangle \leftrightarrow |-\rangle$ are
\begin{equation}
\nu_\pm = D + \mathscr{M}_z+d_\parallel E_z \pm \sqrt{\xi_\perp^2 + \beta_z^2},
\end{equation}
and the coupling strength of transverse strain and electric fields to the NV$^\text{-}$ spin resonance frequencies is given by
\begin{equation}
\frac{\partial \nu_\pm}{\partial \xi_\perp} = {\frac{\pm 1}{\sqrt{1 + \left(\frac{\beta_z}{\xi_\perp}\right)^2}}}.\\
\end{equation}

In the linear Stark regime, characterized by $\beta_z \ll \xi_\perp$, the spin eigenstates become, approximately, equal superpositions of $|\!+\!1\rangle$ and $|\!-\!1\rangle$, and the transition frequencies exhibit maximal sensitivity to variations in $\xi_\perp$:
\begin{equation}
\left|\frac{\partial \nu_\pm}{\partial \xi_\perp}\right|_{\beta_z \ll \xi_\perp}=1-\frac{1}{2}\left(\frac{\beta_z}{\xi_\perp}\right)^2+\mathcal{O}\left[\left(\frac{\beta_z}{\xi_\perp}\right)^4\right].
\end{equation}
In contrast, in the linear Zeeman regime, characterized by $\beta_z \gg \xi_\perp$, the spin eigenstates become, approximately, $|\!+\!1\rangle$ and $|\!-\!1\rangle$, and sensitivity to strain/electric fields is suppressed by the ratio $\frac{\xi_\perp}{\beta_z}$:
\begin{equation}
\left|\frac{\partial \nu_\pm}{\partial \xi_\perp}\right|_{\beta_z \gg \xi_\perp} = \frac{\xi_\perp}{\beta_z}- \frac{1}{2}\left(\frac{\xi_\perp}{\beta_z}\right)^3+\mathcal{O}\left[\left(\frac{\xi_\perp}{\beta_z}\right)^5\right].
\end{equation}

By performing magnetic sensing in the linear Zeeman regime, spatial and temporal variations in transverse electric fields and strain couple less strongly to the NV$^\text{-}$ spin, and thus their contribution to $T_2^*$ is diminished. 
The linear Zeeman regime is best suited for high-sensitivity magnetometry not only because of the $T_2^*$ extension from suppressed sensitivity to variations in $\xi_\perp$, but also because magnetic field changes couple most strongly to $\nu\pm$ in this regime:

\begin{equation}
\left|\frac{\partial \nu_\pm}{\partial \beta_z}\right|_{\beta_z \gg \xi_\perp}
=1-\frac{1}{2}\left(\frac{\xi_\perp}{\beta_z}\right)^2+\mathcal{O}\left[\left(\frac{\xi_\perp}{\beta_z}\right)^4\right].
\end{equation}

Experiments that must operate at near-zero $\vec{B}_0$ for other reasons, such as to protect ferromagnetic samples, should use low-strain diamonds to avoid operating in the unfavorable regime where $\beta_z \ll \xi_\perp$~\cite{Glenn2017,Fu2014,backlund2017diamond,zheng2018zero}. In this linear Stark regime, not only is sensitivity to magnetic signals suppressed by the ratio $\frac{\beta_z}{\xi_\perp}$,
\begin{equation}
\left|\frac{\partial \nu_\pm}{\partial \beta_z}\right|_{\beta_z \ll \xi_\perp}=\frac{\beta_z}{\xi_\perp}-\frac{1}{2}\left(\frac{\beta_z}{\xi_\perp}\right)^3+\mathcal{O}\left[\left(\frac{\beta_z}{\xi_\perp}\right)^5\right],
\end{equation}
but also $T_2^*$ may be shortened by electric field and strain variations.

\subsection{Example annealing calculations}\label{exampleannealing}

We present some calculations to estimate parameters necessary for LPHT annealing to form NV$^\text{-}$ centers. Using $D_0 = 1.6 \times 10^{-3}$ cm$^2$/s~\cite{Fletcher1953} \textcolor{black}{for the diffusion constant}, $E_a = 2.3$ eV \textcolor{black}{for the activation energy}, and $T = 800$~$ ^\circ$C \textcolor{black}{for the annealing temperature}, we expect $D$ = 2.5~nm$^2$/s. When annealing at $T=800~^\circ$C and $t_\text{anneal} = 12 \times 3600$ s, a single vacancy in a perfect lattice is expected (based on the model presented here) to have made $\sim$ $2.7\times10^{7}$ lattice jumps, visited $1.5\times10^7$ distinct lattice sites~\cite{Vineyard1963,Fastenau1982}, and diffused a root-mean-square distance of $\langle r_\text{rms} \rangle \approx .8~\upmu$m, assuming $\langle r_\text{rms} \rangle = \sqrt{6 D t_\text{anneal}}$. The uncertainties in these estimates are dominated by the $\pm 0.3$ eV uncertainty in $E_a$~\cite{Davies1992,Mainwood1999}, which can lead to an order of magnitude variation in $D$ for  $T = 800$ $ ^\circ$C. Ignoring small repulsive forces between substitutional nitrogen and monovacancies~\cite{Davies1992}, a vacancy is expected to visit  $\sim\!10^6/4$ lattice sites in a 1~ppm [N$_\text{S}$] diamond to form an NV. The factor 4 arises from the four closest sites to a substitutional nitrogen while the $10^6$ arises because only 1 out of every $10^6$ lattice sites is occupied by a substitutional nitrogen. Because the number of distinct lattice sites  visited is substantially greater than the number of sites needed to form an NV center (i.e., $\frac{1.5\times 10^7}{10^6/4} \gg 1$), the chosen values of $T$ and $t_\text{anneal}$ are expected to  ensure adequate NV center formation. The simple analysis stated here is complicated by the uncertainty in $D_0$ and $E_a$, as well as the presence of other vacancies, vacancy aggregates, dislocations, surfaces, etc., which which can also trap vacancies, but are beyond the scope of this paper. \textcolor{black}{For additional detail and discussion see Ref.~\cite{alsid2019photoluminescence}.}

\textcolor{black}{The analysis presented above is derived only from first-principles calculations and the measured value of $E_a$. More accurate behavior may be predicted by employing measured values of $D$ at a given temperature, such as $D\approx 1.1$~nm$^2$/s at 750~$^\circ$C~\cite{Martin1999} and $D\approx 1.8$~nm$^2$/s at 850~$^\circ$C~\cite{alsid2019photoluminescence}.}


\begin{table*}
\centering
\begin{center}
\setlength{\tabcolsep}{3pt}
    \begin{tabular}{lccccl}
    \toprule
   Reference  & \cite{Tetienne2012} & \cite{Gupta2016} & \cite{Robledo2011} & \cite{Acosta2010b} & \\ 
     NV$^\text{-}$ centers probed & 4 & 3 & 2 & ensemble & units \\  
     Values reported & avg.~(max, min) & avg.~(max, min) & avg. &   &   \\    \midrule
     $^3$E($m_s\!=\!0)\rightarrow$\,\!$^3$A$_2$($m_s\!=\!0$)  & $67.9$ (63.2, 69.1) & $66.16$ (66.08, 66.43) & 64.2 & - & ~$\upmu$s$^{\text{-}1}$ \\     
    $^3$E($m_s\!=\!\pm 1)\rightarrow$\,\!$^3$A$_2$($m_s\!=\!\pm 1$) & $67.9$ (63.2, 69.1) &  $66.16$ (66.08, 66.43) & 64.9 & - & ~$\upmu$s$^{\text{-}1}$\\
    $^3$E($m_s\!=\!0)\rightarrow ^1$A$_1$  &  $5.7$ (5.2, 10.8) & $11.1$ (10.9, 11.2) & 11.2 & - & ~$\upmu$s$^{\text{-}1}$\\
	$^3$E($m_s\!=\!\pm 1)\rightarrow ^1$A$_1$  & $49.9$ (48.6, 60.7) & $91.8$ (89.3, 92.9) & 80.0 & - & ~$\upmu$s$^{\text{-}1}$\\ 
    $^1$E$\rightarrow ^3$A$_2(m_s\!=\!0$)  & $1.0$ (0.7, 1.5) & $4.87$ (4.75, 4.90) & 3.0 & - & ~$\upmu$s$^{\text{-}1}$\\ 
    $^1$E$\rightarrow ^3$A$_2(m_s\!=\!\pm 1$)  & $0.75$ (0.4, 1.4) & $2.04$ (2.03, 2.13) & 2.6  & - & ~$\upmu$s$^{\text{-}1}$ \\ 
    \textcolor{black}{$^1$E} lifetime & - &  $144.5$ (144.3, 145.3) & $178 \pm 6$ & $219 \pm 3$ & ~ns \\ \bottomrule
   \end{tabular}
\end{center}
\caption{NV$^\text{-}$ decay rates measured at room temperature. Averages over measured NV$^\text{-}$ centers are weighted by reported uncertainties. Dashed lines (-) indicate values not reported. Branching ratios can be derived from the given data. \textcolor{black}{Although not tabulated, vibrational decay within the $^3$E state is fast, with Ref.~\cite{huxter2013vibrational} observing a $\sim\!4$ ps timescale, and Ref.~\cite{ulbricht2018vibrational} observing a $\sim\!50$ fs timescale. The $^1$A$_1$ lifetime was measured to be $\approx 100$ ps at 78K~\cite{ulbricht2018excited} and is likely shorter at room temperature. } }
\label{tab:branchingratios}
\end{table*}

\subsection{The diamond type classification system}\label{diamondtypeclassification}

We briefly overview the ``diamond type'' classification system introduced in the 1930s and outlined in Refs.~\cite{robertson1933two,robertson1936further}. In spite of the system's shortcomings, it has been widely adopted by the gemstone community and is partially used by the scientific community today. In the mid-1930's the authors of Refs.~\cite{robertson1933two,robertson1936further} noted that although the vast majority of natural diamonds exhibited absorption lines in the $225-300$~nm band and near 8~$\upmu$m, these same absorption features were absent in a small minority of diamonds. The authors further observed that diamonds lacking these same absorption features tended to exhibit lower birefringence and higher photoconductivity relative to their peers~\cite{robertson1933two,robertson1936further}. In 1959 the authors of Ref.~\cite{kaiser1959nitrogen} attributed the observed infrared absorption features to carbon-nitrogen molecular vibrations, signaling the presence of nitrogen. Nitrogen was found to be the most common impurity occurring in natural diamonds, which made its presence or absence a logical basis for diamond classification. 

\begin{figure*}[ht]
\centering
\begin{overpic}[height=3 in]{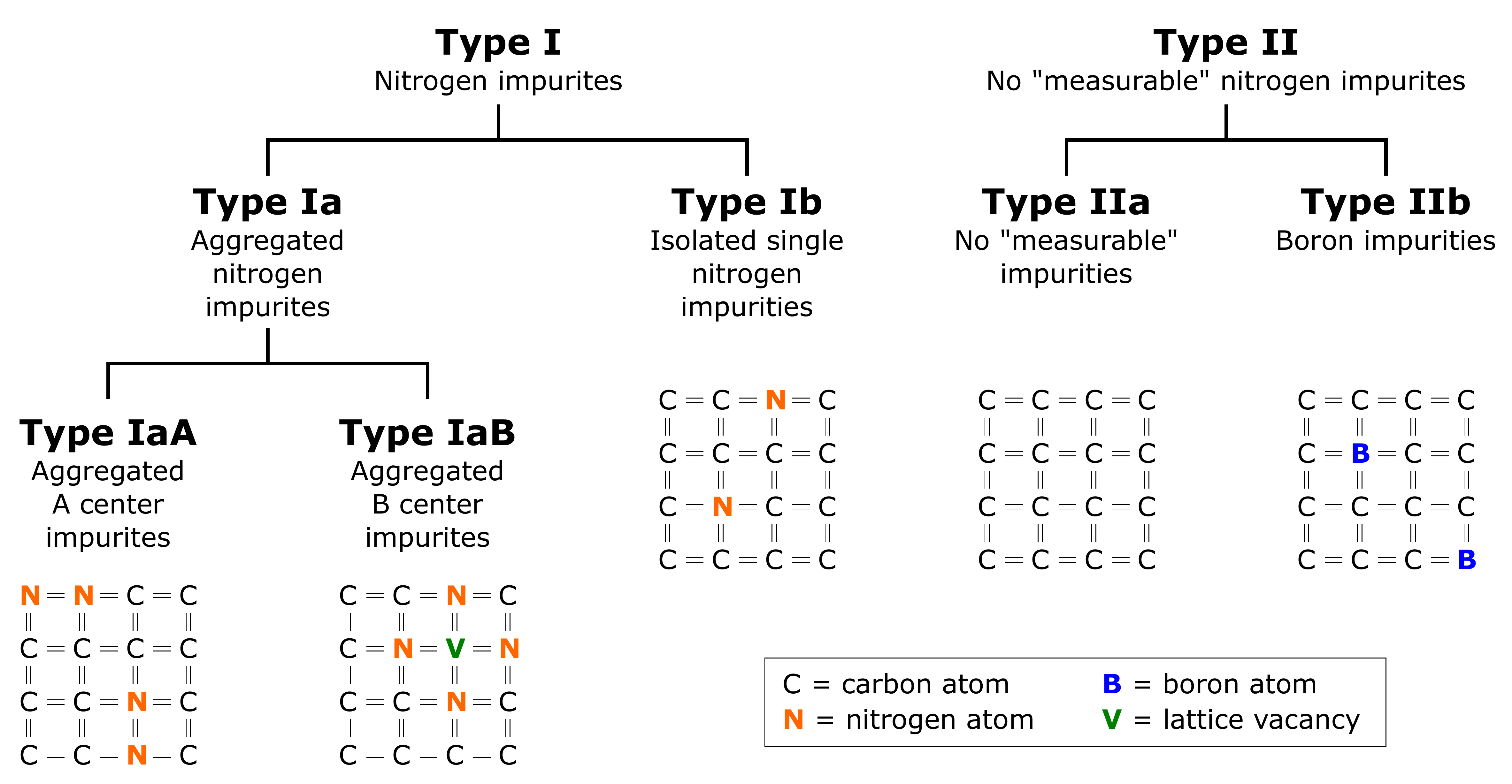} 
\end{overpic}
\caption[diamondtype]{The diamond type classification system as described in the main text. Adapted from Ref.~\cite{breeding2009type}}.  \label{fig:diamondtype}
\end{figure*}

In this nitrogen-based diamond classification system, all diamonds are categorized into one of two primary types: Type I diamonds contain measurable quantities of nitrogen while Type II diamonds do not, as shown in Fig.~\ref{fig:diamondtype}. There is no wide consensus on what constitutes ``measurable'' in an age of ever-advancing characterization tools, although a common definition is a quantity detectable with an FTIR spectrometer~\cite{breeding2009type}. Most sources suggest a delineation somewhere between 0.5~ppm~\cite{Zaitsev2001} and 20~ppm~\cite{dischler2012handbook,gaillou2012boron}. This delineation uncertainty is particularly unfortunate for the NV$^\text{-}$ community, as many diamonds employed for ensemble NV$^\text{-}$ experiments fall in this range. 

Type I diamonds can be further classified by the specific nitrogen complexes incorporated into the carbon lattice. For example, Type Ia diamonds contain aggregated nitrogen impurities, and  describe the vast majority of natural diamonds ($\gtrsim 95\%$, depending on the delineation nitrogen concentration ~\cite{breeding2009type,Zaitsev2001}). Typical nitrogen concentrations in natural Type Ia diamonds are in the hundreds of ppm (e.g., 500~ppm~\cite{Zaitsev2001}) but can be as high as 3000~ppm~\cite{neves2001properties}. If the aggregated nitrogen predominantly forms A centers consisting of two substitutional nitrogens located adjacent in the diamond lattice, the diamond is classified as Type IaA. If the aggregated nitrogen predominantly forms B centers consisting of four substitutional nitrogens surrounding a lattice vacancy, the diamond is classified as Type IaB. In contrast, diamonds containing predominantly isolated single nitrogen impurities are classified as Type Ib and make up about $0.1\%$ of all natural diamonds~\cite{Zaitsev2001}. As higher nitrogen density promotes aggregation, Type Ib diamonds typically exhibit nitrogen concentrations at or below the 100~ppm level~\cite{Zaitsev2001}, less than typical for Type Ia diamonds~\cite{Zaitsev2001}.

Type II diamonds containing no ``measurable'' nitrogen can be additionally classified as well. Type IIa diamonds contain no other measurable impurities and make up the majority of gem-grade diamonds in spite of comprising only $1$ to $2 \%$ of natural diamonds. These diamonds are the most optically transparent diamonds: while Type IIa diamonds with low levels of impurities may exhibit pale shades of yellow, pink, or purple; extremely pure Type IIa diamonds are colorless~\cite{Zaitsev2001}. Nearly all single NV$^\text{-}$ experiments employ Type IIa diamonds. As boron is another common impurity in natural diamond, Type II diamonds with ``measurable'' boron are categorized as Type IIb. These diamonds make up about $0.1 \%$ of all natural diamonds and may exhibit a bluish or greyish hue. 

Although the diamond type classification system was developed for natural diamonds, it appropriately describes synthetic diamonds as well. CVD-grown diamonds without nitrogen doping are Type IIa. Man made HPHT diamonds of Type IaA, Ib, IIa, and IIb have been created. Further diamond types exist: see Ref.~\cite{Zaitsev2001}.

\begin{table}[htbp]
  \centering
    \begin{tabular}{p{5em}p{20.5em}}
    \toprule
    Acronym & Description \\
    \midrule
    CPMG  & Carr-Purcell-Meiboom-Gill (pulse sequences) \\
    CW    & Continuous wave \\
    CVD   & Chemical vapor deposition \\
    DEER  & Double electron-electron resonance \\
    DQ    & Double-quantum \\
    EPR   & Electron paramagnetic resonance \\
    ESLAC & Excited-state level anti-crossing \\
    ESR   & Electron spin resonance \\
    FID   & Free induction decay \\
    GSLAC & Ground state level anti-crossing \\
    HPHT  & High pressure high temperature \\
    LAC   & Level anti-crossing \\
    LPHT  & Low pressure high temperature \\
    MW    & Microwave \\
    NIR   & Near-infrared \\
    NMR   & Nuclear magnetic resonance \\
    NQR   & Nuclear quadrupole resonance \\
    ODMR  & Optically detected magnetic resonance \\
    PDMR  & Photoelectrically detected magnetic resonance \\
    PE    & Photoelectric (readout) \\
    PL    & Photoluminescence \\
    QND   & Quantum non-demolition \\
    RF    & Radiofrequency \\
    SCC   & Spin-to-charge conversion \\
    SQ    & Single quantum (standard basis) \\
    \bottomrule
    \end{tabular}%
  \label{tab:addlabel}%
  \caption{Frequently used acronyms}
\end{table}%

\clearpage\newpage
\begin{table}[htbp]
  \centering
  \setlength{\tabcolsep}{5pt}
  \renewcommand{\arraystretch}{1.2}
    \begin{tabular}{p{30em}p{10em}p{6.5em}p{7em}}
    \toprule
    Quantity & Symbol & Units & Notes \\
    \midrule
    Longitudinal (spin-lattice) relaxation time & $T_1$ & s     & \multicolumn{1}{r}{} \\
    Coherence time (transverse relaxation time) & $T_2$ & s     & \multicolumn{1}{r}{} \\
    Dephasing time (free induction decay time) & $T_2^*$ & s     & \multicolumn{1}{r}{} \\
    Single-NV$^\text{-}$ dephasing time & ${T_2^*}^\text{\{single\}}$ & s     & \multicolumn{1}{r}{} \\[3ex]
    Magnetic field sensitivity & $\eta$ & T$/\sqrt{\text{Hz}}$ & \multicolumn{1}{r}{} \\
    Interrogation time (free-precession time for Ramsey) & $\tau$ & s     & \multicolumn{1}{r}{} \\
    Initialization, readout, and overhead time & $t_I$, $t_R$, $t_O$ & s     & $t_O \equiv t_I+t_R$ \\
    Stretched exponential parameter & $p$   & -     & \multicolumn{1}{r}{} \\
    Static (bias) \& microwave magnetic field & $B_0$ \& $B_1$ & T     & \multicolumn{1}{r}{} \\
    Electronic spin gyromagnetic ratio & $\gamma_e$ & s$^{\text{-}1}$/T & $\equiv g_e \mu_B/\hbar$ \\
    Readout fidelity & $\mathcal{F}$ & -     & $\equiv 1/\sigma_R$ \\
    Factor above spin projection noise & $\sigma_R$ & -     & $\equiv 1/\mathcal{F}$ \\
    Rabi frequency & $\Omega_R$ & s$^{\text{-}1}$ & \multicolumn{1}{r}{} \\
    ODMR center frequency \& linewidth & $\nu$ \&  $\Delta\nu$ & Hz    & \multicolumn{1}{r}{} \\
    Dephasing or decay rate & $\Gamma$ & s$^{\text{-}1}$ & \multicolumn{1}{r}{} \\
    Measurement contrast (fringe visibility) & $C$   & -     & \multicolumn{1}{r}{} \\
    CW-ODMR contrast, pulsed ODMR contrast & $C_\text{CW}$, $C_\text{pulsed}$ & -     & \multicolumn{1}{r}{} \\
    Number of sensors (NV$^\text{-}$ centers in ensemble) & $N$   & - & \multicolumn{1}{r}{} \\
    Average collected photons per readout per NV$^\text{-}$ & $n_\text{avg}$ & -     & \multicolumn{1}{r}{} \\
    Average collected photons per readout from an NV$^\text{-}$ ensemble & $\mathscr{N}$ & -     & \multicolumn{1}{r}{} \\[3ex]
    Concentration of species X & [X]   & cm$^{\text{-}3}$ or ppm & \multicolumn{1}{r}{} \\
    Negative, neutral \& total NV concentration & [NV$^\text{-}$], [NV$^\text{0}$], [NV$^\text{T}$] & cm$^{\text{-}3}$ or ppm & \multicolumn{1}{r}{} \\
    Total nitrogen concentration in the lattice & [N$^\text{T}$] & cm$^{\text{-}3}$ or ppm &  \\
    Neutral, positive, total substitutional nitrogen concentration & [N$_\text{S}^\text{0}$], [N$_\text{S}^\text{+}$], [N$_\text{S}^\text{T}$] & cm$^{\text{-}3}$ or ppm & \multicolumn{1}{r}{} \\[3ex]
    Contribution to $T_2^*$ from mechanism X & $T_2^*\{\text{X}\}$ & s     & \multicolumn{1}{r}{} \\
    Dipolar interaction strength between N$_\text{S}^0$ and NV$^\text{-}$ & $A_{\text{N}_\text{S}^0}$ & s$^{\text{-1}}/$ppm & \multicolumn{1}{r}{} \\
    Dipolar interaction strength between $^{13}$C and NV$^\text{-}$ & $A_{^{13}\text{C}}$ & s$^{\text{-1}}/$ppm & \multicolumn{1}{r}{} \\
    Dipolar interaction strength between NV$^\text{-}$ spins in the same group (same resonance frequency) & $A_{\text{NV}_\parallel^\text{-}}$ & s$^{\text{-1}}/$ppm & \multicolumn{1}{r}{} \\
    Dipolar interaction strength between NV$^\text{-}$ spins in different groups (different resonance frequencies) & $A_{\text{NV}_\nparallel^\text{-}}$ & s$^{\text{-1}}/$ppm & \multicolumn{1}{r}{} \\
    Proportionality factor for N$_\text{S}^0$ contribution to $T_2$ & $B_{\text{N}_\text{S}^0}$ & s$^{\text{-1}}/$ppm & \multicolumn{1}{r}{} \\[3ex]
    Hamiltonian & $H$   & J & \multicolumn{1}{r}{} \\
    Electronic spin, electronic spin projection & $S$, $m_s$   & - & \multicolumn{1}{r}{} \\
    Nuclear spin, nuclear spin projection & $I$, $m_I$   & - & \multicolumn{1}{r}{} \\
    NV$^\text{-}$ ground state spin eigenstates & \{$|\,0\rangle$, $|\!-\!1\rangle$, $|\!+\!1\rangle$\} & - & \multicolumn{1}{r}{} \\
    Zero field splitting parameter & $D$   & Hz    & $\approx 2.87$ GHz \\
    Spin-strain coupling parameters & $\mathscr{M}_z$, $\mathscr{M}_x$, $\mathscr{M}_y$, $\mathscr{N}_x$, $\mathscr{N}_y$ & Hz & \multicolumn{1}{r}{} \\
    Electric field components & $E_x$, $E_y$, $E_z$ & V/m   & \multicolumn{1}{r}{} \\
    NV$^\text{-}$ transverse, axial (longitudinal) electric dipole moment & $d_\perp$, $d_\parallel$ & Hz/(V/m) & \multicolumn{1}{r}{} \\
    Transverse strain and electric field coupling parameter & $\xi_\perp$ & Hz    & \multicolumn{1}{r}{} \\
    Axial magnetic field coupling parameter & $\beta_z$ & Hz    & $\equiv \frac{g \mu_B}{h} B_z$ \\[3ex]
    Total N-to-NV$^\text{-}$ conversion efficiency & E$_\text{conv}$ & -     & $\equiv [\text{NV}^\text{-}]/[\text{N}^\text{T}]$ \\
      N-to-NV conversion efficiency & $\chi$ & - & $\equiv [\text{NV}^{\text{T}}]/[\text{N}^\text{T}]$ \\
    NV-to-NV$^\text{-}$ charge state efficiency & $\zeta$ & -     & $\equiv [\text{NV}^{\text{-}}]/[\text{NV}^\text{T}]$ \\
    \bottomrule
    \end{tabular}%
      \label{tab:addlabel}%
    \caption{Frequently used symbols}
\end{table}%

\begin{table*}
\begin{center}
    \begin{tabular}{ l  l  l  l}
    \toprule
    \parbox[t]{8em}{\raggedright Const.} & \parbox[t]{8em}{\raggedright Description} & 
    \parbox[t]{8em}{\raggedright Value} & \parbox[t]{8em}{\raggedright Ref.} \\ \midrule
          $g_\parallel$ & Axial g-factor & $2.0028 \pm 0.0003$ & \cite{Loubser1978}    \\ 
     & & $2.0029 \pm 0.0002$ & \cite{Felton2009}    \\ 
          & & & \\   
   $g_\perp$ & Transverse g-factor & $2.0028 \pm 0.0003$ & \cite{Loubser1978}    \\ 
    & & $2.0031 \pm 0.0002$ & \cite{Felton2009}    \\ 
    & & & \\

    $A_\parallel$ & $^{14}$N axial magnetic hyperfine constant & $\pm 2.32 \pm 0.01$~MHz &\cite{Loubser1978}   \\ 
     &  & $2.30 \pm 0.02$~MHz &\cite{he1993paramagnetic}    \\ 
     &  & $-2.14 \pm 0.07$~MHz &\cite{Felton2008}   \\ 
     &  & $-2.166 \pm 0.01$~MHz &\cite{Steiner2010}    \\ 
     &  & $-2.162 \pm 0.002$~MHz &\cite{smeltzer2009robust} \\ 
    & & & \\ 
    & $^{15}$N axial magnetic hyperfine constant & $-3.1$~MHz &\cite{rabeau2006implantation}    \\ 
     & & $3.01  \pm  0.05$~MHz &\cite{Fuchs2008}   \\ 
     & & $3.03  \pm  0.03$~MHz&\cite{Felton2009}   \\ 
    & & & \\ 
    $A_\perp$ & $^{14}$N transverse magnetic hyperfine constant & $+2.10  \pm  0.10$~MHz &\cite{he1993paramagnetic}   \\ 
     &  & $-2.70  \pm  0.07$~MHz &\cite{Felton2008}  \\ 
    & & & \\ 
    & $^{15}$N transverse magnetic hyperfine constant & $-3.1$~MHz &\cite{rabeau2006implantation}   \\ 
     &  & $3.01  \pm  0.05$~MHz &\cite{Fuchs2008}   \\ 
     &  & $3.65  \pm  0.03$~MHz &\cite{Felton2009}   \\ 
    & & & \\ 
    $P$ & $^{14}$N nuclear electric quadrupole parameter & $-5.04 \pm  0.05$ &\cite{he1993paramagnetic} \\ 
     &  & $-5.01 \pm  0.06$ &\cite{Felton2008} \\ 
     &  & $-4.945 \pm  0.01$ &\cite{Steiner2010} \\ 
     &  & $-4.945 \pm  0.005$ &\cite{smeltzer2009robust} \\ 
     & & & \\   

    $d_\parallel$ & Axial dipole moment & $3.5 \pm 0.02\times10^{-3}$~Hz/(V/m) &\cite{VanOort1990} \\ 
         & & & \\   

        $d_\perp$ & Transverse dipole moment & $0.165 \pm 0.007$ Hz/(V/m) &\cite{michl2019robust}   \\ 
     &  & $0.175 \pm 0.030$ Hz/(V/m) &\cite{Dolde2011}   \\ 
     &  & $0.17 \pm 0.025$ Hz/(V/m) &\cite{VanOort1990}   \\ 
    \bottomrule
    \end{tabular}
\end{center}
\caption{\textcolor{black}{Compiled constants for the electronic ground state of the NV$^\text{-}$ center in diamond. Data are  reproduced in part from Ref.~\cite{Doherty2013}.}}\label{tab:hamconstants}
\end{table*}

\begin{table*}
\begin{center}
    \begin{tabular}{ l  l l  }
    \toprule
    \parbox[t]{8em}{\raggedright Cross section} & \parbox[t]{8em}{\raggedright Value}  & Reference\\ \midrule
      $\sigma_{^3\text{A}_2\rightarrow^{3}\text{E}}(\lambda = 532 \text{ nm})$ &  $(3.1\pm 0.8)\times10^{-17}\text{ cm}^2$ & \cite{Wee2007}     \\ 
      $\sigma_{^3\text{A}_2\rightarrow^{3}\text{E}}(\lambda = 532 \text{ nm})$ &  $(9.5\pm 2.5)\times10^{-17}\text{ cm}^2$ & \cite{Chapman2011}     \\ 
      $\sigma_{^3\text{A}_2\rightarrow^{3}\text{E}}(\lambda = 532 \text{ nm})$ &  $(2.6\pm 0.5)\times10^{-17}\text{ cm}^2$ & \cite{Fraczek2017}     \\ 
      $\sigma_{^3\text{A}_2\rightarrow^{3}\text{E}}(\lambda = 532 \text{ nm})$ &  $2.4\times10^{-17}\text{ cm}^2$ & \cite{subedi2019laser} \\     
  \bottomrule
    \end{tabular}
\end{center}
\caption{\textcolor{black}{Absorption cross section at 532 nm for the NV$^\text{-}$ $^3\text{A}_2\rightarrow^{3}$E transition. The value from Ref.~\cite{Fraczek2017} was calculated from their data under the assumption that the NV$^\text{-}$ and NV$^0$ absorption cross sections are equal to within 2$\times$.}}\label{tab:532nmabsorptioncrosssectioncompiled}
\end{table*}